\documentclass[school=estg,
    language=english,
    chapterstyle=fancy,
    coverstyle=classic
]{IPLeiriaThesis}

\usepackage[
    b5paper,
    left=2.1cm, 
    right=1.75cm,
    top=2cm,
    bottom=1.5cm,
    headsep=0.4cm,
    footskip=5cm
  ]{geometry}

\usepackage{graphicx}
\usepackage{amsmath,amssymb,amsthm,textcomp}
\usepackage{mathrsfs}
\usepackage{comment}
\usepackage{subcaption}
\usepackage{dsfont}
\usepackage{bbm,bbold}
\usepackage{mathtools}
\usepackage{physics}
\usepackage{bm}
\usepackage{multicol}

\newcommand{\bfk}{\textbf{k}}
\newcommand{\bfx}{\textbf{x}}

\usepackage{xcolor}

\FirstAuthor{Joe Smith}
\FirstAuthorNumber{2230455}

\Supervisor{John Smith}
\SupervisorMail{joe.smith@ipleiria.pt}
\SupervisorTitle{Full Professor, Polytechnic of Leiria} 

\CoSupervisor{Steve Smith}
\CoSupervisorMail{steve.smith@ipleiria.pt}
\CoSupervisorTitle{Associate Professor, Polytechnic of Leiria}

\SecCoSupervisor{Shak Smith}
\SecCoSupervisorMail{shak.smith@ipleiria.pt}
\SecCoSupervisorTitle{Associate Researcher, Computer Science \& Communication Research Centre}

\Title{Improving Machine Learning Efficiency Against Noisy Data Sources}

\Subtitle{Investigating Advanced Strategies to Mitigate Adverse Effects of Noisy Data}

\University{Polytechnic of Leiria}

\School{School of Management and Technology}

\Department{Department of Computer Engineering}

\Degree{Master in Cybersecurity \& Digital Forensics}

\ThesisType{Dissertation/Project/Internship \\ \textcolor{blue}{(Erase the Non-Essential)}}

\Date{Leiria, \DTMmonthname{\month} \number\year}

\AcademicYear{2024/25}

\makeglossaries
\newglossaryentry{KGEquation}
{
    name=KG equation,
    description={}
}

\newglossaryentry{CanonicalPhaseSpace}
{
    name=canonical phase space,
    description={}
}

\newglossaryentry{CovariantPhaseSpace}
{
    name=covariant phase space,
    description={}
}

\newglossaryentry{KGProduct}
{
    name=KG product,
    description={}
}

\newglossaryentry{ComplexStructure}
{
    name=complex structure,
    description={}
}

\newglossaryentry{OneParticleHilbertSpace}
{
    name=one-particle Hilbert space,
    description={}
}

\newglossaryentry{OneAntiparticleHilbertSpace}
{
    name=one-antiparticle Hilbert space,
    description={}
}

\newglossaryentry{AnnihilationCreationVariables}
{
    name=annihilation and creation variables,
    description={}
}

\newglossaryentry{SymmetricFockSpace}
{
    name=symmetric Fock space,
    description={}
}

\newglossaryentry{AnnihilationOperators}
{
    name=annihilation operators,
    description={}
}

\newglossaryentry{CreationOperators}
{
    name=creation operators,
    description={}
}

\newglossaryentry{QuantumVacuum}
{
    name=quantum vacuum,
    description={}
}

\newglossaryentry{QuantumFieldOperator}
{
    name=quantum field operator,
    description={}
}

\newglossaryentry{BogoliubovCoefficients}
{
    name=Bogoliubov coefficients,
    description={}
}

\newglossaryentry{UnitarilyEquivalent}
{
    name=unitarily equivalent,
    description={}
}

\newglossaryentry{UnitarilyImplementable}
{
    name=unitarily implementable,
    description={}
}

\newglossaryentry{DiracEquation}
{
    name=Dirac equation,
    description={}
}

\newglossaryentry{DiracProduct}
{
    name=Dirac product,
    description={}
}

\newglossaryentry{MinkowskiQuantumVacuum}
{
    name=Minkowski quantum vacuum,
    description={}
}

\newglossaryentry{InQuantumVacuum}
{
    name=`in' quantum vacuum,
    description={}
}

\newglossaryentry{OutQuantumVacuum}
{
    name=`out' quantum vacuum,
    description={}
}

\newglossaryentry{ILESt0}
{
    name=instantaneous lowest-energy state at~$t_0$,
    description={}
}

\newglossaryentry{AdiabaticApproximation}
{
    name=adiabatic approximation of order~$n$,
    description={}
}

\newglossaryentry{AdiabaticQuantumVacuum}
{
    name=adiabatic quantum vacuum of order~$n$ at time~$t_0$,
    description={}
}

\newglossaryentry{geons}
{
    name=geons,
    description={}
}

\newglossaryentry{kugelblitz}
{
    name=kugelblitz,
    description={}
}

\newglossaryentry{TransmissionAndReflectionCoefficients}
{
    name=transmission and reflection coefficients,
    description={}
}

\newglossaryentry{SmearingFunction}
{
    name=smearing function,
    description={}
}

\newglossaryentry{SmearedEnergyDensity}
{
    name=smeared energy density,
    description={}
}

\newglossaryentry{StateOfMinimalEnergy}
{
    name=state of minimal energy,
    description={}
}

\newglossaryentry{PowerSpectrum}
{
    name=power spectrum,
    description={}
}

\newglossaryentry{InitialConditionDistributors}
{
    name=initial condition distributors,
    description={}
}

\newglossaryentry{SpectralNumberOfCreatedExcitationsBetweenTau0AndTau}
{
    name=spectral number of created excitations between~$\tau_0$ and~$\tau$,
    description={}
}

\newglossaryentry{GeneralizedQuantumVlasovEquation}
{
    name=generalized quantum Vlasov equation,
    description={}
}

\newglossaryentry{QuantumVlasovEquation}
{
    name=quantum Vlasov equation,
    description={}
}

\newglossaryentry{Superradiance}
{
    name=superradiance,
    description={}
}

\newglossaryentry{ReissnerNordstromBlackHole}
{
    name=Reissner-Nordstr\"om black hole,
    description={}
}

\newglossaryentry{InUpBasis}
{
    name=`in-up' basis,
    description={}
}

\newglossaryentry{OutDownBasis}
{
    name=`out-down' basis,
    description={}
}

\newglossaryentry{QuantumChargeCurrentOperator}
{
    name=quantum charge current operator,
    description={}
}

\newglossaryentry{QuantumStressEnergyMomentumTensorOperator}
{
    name=quantum stress-energy momentum tensor operator,
    description={}
}

\newglossaryentry{EffectiveErgosphere}
{
    name=effective ergosphere,
    description={}
}

\newacronym{QFT}{QFT}{quantum field theory}
\newacronym{QFTCS}{QFTCS}{Quantum field theory in curved spacetimes}
\newacronym{KG}{KG}{Klein-Gordon}
\newacronym{QVE}{QVE}{quantum Vlasov equation}
\newacronym{SLEs}{SLEs}{states of low energy}
\newacronym{FLRW}{FLRW}{Friedmann-Lema\^itre-Robertson-Walker}
\newacronym{ILES}{ILES}{instantaneous lowest-energy state}
\newacronym{BECs}{BECs}{Bose-Einstein condensates}
\newacronym{RN}{RN}{Reissner-Nordstr\"om}

\begin{document}

\pagenumbering{roman}

%%% Roman Numeration %%%

%%%%%%Portada%%%%%%%
\begin{titlepage}
\centering
{\bfseries \Large Universidad Complutense de Madrid}

{\Large Facultad de Ciencias Físicas} 

{\large Programa de Doctorado en Física}
\vspace{0.8cm}

%%%%Logo Complutense%%%%%
{\includegraphics[width=0.5\textwidth]{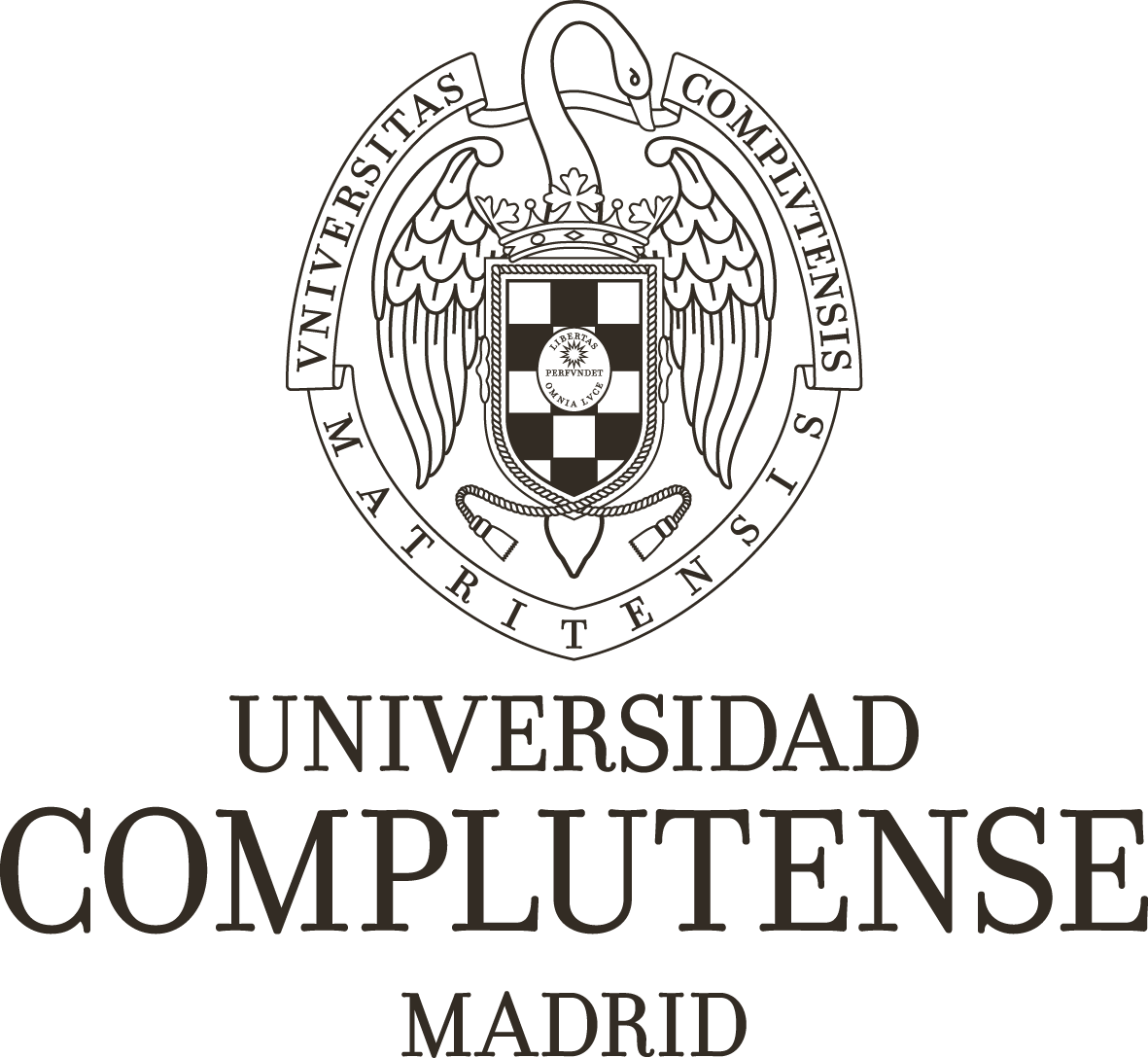}}

\vspace{0.8cm}

%%PORTADA

{\large PhD Thesis}
\vspace{0.5cm} 

{\color{darkred} \scshape \huge Cuantizaci\'on de campos cargados en presencia de campos electromagn\'eticos intensos }
\vspace{0.15cm}

\rule{\textwidth}{0.1mm}

\vspace{0.3cm}
{\color{darkred} \scshape \huge Quantization of charged fields in the presence of intense electromagnetic fields}
\vspace{1cm} 

{\large Dissertation submitted for the degree of Doctor of Phylosophy in Physics by}\vspace{0.5cm} 

{\color{chapterdarkred} \scshape \bfseries \Large \'Alvaro \'Alvarez Dom\'inguez}\vspace{1cm} 

{\large under the supervision of}\vspace{0.5cm} 

{\color{chapterdarkred} \scshape \Large Luis Javier Garay Elizondo}

{\color{chapterdarkred} \scshape \Large Mercedes Mart\'in Benito}

\end{titlepage}
\plainblankpage
%\plainblankpage
%\plainblankpage

\thispagestyle{empty}
\vspace*{\fill} % Empuja el contenido hacia el centro verticalmente
\begin{minipage}{0.8\textwidth}

\textit{A mi mam\'a y a mi pap\'a. \\ A mi hermano. \\ Y a mi compañero de vida. \\ Gracias.}

\end{minipage}
\vspace*{\fill} % Empuja el contenido restante hacia abajo para equilibrar
\plainblankpage

%%% Acknowledgements %%%
\chapter*{Agradecimientos}

Yo siempre lo he tenido bastante claro. Yo, de mayor, quería ser científico, de esos que investigan y dan clases. Y aquí estoy. Cerrando una bonita etapa de mi vida. O, como mejor me gusta verlo a mí: dando mis primeros pasos en esta montaña rusa de la academia.

Hace cinco años ya que me vine a Madrid para estudiar el máster en física teórica de la Universidad Complutense de Madrid. Fue así como conocí a Luis y a Merce, mis por aquel entonces directores del trabajo fin de máster. Eran mundialmente conocidos por ser los \textit{duros} de la facultad. Y a mí (antes mucho más que ahora) me encantaba que me metieran caña. Solo un mes después, les propuse hacer la tesis doctoral con ellos. Diría que fue fácil convencerlos, aunque la realidad es que me costó tres días hasta que me dieron el \textit{sí quiero}. Qué tres días aquellos más angustiosos. Espero que no os hayáis arrepentido. Sé que no.

Gracias, Merce, por guiarme y por aconsejarme en esta aventura de la que tengo todavía tanto por aprender. Es una gozada tener a alguien a tu lado que de verdad mira por tu bien. Y a ti, Luis, gracias. Hemos discutido lo que no está escrito. Y es que si a mí me gusta pincharte, a ti te gusta aún más. Gracias por escucharme cuando lo he necesitado, y por enseñarme que un \textit{señor} catedrático como tú puede tener una capacidad de autocrítica tan exquisita. Os preocupáis por los vuestros. Vuestro interés por formar un grupito \textit{apañao'} de \textit{gravillorones} va mucho más allá de lo puramente científico. Para mí sois un ejemplo de motivación y de pasión por lo que hacéis.

In the second year of my PhD, I met Elizabeth Winstanley at a conference in Granada. Nervous---perhaps more than a little---I asked her if I could do a research stay at the University of Sheffield under her guidance. She welcomed me with incredible warmth and generosity. Working with you, Elizabeth, has been a joy: easy, inspiring, and genuinely enjoyable. You bring a smile to your work and effortlessly share your passion and dedication with your students. I am certain this is just the beginning of our collaboration. Thank you.

Y qué hubieran sido estos años sin mi \textit{despachito 15}. Mercè, mi primera compañera. Qué pena que acabases tan pronto tu tesis. Hemos tenido unas conversaciones tan sinceras, tan bonitas. Gracias por ser una de las personas más sensibles y empáticas que conozco. Eva: cuántas charlas y cuántas risas. E cosa sarebbe stato il \textit{despachito} senza Sara, la mia Sara Datas. Sempre così generosa, sempre attenta agli altri in modo così sincero. Y sin Darío, la única persona en este mundo que no sabe enfadarse. No sabe. Aunque lo intente, no es capaz. Le sale cariño y bondad por los poros. Y qué seríamos todos sin Álvaro, la persona más resolutiva, organizativa y propositiva que se puede parir. Cuánto me alegro de haber compartido mis días con una persona como tú. Gracias a todas las personas que habéis pasado por el \textit{despachito}. A ti en especial, Sofía. Qué pena que tu tesis no la hagas en Madrid, porque qué energía más bonita tienes. Qué suerte tengo de teneros. Así, el doctorado es coser y cantar.

Más allá del \textit{despachito}, el departamento de física teórica encierra a mucha gente especial. En el caso de Dani, lo de especial se queda muy corto. Gracias por preguntarme cada vez que me ves por cómo estoy. Gracias por ese interés tan grande en todo el mundo que te rodea. Eres un osito amoroso. A ti, Parra. Porque he disfrutado tanto de la física contigo... Han sido tantas horas intensas de trabajo mano a mano... Y es que no hay nada mejor que poder equivocarse sin filtros, sin pensar que vas a quedar mal. Contigo he aprendido cómo quiero investigar. Gracias a Patri por siempre sonreírme en los pasillos, y por esos guantes tan calentitos. Gracias, Diego, por nuestras conversaciones eurovisivas día sí y día también en un espacio lleno de \textit{haters} de Eurovisión. Gracias, Sarita, por tu sensibilidad. Gracias a Lucía, Óscar, Toni, Alba, Teo, Pablos, Guille, Andrea, Diego, Joan, Alfredo, Clara, y un largo etcétera, por permitirme sentirme tan yo. Thanks also to all the colleagues I met in Sheffield, who made me feel at home during those three months. Obrigada, Rita, por seres sempre tão generosa comigo. A minha estadia em Sheffield não teria sido a mesma sem ti. Trabalhar com uma amiga como tu é um prazer imenso.

Ahora os animo a coger la maleta y viajar unos kilómetros desde Madrid hasta Sevilla. La que tiene un color \textit{epesiá}. Allí están mis amigas del alma. Y es que tengo una suerte que no se puede \textit{aguantá}. Las tengo lejos y cerca a la vez. Lara, en el \textit{top 1} de nuestras visitantes en casa. ?`Cuántas veces has venido? He perdido la cuenta. Y pocas son para las que me gustarían. Ya sabes que eres un pilar fundamental en mi vida. María Isabel, mi alma gemela desde antes de nacer. Mi hermana. Entre nosotros no hacen falta palabras para que nos entendamos. Espe, Laura: con ustedes los días están llenos de carcajadas. El año que viene volvemos al Benidorm Fest, ?`no? Y Marina, Samuel, Bea, Montse: qué gusto es volver a Sevilla y ver cómo los años van pasando, pero entre nosotras todo sigue igual. 

En Sevilla también he dejado a otra familia: mis amigos de la carrera. Fran, Abraham, Sara, Pena, Silvia... Ellos también son muy culpables de que yo esté aquí. Y es que sin la hermandad que formamos durante aquellos años tan intensos, todo hubiese sido mucho más complicado. Tan bien rodeado, tan alejados de la competitividad tóxica, fue mucho más fácil acabarme de enamorar de la física y de las matemáticas. Alexander: a ti tengo la suerte de tenerte más cerca. Gracias por hacerme sentir en Madrid tan a gusto. Eres un grande. Y detrás de mí, en una clase de análisis funcional, estaba Marta. Qué maravilla las casualidades. Desde entonces, Marta me abrió las puertas de su casa, y ahora no hay quien nos separe. Y es que cuando digo que soy un afortunado, no lo digo por decir.

Aunque ya os he presentado a gran parte de mi familia, todavía queda la parte que me corre por las venas. Mi familia. Mis tías, mis tíos. Mi abuela Pepa, que aún la siento conmigo. Mi prima Carmina, que tanto me quería. Nos dejasteis demasiado pronto, pero sé que hoy sois las personas más orgullosas de vuestro Álvaro. 

A mi hermano, por ser un hermano tan especial. Con él he aprendido que la sensibilidad es un deber. A mamá. A papá. Que lo dais todo por mí. Sois los mejores padres que existen. Me habéis enseñado a ser, siempre desde la comprensión y desde el amor. Soy como soy por ustedes. No puedo estar más orgulloso de los padres que tengo.

A Jesús, mi compañero de vida, de aventuras, quien comparte mis sueños, el que hace de nuestra casa en Madrid (con wáter rosa) nuestro hogar. Llegar a casa y darte un abrazo es el mejor momento del día. Me escuchas, me entiendes, me apoyas. Me regañas cuando digo que soy \textit{pesao'} porque para ti nunca lo soy. Me has enseñado las cosas bonitas de la vida, a mirar mucho más allá de la intensidad del trabajo. Gracias a ti he ampliado mi familia. Tu familia (mi familia) me ha acogido desde el minuto cero. Tu madre, tus tías, tus primos. Ahora incluso tengo dos abuelos maravillosos a los que no puedo tener más cariño. Gracias por traer a mi vida a amigas tan especiales como Tania y Lara. Incluso hemos parido a nuestro querido Benito, \textit{miau}. Contigo he aprendido cómo construir una relación desde el entendimiento, el cariño, el respeto y el amor. Contigo soy mejor. 

Y es que, esté donde esté, siempre me he sentido arropado por los míos. Así, todo resulta mucho más fácil, incluso escribir las páginas que siguen. Ojalá las disfruten tanto como yo he disfrutado escribiéndolas.

\vspace{1cm}

\textit{This work is part of the R+D+I Projects PID2022-139841NB-I00 and PID2023\\-149018NB-C44, funded by MICIU/AEI/10.13039/501100011033 and by ERDF/EU.}

\plainblankpage
%%% Table of Contents, List of Figures and List of Tables %%%
\bookmarktocentry
{
\hypersetup{
    linkcolor=black,
    urlcolor=black,
    citecolor=darkred,
}
\tableofcontents
}
%\plainblankpage
{
\hypersetup{
    linkcolor=black,
    urlcolor=black,
    citecolor=darkred,
}
\plainblankpage
\listoffigures
}
%\plainblankpage

%%% Print: Glossary and Acronyms %%%
\plainblankpage
\glossarytoc\printglossary
\acronymtoc\printglossary[type=\acronymtype]

\plainblankpage
%%% Abstract %%%
\addcontentsline{toc}{frontmatter}{Abstract}
\chapter*{Abstract}

This thesis applies techniques from quantum field theory in curved spacetimes to investigate various particle creation processes, with a special focus on pair production in strong electromagnetic backgrounds. The Schwinger effect, the central phenomenon explored here, occurs when an extremely strong electric field generates particle-antiparticle pairs out of the vacuum. Its experimental verification remains elusive due to the extreme electric field strengths required---on the order of $10^{18} \, \mathrm{V/m}$---which current laser technologies have yet to achieve.

The thesis is organized into five main parts.
Part I establishes the theoretical framework for the canonical quantization of charged fields in non-trivial backgrounds, including both curved spacetimes and external electromagnetic fields. It addresses the ambiguities arising in the canonical quantization procedure and explores different choices of quantum vacua. In addition, it extends the definition of states of low energy, originally developed in cosmology, to the context of the Schwinger effect. These states minimize the smeared energy density of the test field over a finite time window, offering a physically motivated and mathematically well-posed choice of vacua in non-trivial backgrounds.

Part II examines the time evolution of quantum theories and the impact of different quantization schemes. Only certain quantizations admit unitary dynamics, a desirable feature for the physical consistency of the theory. By identifying criteria that favour unitary time evolution, this part narrows the class of physically viable quantum theories applicable to the Schwinger effect. It also analyses how the number of created particle-antiparticle pairs evolves over time and how quantum ambiguities affect their production rates. In particular, it generalizes the standard quantum Vlasov equation---traditionally derived under restrictive assumptions regarding the choice of vacuum---to a framework that accommodates general quantization choices and a more flexible physical interpretation.

Part III adopts an operational perspective to address whether quantum ambiguities are fundamentally physical or mere mathematical artifacts. By proposing an operational realization of quantum vacuum ambiguities, this part confirms their physical nature, establishing a connection between the theoretical infinite freedom in vacuum choice and the infinite possibilities for interacting with and measuring the system. It also shows that interactions with the system inevitably induce `on' and `off' transitions out of and into static regimes, which can significantly influence experimental outcomes. This highlights the need for careful interpretation of results, particularly in analogue gravity experiments. The study is carried out both for Bose-Einstein condensates simulating cosmological pair production in a homogeneous and isotropic expanding universe and for the Schwinger effect, revealing fundamental differences in the behaviour of quantum fields across different backgrounds.

Part IV explores the broader physical implications of pair creation in strong electromagnetic backgrounds, particularly in black hole physics. It demonstrates that the Schwinger effect prevents the formation of black holes from pure light in the present-day universe, whether artificially or naturally---a result that contrasts with classical general relativity predictions. Moreover, it examines fermionic charge superradiance in charged black holes, where quantum effects lead to black hole discharge and energy loss through pair production. This provides a striking example of a purely quantum phenomenon with no classical counterpart: fermionic fields do not exhibit classical superradiance, unlike scalar fields.

Part V concludes by summarizing the main contributions of the thesis and outlining possible directions for future research.

This work highlights the fundamental challenge of defining particles and vacua in non-trivial backgrounds, confronting the standard intuitions derived from flat spacetime quantum field theory. It shows how external fields---gravitational or electromagnetic---can fundamentally alter the quantum structure of spacetime, leading to spontaneous particle creation and revealing inherent ambiguities in the choice of quantum vacuum. It also emphasizes the importance of purely quantum phenomena, with no classical analogues, in shaping the nature of physical processes. Addressing these issues is crucial for advancing our understanding of the quantum nature of spacetime and bridging the longstanding gap between quantum field theory and general relativity.

\addcontentsline{toc}{frontmatter}{Resumen}
\chapter*{Resumen}

Esta tesis aplica técnicas de la teoría cuántica de campos en espaciotiempos curvos para investigar diversos procesos de creación de partículas, con especial atención a la producción de pares en campos electromagnéticos intensos. El efecto Schwinger, el fenómeno central explorado aquí, se produce cuando un campo eléctrico extremadamente intenso genera pares partícula-antipartícula del vacío. Su verificación experimental sigue siendo difícil debido a las extremas intensidades de campo eléctrico que se requieren, del orden de $10^{18} \, \mathrm{V/m}$, que las tecnologías láser actuales aún no han alcanzado.

La tesis está organizada en cinco partes principales. 
La Parte I establece el marco teórico para la cuantización canónica de campos cargados en entornos no triviales, incluyendo tanto espaciotiempos curvos como campos electromagnéticos externos. Aborda las ambigüedades que surgen en el procedimiento de cuantización canónica y explora diferentes elecciones de vacíos cuánticos. Además, extiende la definición de estados de baja energía, desarrollada originalmente en cosmología, al contexto del efecto Schwinger. Estos estados minimizan la densidad de energía suavizada del campo de prueba en una ventana temporal finita, ofreciendo una elección de vacío físicamente motivada y con robustas propiedades matemáticas.

La Parte II estudia la evolución temporal de las teorías cuánticas, enfatizando en el impacto de las distintas elecciones de los esquemas de cuantización. Solo determinadas cuantizaciones admiten una dinámica unitaria. Esta característica es deseable para la consistencia física de la teoría. Al identificar los criterios necesarios para imponer una evolución temporal unitaria, se restringe la clase de teorías cuánticas físicamente viables aplicables al efecto Schwinger. Esta parte también analiza cómo evoluciona en el tiempo el número creado de pares partícula-antipartícula, y cómo afectan las ambigüedades cuánticas al ritmo de producción. En particular, generaliza la ecuación cuántica estándar de Vlasov, derivada tradicionalmente bajo condiciones muy restrictivas en la elección del vacío cuántico, a un marco más amplio que da cabida a opciones generales de cuantización y a una interpretación física más flexible.

La Parte III adopta una perspectiva operativa para abordar si las ambigüedades cuánticas son fundamentalmente físicas o meros artefactos matemáticos. Al proponer una realización operativa de las ambigüedades cuánticas del vacío, confirmamos su naturaleza física, estableciendo una conexión entre la infinita libertad teórica que existe en la elección del vacío y las infinitas posibilidades de interactuar con el sistema y medir. Además, se muestra que las interacciones con el sistema inducen inevitablemente transiciones de encendido y apagado desde y hacia regímenes estáticos, que pueden influir significativamente en los resultados experimentales. Esto enfatiza la necesidad de una interpretación cuidadosa de los resultados, especialmente en experimentos gravitatorios análogos. El estudio se lleva a cabo tanto para condensados de Bose-Einstein que simulan la producción cosmológica de pares en un universo homogéneo e isótropo en expansión como para el efecto Schwinger, revelando diferencias fundamentales en el comportamiento de los campos cuánticos en distintos contextos.

La Parte IV explora las implicaciones físicas de la creación de pares en la física de los agujeros negros. Se demuestra que el efecto Schwinger impide la formación de agujeros negros a partir de luz en el universo actual, ya sea de forma artificial o natural; un resultado que contrasta con las predicciones de la relatividad general clásica. Además, se examina la superradiancia de carga fermiónica en agujeros negros cargados, donde los efectos cuánticos conducen a la descarga del agujero negro y a la pérdida de energía por producción de pares. Esto proporciona un ejemplo interesante de un fenómeno puramente cuántico sin análogo clásico: los campos fermiónicos no exhiben superradiancia clásica, a diferencia de los campos escalares.

La Parte V concluye resumiendo las principales contribuciones de la tesis e indicando posibles ideas y direcciones para investigaciones futuras.

Este trabajo pone de relieve el reto fundamental de definir partículas y vacíos en contextos no triviales, confrontando las intuiciones usuales derivadas de la teoría cuántica de campos en espaciotiempo plano. La tesis muestra cómo los campos externos, gravitatorios o electromagnéticos, pueden alterar de forma fundamental la estructura cuántica del espaciotiempo, dando lugar a la creación espontánea de partículas y revelando ambigüedades inherentes a la elección del vacío cuántico. También se estudian fenómenos puramente cuánticos, sin análogos clásicos, que tienen consecuencias interesantes en la naturaleza de los agujeros negros. Abordar estas cuestiones es crucial para avanzar en nuestra comprensión de la naturaleza cuántica del espaciotiempo y salvar la brecha existente entre la teoría cuántica de campos y la relatividad general.

%%% Bibliography (my papers) %%%
\begin{refsection}

\newrefcontext[labelprefix=A]
\renewcommand*{\bibname}{Publications} 
\renewcommand*{\refname}{Publications}

\defbibnote{mainbibnote}{The content of this thesis is primarily based on my works:}
\printbibliography[
  title={Publications},
  heading=bibintoc,   % Makes it appear in the TOC with correct level (section/chapter)
  prenote=mainbibnote,
  keyword={Alvarez},
  resetnumbers=true]

\plainblankpage

%%% Arabic Numeration %%%
\pagenumbering{arabic}

%%% Chapters %%%
\chapter[Introduction]{Introduction}
\label{chap:Introduction}

Throughout the 20th century, \acrfull{QFT} emerged as a basis of modern particle physics, providing an exceptionally precise and experimentally validated framework, mainly through the study of perturbative effects observed for example in particle-particle collisions. The foundations of QFT rest on three key principles: quantum mechanics, classical field theory, and special relativity. However, QFT does not inherently incorporate the effects of gravity, making it insufficient to describe phenomena where gravitational effects play a dominant role.

General relativity, on the other hand, extends special relativity to incorporate gravity, offering a well-established and experimentally verified description of gravitational interactions. Over the past century, general relativity has withstood numerous tests, including the recent direct detection of gravitational waves~\cite{Collaboration2016}, a century after their theoretical prediction by Einstein.

Despite their successes, QFT and general relativity remain fundamentally incompatible in extreme physical regimes---such as those inside black holes or in the early universe---where both quantum effects and gravity are significant. Despite significant progress over the last century, the quest for a self-consistent theory of quantum gravity that unifies these two frameworks remains one of the greatest unsolved challenges in physics.

\acrfull{QFTCS} is an effective approach that lies in the interface between QFT and general relativity~\cite{Birrell1982,Wald1994,Fabbri2005,Mukhanov2007, Calzetta2008, Parker2009}. This framework allows for the study of quantum fields propagating on a classical gravitational background, such as black holes or an expanding universe. QFTCS provides valuable insights into scenarios where gravity is strong enough to curve spacetime but not so extreme as to necessitate full quantization. One of the most profound achievements of QFTCS is its prediction of non-perturbative quantum particle creation phenomena. In 1974, Stephen Hawking demonstrated that black holes emit pairs of particles and antiparticles, leading to Hawking radiation, which causes black holes to lose energy and potentially evaporate over time~\cite{Hawking1974,Hawking1975}. Similar particle creation processes also arise in other curved backgrounds, such as during the cosmological expansion of the universe~\cite{Parker1968,Parker1969,Ford1987}.

In this thesis, we use techniques from QFTCS to cover a wide range of particle creation phenomena. Our central example will be the pair creation phenomenon that happens when we have, instead of a strong gravitational background, a strong electromagnetic background. This phenomenon was first suggested by F. Sauter~\cite{Sauter1931}, although it carries the name of Schwinger as he was the one who first explained it in the context of quantum electrodynamics for slowly varying fields~\cite{Schwinger1951}. In addition, the perturbative contribution counterpart of the particle and antiparticle creation in electromagnetic backgrounds takes the name of multiphoton Breit-Wheeler scattering~\cite{Brodsky1970,Jaccarini1970,Budnev1975}, and involves high-frequency modes. Here, we will use the term `Schwinger effect' to encompass both phenomena, and the formalism that we develop in this thesis takes into account both non-perturbative and perturbative contributions.

Empirically verifying the Schwinger effect in the case of a constant electric field requires generating field strengths that exceed the so-called Schwinger limit~\cite{Schwinger1951}:
\begin{equation}
    E_{\text{c}} = \frac{m^2c^3}{\hbar q},
\end{equation}
where~$m$ and~$q$ are the mass and the charge of the created particles and antiparticles. For electron-positron production, this critical field strength is approximately of the order of $10^{18} \ \mathrm{V/m}$. When the electric field~$E$ is below this threshold, the probability of pair creation is exponentially suppressed~\cite{Schwinger1951}:
\begin{equation}
    e^{-\pi E_\text{c}/E},
    \label{eq:ExponentialSuppression}
\end{equation}
making direct observation extremely challenging. Achieving such extreme field strengths poses major technical and engineering challenges and has not yet been realized. Current state-of-the-art laser systems can reach intensities up to \mbox{$10^{27} \ \text{W/m}^2$}, which correspond to electric field strengths still about three orders of magnitude below the Schwinger limit~\cite{Yoon2021}. Despite these promising technological advances, such experiments involve ultraintense lasers operating at very high frequencies, where non-perturbative contributions from the Schwinger effect remain largely inaccessible with current capabilities.

Studying the Schwinger effect reveals fundamental features shared by particle creation processes in curved spacetimes. One crucial concept that arises is the inherent ambiguity in defining the quantum vacuum, and consequently, the very notions of what we call \textit{particles} and \textit{antiparticles}~\cite{Fulling1973}. This contrasts with our usual intuition in standard QFT in flat spacetime, where we typically assume well-defined notions of particles and antiparticles. Indeed, in the canonical quantization of a free field in Minkowski spacetime, one usually selects the so-called Minkowski quantum vacuum, which preserves Poincar\'{e} invariance. However, even in flat spacetime, alternative quantization schemes exist that do not preserve classical symmetries, leading to different choices of quantum vacuum states. A striking example is the Unruh effect, where an accelerating observer perceives a thermal bath of particles in a vacuum state that an inertial observer would consider empty~\cite{Davies1975,Unruh1976}. In this case, the Rindler vacuum---adapted to the accelerating observer’s frame---is not Poincar\'{e} invariant, demonstrating how observer-dependent quantization schemes can lead to physically distinct particle interpretations.

In the presence of an intense external agent---such as a curved background or a strong electromagnetic field---classical symmetries, particularly time-translational invariance, can be broken. The absence of a preferred time symmetry results in multiple possible vacuum choices, potentially leading to nonequivalent quantum theories. In the case of the Schwinger effect, the presence of a strong electromagnetic field explicitly breaks the Poincar\'{e} symmetry of flat spacetime. This symmetry breaking prevents the existence of a preferred vacuum state, even if we require the preservation of all the classical symmetries, causing the quantum vacuum to evolve dynamically. As a consequence, particle-antiparticle pairs are spontaneously created throughout the evolution of the quantum state.

This thesis is divided into four main parts:

\begin{itemize}
    \item \textbf{Part I.} We set the theoretical framework for the canonical quantization of charged fields in non-trivial backgrounds. As classical background, we consider a general curved background in addition to an external electromagnetic field. For the charged field, we analyse both scalar and fermionic fields. 
    
    After establishing a general formalism applicable to a wide range of scenarios, we focus on a particular yet physically relevant background: a time-dependent electric field in flat spacetime, which allows us to explore the Schwinger effect in detail. A central aspect of our analysis is the study of ambiguities arising in the canonical quantization process. Depending on the physical criteria we aim to impose to the resulting quantum theory, different quantization choices can be made. These different prescriptions may result in distinct physical predictions, emphasizing the fundamental importance of selecting an appropriate quantization scheme.
    
    We review the family of states of low energy, originally proposed in the context of homogeneous and isotropic cosmologies~\cite{Fewster2000,Olbermann2007}, and extend their definition to the Schwinger effect~\cite{AlvarezSLEs}. These states are designed to minimize the smeared energy density of the test field, providing a physically motivated and mathematically well-behaved choice of quantum vacuum. Notably, this family of states of low energy encompasses many well-known quantum vacua from the literature, offering a unifying approach to defining quantum vacuum states in non-trivial backgrounds.
    \item \textbf{Part II.} We analyse the time evolution of quantum theories, examining how different quantization choices impact unitarity. While some quantizations allow for unitary dynamics, others do not. With the aim of reducing the quantum ambiguities in the choice of vacuum, it is desirable to identify physically reasonable criteria for quantization. One such criterion is requiring that the quantum theory admits a unitary implementation of time evolution, which effectively narrows down the range of viable quantizations in various backgrounds~\cite{Corichi2006,Cortez2012,Cortez2015,Garay2020}. For these unitary quantizations, we investigate how the number of created particle-antiparticle pairs evolves over time, exploring its dependence on quantum ambiguities. This part is developed based on the results presented in~\cite{AlvarezUnitary,AlvarezGQVE}.
    \item \textbf{Part III.} Building on~\cite{AlvarezOperational}, we confirm through an operational approach that quantum ambiguities are fundamentally physical, addressing an open debate in the literature concerning the interpretation of the time evolution of pair creation~\cite{Ilderton2022,Dabrowski2014,Dabrowski2016a,Yamada2021,Domcke2022,Diez2023}. Moreover, this perspective enables us to bridge the gap between theoretical predictions and potential experimental realizations. Expanding on this connection, we further investigate the Schwinger effect and gravitational analogue experiments, demonstrating how the interpretation of experimental outcomes can be significantly obscured by our unavoidable interaction with the system~\cite{AlvarezInOut}.
    \item \textbf{Part IV.} All the phenomena we have explored in the context of strong electromagnetic backgrounds throughout this thesis have profound implications in nature, particularly in the study of black holes. For instance, revisiting our works~\cite{AlvarezKugelblitz,AlvarezKugelblitzComment}, we demonstrate that the Schwinger effect prevents the formation of black holes from pure light---an outcome otherwise permitted by general relativity~\cite{Robinson1962,Senovilla2014}. Additionally, particle creation processes also occur in charged black holes, giving rise to a phenomenon known as charge superradiance~\cite{Bekenstein1973,Nakamura1976,DiMenza2015,Benone2016,DiMenza2020}. Following~\cite{AlvarezSuperradiance}, we show how this quantum effect manifests for fermions, ultimately leading to both the discharge and energy loss of the black hole.
    \item \textbf{Part V.} We conclude with some final remarks, summarizing the key findings of this thesis and highlighting the most significant contributions. Additionally, we discuss open questions and potential future research directions that I aim to explore in the near future.
\end{itemize} 

\textit{Notation.} Unless explicitly stated otherwise, this thesis follows natural units $\hbar = c = G = 1$. The chosen metric signature is~$(-,+,+,+)$.

\part[Ambiguities in the canonical quantization of charged fields]{Ambiguities in the canonical quantization of charged fields}
\label{part:Ambiguities}

\plainblankpage

\vspace*{\fill}
\begin{minipage}{0.8\textwidth}

In this first part of the thesis, we investigate the ambiguities that arise in the canonical quantization of test fields in non-trivial backgrounds. This formalism is crucial for studying pair creation phenomena. In particular, we focus primarily on the Schwinger effect and explore how different choices of quantization schemes can lead to different theoretical predictions.

In~\autoref{chap:QuantizationScalars}, we review the fundamental concepts of canonical quantization for both scalar and fermionic fields in the presence of an electromagnetic background in a generic curved spacetime. Particular attention is given to Bogoliubov transformations, which allow us to compare different quantization schemes. In~\autoref{chap:ChoiceVacuum}, we apply this framework to a quantum scalar field in a homogeneous electric field in flat spacetime. We present the most well-known quantum vacua in the literature, each based on different physical criteria, which will be useful throughout the rest of the thesis. Finally, in~\autoref{chap:SLEs}, we study a particular family of quantum vacua in the Schwinger effect: the so-called states of low energy, which minimize the energy density of the test field over a finite time interval. Originally introduced in cosmological settings, we extend their definition to anisotropic electric backgrounds.

Although most of the results in \autoref{chap:QuantizationScalars} and \autoref{chap:ChoiceVacuum} are well-established in the literature, they have been carefully rewritten to adapt the formalism and ensure consistency with the rest of the thesis. As far as I know, the canonical quantizations of scalars and fermions have not been explicitly reviewed in the presence of both a generic curved background and a generic electromagnetic field. In this sense, I hope the general treatment developed in \autoref{chap:QuantizationScalars} provides a solid foundation for handling both contributions simultaneously. On the other hand, \autoref{chap:SLEs} is primarily based on~\cite{AlvarezSLEs}.

\end{minipage}
\vspace*{\fill} 
\plainblankpage
\chapter[Canonical quantization of charged fields]{Canonical quantization of \\ charged fields}
\label{chap:QuantizationScalars}

\chaptermark{Canonical quantization of charged fields}

In this chapter, we establish the foundational elements necessary for the rest of this thesis. Our approach is primarily based on the treatments found in~\cite{Birrell1982, Wald1994, Fabbri2005, Mukhanov2007}. In~\autoref{sec:ClassicalScalars}, we examine the canonical classical theory of a charged scalar field in a general curved and electromagnetic background, introducing the \acrfull{KG} inner product. Since this product is not positive-definite, constructing a proper Hilbert space of solutions to the KG equation requires defining a positive-definite inner product through the introduction of a complex structure. This step is crucial for distinguishing particles and antiparticles in the quantum theory. However, the presence of a background electromagnetic field introduces additional challenges, which we discuss in~\autoref{sec:ComplexStructure}.

In~\autoref{sec:QuantumTheory}, we proceed with the quantization of the scalar field while treating the curved and electromagnetic background as classical. We find that, in general, a single classical theory can give rise to infinitely many different quantum theories, each associated with its own notion of particles and antiparticles. Understanding these ambiguities in the canonical quantization is a central theme throughout this thesis. A key tool in analysing these ambiguities involves the Bogoliubov transformations, introduced in~\autoref{sec:Bogoliubov}. These transformations allow us to compare different quantum theories and define a crucial observable in particle creation phenomena: the number of created particles and antiparticles. 

In~\autoref{sec:Fermions}, we extend the canonical quantization framework from scalar fields to fermionic fields, highlighting the key differences between both cases. Finally, in \autoref{sec:ConclusionsQuantization}, we summarize the main concepts discussed throughout this chapter.

\section{Classical scalar theory}\label{sec:ClassicalScalars}

Let us consider a complex scalar field~$\Phi$ with mass~$m$ and charge~$q$, minimally coupled to a globally hyperbolic spacetime in the presence of an electromagnetic background characterized by the four-vector potential~$A_{\mu}$. An action for this field is 
\begin{equation}
    S = - \int \text{d}^4 x \ \sqrt{-\mathfrak{g}} \left[ \left( D_\mu \Phi \right)^* \left( D^\mu \Phi \right) + m^2 \Phi^* \Phi \right],
    \label{eq:ActionScalar}
\end{equation}
where~$\mathfrak{g}$ is the determinant of the metric tensor~$g_{\mu\nu}$, 
\begin{equation}
    D_\mu = \nabla_\mu + iqA_\mu
\end{equation}
is the covariant derivative, and~$*$ denotes complex conjugation. From here, the equations of motion for the charged field can be derived, yielding the \gls{KGEquation}:
\begin{equation} 
    \left( D_\mu D^\mu - m^2 \right)  \Phi = 0.
    \label{eq:KG}
\end{equation}

Since we are working in a globally hyperbolic spacetime, the spacetime admits a global time function~$t$ and can be foliated into Cauchy hypersurfaces~$\Sigma_t$ defined by constant values of~$t$. The conjugate momentum field corresponding to the scalar field~$\Phi$ is given by
\begin{equation}
    \Pi = \frac{\delta S}{\delta (\partial_t \Phi)} = \sqrt{\mathfrak{h}} n^\mu \left( D_\mu \Phi \right)^*,
\end{equation}
where~$\mathfrak{h}$ is the determinant of the induced metric on~$\Sigma_t$, and~$n^\mu$ is the unit normal vector to the Cauchy hypersurfaces. The only non-vanishing Poisson brackets is
\begin{equation}
    \{\Phi(t,\bfx),\Pi(t,\textbf{y})\} = \delta(\bfx-\textbf{y}),
    \label{eq:PoissonPhi}
\end{equation}
where~$\bfx$ and~$\textbf{y}$ are points in~$\Sigma_t$, and~$\delta$ denotes the Dirac delta distribution.

One way to proceed in the description of the classical theory is the canonical approach. The \gls{CanonicalPhaseSpace} is defined as the set of pairs composed of a field and its conjugate momentum \mbox{$(\Phi_{t_0}(\bfx),\Pi_{t_0}(\bfx))$} on a given Cauchy hypersurface~$\Sigma_{t_0}$ at a fixed time~$t_0$. The KG equation~\eqref{eq:KG} admits a well-posed initial value formulation. For any given pair of initial data~\mbox{$(\Phi_{t_0}(\bfx),\Pi_{t_0}(\bfx))$}, there exists a unique smooth solution~$\Phi$ on the whole spacetime manifold satisfying the initial conditions
\begin{equation}
    \Phi|_{\Sigma_{t_0}}(t,\bfx)= \Phi_{t_0}(\bfx) \qquad \text{and} \qquad \Pi|_{\Sigma_{t_0}}(t,\bfx)=\Pi_{t_0}(\bfx),
\end{equation}
when restricted to the Cauchy hypersurface~$\Sigma_{t_0}$~\cite{Wald1994}. Therefore, the canonical phase space can be identified with the \gls{CovariantPhaseSpace}~$\mathcal{S}$: the vector space of all smooth solutions~$\Phi$ of the KG equation with smooth initial data.

The \gls{KGProduct} is defined on the covariant phase space~$\mathcal{S}$. For two solutions~$\Phi_1$ and~$\Phi_2$ of the KG equation~\eqref{eq:KG}, it is given by
\begin{equation}
    (\Phi_1,\Phi_2)_{\text{KG}} = -i \int_{\Sigma_t} \text{d}^3\bfx \ \sqrt{\mathfrak{h}} n^\mu \left[ \Phi_1^* D_\mu \Phi_2 - (D_\mu \Phi_1)^* \Phi_2 \right].
    \label{eq:KGproduct}
\end{equation}
An important property of the KG product is its independence of the choice of the Cauchy hypersurface~$\Sigma_t$ on which it is evaluated. This invariance can be shown using integration by parts and assuming the condition that the fields vanish in the boundary of~$\Sigma_t$. Thus, the KG product is conserved under time evolution.

The KG product satisfies most of the properties of an inner product: it is antilinear in the first argument, linear in the second, and hermitian; i.e., \mbox{$(\Phi_1,\Phi_2)_{\text{KG}}^* = (\Phi_2,\Phi_1)_{\text{KG}}$}. However, it fails to be positive-definite, as we now explain.

If~$\Phi$ is a solution to the KG equation, its complex conjugate~$\Phi^*$ is not a solution to the same KG equation, but of its complex conjugate. This means that if~$\Phi$ is a KG field with charge~$q$, $\Phi^*$ is a KG field with charge~$-q$. Consequently, the KG product for fields with charge~$-q$, $(\cdot,\cdot)_{\text{KG}^*}$, is defined differently, incorporating the complex-conjugate covariant derivative, \mbox{$D_\mu^* = \nabla_\mu - iqA_\mu$}. Specifically, if~$\Phi_1^*$ and~$\Phi_2^*$ are fields with charge~$-q$, their KG product is given by
\begin{equation}
    (\Phi_1^*,\Phi_2^*)_{\text{KG}^*} = -i \int_{\Sigma_t} \text{d}^3\bfx \ \sqrt{\mathfrak{h}} n^\mu \left[ \Phi_1 (D_\mu \Phi_2)^* - (D_\mu \Phi_1) \Phi_2^* \right] = - (\Phi_1,\Phi_2)_{\text{KG}}^*.
    \label{eq:KGproduct*}
\end{equation}
From this, it follows that if~$\Phi$ has positive norm with respect to the original KG product $(\cdot,\cdot)_{\text{KG}}$ defined in~\eqref{eq:KGproduct}, then its complex conjugate~$\Phi^*$ has negative norm with respect to the KG product $(\cdot,\cdot)_{\text{KG}^*}$ introduced in~\eqref{eq:KGproduct*}. In particular, for real solutions~$\Phi$ of the KG equation (which have no charge), the norm always vanishes, i.e., \mbox{$(\Phi, \Phi)_{\text{KG}} = (\Phi, \Phi)_{\text{KG}^*} = 0$.}
\begin{block}[note]
If there were no electromagnetic interaction, the KG equation would be real. In this case, both~$\Phi$ and~$\Phi^*$ would satisfy the same equation and the covariant phase space~$\mathcal{S}$ would coincide with its complex conjugate~$\mathcal{S}^*$. Consequently, there would be no need to extend the definition of the KG product to accommodate complex-conjugate solutions.
\end{block}

This lack of positive-definiteness renders the KG product unsuitable as a proper inner product, which prevents us from directly constructing a Hilbert space of solutions. This presents an obstacle to formulating the quantum theory. However, as we will discuss in the following section, this issue can be addressed and resolved, allowing for a consistent quantization framework.

\section{One-particle and one-antiparticle Hilbert spaces}
\label{sec:ComplexStructure}

Our goal is to construct a Hilbert space with a proper positive-definite inner product from the covariant phase space~$\mathcal{S}$ of solutions to the KG equation. To achieve this, we will select a subspace of~$\mathcal{S}$ consisting of solutions with positive KG norm. On this subspace, we define a positive-definite inner product that coincides with the KG product. Similarly, we construct a complementary subspace of solutions with negative KG norm and redefine the inner product by changing the sign of the KG product for these solutions.

To formalize this construction, we introduce a \gls{ComplexStructure}~$J$ on the covariant phase space~$\mathcal{S}$. By definition, $J$ is an antihermitian linear operator satisfying~$J^2=-\mathds{1}$ such that 
\begin{equation}
    (\cdot,\cdot) \equiv i(J \cdot, \cdot)_{\text{KG}}
    \label{eq:InnerProduct}
\end{equation}
is a positive-definite inner product on~$\mathcal{S}$. It is straightforward to verify that~$iJ$ is self-adjoint with respect to this inner product, and that its eigenvalues are~$\pm i$. 

Using the spectral theorem, we decompose the covariant phase space~$\mathcal{S}$ into two orthogonal eigenspaces corresponding to the eigenvalues~$+i$ and~$-i$. The orthogonal projection operators onto these eigenspaces are defined as:
\begin{equation}
    P^{\pm}=\frac{1}{2}(\mathds{1} \mp iJ).
\end{equation}
This allows us to write the decomposition of~$\mathcal{S}$ as
\begin{equation}
    \mathcal{S} = P^+\mathcal{S} \oplus P^-\mathcal{S},
\end{equation}
where~$P^+\mathcal{S}$ and $P^-\mathcal{S}$ are the eigenspaces associated with the eigenvalues~$+i$ and~$-i$, respectively.

The \textit{particle} states are identified as solutions in~$P^+\mathcal{S}$. As solutions to the KG equation~\eqref{eq:KG}, they have charge~$q$. The \gls{OneParticleHilbertSpace}~$\mathcal{H}^+$ is then defined as the Cauchy completion of~$P^+\mathcal{S}$ with respect to the positive-definite inner product~\eqref{eq:InnerProduct}. If~$\Phi^+$ is in $\mathcal{H}^+$, we have that~$J\Phi^+ = i\Phi^+$, and then its KG norm is positive:
\begin{equation}
(\Phi^+, \Phi^+) = i(J\Phi^+, \Phi^+)_{\text{KG}} = (\Phi^+, \Phi^+)_{\text{KG}} > 0.
\label{eq:Phi+KG}
\end{equation}

The solutions~$\Phi^-$ in~$P^-\mathcal{S}$, associated with the eigenvalue~$-i$, could be interpreted as \textit{holes}. However, it is standard to describe these states in terms of \textit{antiparticles} (with charge~$-q$) rather than \textit{holes} (with charge $q$). To formalize this interpretation, we consider the complex conjugates~$(\Phi^-)^*$ living in the eigenspace \mbox{$(P^-\mathcal{S})^*=P^+\mathcal{S}^* \subset \mathcal{S}^*$}. To extend the definition of the positive-definite inner product~\eqref{eq:InnerProduct}, which is defined only on~$\mathcal{S}$, to the complex-conjugate space of solutions~$\mathcal{S}^*$, we define it in terms of the KG product~$(\cdot,\cdot)_{\text{KG}^*}$ introduced in~\eqref{eq:KGproduct*}:
\begin{equation}
    (\cdot,\cdot) \equiv i(J\cdot,\cdot)_{\text{KG}^*}.
    \label{eq:InnerProduct*}
\end{equation}

The \gls{OneAntiparticleHilbertSpace}~$\mathcal{H}^-$ is then defined as the Cauchy completion of $(P^-\mathcal{S})^*$ with respect to this inner product. For any solution~$(\Phi^-)^*$ in $\mathcal{H}^-$, its norm in the inner product~\eqref{eq:InnerProduct*} is positive, and as a consequence, $\Phi^-$ has negative KG norm. Indeed, since~\mbox{$J(\Phi^-)^*=i(\Phi^-)^*$}: 
\begin{equation}
    ((\Phi^-)^*,(\Phi^-)^*) = i(J(\Phi^-)^*,(\Phi^-)^*)_{\text{KG}^*} = -(\Phi^-, \Phi^-)_{\text{KG}} > 0.
    \label{eq:Phi-KG}
\end{equation}
Finally, the complete Hilbert space~$\mathcal{H}$ is constructed as the direct sum:
\begin{equation}
\mathcal{H} = \mathcal{H}^+ \oplus \mathcal{H}^-.
\end{equation}
\begin{block}[note]
The complete Hilbert space~$\mathcal{H}$ is not the same as the covariant phase space~$\mathcal{S}$. While the one-particle Hilbert space~$\mathcal{H}^+$ is a subset of the covariant phase space~$\mathcal{S}$, the one-antiparticle Hilbert space~$\mathcal{H}^-$ is a subset of the complex conjugate of the covariant phase space~$\mathcal{S}^*$. This distinction reflects the role of the complex conjugate space in the proper treatment of antiparticle states when there is an electromagnetic background.
\end{block}

With the inner product defined in~\eqref{eq:InnerProduct} for~$\mathcal{S}$ and its extension~\eqref{eq:InnerProduct*} for~$\mathcal{S}^*$, we resolve the issue of the non-positive definiteness of the KG product. Unlike the KG product, which satisfies~\eqref{eq:KGproduct*} for complex-conjugate solutions, the redefined inner product ensures that:
\begin{equation}
(\Phi_1^*, \Phi_2^*) = (\Phi_1, \Phi_2),
\end{equation}
for any two solutions~$\Phi_1$ and~$\Phi_2$ of the KG equation. This property guarantees that the inner product on the Hilbert space~$\mathcal{H}$ is positive-definite and consistent, enabling a well-defined quantum framework.

We can now choose an orthonormal basis~$\{\Phi_n^+\}$ for the one-particle Hilbert space~$\mathcal{H}^+$, as well as an orthonormal basis~$\{(\Phi_n^-)^*\}$ for the one-antiparticle Hilbert space~$\mathcal{H}^-$, with respect to the positive-definite inner product~$(\cdot,\cdot)$. Then, $\{\Phi_n^+,(\Phi_n^-)^*\}$ is an orthonormal basis of~$\mathcal{H}$. Consequently, for every solution $\Phi$ in the covariant phase space~$\mathcal{S}$ there exist unique complex coefficients $a_n$ and $b_n^*$ such that
\begin{equation}
    \Phi=\sum_n \left( a_n\Phi_n^+ +b_n^*\Phi_n^- \right).
    \label{eq:PhiExpansion}
\end{equation} 
These coefficients, which are associated with the complex structure~$J$, are called \gls{AnnihilationCreationVariables}, respectively. 

The Poisson bracket structure~\eqref{eq:PoissonPhi} induces the following algebra for the creation and annihilation variables:
\begin{equation}
    \{a_n,a_m^*\} = \{b_n,b_m^*\} = -i\delta_{n,m},
    \label{eq:PoissonVariables}
\end{equation}
the rest of Poisson brackets among them being zero. Here we consider the label~$n$ to be discrete. In the case that~$n$ were a continuous index, all equations would be naturally written with integrals instead of summations, and $\delta_{n,m}$ being the Dirac delta instead of the Kronecker delta.

\section{Quantum scalar theory}
\label{sec:QuantumTheory}

To define the quantum theory, the full Hilbert space is chosen to be the \gls{SymmetricFockSpace}:
\begin{equation}
    \mathcal{F}_{\text{S}} = \oplus_{n=0}^{\infty}\left( \otimes^n_{\text{S}} \mathcal{H} \right) = \mathbb{C} \oplus \mathcal{H} \oplus ( \mathcal{H} \otimes_{\text{S}} \mathcal{H} ) \oplus ...,
    \label{eq:SymmetricFock}
\end{equation}
where~$\oplus$ denotes the direct sum, and~$\otimes_{\text{S}}$ represents the symmetric tensor product. 

The annihilation coefficients~$a_n$ and~$b_n$ are promoted to \gls{AnnihilationOperators}~$\hat{a}_n$ and~$\hat{b}_n$ acting on the Fock space. Similarly, their complex conjugates~$a_n^*$ and~$b_n^*$ are mapped to \gls{CreationOperators}~$\hat{a}_n^\dagger,\hat{b}_n^\dagger$, where $\dagger$ denotes hermitian conjugation. While~$\hat{a}_n$ and~$\hat{a}^\dagger_n$ annihilate and create particles, respectively, $\hat{b}_n$ and~$\hat{b}^\dagger_n$ play the same role for antiparticles. The commutation relations of these operators are derived from the classical Poisson bracket algebra via the quantization prescription:
\begin{equation}
    \{\cdot,\cdot\} \rightarrow [\hat{\cdot},\hat{\cdot}]=i\widehat{\{{\cdot},{\cdot}\}},
\end{equation}
where~$[\hat{\cdot},\hat{\cdot}]$ is the commutator. From the Poisson algebra of the classical variables~\eqref{eq:PoissonVariables}, the only non-vanishing commutators are
\begin{equation} 
    [\hat{a}_n,\hat{a}_m^{\dagger}]=[\hat{b}_n,\hat{b}_m^{\dagger}]=\delta_{n,m}.
    \label{eq:Commutatorsab}
\end{equation}

The Fock \gls{QuantumVacuum}~$|0\rangle$ is defined as the state annihilated by all annihilation operators:
\begin{equation}
    \hat{a}_n |0\rangle = \hat{b}_n |0\rangle = 0,
\end{equation}
for all~$n$. Physically, the quantum vacuum represents the state with no particles or antiparticles. 

The \gls{QuantumFieldOperator}~$\hat{\Phi}$ on the Fock space is constructed by simply replacing the coefficients~$a_n$ and~$b_n^*$ with the corresponding operators in the expansion~\eqref{eq:PhiExpansion}:
\begin{equation}
    \hat{\Phi}=\sum_n \left( \hat{a}_n\Phi_n^+ + \hat{b}_n^\dagger \Phi_n^- \right).
    \label{eq:FieldOperator}
\end{equation} 

In summary, every complex structure~$J$ defines a split of the space of solutions~$\mathcal S$ which leads to two Hilbert spaces, namely~$\mathcal{H}^+$ and~$\mathcal{H}^-$. Elements of~$\mathcal{H}^+$ are interpreted as particles with charge~$q$, while elements of~$\mathcal{H}^-$ are interpreted as antiparticles with charge~$-q$. The complex structure~$J$ encodes all the information about the particular choice of quantization, and therefore determines the notions of particles and antiparticles.

\section{Bogoliubov transformations}
\label{sec:Bogoliubov}

In the definition of the one-particle Hilbert space~$\mathcal{H}^+$---and consequently in the selection of annihilation and creation operators in the quantum theory---there exists an inherent ambiguity due to the choice of the complex structure~$J$. Since different choices of~$J$ can lead to distinct quantum theories, each with its own set of observable predictions, it is important to compare different complex structures systematically. To formulate this problem mathematically, we consider canonical transformations of the fields.

\subsection*{Classical Bogoliubov transformations}

Let~$\mathcal{S}$ be the vector space of classical solutions to the KG equation~\eqref{eq:KG}, endowed with two different complex structures, $J$ and~$\widetilde{J}$. These structures define the corresponding Hilbert spaces~$\mathcal{H}$ and~$\widetilde{\mathcal{H}}$, each equipped with a positive-definite inner product, denoted here as~$(\cdot,\cdot)$ and~$\widetilde{(\cdot,\cdot)}$, respectively. These spaces have corresponding orthonormal bases \mbox{$\{\Phi^+_n,(\Phi^-_n)^*\}$} and \mbox{$\{\widetilde{\Phi}^+_n,(\widetilde{\Phi}^-_n)^*\}$}. The solutions~$\widetilde{\Phi}_n^\pm$ can be expressed as linear combinations of the basis solutions~$\Phi_n^\pm$:
\begin{equation}
    \begin{pmatrix} \widetilde{\Phi}^+_n \\ \widetilde{\Phi}^-_n \end{pmatrix}
    =
    \sum_m 
    \begin{array}{c}
        \begin{pmatrix} \alpha_{nm}^+ & \beta_{nm}^+ \\ \beta_{nm}^- & \alpha_{nm}^- \end{pmatrix}
        \begin{pmatrix} \Phi^+_m \\ \Phi^-_m \end{pmatrix}
    \end{array}.
    \label{eq:BogoliubovPhis}
\end{equation}
The coefficients~$\alpha^\pm_{nm}$ and~$\beta^\pm_{nm}$ in this expansion are known as the \gls{BogoliubovCoefficients}. They can be computed directly from the orthonormality conditions of the bases:
\begin{equation}
    \alpha_{nm}^\pm = \left(\Phi_m^\pm, \widetilde{\Phi}_n^\pm\right) = \pm\left(\Phi_m^\pm, \widetilde{\Phi}_n^\pm\right)_{\text{KG}}, \qquad \beta_{nm}^\pm = \left(\Phi_m^\mp, \widetilde{\Phi}_n^\pm\right) = \mp \left(\Phi_m^\mp, \widetilde{\Phi}_n^\pm\right)_{\text{KG}}.
    \label{eq:BogoliubovCoeffs}
\end{equation}

Since the classical Klein-Gordon field~$\Phi$ can be expanded in terms of both sets~${\Phi_n^\pm}$ and~${\widetilde{\Phi}_n^\pm}$:
\begin{equation}
    \Phi=\sum_n \left( a_n\Phi_n^+ +b_n^*\Phi_n^- \right)
    =\sum_n \left( \widetilde{a}_n\widetilde{\Phi}_n^+ +\widetilde{b}_n^*\widetilde{\Phi}_n^-\right),
\end{equation} 
it follows that the annihilation and creation variables transform via a Bogoliubov transformation:
\begin{equation}
    \begin{pmatrix} a_m \\ b^*_m \end{pmatrix}
    =
    \sum_n 
    \begin{array}{c}
        \begin{pmatrix} \alpha_{nm}^+ & \beta_{nm}^- \\ \beta_{nm}^+ & \alpha_{nm}^- \end{pmatrix}
        \begin{pmatrix} \widetilde{a}_n \\ \widetilde{b}^*_n \end{pmatrix}
    \end{array}.
    \label{eq:BogoliubovTransCoeffs}
\end{equation}

The inverse Bogoliubov transformation of~\eqref{eq:BogoliubovPhis} follows from the orthonormality of the bases, the hermitian properties of the KG inner product and the relations~\eqref{eq:BogoliubovCoeffs}:
\begin{equation}
    \begin{pmatrix} \Phi^+_m \\ \Phi^-_m \end{pmatrix}
    =
    \sum_n 
    \begin{array}{c}
        \begin{pmatrix} (\alpha_{nm}^+)^* & -(\beta_{nm}^-)^* \\ -(\beta_{nm}^+)^* & (\alpha_{nm}^-)^* \end{pmatrix}
        \begin{pmatrix} \widetilde{\Phi}^+_n \\ \widetilde{\Phi}^-_n \end{pmatrix}
    \end{array}.
    \label{eq:BogoliubovPhisInverse}
\end{equation}
Then, the inverse Bogoliubov transformation of~\eqref{eq:BogoliubovTransCoeffs}, relating the creation and annihilation coefficients, is
\begin{equation}
    \begin{pmatrix} \widetilde{a}_n \\ \widetilde{b}^*_n \end{pmatrix}
    =
    \sum_m 
    \begin{array}{c}
        \begin{pmatrix} (\alpha_{nm}^+)^* & -(\beta_{nm}^+)^* \\ -(\beta_{nm}^-)^* & (\alpha_{nm}^-)^* \end{pmatrix}
        \begin{pmatrix} a_m \\ b^*_m \end{pmatrix}
    \end{array}.
    \label{eq:BogoliubovTransCoeffsInverse}
\end{equation}

Finally, the Bogoliubov coefficients satisfy non-trivial constraints among themselves. By substituting~\eqref{eq:BogoliubovPhis} into its inverse transformation~\eqref{eq:BogoliubovPhisInverse}, and vice versa, we obtain the following relations:
\begin{align}
    \sum_i \left[ \alpha^\pm_{ni} (\alpha^\pm_{mi})^* - \beta^\pm_{ni} (\beta^\pm_{mi})^* \right] = \sum_i \left[ \alpha^\pm_{in} (\alpha^\pm_{im})^* - \beta^\mp_{in} (\beta^\mp_{im})^* \right] &= \delta_{nm}, \nonumber \\
    \sum_i \left[ \alpha^+_{ni}(\beta_{mi}^-)^* - \beta_{ni}^+(\alpha_{mi}^-)^* \right] = \sum_i \left[ \alpha^+_{in}(\beta_{im}^+)^* - \beta_{in}^-(\alpha_{im}^-)^* \right] &= 0.
    \label{eq:BogoliubovConstraints}
\end{align}
These constraints ensure the preservation of the Poisson algebra of the annihilation and creation variables~\eqref{eq:PoissonVariables}, thereby guaranteeing that the Bogoliubov transformation remains canonical.

\begin{comment}
\begin{align}
    \sum_i \left[ \alpha^\pm_{ni} (\alpha^\pm_{mi})^* - \beta^\mp_{ni} (\beta^\mp_{mi})^* \right] = \sum_i \left[ \alpha^\pm_{in} (\alpha^\pm_{im})^* - \beta^\mp_{in} (\beta^\mp_{im})^* \right] &= \delta_{nm}, \nonumber \\
    \sum_i \left[ \alpha^+_{ni}(\beta_{mi}^+)^* - \beta_{ni}^-(\alpha_{mi}^-)^* \right] = \sum_i \left[ \alpha^+_{in}(\beta_{im}^+)^* - \beta_{in}^-(\alpha_{im}^-)^* \right] &= 0.
    \label{eq:BogoliubovConstraints}
\end{align}
\end{comment}

\subsection*{Quantum Bogoliubov transformations}

Until now, our discussion regarding Bogoliubov transformations has been entirely classical. We now examine how Bogoliubov transformations manifest in the quantum theory.

Let~$\hat{\Phi}$ and~$\hat{\widetilde{\Phi}}$ denote the field operators associated with the complex structures~$J$ and~$\widetilde{J}$, respectively. These operators are obtained by promoting their corresponding annihilation and creation variables to operators, as defined in~\eqref{eq:FieldOperator}. Notably, their respective symmetric Fock spaces, $\mathcal{F}_S$ and~$\widetilde{\mathcal{F}}_S$, are not necessarily the same. The relation between these two quantum field operators is mediated by an operator~$\hat{B}: \mathcal{F}_S \rightarrow \widetilde{\mathcal{F}}_S$, satisfying
\begin{equation}
    \hat{\widetilde{\Phi}} = \hat{B} \hat{\Phi} \hat{B}^{-1}.
    \label{eq:BPhiB-1}
\end{equation}

This relation extends naturally to the annihilation and creation operators associated with both complex structures. From~\eqref{eq:BPhiB-1} and the definition of the quantum field operator~\eqref{eq:FieldOperator}, we obtain
\begin{equation}
    \hat{\widetilde{\Phi}} = \sum_n \left( \hat{\widetilde{a}}_n\widetilde{\Phi}_n^+ + \hat{\widetilde{b}}_n^\dagger \widetilde{\Phi}_n^- \right) = \sum_n \left( \hat{B}\hat{a}_n\hat{B}^{-1} \Phi^+_n + \hat{B}\hat{b}_n^\dagger \hat{B}^{-1} \Phi_n^- \right).
\end{equation}
Using the expressions for the Bogoliubov coefficients in terms of the basis elements~\eqref{eq:BogoliubovCoeffs}, we deduce the transformation:
\begin{equation}
    \begin{pmatrix} \hat{B} \hat{a}_m \hat{B}^{-1} \\ \hat{B} \hat{b}^\dagger_m \hat{B}^{-1} \end{pmatrix}
    =
    \sum_n 
    \begin{array}{c}
        \begin{pmatrix} \alpha_{nm}^+ & \beta_{nm}^- \\ \beta_{nm}^+ & \alpha_{nm}^- \end{pmatrix}
        \begin{pmatrix} \hat{\widetilde{a}}_n \\ \hat{\widetilde{b}}^\dagger_n \end{pmatrix}
    \end{array}.
    \label{eq:BogoliubovTransOperators}
\end{equation}

From these relations, it is evident that the notions of particle and antiparticle are, in general, different for the two complex structures. Specifically:
\begin{itemize}
    \item The coefficients~$\beta_{nm}^+$ mix antiparticle states of~$J$ with particle states of~$\widetilde{J}$.
    \item The coefficients~$\beta_{nm}^-$ mix particle states of~$J$ with antiparticle states of~$\widetilde{J}$.
\end{itemize}
Only if these~$\beta$-coefficients vanish do the definitions of particle and antiparticle remain the same for both~$J$ and~$\widetilde{J}$. In such a case, the transformation reduces to an independent change of basis within the one-particle and one-antiparticle Hilbert spaces, leaving the quantization of the classical theory unaffected.

\subsection*{Unitary equivalence}

In the special case where the quantum field operators~$\hat{\Phi}$ and~$\hat{\widetilde{\Phi}}$ are related by a unitary operator~$\hat{B}$, according to the relation~\eqref{eq:BPhiB-1}, the corresponding quantum theories are said to be \gls{UnitarilyEquivalent}. In this case, the associated Bogoliubov transformation is said to be \gls{UnitarilyImplementable} in the quantum theory.

When unitary equivalence holds, the states $|\phi_i\rangle\in \mathcal{F}_S$ and their transformed counterparts $|\widetilde{\phi}_i\rangle=\hat{B}|\phi_i\rangle\in \widetilde{\mathcal{F}}_S$ yield identical transition amplitudes. This follows from the unitarity of~$\hat{B}$, which ensures that~$\hat{B}^{-1} = \hat{B}^\dagger$, leading to
\begin{equation}
    \langle\widetilde{\phi}_1 | \hat{\widetilde{\Phi}}|\widetilde{\phi}_2\rangle=\langle\phi_1|\hat{\Phi}|\phi_2\rangle.
\end{equation}

However, even when two quantizations are unitarily equivalent, the vacuum state~$| \widetilde{0} \rangle$ may contain excitations relative to the vacuum state~$|0\rangle$. The total number of these excitations is given by
\begin{equation}
    \mathcal{N} = \sum_n \langle 0 | \hat{B} ( \hat{\widetilde{a}}_n^\dagger \hat{\widetilde{a}}_n + \hat{\widetilde{b}}_n^\dagger \hat{\widetilde{b}}_n ) \hat{B}^{-1} | 0 \rangle = \sum_{n,m} \left( |\beta_{nm}^+|^2 + |\beta_{nm}^-|^2 \right),
    \label{eq:Num}
\end{equation}
where the last equality follows from the Bogoliubov transformation~\eqref{eq:BogoliubovTransOperators} and the commutation relations~\eqref{eq:Commutatorsab}. 
The Bogoliubov coefficients~$\beta_{nm}^+$ and~$\beta_{nm}^-$ encode the contribution from each mode to the total number of particles and antiparticles, respectively.

A necessary and sufficient condition for two quantizations to be unitarily equivalent is that the sum in~\eqref{eq:Num} remains finite, provided the norms associated with their respective inner products are equivalent~\cite{Wald1994}.\footnote{See also~\cite{Wald1994} for a detailed discussion of norm equivalence in this context.} For systems with a finite number of degrees of freedom, the Stone-von Neumann theorem~\cite{Stone1932,vonNeumann1932} guarantees the uniqueness of the quantum representation. In particular, the sum in~\eqref{eq:Num} is trivially finite. However, in the infinite-dimensional case, as we will explore in~\autoref{chap:QuantumUnitary}, there exist Bogoliubov transformations that cannot be implemented as unitary operators. As a result, unitarily nonequivalent quantizations emerge. This highlights the crucial role of the choice of complex structure in the quantization process. Careful consideration must therefore be given to selecting the most appropriate complex structure for each specific physical scenario.

\section{Fermions}
\label{sec:Fermions}

Having thoroughly studied the canonical quantization of a charged scalar field, we now extend the formalism to the case of fermions. We examine a Dirac fermionic field~$\Psi$ with mass~$m$ and charge~$q$, propagating in a globally hyperbolic spacetime and interacting with an electromagnetic background represented by the four-vector potential~$A_{\mu}$. The dynamics of this field, minimally coupled to the gravitational background, is governed by the action:
\begin{equation}
    S = i\int \text{d}^4 x \ \sqrt{-\mathfrak{g}} \left[ \frac{1}{2} \overline{\Psi} \gamma^\mu D_\mu\Psi - \frac{1}{2} \left( D_\mu^* \overline{\Psi} \right) \gamma^\mu \Psi - m\overline{\Psi}\Psi \right].
    \label{eq:ActionFermion}
\end{equation}
To fully understand this equation, let us clarify the key components involved:
\begin{itemize}
    \item The Dirac matrices~$\gamma ^{\mu }$ satisfy the anticommutation relations~$\{\gamma^\mu,\gamma^\nu\} = 2g^{\mu \nu }$. The chosen representation for the flat-space gamma matrices~$\widetilde{\gamma}^\mu$ in this thesis is:
\begin{equation}
{ \widetilde {\gamma }}^{0} =
\left(
\begin{array}{cc}
iI_{2} & 0  \\
0 & -iI_{2}
\end{array}
\right) ,
\qquad
{ \widetilde {\gamma }}^{j} =
\left(
\begin{array}{cc}
0 & i\sigma _{j} \\
-i\sigma _{j} & 0
\end{array}
\right) ,
\label{eq:flatspacegamma}
\end{equation}
with $I_{2}$ the $2\times 2$ identity matrix and $\sigma _{i}$  the usual Pauli matrices
\begin{equation}
\sigma _{1} =
\left(
\begin{array}{cc}
0 & 1 \\
1 & 0
\end{array}
\right) ,
\quad
\sigma _{2} =
\left(
\begin{array}{cc}
0 & -i \\
i & 0
\end{array}
\right) ,
\quad
\sigma _{3}=
\left(
\begin{array}{cc}
1 & 0 \\
0 & -1
\end{array}
\right) .
\label{eq:FlatGammaMatrices}
\end{equation}
In this representation, the flat-space gamma matrices verify the relations $(\widetilde{\gamma}^0)^2 = -I$ and~$(\widetilde{\gamma}^j)^2 = I$.
    \item The spinor connection matrices~$\Gamma_{\mu }$ are defined in terms of covariant derivatives of the Dirac matrices $\gamma ^{\mu }$:
\begin{equation}
\nabla_\nu \gamma^\mu = \partial _{\nu }\gamma ^{\mu } + \Gamma _{\nu \kappa }^{\mu } \gamma ^{\kappa }
- \Gamma _{\nu }\gamma ^{\mu } +
\gamma ^{\mu } \Gamma _{\nu }=0,
\end{equation}
where $\Gamma _{\nu \kappa }^{\mu }$ are the standard Christoffel symbols. They allow to define the spinor covariant derivatives~$\nabla _{\mu }$ according to~\cite{Unruh1974}
\begin{equation}
\nabla _{\mu } \Psi = \partial_\mu \Psi - \Gamma _{\mu }\Psi .
\label{eq:SpinorCovariantDerivative}
\end{equation}
In terms of the vierbein components~$e^{\mu }_{a}$, satisfying \mbox{$\gamma ^{\mu } = e^{\mu }_{a} {\tilde {\gamma }}^{a}$}, the spinor connection matrices can be easily computed as follows \cite{Brill1957,Iyer1982}:
\begin{equation}
\Gamma _{\nu } = - \frac {1}{4} g_{\sigma \rho }
e^{\sigma }_{a} e^{\rho }_{b ;\nu } {\tilde {\gamma }}^{a}
{\tilde {\gamma }}^{b},
\label{eq:SpinorConnectionMatrices}
\end{equation}
where $e^{\rho }_{b ;\nu } = \partial_{\nu} e^{\rho }_b + \Gamma^\rho_{\nu \kappa} e^\kappa_b$.
\item The conjugate spinor ${\overline {\Psi }}$ is given by
${\overline {\Psi }} = \Psi ^{\dagger }\widetilde{\gamma}^0 $, with~$\Psi ^{\dagger }$ the usual
hermitian conjugate of~$\Psi $ considered as a matrix. The spinor covariant derivative of ${\overline {\Psi }}$ is
\begin{equation}
    \nabla_{\mu}\overline{\Psi} = \partial_\mu \overline{\Psi}+\overline{\Psi}\Gamma_\mu .
\end{equation}
\end{itemize}

From this, we can derive the equations of motion for the fermionic field, leading to the \gls{DiracEquation}:
\begin{equation} 
    \left( \gamma^\mu D_\mu - m \right)\Psi = 0.
    \label{eq:Dirac}
\end{equation}

\begin{block}[note]
We might be more familiar with the standard form of the Dirac equation:
\begin{equation}
    \left( i\gamma^\mu D_\mu - m \right) \Psi = 0,
\end{equation}
where the gamma matrices satisfy the anticommutation relations~\mbox{$\{\gamma^\mu, \gamma^\nu\} = 2g^{\mu\nu}$}. With the signature convention~$(+,-,-,-)$, this implies that the flat-space gamma matrices satisfy~\mbox{$(\widetilde{\gamma}^0)^2=I$} and~\mbox{$(\widetilde{\gamma}^j)^2=-I$}. However, with our chosen signature~$(-,+,+,+)$, maintaining the same anticommutation relations~$\{\gamma^\mu, \gamma^\nu\} = 2g^{\mu\nu}$ requires the flat-space gamma matrices to satisfy~ $(\widetilde{\gamma}^0)^2 = -I$ and~$(\widetilde{\gamma}^j)^2 = I$. To achieve this, the flat-space gamma matrices must absorb a factor of~$i$ in their definition. In particular, our choice for the flat-space gamma matrices in this thesis~\eqref{eq:flatspacegamma} is based on the Dirac representation, but multiplied by~$i$.

Alternatively, as used for instance in some of my works~\cite{AlvarezUnitary,AlvarezKugelblitz}, one can choose a representation of the Dirac matrices satisfying~ $(\widetilde{\gamma}^0)^2 = -I$ and~$(\widetilde{\gamma}^j)^2 = I$. However, for the signature~$(-,+,+,+)$, this choice modifies the anticommutation relations to~$\{\gamma^\mu, \gamma^\nu\} = -2g^{\mu\nu}$.
\end{block}

By foliating the globally hyperbolic spacetime into Cauchy hypersurfaces~$\Sigma_t$ of constant~$t$, we can define the conjugate momentum field corresponding to~$\Psi$ as
\begin{equation}
    \Pi = \frac{\delta S}{\delta (\partial_t \Psi)} = \frac{i}{2} \sqrt{-\mathfrak{g}} \overline{\Psi}\gamma^t.
\end{equation}

Analogous to the scalar case, the Dirac equation~\eqref{eq:Dirac} admits a well-posed initial value formulation, allowing us to identify the canonical phase space with the covariant phase space associated with the equation. The \gls{DiracProduct} is an inner product defined on this covariant phase space, given by
\begin{equation}
    \left( \Psi_1, \Psi_2 \right) =  - \int_{\Sigma_t} \text{d}^3\bfx \  \sqrt{\mathfrak{h}} \overline{\Psi}_1 \gamma^{\mu} n_\mu \Psi_2,
    \label{eq:DiracProduct}
\end{equation}
where $\Psi_1$ and $\Psi_2$ are two Dirac solutions.

Similar to the KG product, the Dirac product is independent of the choice of hypersurface~$\Sigma_t$. However, a key distinction exists between the scalar and fermionic cases: the Dirac product is positive-definite, naturally endowing the space of Dirac solutions with a Hilbert space structure. Indeed, 
\begin{equation}
    (\Psi,\Psi) = \frac{1}{2}[(\Psi,\Psi) + (\Psi,\Psi)^*] = \int_{\Sigma_t} \text{d}^3\bfx \ \sqrt{\mathfrak{h}} \Psi^\dagger \Psi e^t_0 \geq 0,
\end{equation}
where we used that in adapted coordinates $n_\mu = \delta_\mu^t$, along with the antihermitian or hermitian properties of the flat-space gamma matrices: $(\widetilde{\gamma}^0)^{\dagger} = -\widetilde{\gamma}^0$ and $(\widetilde{\gamma}^j)^{\dagger} = \widetilde{\gamma}^j$. In contrast, as we saw above, the KG product is not positive-definite, necessitating the introduction of a complex structure to define a proper inner product. This means that while charged scalars can have positive or negative KG norm, charged fermions always have a positive Dirac norm.

Although the covariant phase space in the fermionic case already forms a Hilbert space of solutions, we will still introduce a complex structure~$J$ to construct one-particle and one-antiparticle Hilbert spaces, $\mathcal{H}^+$ and $\mathcal{H}^-$. Consequently, the full Hilbert space of the quantum theory is not merely the covariant phase space but rather the direct sum of these two subspaces: $\mathcal{H} = \mathcal{H}^+ \oplus \mathcal{H}^-$. This approach ensures a clear distinction between particles and antiparticles in the quantum theory.

The canonical quantization procedure follows a similar framework to the scalar case but differs in key aspects due to the symmetric nature of the algebra satisfied by the field~$\Psi$ and its conjugate momentum~$\Pi$---unlike the antisymmetric Poisson algebra in the scalar case. The main differences are:
\begin{itemize}
    \item The full Hilbert space is now constructed as an antisymmetric Fock space~$\mathcal{F}_\text{A}$. Instead of using the symmetric tensor product~$\otimes_\text{S}$ of the scalar case, we now employ an antisymmetric tensor product~$\otimes_\text{A}$.
    \item We choose an orthonormal basis~$\{\Psi_n^+\}$ for~$\mathcal{H}^+$ and another basis~$\{(\Psi_n^-)^*\}$ for~$\mathcal{H}^-$, allowing us to expand the original fermionic field~$\Psi$ as
\begin{equation}
    \Psi = \sum_n \left( c_n \Psi_n^+ + d_n^* \Psi_n^- \right).
\end{equation}
    Here, the annihilation and creation variables~$c_n$ and~$d_n^*$ satisfy an algebra similar to~\eqref{eq:PoissonVariables}, but now with respect to a symmetric bracket structure. 
    \item These variables are promoted to annihilation and creation operators~$\hat{c}_n$ and~$\hat{d}_n^\dagger$ satisfying the anticommutator relations:
    \begin{equation}
        \{\hat{c}_n,\hat{c}_m^{\dagger}\} = \{\hat{d}_n,\hat{d}_m^{\dagger}\} = \delta_{n,m},
    \end{equation}
    where all the other anticommutators between these operators vanish.
    \item Finally, the quantum field operator is defined as
    \begin{equation}
        \hat{\Psi} = \sum_n \left( \hat{c}_n \Psi_n^+ + \hat{d}_n^\dagger \Psi_n^- \right).
        \label{eq:DiracOperator}
    \end{equation}
\end{itemize}

Regarding Bogoliubov transformations, the fundamental differences between scalars and fermions stem from two main factors: the KG product is not positive-definite, whereas the Dirac product is, and the Poisson algebra for creation and annihilation variables is symmetric for fermions rather than antisymmetric as in the bosonic case. As a result, the minus signs that appear in the inverse relations~\eqref{eq:BogoliubovPhisInverse} for bosonic fields are replaced by plus signs in the fermionic case:
\begin{equation}
    \begin{pmatrix} \widetilde{\Psi}^+_n \\ \widetilde{\Psi}^-_n \end{pmatrix}
    =
    \sum_m 
    \begin{array}{c}
        \begin{pmatrix} \alpha_{nm}^+ & \beta_{nm}^+ \\ \beta_{nm}^- & \alpha_{nm}^- \end{pmatrix}
        \begin{pmatrix} \Psi^+_m \\ \Psi^-_m \end{pmatrix}
    \end{array},
    \qquad
    \begin{pmatrix} \Psi^+_m \\ \Psi^-_m \end{pmatrix}
    =
    \sum_n 
    \begin{array}{c}
        \begin{pmatrix} (\alpha_{nm}^+)^* &(\beta_{nm}^-)^* \\ (\beta_{nm}^+)^* & (\alpha_{nm}^-)^* \end{pmatrix}
        \begin{pmatrix} \widetilde{\Psi}^+_n \\ \widetilde{\Psi}^-_n \end{pmatrix}
    \end{array}.
    \label{eq:BogoliubovPsis}
\end{equation}
Similarly, for the annihilation and creation variables:
\begin{equation}
    \begin{pmatrix} c_m \\ d^*_m \end{pmatrix}
    =
    \sum_n 
    \begin{array}{c}
        \begin{pmatrix} \alpha_{nm}^+ & \beta_{nm}^- \\ \beta_{nm}^+ & \alpha_{nm}^- \end{pmatrix}
        \begin{pmatrix} \widetilde{c}_n \\ \widetilde{d}^*_n \end{pmatrix}
    \end{array}, \qquad
        \begin{pmatrix} \widetilde{c}_n \\ \widetilde{d}^*_n \end{pmatrix}
    =
    \sum_m 
    \begin{array}{c}
        \begin{pmatrix} (\alpha_{nm}^+)^* & (\beta_{nm}^+)^* \\ (\beta_{nm}^-)^* & (\alpha_{nm}^-)^* \end{pmatrix}
        \begin{pmatrix} c_m \\ d^*_m \end{pmatrix}
    \end{array}.
    \label{eq:BogoliubovTransCoeffsFermions}
\end{equation}
This modifications lead to Bogoliubov coefficient constraints that differ in sign from those for scalar fields~\eqref{eq:BogoliubovConstraints}
\begin{align}
    \sum_i \left[ \alpha^\pm_{ni} (\alpha^\pm_{mi})^* + \beta^\pm_{ni} (\beta^\pm_{mi})^* \right] = \sum_i \left[ \alpha^\pm_{in} (\alpha^\pm_{im})^* + \beta^\mp_{in} (\beta^\mp_{im})^* \right] &= \delta_{nm}, \nonumber \\
    \sum_i \left[ \alpha^+_{ni}(\beta_{mi}^-)^* + \beta_{ni}^+(\alpha_{mi}^-)^* \right] = \sum_i \left[ \alpha^+_{in}(\beta_{im}^+)^* + \beta_{in}^-(\alpha_{im}^-)^* \right] &= 0.
    \label{eq:BogoliubovConstraintsFermions}
\end{align}

\section{Conclusions}
\label{sec:ConclusionsQuantization}

In this chapter, we have established the fundamental framework that will be used throughout the rest of the thesis. We have carefully examined the difficulties in quantizing a charged scalar field in a general curved background with an external electromagnetic field. This general formalism will allow us to address various settings explored in subsequent chapters, including flat spacetime with a homogeneous electric field (the main focus of this thesis), cosmological expansion in~\autoref{chap:InOut}, and charged black holes in~\autoref{chap:Superradiance}.

A crucial aspect of this framework involves constructing a proper inner product from the conventional Klein-Gordon product, which is not positive-definite. The construction of a quantum theory fundamentally depends on the choice of a complex structure, which in turn relies on two key elements: 1) the basis of solutions used to expand the test field, and 2) the way this basis is split to define particles and antiparticles.

Different choices in the quantization procedure can result in different quantum theories, each defining its own quantum vacuum. To compare these quantizations, we introduced the concept of Bogoliubov transformations. These transformations provide a powerful tool for assessing whether two quantizations are unitarily equivalent. Moreover, the squared modulus of the $\beta$-Bogoliubov coefficients offers a direct observable indicating the number of excitations of one quantum vacuum relative to another.

Finally, we extended the canonical quantization procedure to Dirac fields in a classical curved and electromagnetic background, establishing a foundational framework that will be crucial in later chapters. Although most of this thesis focuses on scalar fields, the treatment of Dirac fields will be essential for studying significant physical implications of pair creation in nature in~\autoref{chap:Kugelblitz} and~\autoref{chap:Superradiance}.
\chapter[Choice of quantum vacuum]{Choice of quantum vacuum}
\label{chap:ChoiceVacuum}

\chaptermark{Choice of quantum vacuum}

In the previous chapter, we established that the definitions of particles and antiparticles are inherently tied to the quantization procedure, which is in turn, dictated by the choice of a complex structure. As a consequence, different quantum theories can be constructed from the same classical system, each leading to distinct observable predictions.

In~\autoref{sec:HomogeneousElectricFlat}, we apply the framework developed in~\autoref{chap:QuantizationScalars} to study the Schwinger effect. Specifically, we analyse the canonical quantization of a charged scalar field in flat spacetime under the influence of a strong homogeneous electric field. This will serve as our primary reference setting throughout this thesis. In~\autoref{sec:ParametrizationVacua}, we systematically parametrize the different choices of quantum vacua in the Schwinger effect and examine the most well-known quantum vacua discussed in the literature. Finally, \autoref{sec:ConclusionsChoiceQuantumVacuum} provides a brief summary of the results discussed in the chapter.

\section{Charged scalars in flat spacetime with a homogeneous electric background} 
\label{sec:HomogeneousElectricFlat}

As an illustrative example, we focus on the case of a charged scalar field in flat spacetime coupled to a homogeneous, time-dependent electric field. While this thesis also explores more complex scenarios, including inhomogeneous electric field configurations in black hole backgrounds in~\autoref{chap:Superradiance}, this particular case serves as the central topic of interest for the majority of the analysis.

Let us consider a homogeneous, time-dependent electric field~$\textbf{E}(t)$ in flat spacetime. In this scenario, a natural choice of gauge is the temporal gauge, 
\begin{equation}
    A_{\mu}(t,\textbf{x}) = (0, \textbf{A}(t)),
\end{equation}
which ensures that the equation of motion becomes explicitly spatially homogeneous. In this gauge, the electric field in terms of the potential is given by
\begin{equation}
    \textbf{E}(t) = -\dot{\textbf{A}}(t).
\end{equation}
In addition, we consider that the direction of the electric field remains fixed over time. Without loss of generality, we align it along the $z$-axis.

As we described in~\autoref{chap:QuantizationScalars}, the first step to quantize our theory is to look for a separable basis of solutions $\{\Phi_\bfk\}$ of the KG equation~\eqref{eq:KG}, where~$\bfk$ denotes the continuous index identifying each solution. Due to the invariance of the equations of motion under spatial translations, we propose solutions of the form:
\begin{equation}
    \Phi_\bfk(t,\bfx) = (2\pi)^{-\frac{3}{2}} \phi_\bfk(t) e^{i\bfk\cdot \bfx},
    \label{eq:AnsatzPhi}
\end{equation}
where~$\phi_\bfk(t)$ is a time-dependent function. Introducing this ansatz into the KG equation~\eqref{eq:KG}, we deduce that the modes~$\phi_\bfk(t)$ satisfy harmonic oscillator equations of the form
\begin{equation} 
    \ddot{\phi}_{\bfk}(t)+\omega_{\bfk}(t)^2\phi_{\bfk}(t)=0,
    \label{eq:Harmonic} 
\end{equation}
where the time-dependent frequency $\omega_{\bfk}(t)$ is given by
\begin{equation} 
    \omega_{\bfk}(t)=\sqrt{[\bfk+q\textbf{A}(t)]^2+m^2}=k^2 + 2qA(t)k\cos{\theta} + q^2A(t)^2 + m^2.
    \label{eq:Frequency}
\end{equation}
Here, we have introduced the magnitudes~$k=|\bfk|$ and~$A(t)=|\textbf{A}(t)|$. The anisotropic nature of the system becomes evident in the form of the frequency~$\omega_{\bfk}(t)$, as it depends on the angle~$\theta$ between the wavevector~$\bfk$ and the direction of the vector potential~$\textbf{A}(t)$ through a linear term in~$k$. 

It is important to note that the influence of the external electric field on the dynamics of the scalar field is entirely encoded in the time-dependent frequency~$\omega_{\bfk}(t)$. Since the electromagnetic background is treated as an external, fixed agent, the frequencies $\omega_{\bfk}(t)$ are determined solely by the external field configuration and remain unaffected by the dynamics of the modes $\phi_{\bfk}(t)$. In other words, we consider a regime where backreaction effects can be neglected and focus exclusively on solving the harmonic oscillator equations~\eqref{eq:Harmonic}, disregarding the equations of motion for the electric field.

To quantize the classical theory, the following step is to choose the complex structure. As we saw in the previous section, this is equivalent to choosing orthonormal bases for the one-particle and one-antiparticle Hilbert spaces. To better understand and motivate the following, let us consider first a simpler case: when there is no electric field.

\begin{block}[example]
In the special case where there is no electric field, we can set the vector potential~$\textbf{A}(t)$ to zero. In this scenario, the frequency~$\omega_\bfk = \sqrt{k^2 + m^2}$ becomes constant. A basis of solutions to the harmonic oscillator equation~\eqref{eq:Harmonic} is then given by the positive and negative frequency modes
\begin{equation}
    \phi_\bfk^\pm(t) = \frac{1}{\sqrt{2\omega_\bfk}} e^{\mp i\omega_\bfk t}.
    \label{eq:phiE=0}
\end{equation}
According to~\eqref{eq:AnsatzPhi}, these modes lead to positive and negative plane wave solutions to the whole KG equation:
\begin{equation}
    \Phi_\bfk^\pm(t,\bfx) = (2\pi)^{-\frac{3}{2}} \phi_\bfk^\pm(t) e^{i\bfk\cdot \bfx}.
\end{equation}

The terminology of positive and negative frequency modes originates from the historical interpretation of fields as wavefunctions of relativistic particles. Specifically, since
\begin{equation}
    i\partial_t \Phi_\bfk^\pm(t,\bfx) = \pm \omega_\bfk \Phi_\bfk^\pm(t,\bfx),
    \label{eq:EnergyPlaneWaves}
\end{equation}
the mode~$\Phi_\bfk^+$ was traditionally associated with a particle of positive energy~$\omega_\bfk$, while~$\Phi_\bfk^-$ was associated with a hole of negative energy~$-\omega_\bfk$. 
\end{block}

The selection of plane wave solutions as the standard basis for constructing the quantum theory of a charged scalar field in the absence of an electric field is not merely a historical convention but is motivated by fundamental physical principles. Specifically, this choice of complex structure preserves all the classical symmetries of the system---the full Poincar\'{e} group---ensuring that these symmetries remain intact in the quantum theory. As a result, this defines a preferred notion of quantum vacuum, known as the Minkowski quantum vacuum, which we will analyse in detail later.

However, it is crucial to recognize that QFT is fundamentally a theory of fields rather than particles. In generic curved spacetimes, a clear particle interpretation may not even exist, since the classical system may lack the necessary symmetries to single out a preferred vacuum state. In the case of the Schwinger effect, the presence of an electric field explicitly breaks part of the Poincar\'{e} symmetry of flat spacetime. In particular, it breaks time translation invariance, as reflected in the time dependence of the frequency~$\omega_\bfk(t)$. Note that even if the electric field is constant, $\omega_\bfk(t)$ remains time-dependent. This is because the frequency~\eqref{eq:Frequency} depends on the vector potential~$\textbf{A}(t)$ rather than on the electric field~$\textbf{E}(t)$. While we can still impose the condition that our choice of complex structure should preserve the remaining classical symmetries into the quantum theory, this condition alone is insufficient to uniquely determine a preferred complex structure. Consequently, a residual ambiguity remains in the choice of quantization.

In the following, we only consider complex structures that preserve the symmetries of the equations of motion when an electric field is present in the background. To split the basis of solutions into~$\Phi_\bfk^+$ and~$\Phi_\bfk^-$, we compute the KG product between the solutions:
\begin{equation}
    (\Phi_\bfk,\Phi_{\bfk^\prime})_{\text{KG}} = -i \delta(\bfk-\bfk^\prime) \left[ \dot{\phi}_\bfk^*(t)\phi_\bfk(t) - \phi_\bfk^*(t)\dot{\phi}_\bfk(t) \right],
\end{equation}
where we used that the Dirac delta distribution satisfies
\begin{equation}
    \delta(\bfk-\bfk^\prime) = \int \frac{\text{d}^3\bfx}{(2\pi)^3} \ e^{i(\bfk-\bfk^\prime)\cdot \bfx}.
\end{equation}
Then, the solutions~$\Phi_\bfk$ form an orthonormal basis with respect to the KG product and have positive KG norm if and only if, for all time~$t$, the modes~$\phi_\bfk$ satisfy the normalization condition:
\begin{equation}
    \dot{\phi}_\bfk^*(t) \phi_\bfk(t) - \phi_\bfk^*(t) \dot{\phi}_\bfk(t) = i.
    \label{eq:Normalization}
\end{equation}
If this condition holds, the solutions $(2\pi)^{-\frac{3}{2}} \phi^*_\bfk(t) e^{i\bfk\cdot \bfx}$ are also normalized but carry negative KG norm. Accordingly, we propose a splitting of the solutions corresponding to:
\begin{equation}
    \Phi_\bfk^+(t,\bfx) = (2\pi)^{-\frac{3}{2}} \phi_\bfk(t) e^{i\bfk\cdot \bfx}, \qquad \Phi_\bfk^-(t,\bfx) = (2\pi)^{-\frac{3}{2}} \phi_\bfk^*(t) e^{i\bfk\cdot \bfx},
    \label{eq:Phi+-Schwinger}
\end{equation}
where modes~$\phi_\bfk$ satisfy the harmonic oscillator equations with frequencies that depend on time~\eqref{eq:Harmonic} and are normalized according to~\eqref{eq:Normalization}. To fully specify a particular quantization, we must still determine the explicit form of the modes~$\phi_\bfk$. This choice is motivated by two key considerations:
\begin{itemize}
    \item The factor~$e^{i\bfk\cdot\bfx}$ exploits the homogeneity of the background, ensuring that modes with different wavenumbers~$\bfk$ remain dynamically decoupled. This guarantees that the system of harmonic oscillator equations does not couple different wavevectors~$\bfk$.
    \item If~$\phi_\bfk$ is a solution to the harmonic oscillator equation, its complex conjugate~$\phi_\bfk^*$ is also a solution. By structuring the complex structure so that~$\Phi^+_\bfk$ is proportional to~$\phi_\bfk$ and~$\Phi^-_\bfk$ is proportional to~$\phi_\bfk^*$, we ensure that the quantum theory reflects this symmetry. As a consequence, particles and antiparticles are always created in pairs, maintaining equal numbers of both.
\end{itemize}

Despite these physically motivated criteria, which aim to preserve classical symmetries in the quantum theory, they do not uniquely determine a single preferred complex structure. Different choices of modes~$\phi_\bfk$ lead to different quantizations, some of which may even be unitarily nonequivalent. In the following section, we will explore specific examples of different quantizations. In~\autoref{chap:QuantumUnitary} and~\autoref{chap:GQVE}, we will introduce further physically motivated criteria to reduce the ambiguities in the construction of the quantum theory. Later, in~\autoref{chap:OperationalRealization}, we will discuss how these ambiguities are not merely theoretical but are inherently physical.

\begin{comment}
\begin{block}[note]
In contrast to the case without an electric field, the system becomes non-static as soon as an electric field is introduced. Even if the electric field is constant, the frequency~$\omega_\bfk(t)$ remains time-dependent. This is because the frequency~\eqref{eq:Frequency} depends on the vector potential~$\textbf{A}(t)$ rather than directly on the electric field~$\textbf{E}(t)$. This time dependence complicates the identification of an explicit formulation for the complex structure~$J$ analogous to the static case without an electric field~\eqref{eq:JNoElectricField}. However, the complex structure is implicitly defined by the splitting of the space of solutions, which is determined by our choice of the functions~$\Phi_\bfk^\pm$, and, specifically, of the modes~$\phi_\bfk$.
\end{block}
\end{comment}

We now proceed with the canonical quantization procedure outlined in the previous sections. The quantum field operator is defined via~\eqref{eq:FieldOperator} as:
\begin{equation}
    \hat{\Phi}(t,\bfx) = \int \frac{\text{d}^3\bfk}{(2\pi)^{\frac{3}{2}}} \ \left[ \hat{a}_\bfk \phi_\bfk(t) + \hat{b}_\bfk^\dagger \phi_{\bfk}^*(t) \right]  e^{i\bfk\cdot \bfx}.
    \label{eq:FieldOperatorSchwinger}
\end{equation}
The annihilation and creation operators~$\hat{a}_\bfk$,~$\hat{b}_\bfk^\dagger$ satisfy the commutation relations~\eqref{eq:Commutatorsab}.

Since we focus on bases of solutions that do not mix different~$\bfk$ modes, the Bogoliubov coefficients in~\eqref{eq:BogoliubovPhis} take a diagonal form:
\begin{equation}
    \alpha^\pm_{\bfk\bfk^\prime} = \alpha^\pm_\bfk \delta(\bfk-\bfk^\prime), \qquad \beta^\pm_{\bfk\bfk^\prime} = \beta^\pm_\bfk \delta(\bfk-\bfk^\prime).
    \label{eq:BogoliubovDiagonal}
\end{equation}
The constraints imposed by the Bogoliubov relations~\eqref{eq:BogoliubovConstraints} determine two of these coefficients in terms of the others: \mbox{$\alpha_\bfk^- = (\alpha_\bfk^+)^* \equiv \alpha_\bfk^*$} and \mbox{$\beta_\bfk^- = (\beta_\bfk^+)^* \equiv \beta_\bfk^*$}. With this notation, the full set of Bogoliubov constraints~\eqref{eq:BogoliubovConstraints} reduces to a single equation:
\begin{equation}
    |\alpha_\bfk|^2 - |\beta_\bfk|^2 = 1.
    \label{eq:|alpha||beta|}
\end{equation}

The Bogoliubov transformation~\eqref{eq:BogoliubovPhis} relating the solutions~$\Phi_\bfk^\pm$ and~$\widetilde{\Phi}_\bfk^\pm$ of the full KG equation translates into a Bogoliubov transformation between the corresponding mode functions~$\phi_\bfk$ and~$\widetilde{\phi}_\bfk$, as defined in~\eqref{eq:Phi+-Schwinger}:
\begin{equation}
    \begin{pmatrix} \widetilde{\phi}_\bfk \\ \widetilde{\phi}^*_\bfk \end{pmatrix}
    =
    \begin{array}{c}
        \begin{pmatrix} \alpha_\bfk & \beta_\bfk \\ \beta_\bfk^* & \alpha_\bfk^* \end{pmatrix}
        \begin{pmatrix} \phi_\bfk \\ \phi^*_\bfk \end{pmatrix}
    \end{array}.
\end{equation}
From these relations, we can express the Bogoliubov coefficients in terms of the mode functions as
\begin{equation}
    \alpha_\bfk = i\left( \phi_\bfk^*\dot{\widetilde{\phi}}_\bfk - \widetilde{\phi}_\bfk \dot{\phi}_\bfk^* \right), \qquad \beta_\bfk = i\left( \widetilde{\phi}_\bfk \dot{\phi}_\bfk - \phi_\bfk \dot{\widetilde{\phi}}_\bfk \right).
    \label{eq:BogoliubovCoeffsphis}
\end{equation}
It is important to note that~$\alpha_\bfk$ and~$\beta_\bfk$ are time-independent. Consequently, they can be evaluated at any convenient time.

According to~\eqref{eq:Num}, the total number of particles plus antiparticles is given by
\begin{equation}
    \mathcal{N} = \int \text{d}^3\bfk \int \text{d}^3\bfk^\prime \ ( |\beta_{\bfk\bfk^\prime}^+|^2 + |\beta_{\bfk\bfk^\prime}^-|^2 ) = \delta(0) \int \text{d}^3\bfk \ \mathcal{N}_\bfk,
\end{equation}
where 
\begin{equation}
    \mathcal{N}_\bfk = |\beta_{\bfk}^+|^2 + |\beta_{\bfk}^-|^2 = 2|\beta_\bfk|^2
    \label{eq:NkSchwinger}
\end{equation}
represents the number of excitations per mode. The factor of~$2$ accounts for the fact that the particle creation process produces the same number of particles and antiparticles. The divergent factor~$\delta(0)=(2\pi)^{-3} \int \text{d}^3\bfx$ arises from squaring the Dirac delta~$\delta(\bfk-\bfk^\prime)$ appearing in the diagonal Bogoliubov coefficients~\eqref{eq:BogoliubovDiagonal}, reflecting the fact that we are computing the total number over an infinite spatial volume. Therefore, the quantity~$\int \text{d}^3\bfk \ \mathcal{N}_\bfk$ should be interpreted as a number density (i.e., number of pairs per unit volume).

\section{Parametrization of quantum vacua}
\label{sec:ParametrizationVacua}

Given the natural choices established in~\autoref{sec:HomogeneousElectricFlat}, ensuring the preservation of the classical symmetries into the resulting quantum theory, the remaining freedom in defining the quantum theory lies entirely in the choice of the time-dependent modes~$\phi_\bfk$. Different choices of functions~$\phi_\bfk$ translate into different annihilation and creation operators via~\eqref{eq:FieldOperatorSchwinger}, which, in turn, define different quantum theories, each characterized by its own notion of quantum vacuum. 

Constructing a specific quantum theory, therefore, reduces to selecting a set of solutions~$\phi_\bfk$ to the harmonic oscillator equation with time-dependent frequency~\eqref{eq:Harmonic} for every wavenumber~$\bfk$. To uniquely specify these solutions, we need to impose initial conditions~\mbox{$(\phi_\bfk(t_0),\dot{\phi}_\bfk(t_0))$} at some reference time~$t_0$. The value of~$\phi_\bfk(t_0)$ can be any complex number and is conventionally parametrized as~\cite{Habib2000},
\begin{equation}
    \phi_\bfk(t_0) = \frac{1}{\sqrt{2W_\bfk(t_0)}}e^{-i\varphi_\bfk(t_0)}, 
    \label{eq:ParametrizationICphi}
\end{equation}
where~$W_\bfk(t_0)>0$ and~$\varphi_\bfk(t_0)$ are real quantities associated with the magnitude and phase, respectively, of~$\phi_\bfk(t_0)$. The normalization condition~\eqref{eq:Normalization} imposes a constraint that reduces the degrees of freedom in the choice of~$\dot{\phi}_\bfk(t_0)$ to a real quantity~$Y_\bfk(t_0)$ such that 
\begin{equation}
    \dot{\phi}_\bfk(t_0) = \sqrt{\frac{W_\bfk(t_0)}{2}}\left[ Y_\bfk(t_0) - i \right]e^{-i\varphi_\bfk(t_0)}.
    \label{eq:ParametrizationICDotphi}
\end{equation}

Consider two different quantizations, each defined by a set of modes~$\phi_\bfk$ and~$\widetilde{\phi}_\bfk$. Using the above parametrization, the number of created excitations per mode~\eqref{eq:NkSchwinger} can be rewritten as:
\begin{equation}
    \mathcal{N}_\bfk = 2|\beta_\bfk|^2 = \frac{W_\bfk(t_0)}{2\widetilde{W}_\bfk(t_0)}\left[ Y_\bfk(t_0)^2 +1 \right] + \frac{\widetilde{W}_\bfk(t_0)}{2W_\bfk(t_0)}\left[ \widetilde{Y}_\bfk(t_0)^2 +1 \right] - Y_\bfk(t_0) \widetilde{Y}_\bfk(t_0) - 1,
    \label{eq:NumWY}
\end{equation}
where~$\widetilde{W}_\bfk(t_0)$ and~$\widetilde{Y}_\bfk(t_0)$ are the real parameters associated with the mode functions~$\widetilde{\phi}_\bfk$, defined through the parametrization~\eqref{eq:ParametrizationICphi} and~\eqref{eq:ParametrizationICDotphi}.

Although, in principle, three independent real quantities---$W_\bfk(t_0)$, $\varphi_\bfk(t_0)$, and~$Y_\bfk(t_0)$---determine the specific quantum vacuum selected for the theory, $\mathcal{N}_{\bfk}$ does not depend on the phase $\varphi_{\bfk}(t_0)$. This phase independence also extends to the power spectrum that will be introduced in~\autoref{chap:SLEs}. This result can be understood by noting that multiplying the mode functions~$\phi_\bfk(t_0)$ and~$\dot{\phi}_\bfk(t_0)$ in~\eqref{eq:ParametrizationICphi} and~\eqref{eq:ParametrizationICDotphi} by a time-dependent phase corresponds to a trivial Bogoliubov transformation, i.e., a transformation with vanishing $\beta$-coefficients. Then, the parameters~$W_\bfk(t_0)$ and~$Y_\bfk(t_0)$ encapsulate the physical choice of initial conditions for the mode functions~$\phi_\bfk(t)$, which ultimately define the quantum theory.

Now, we introduce several well-known and widely used quantum vacua that appear frequently in the literature, each motivated by specific physical properties that authors aim to imprint on the quantum theory. However, the choices presented here are by no means exhaustive. The literature offers a vast array of physically motivated quantum vacua that we do not cover in this discussion. To briefly mention a few examples: the Bunch-Davies vacuum in de Sitter spacetime is invariant under the de Sitter group~\cite{Chernikov1968,Bunch1978}; Hadamard states are characterized by a precise ultraviolet behaviour that resembles that of Minkowski vacuum states in the absence of external fields~\cite{DeWitt1960,Fulling1981,Radzikowski1996,Junker2002}; some approaches focus on the instantaneous minimization of the renormalized stress-energy tensor~\cite{Agullo2015,Handley2016}; others aim to suppress oscillations in the primordial power spectrum in cosmological scenarios~\cite{MartindeBlas2016,ElizagaNavascues2021}, or to minimize the oscillations in the time evolution of the particle number in the Schwinger effect~\cite{Dabrowski2014,Dabrowski2016a}, among many others.

\subsection*{Minkowski quantum vacuum}

In~\autoref{sec:HomogeneousElectricFlat}, we reviewed the well-known case of quantization in the absence of an electromagnetic background. In that scenario, the standard quantization relies on the positive and negative frequency modes given in~\eqref{eq:phiE=0}, with constant frequency $\omega_\bfk^\text{M}=\sqrt{k^2 + m^2}$. This choice preserves Poincar\'{e} invariance into the quantum theory. By identifying these modes with the parametrization of the initial conditions~\eqref{eq:ParametrizationICphi} and~\eqref{eq:ParametrizationICDotphi}, the \gls{MinkowskiQuantumVacuum} is characterized by:
\begin{equation}
    W^{\text{M}}_\bfk(t_0) = \omega_\bfk = \sqrt{k^2 + m^2}, \qquad Y^{\text{M}}_\bfk(t_0) = 0.
    \label{eq:WYMinkowski}
\end{equation}
In the absence of an electric field, the initial conditions remain independent of the choice of reference time~$t_0$ at which they are imposed, reflecting the time-translation symmetry of Minkowski spacetime. However, even when an electric field is present, we can still impose the initial conditions~\eqref{eq:WYMinkowski}. In this case, plane waves no longer satisfy the harmonic oscillator equations with the time-dependent frequency~$\omega_\bfk(t)$. Nevertheless, it is always possible to select a solution that locally behaves as a plane wave with frequency~$\omega_\bfk^{\text{M}} = \sqrt{k^2 + m^2}$ around a specific time~$t_0$.

\subsection*{`In' and `out' quantum vacua}

We now consider a physically relevant scenario in which an electric field is switched on and later switched off. In this case, there exist well-defined asymptotic past and future times where the electric field vanishes, and the frequency~$\omega_\bfk(t)$ in~\eqref{eq:Harmonic} asymptotically approaches a constant value. This enables us to construct two distinguished bases of solutions:
\begin{itemize}
    \item `In' solutions, which behave as plane waves in the asymptotic past and evolve non-trivially as the electric field is applied. They define a global notion of quantum vacuum, known as the \gls{InQuantumVacuum}.
    \item `Out' solutions, which behave as plane waves in the asymptotic future after the field is switched off. The globally defined quantum vacuum in this case is the \gls{OutQuantumVacuum}.
\end{itemize}
Since the electric field vanishes in both asymptotic regimes, the system locally recovers Poincar\'{e} symmetry in these limits. The `in' and `out' solutions are preferred because they allow for a restoration of this symmetry in the quantum theory, although only locally in the asymptotic past and future, respectively.

\begin{figure}
    \centering
    \includegraphics[width=0.6\linewidth]{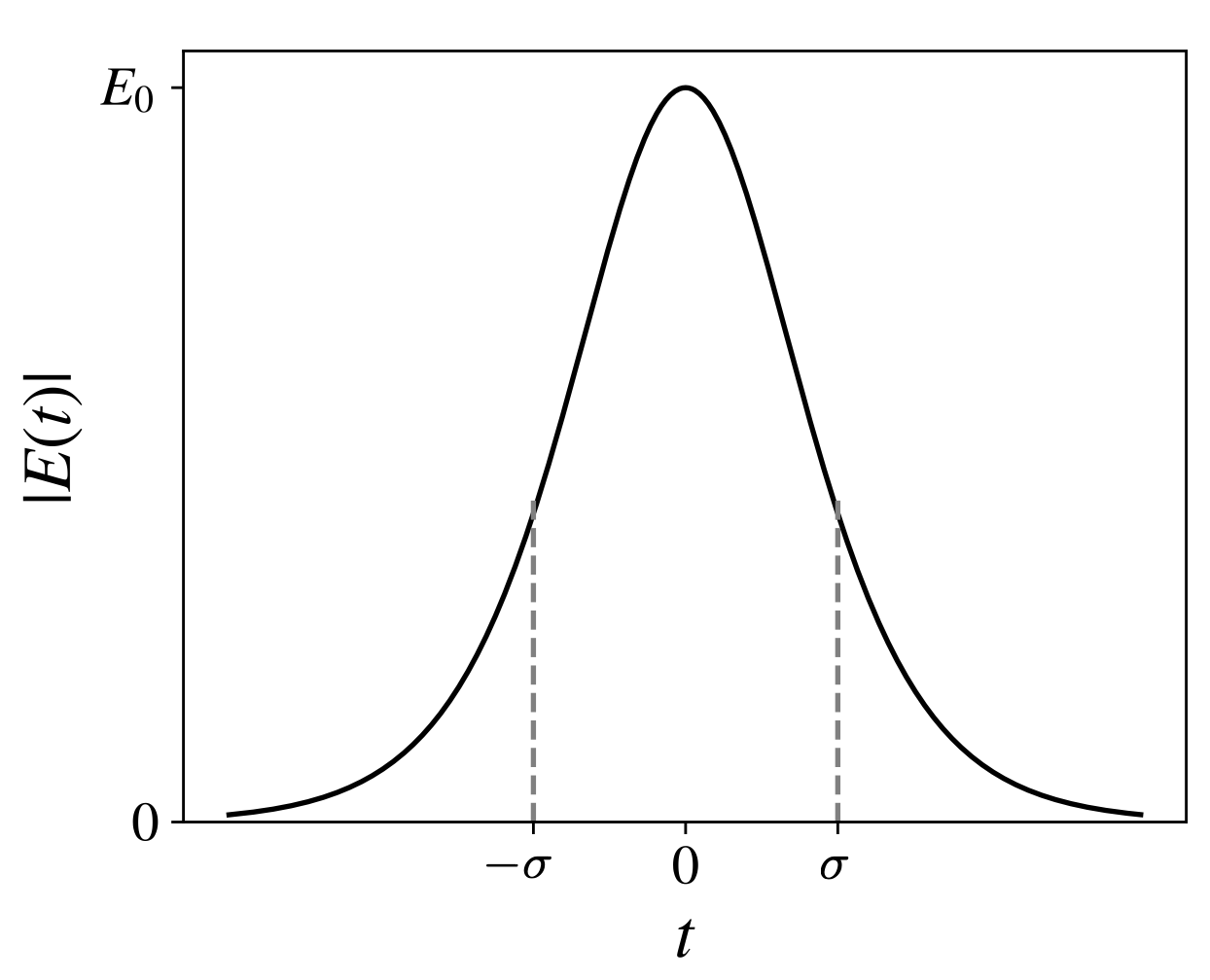}
    \caption{Sauter-type electric field of time width~$\sigma$ and maximum amplitude~$E_0$, corresponding to the vector potential given in~\eqref{eq:Sauter}.}
    \label{fig:ElectricSauter}
\end{figure}

A particularly useful example, which we will refer to multiple times in this thesis, is the Sauter-type potential~\cite{Sauter1931}. This corresponds to an electric field potential of the form
\begin{equation}
    \textbf{A}(t)=E_0\sigma\left[\tanh\left(t/\sigma\right)+1 \right] \ \textbf{e}_3.
\label{eq:Sauter}
\end{equation}
As shown in~\autoref{fig:ElectricSauter}, it models a P{\"o}schl-Teller electric pulse~\cite{Poschl1933},
\begin{equation}
    \textbf{E}(t)=-\frac{E_0}{\cosh^2\left(t/\sigma\right)} \ \textbf{e}_3,
    \label{eq:PoschlTeller}
\end{equation}
of maximum amplitude~$E_0$ at time~$t=0$. It vanishes asymptotically, and the characteristic width of the pulse is given by~$\sigma$.

This potential allows us to find an analytic expression for the `in' solution $\phi_{\bfk}^{\text{in}}$ to \eqref{eq:Harmonic}, which behaves asymptotically in the past as a plane wave of frequency
\begin{equation}
    \omega_{\bfk}^{\text{in}} = \sqrt{k^2+m^2}.
    \label{eq:InFrequency}
\end{equation}
Following~\cite{BeltranPalau2019}, the `in' solution can be written in terms of hypergeometric functions \cite{NISTDLMF} as
\begin{equation}
    \phi_{\bfk}^{\text{in}}(t)=\frac{1}{\sqrt{2\omega_{\bfk}^{\text{in}}}}e^{-i\omega_{\bfk}^{\text{in}}t}\left( 1+e^{\frac{2t}{\sigma}} \right)^{\frac{1-i\delta}{2}} {}_2F_1\left( \rho_{\bfk}^+,\rho_{\bfk}^-,1-i\sigma\omega_{\bfk}^{\text{in}};-e^{\frac{2t}{\sigma}} \right), 
    \label{eq:InSolution}
\end{equation}
where 
\begin{equation}
    \delta=\sqrt{(2qE_0\sigma^2)^2-1}, \qquad \rho_{\bfk}^{\pm}=\frac{1}{2}\left[ 1-i\sigma\left(\omega_{\bfk}^{\text{in}}\pm \omega_{\bfk}^{\text{out}}\right)-i\delta \right].
\end{equation}
Here, the `out' frequency, defined as the asymptotic limit of~$\omega_{\bfk}(t)$ for \mbox{$t\rightarrow +\infty$}, is
\begin{equation}
    \omega_{\bfk}^{\text{out}}=\sqrt{(k_3+2qE_0\sigma)^2+k_1^2+k_2^2+m^2}.
    \label{eq:OutFrequency}
\end{equation}
The `out' solutions~$\phi_\bfk^{\text{out}}$ are defined by their plane wave behaviour in the asymptotic future:
\begin{equation}
    \phi_\bfk^{\text{out}}(t) \sim \frac{1}{\sqrt{2\omega_\bfk^{\text{out}}}} e^{-i\omega_\bfk^{\text{out}} t} \qquad \text{when}\quad t \rightarrow +\infty.
\end{equation}

However, the `out' solutions do not coincide with the `in' solutions. In fact, the Bogoliubov coefficient~$\beta_\bfk$ that relates the two bases of solutions is nonzero, indicating particle production. Indeed, asymptotically in the future, the `in' solutions evolve as a linear combination of positive and negative frequency `out' plane waves:
\begin{equation}
    \phi_\bfk^{\text{in}}(t) \sim \frac{\alpha_\bfk}{\sqrt{2\omega_\bfk^{\text{out}}}} e^{-i\omega_\bfk^{\text{out}} t} + \frac{\beta_\bfk}{\sqrt{2\omega_\bfk^{\text{out}}}} e^{i\omega_\bfk^{\text{out}}t } \qquad \text{when}\quad t \rightarrow +\infty.
\end{equation}
To compute the $\beta$-Bogoliubov coefficient, we can evaluate the expression~\eqref{eq:BogoliubovCoeffsphis} at any convenient time. For instance, 
\begin{equation}
    \beta_\bfk = \lim_{t\rightarrow +\infty} i\left[ \phi^{\text{in}}_\bfk(t) \dot{\phi}^{\text{out}}_\bfk(t) - \phi^{\text{out}}_\bfk(t) \dot{\phi}^{\text{in}}_\bfk(t) \right].
\end{equation}
By analysing the asymptotic behaviour of the hypergeometric functions, one finds the expression for the number of created particles and antiparticles~\cite{BeltranPalau2019}:
 \begin{equation}
    \mathcal{N}_\bfk = 2|\beta_\bfk|^2 = \frac{\cosh \left[ \pi(\omega_\bfk^{\text{out}} - \omega_\bfk^{\text{in}}) \sigma \right] - \cosh\left( \pi \delta \right)}{\sinh \left( \pi \omega_\bfk^{\text{in}} \sigma \right) \sinh \left( \pi\omega_\bfk^{\text{out}} \sigma \right)}.
    \label{eq:nkSauterScalar}
\end{equation}
This result explicitly demonstrates that the Sauter-type electric field induces vacuum pair production via the Schwinger effect.

\subsection*{Instantaneous lowest energy vacua}

A natural approach to defining the quantum vacuum would be to select the state that minimizes the energy of the system. However, in our case, the presence of an external agent introduces an explicit time dependence in the Hamiltonian, preventing the existence of a universal state of minimal energy (see~\autoref{sec:ConstructionsSLEs}). Nevertheless, we can still determine the state that minimizes the energy at a specific time~$t_0$. However, since the Hamiltonian evolves with time, the state that minimizes the energy at one instant will generally differ from the state that minimizes it at another.

From its classical definition, obtained from the action~\eqref{eq:ActionScalar}, we define the Hamiltonian for the charged field~$\Phi$ as
\begin{align}
H(t) &= \int \text{d}^3\bfx \ \left[ (\partial_t\Phi) \Pi + (\partial_t\Phi^*) \Pi^* - L \right] \nonumber \\
&= \int \text{d}^3\bfx \ \left[ (\partial_t\Phi^*) (\partial_t\Phi) + (\partial_i - iqA_i)\Phi^* ( \partial_i +iqA_i)\Phi + m^2\Phi^* \Phi \right],
\label{eq:HamiltonianClassical}
\end{align}
where~$L$ denotes the Lagrangian density. To define a symmetric Hamiltonian quantum operator, we need to introduce anticommutators:
\begin{equation}
    \hat{H}(t) = \frac{1}{2} \int \text{d}^3\bfx \ \left[ \left\{ \partial_t\hat{\Phi}^\dagger, \partial_t\hat{\Phi} \right\} + \left\{ (\partial_i - iqA_i) \hat{\Phi}^\dagger, (\partial_i + iqA_i)\hat{\Phi} \right\} + m^2\left\{ \hat{\Phi}^\dagger, \hat{\Phi} \right\} \right].
    \label{eq:HamiltonianOperator}
\end{equation}

\begin{block}[note]
To further motivate the introduction of anticommutators, let us consider, as an illustrative example, the last term in~\eqref{eq:HamiltonianClassical}, which is proportional to~$\Phi^*\Phi$. If we were to define the corresponding quantum operator as~$\hat{\Phi}^\dagger \hat{\Phi}$, then its expectation value in the quantum vacuum~$|0\rangle$, given by~$\langle 0 | \hat{\Phi}^\dagger \hat{\Phi} | 0 \rangle$, would only involve terms associated with the creation and annihilation operators of antiparticles, specifically~$\langle 0 | \hat{b}_{\bfk^\prime}^\dagger \hat{b}_\bfk | 0 \rangle$.

However, we must also account for the contribution from particles, which arises from the operator~$\hat{\Phi} \hat{\Phi}^\dagger$, leading to terms of the form~$\langle 0 | \hat{a}_{\bfk} \hat{a}_{\bfk^\prime}^\dagger | 0 \rangle$. To incorporate both contributions symmetrically, the most natural definition is to use the anticommutator:
\begin{equation}
     \Phi^*\Phi \rightarrow \frac{1}{2}\left\{ \hat{\Phi}^\dagger, \hat{\Phi} \right\} = \frac{1}{2}\left( \hat{\Phi}^\dagger\hat{\Phi} + \hat{\Phi}\hat{\Phi}^\dagger \right).
\end{equation}
\end{block}

The expectation value of the Hamiltonian operator in a quantum vacuum~$|0\rangle$ can be computed by substituting the field operator~$\hat{\Phi}$ in terms of the annihilation and creation operators from~\eqref{eq:FieldOperatorSchwinger}:
\begin{equation}
\langle 0 | \hat{H}(t) |0\rangle = \delta(0) \int \text{d}^3\bfk \ E[\phi_\bfk](t),
\label{eq:HamiltonianExpectation}
\end{equation}
where the contribution from each mode is given by
\begin{equation}
E[\phi_\bfk](t) = |\dot{\phi}_{\bfk}(t)|^2 + \omega_{\bfk}(t)^2 |\phi_{\bfk}(t)|^2.
\label{eq:EnergyILES}
\end{equation}
The delta factor~$\delta(0)$ in~\eqref{eq:HamiltonianExpectation} reflects the fact that the total spatial volume is infinite. Note that although the energy per mode generally remains finite, the total energy---obtained by summing over all modes---requires renormalization. In particular, this imposes ultraviolet convergence conditions on the quantum state. More precisely, the state must exhibit sufficiently high-order adiabatic behavior, approaching the Hadamard ultraviolet structure~\cite{Cooper1989,Kluger1991,Ferreiro2018}. The adiabatic regularization of this quantity, specifically for Dirac fields, will be carried out in~\autoref{chap:Kugelblitz}.

We seek to minimize the energy density per mode at a particular time~$t_0$, $E[\phi_\bfk](t_0)$. Substituting the parametrizations~\eqref{eq:ParametrizationICphi} and~\eqref{eq:ParametrizationICDotphi}, we obtain
\begin{equation}
E[\phi_\bfk](t_0) = \frac{1}{2} \left[ W_\bfk(t_0) Y_\bfk(t_0)^2 + W_\bfk(t_0) + \frac{\omega_\bfk(t_0)^2}{W_\bfk(t_0)} \right].
\label{eq:EnergyILESWY}
\end{equation}
Thus, minimizing the energy reduces to finding the real coefficients~$W_\bfk(t_0)$ and~$Y_\bfk(t_0)$ that minimize this expression. Since~$W_\bfk(t_0) > 0$, all terms are positive. The first term is minimized when~$Y_\bfk(t_0) = 0$, while the remaining two terms reach their minimum when \mbox{$W_\bfk(t_0) = \omega_\bfk(t_0)$}. Therefore, the \gls{ILESt0} (\acrshort{ILES}) is defined by the initial conditions:
\begin{equation}
W^{\text{ILES}}_\bfk(t_0) = \omega_\bfk(t_0), \qquad Y^{\text{ILES}}_\bfk(t_0) = 0.
\label{eq:ILES}
\end{equation}

Another key property of the ILES at time~$t_0$ is that it instantaneously diagonalizes the Hamiltonian operator. Indeed, for the ILES at~$t_0$, the energy per mode satisfies 
\begin{equation}
    E[\phi_\bfk^{\text{ILES}}](t_0) = \omega_\bfk(t_0),
\end{equation}
and all terms that mix different wavevectors~$\bfk$ in the Hamiltonian operator~\eqref{eq:HamiltonianOperator} vanish. As a result, the Hamiltonian takes the diagonal form:
\begin{equation}
    \hat{H}^{\text{ILES}}(t_0) = \frac{1}{2} \int \text{d}^3 \bfk \ \omega_\bfk(t_0) \left( \left\{ \hat{a}_\bfk^\dagger, \hat{a}_\bfk \right\} + \left\{ \hat{b}_\bfk^\dagger, \hat{b}_\bfk \right\} \right).
\end{equation}

In~\autoref{chap:SLEs}, we will discuss the definition of states of low energy, which generalize the concept of ILESs. Unlike ILESs, which minimize the energy density at a single instant, states of low energy are defined by minimizing the energy density over a finite time interval~\mbox{$[t_1, t_2]$}.

\subsection*{Adiabatic quantum vacua}

Adiabatic states, proposed by Parker in~\cite{Parker1969} and formalized by L\"uders and Roberts in~\cite{Luders1990}, are among the most widely used choices of quantum vacua. These states are constructed based on the Wentzel-Kramers-Brillouin (WKB) approximation and naturally generalize the concept of plane waves, which define the Minkowski vacuum in flat spacetime, to scenarios with a slowly varying external agent---specifically, a time-dependent frequency~$\omega_\bfk(t)$.

Let us seek an approximate solution to the harmonic oscillator equation~\eqref{eq:Harmonic} under the assumption that the time-dependent frequency~$\omega_\bfk(t)$ varies slowly. First, we introduce the adiabatic parameter~$T$, which quantifies the timescale of variation, and perform a change of variables to the dimensionless time~$\tilde{t} = t/T$. The equation of motion then takes the form:
\begin{equation}
    \frac{\text{d}^2 \phi_\bfk}{\text{d}\tilde{t}^2} + T^2 \omega_\bfk(\tilde{t})^2 \phi_\bfk(\tilde{t}) = 0.
    \label{eq:HarmonicTilde}
\end{equation}
We now write the exact solution~$\phi_\bfk$ in polar form, analogous to the parametrization of initial conditions in~\eqref{eq:ParametrizationICphi}. Similarly, we write its derivative in a form analogous to~\eqref{eq:ParametrizationICDotphi}:
\begin{equation}
    \phi_\bfk(\tilde{t}) = \frac{1}{\sqrt{2TW_\bfk(\tilde{t})}}e^{-iT\varphi_\bfk(\tilde{t})}, \qquad \frac{\text{d}\phi_\bfk}{\text{d}\tilde{t}} = \sqrt{\frac{TW_\bfk(\tilde{t})}{2}} \left[ \frac{Y_\bfk(\tilde{t})}{T} -i \right] e^{-iT\varphi_\bfk(\tilde{t})}.
\end{equation}
Substituting this ansatz into the harmonic oscillator equation~\eqref{eq:HarmonicTilde} and using the normalization condition~\eqref{eq:Normalization}, we obtain a dynamical equation for~$W_\bfk(\tilde{t})$:
\begin{equation}
    W_\bfk^2 = \omega_\bfk^2 - \frac{1}{2T^2} \left[ \frac{1}{W_\bfk} \frac{\text{d}^2 W_\bfk}{\text{d}\tilde{t}^2} - \frac{3}{2W_\bfk^2} \left(\frac{\text{d} W_\bfk}{\text{d}\tilde{t}} \right)^2  \right],
    \label{eq:WkEquation}
\end{equation}
while the phase~$\varphi_\bfk(\tilde{t})$ and~$Y_\bfk(\tilde{t})$ are determined entirely by~$W_\bfk(\tilde{t})$:
\begin{equation}
    \varphi_\bfk(\tilde{t}) = \int \text{d}\tilde{t} \ W_\bfk(\tilde{t}), \qquad Y_\bfk(\tilde{t}) = -\frac{1}{2W_\bfk(\tilde{t})^2} \frac{\text{d}W_\bfk}{\text{d}\tilde{t}}.
\end{equation}

At this stage, the equations for~$W_\bfk(\tilde{t})$, $\varphi_\bfk(\tilde{t})$ and~$Y_\bfk(\tilde{t})$ are exact. To proceed with the adiabatic approximation, we expand~$W_\bfk(\tilde{t})$ as a power series in~$T^{-1}$, assuming that~$T^{-1}$ is small (in the adiabatic limit, $T\rightarrow \infty$). Keeping only terms of order smaller or equal to~$T^{-n}$, the \gls{AdiabaticApproximation} is given by
\begin{equation}
    W_\bfk^{(n)} = \sum_{i=0}^n T^{-i} W_{i,\bfk}.
    \label{eq:WKBSum}
\end{equation}

Substituting this last expansion into~\eqref{eq:WkEquation}, the zeroth-order adiabatic approximation is defined by
\begin{equation}
    W_\bfk^{(0)}(t) = \omega_\bfk(t), \qquad \varphi_\bfk^{(0)}(t) = \int \text{d}t \ W_\bfk(t), \qquad Y_\bfk^{(0)}(t) = - \frac{\dot{\omega}_\bfk(t)}{2\omega_\bfk(t)^2},
    \label{eq:0thWKB}
\end{equation}
where we have reverted to the original time variable~$t$. 

The $n$th-order WKB approximation can be obtained in the standard way~\cite{Birrell1982} recursively introducing the previous order in~\eqref{eq:WkEquation}. Note that, since in the equation~\eqref{eq:WkEquation} the parameter~$T$ appears squared, odd WKB orders vanish. For instance, the following non-vanishing WKB order approximation is the second, defined from
\begin{equation}
    \left( W_\bfk^{(2)} \right)^2 = \omega_\bfk^2 - \frac{1}{2} \left[ \frac{\ddot{\omega}_\bfk}{\omega_\bfk}  - \frac{3}{2} \left( \frac{\dot{\omega}_\bfk}{\omega_\bfk} \right)^2  \right].
\end{equation}

\begin{block}[note]
The WKB series, as defined by the partial sum~\eqref{eq:WKBSum}, does not generally converge. This implies that higher-order approximations do not necessarily yield better accuracy than lower-order ones. Nevertheless, it is well established that, despite its formal divergence, the WKB approximation can provide an extremely accurate numerical approximation to the exact solution in many practical scenarios~\cite{Bender1999}.
\end{block}

Motivated by the WKB approximation, we can define physically meaningful quantum vacua. Specifically, the \gls{AdiabaticQuantumVacuum} is constructed using the exact solutions~$\phi_\bfk(t)$ of the harmonic oscillator equations~\eqref{eq:Harmonic}, with initial conditions at time~$t_0$ chosen to match the adiabatic approximation of order~$n$ at that instant. For example, the zeroth-order adiabatic quantum vacuum at time~$t_0$ is defined by the initial conditions:
\begin{equation}
    W_\bfk^{(0)}(t_0) = \omega_\bfk(t_0),  \qquad Y_\bfk^{(0)}(t_0) = -\frac{\dot{\omega}_\bfk(t_0)}{2\omega_\bfk(t_0)^2},
    \label{eq:0thAdiabaticVacuum}
\end{equation}
which are motivated by the zeroth-order WKB approximation~\eqref{eq:0thWKB}. It is important to emphasize that the modes~$\phi_\bfk(t)$ are exact solutions, and that the adiabatic approximation is used only to determine their initial conditions.

\section{Conclusions}
\label{sec:ConclusionsChoiceQuantumVacuum}

In this chapter, we have specialized the formalism introduced in~\autoref{chap:QuantizationScalars} to the case of a charged scalar field in Minkowski spacetime coupled to a homogeneous electric field. This setup will serve as the foundational framework for most of the chapters in this thesis.

For this particular scenario, we have explicitly identified the ambiguities inherent in the quantization procedure and parametrized the various possibilities. The specific choice of quantization is usually guided by the desired physical properties to be imprinted on the quantum theory. We introduced several common quantization schemes, each defining its own notion of quantum vacuum. For example, when there is no external field in flat spacetime, the standard choice of quantum vacuum is the Minkowski vacuum, which preserves Poincar\'e invariance. 

The instantaneous lowest energy vacuum at a given time is defined as the quantum vacuum that minimizes the energy density per mode locally at that specific time. In~\autoref{chap:SLEs}, we will generalize this concept to define states of low energy, which minimize the energy density over a finite time interval instead of instantaneously.

Another important class of vacua is the adiabatic quantum vacua, constructed via the WKB approximation, which generalizes the Minkowski vacuum to scenarios where the external field evolves slowly. This family of vacua will play a significant role in our study of the generalized quantum Vlasov equation in~\autoref{sec:GQVE}, as well as in the adiabatic regularization procedure for the stress-energy tensor used in~\autoref{chap:Kugelblitz} to compute the energy dissipated via the Schwinger effect.
\chapter[States of low energy]{States of low energy}
\label{chap:SLEs}

\chaptermark{States of low energy}

In this chapter, based on~\cite{AlvarezSLEs}, we explore the so-called \acrfull{SLEs} in the context of the Schwinger effect. Their definition in cosmology was originally motivated by the work of~\cite{Fewster2000}, which showed that the renormalized energy density, when smeared along a timelike curve, is bounded from below as a function of the state. This result was later applied in~\cite{Olbermann2007} to general cosmological models considering smearing functions supported on the worldline of an isotropic observer. A systematic procedure was then developed to explicitly construct the vacuum states that minimize this smeared energy density: the SLEs.

H. Olbermann proved in~\cite{Olbermann2007} one appealing property of SLEs in \acrfull{FLRW} cosmological backgrounds: they satisfy the Hadamard condition. This relates to the ultraviolet behaviour of the two-point function, and guarantees that computations such as that of the stress-energy tensor are well defined~\cite{Fewster2013}. While the Hadamard condition has been studied in the context of static electric backgrounds~\cite{Wrochna2012} and time-dependent external potentials~\cite{Finster2016}, its validity in the Schwinger effect remains an open question. In this chapter, we will show that the ultraviolet behaviour of SLEs in the Schwinger effect is consistent with the Hadamard condition, though a rigorous proof is still missing.

The properties of SLEs for cosmological models were further investigated in~\cite{Banerjee2020}. They were found to have the same infrared behaviour up to a constant factor for any smearing function and the same ultraviolet behaviour independently of the smearing function. In the context of cosmological perturbations, these authors found that these states are suitable candidates for vacua in models with a period of kinetic dominance prior to inflation, as they provide the correct infrared and ultraviolet behaviours for perturbations at the end of inflation. This prompted the proposal of SLEs as vacua of cosmological perturbations in the context of loop quantum cosmology~\cite{MartinBenito2021,MartinBenito2021a}. These two works have shown an interesting dependence of SLEs on the smearing function: they are independent of it as long as it is wide enough around the bounce of loop quantum cosmology~\cite{MartinBenito2021}, but very sensitive to whether the moment of the bounce is included in the support of the smearing function~\cite{MartinBenito2021a}. 

The concept of SLEs has also been extended to fermionic fields in~\cite{NadalGisbert2023}, where they are applied in a radiation-dominated, CPT-invariant Universe. In addition, following our study of SLEs in an anisotropic electric field background~\cite{AlvarezSLEs}, they were rigorously formulated in another anisotropic setting: the Bianchi I cosmological model~\cite{Banerjee2023}.

In~\autoref{sec:ConstructionsSLEs}, we generalize the construction of SLEs to arbitrary homogeneous settings, with a particular focus on anisotropic scenarios such as the Schwinger effect. In~\autoref{sec:RoleSmearing}, we analyse how different choices of the smearing function lead to distinct quantum vacua within the family of SLEs. \autoref{sec:Anisotropies} is dedicated to studying anisotropies. We extend the conventional notion of the power spectrum, commonly used in cosmology, to the context of the Schwinger effect, and we examine the multipolar contributions predicted by SLEs. Finally, in~\autoref{sec:NumberSLEs}, we investigate the number of created particles for different choices of smearing functions, and assess the compatibility of SLEs with the Hadamard condition. In \autoref{sec:ConclusionsSLEs} we present the main conclusions of the chapter.

\section{Construction of SLEs}
\label{sec:ConstructionsSLEs}

Here we propose a direct generalization of Olbermann's procedure in~\cite{Olbermann2007} to systems characterized by modes~$\phi_{\bfk}(t)$ satisfying harmonic oscillator equations with time-dependent frequencies~\eqref{eq:Harmonic}. In this construction, we do not assume the explicit expression for the frequency~$\omega_\bfk(t)$ in a homogeneous electric background given in~\eqref{eq:Frequency}. 

More broadly, matter fields coupled to other external, time-dependent, spatially homogeneous backgrounds---beyond just an electric field---are also governed by harmonic oscillator equations with time-dependent frequencies of the form~\eqref{eq:Harmonic}. Thus, the construction of SLEs that we are presenting here remains entirely valid for these more general models. A notable example is the case of scalar and tensor gauge-invariant perturbations in FLRW backgrounds, where the gravitational field plays a role analogous to that of the electric field. The formalism developed here can be applied not only to these models but also to broader scenarios involving particle creation, provided they can be described within this framework.

Let~$f(t)$ be a \gls{SmearingFunction} of compact support~$[t_1,t_2]$. By smearing the energy density given in~\eqref{eq:EnergyILES}, the contribution of each mode~$\phi_{\bfk}(t)$ to the total \gls{SmearedEnergyDensity} is given by\footnote{In the original paper~\cite{AlvarezSLEs}, the definition of~$E_f[\phi_\bfk]$ includes a factor of~$1/2$ because in~\cite{AlvarezSLEs} $\phi_\bfk$ refers to real and imaginary parts of the complex mode, each contributing equally to the total smeared energy.}
\begin{equation}
    E_f[\phi_{\bfk}]=\int \text{d}t \ f(t)^2\left[ |\dot{\phi}_{\bfk}(t)|^2+\omega_{\bfk}(t)^2|\phi_{\bfk}(t)|^2 \right].
     \label{eq:Ek}
\end{equation}
The aim is to find, for each~$\bfk$, the mode~$\phi_\bfk^{\text{SLE}}(t)$ which minimizes this energy density. 

The strategy is as follows. First, we provide a fiducial solution~$F_{\bfk}(t)$ to the equation of motion~\eqref{eq:Harmonic}. Then, since this differential equation has real coefficients, the complex conjugate of the fiducial solution, $F_\bfk^*(t)$, is also a solution, and the problem translates into finding complex constants~$\lambda_{\bfk}$ and~$\mu_{\bfk}$ such that the solution~$\phi_\bfk^{\text{SLE}}(t)$ is written as the linear combination
\begin{equation} 
    \phi_\bfk^{\text{SLE}}(t)=\lambda_{\bfk}F_{\bfk}(t)+\mu_{\bfk}F_{\bfk}(t)^*.
    \label{eq:TS}
\end{equation}
Note that this is actually a Bogoliubov transformation, so in order to preserve the Poisson algebra of the corresponding annihilation and creation operators, the Bogoliubov coefficients should satisfy ~\mbox{$|\lambda_{\bfk}|^2-|\mu_{\bfk}|^2=1$}. On the other hand, the phase of the solution $\phi_\bfk^{\text{SLE}}(t)$ is irrelevant, so without loss of generality we can assume that~$\mu_{\bfk}$ is a positive real constant. Substituting~\eqref{eq:TS} in the smeared energy density~\eqref{eq:Ek}, we can write
\begin{equation} \label{eq:ETk}
    E_f[\phi_\bfk^{\text{SLE}}]=\left( 1+2\mu_{\bfk}^2\right)E_f[F_{\bfk}]+2\mu_{\bfk}\Re \left( \lambda_{\bfk}C[F_{\bfk}]\right).
\end{equation}
Here, $\Re$ denotes the real part of complex quantities, and the complex constant $C[F_{\bfk}]$ depends on the fiducial solution $F_{\bfk}(t)$ as
\begin{equation} \label{eq:ck}
    C[F_{\bfk}]=\int \text{d}t \ f(t)^2\left[ \dot{F}_{\bfk}(t)^2+\omega_{\bfk}(t)^2F_{\bfk}(t)^2 \right].
\end{equation}

Since both~$\mu_\bfk$ and~$E_f[F_\bfk]$ are non-negative, direct inspection of~\eqref{eq:ETk} reveals that the minimum of~$E_f[\phi_\bfk^{\text{SLE}}]$ is reached for the most negative value that the quantity~\mbox{$\Re \left(\lambda_{\bfk}C[F_{\bfk}]\right)$} can attain. This is achieved when the principal arguments satisfy~\mbox{$\text{Arg\,}\lambda_{\bfk}+\text{Arg\,}C[F_{\bfk}]=\pi$}. Then, using the relation~\mbox{$|\lambda_{\bfk}|^2-|\mu_{\bfk}|^2=1$} we can write~$E_f[\phi_{\bfk}^{\text{SLE}}]$ in~\eqref{eq:ETk} only in terms of the Bogoliubov coefficient~$\mu_{\bfk}$. Finally, we minimize~$E_f[\phi_{\bfk}^{\text{SLE}}]$ with respect to~$\mu_{\bfk}$ and obtain
\begin{equation} 
\mu_{\bfk}=\sqrt{\frac{E_f[F_{\bfk}]}{2\sqrt{E_f[F_{\bfk}]^2-|C[F_{\bfk}]|^2}}-\frac{1}{2}}, \qquad
\lambda_{\bfk}=-e^{-i\arg{C[F_{\bfk}]}}\sqrt{\mu_{\bfk}^2+1}.
\label{eq:mulambda} 
\end{equation}
These two coefficients define the SLE~$\phi_\bfk^{\text{SLE}}(t)$ through the Bogoliubov transformation~\eqref{eq:TS}.

The construction of SLEs is strongly dependent on the choice of the smearing function~$f(t)$, which defines the smeared energy in~\eqref{eq:Ek}. Consequently, rather than yielding a unique vacuum, this framework gives rise to a family of SLEs, each corresponding to a different choice of~$f(t)$. The implications of this dependence, as well as the role of the smearing function in shaping the resulting quantum vacua, are examined in detail in~\autoref{sec:RoleSmearing}. On the other hand, the construction presented here seems to depend explicitly on the fiducial solution~$F_{\bfk}(t)$. However, the SLE is independent of this choice~\cite{Banerjee2020}.

\subsection*{States of minimal energy}
A natural question that arises is whether there exists a particular SLE that minimizes the smeared energy density for all choices of smearing functions. Reference~\cite{Olbermann2007} investigated this question in the context of FLRW spacetimes and demonstrated that such a state exists only when the scale factor remains constant; otherwise, no universal minimal-energy state can be defined. 

We now extend this result by proving that, for a \gls{StateOfMinimalEnergy} to exist in a general homogeneous background, the frequencies~$\omega_{\bfk}(t)$ in the harmonic oscillator equations must be time-independent. In the case of the Schwinger effect, the frequency~\eqref{eq:Frequency} remains constant only in the absence of an electric field. This implies that when an electric field is applied, a notion of state of minimal energy does not exist.

Indeed, let us consider a solution~$\phi_{\bfk}(t)$ to the harmonic oscillator equation with time-dependent frequency~\eqref{eq:Harmonic}. Analogous to~\eqref{eq:TS}, there exist Bogoliubov coefficients~$\mu_\bfk$ and~$\lambda_\bfk$ such that \mbox{$\phi_\bfk^{\text{SLE}}(t)=\lambda_{\bfk}\phi_{\bfk}(t)+\mu_{\bfk}\phi_{\bfk}^*(t)$}. Then, $\phi_\bfk(t)$ is a SLE if and only the Bogoliubov coefficient~$\mu_\bfk$ vanishes. According to~\eqref{eq:mulambda}, this occurs if and only if the coefficient~$C[\phi_{\bfk}]$ vanishes. Moreover, from~\eqref{eq:ck}, if we require~$\phi_{\bfk}(t)$ to be a SLE for all smearing functions, then it must satisfy the equation \mbox{$\dot{\phi}_{\bfk}(t)^2+\omega_{\bfk}(t)^2\phi_{\bfk}(t)^2=0$}. Differentiating this equation yields: \mbox{$2\omega_\bfk(t)\dot{\omega}_\bfk(t)\phi_\bfk(t)^2=0$}, which is compatible with the equation of motion if and only if the frequency is constant.

\subsection*{Computation of SLEs}
For the numerical computations, we focus on an electric background modelled by the Sauter potential~\eqref{eq:Sauter}, which provides a smooth, time-dependent electric field pulse that vanishes asymptotically. This choice allows us to derive an analytic expression for the fiducial solution~$F_{\bfk}(t)$, which we can set to be the `in' solution~\eqref{eq:InSolution}. Consequently, the SLEs can be expressed in terms of integrals involving hypergeometric functions. However, remember that the construction of the SLEs is independent of the specific fiducial solution chosen. Any convenient solution may be used.

The numerical computations required to obtain the fiducial solution and generate the figures presented in this chapter were carried out using Python. In particular, we used the \texttt{scipy.integrate.odeint} function to solve, for each~$k$ and each~$\theta$, the harmonic oscillator equation with zeroth-order adiabatic initial conditions~\eqref{eq:0thWKB} at~$t=0$. Given the large size of the data and the computational demands, we relied on the High Performance Computing cluster resources provided by the Universidad Complutense de Madrid. In addition, to compute the `in' solution~\eqref{eq:InSolution}, we imported into \textsc{Python} the hypergeometric functions already implemented analytically in \textsc{Mathematica}, as these specific functions are not natively available in Python.

In the following, time will always be expressed in units of the time width~$\sigma$ of the Sauter-type pulse, and frequencies in units of~$\sigma^{-1}$. On the other hand, in our plots, we fix the value of the mass to~$m=\sigma^{-1}$ and the maximum amplitude of the electric field to~\mbox{$qE_0=\sigma^{-2}$}. This corresponds to the critical Schwinger limit~\mbox{$qE_0=m^2$}~\cite{Schwinger1951}. For lower strengths, the probability of pair production is exponentially suppressed. Only for electric fields of this order does the Schwinger effect become physically relevant. In practice, the qualitative behaviour of the system for stronger electric fields is the same, and provides no additional information that is relevant to this work. Furthermore, we are interested in studying the physical differences between choices of vacua. Considering larger intensities than the Schwinger limit makes these differences less clear.

\section{Role of the smearing function}
\label{sec:RoleSmearing}

\begin{figure}
    \centering
    \includegraphics[width=0.6\linewidth]{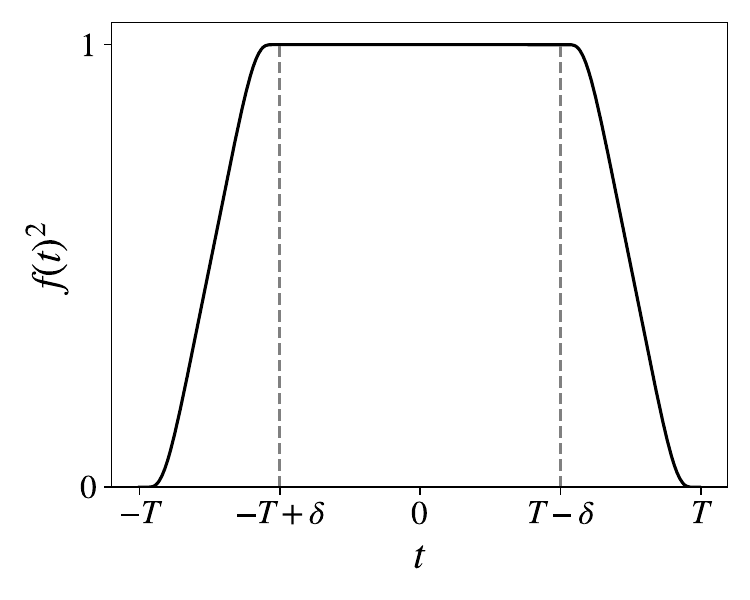}
    \caption{Smearing function~\eqref{eq:SmearingFunction} of compact support~$[-T,T]$ and slope of length~$\delta$.}
    \label{fig:SmearingFunction}
\end{figure}

In~\autoref{sec:ConstructionsSLEs} we saw that each SLE minimizes the energy density smeared with a certain compact support function~$f(t)$. We are interested in studying the physical interpretation of choosing different supports for the smearing function, each defining a particular notion of SLE. As the Sauter potential~\eqref{eq:Sauter} is symmetric around its maximum at $t=0$, it will be useful to consider smearing functions with compact support~$[-T,T]$, where~$T>0$. In particular, we are going to use smooth window functions as shown in~\autoref{fig:SmearingFunction}. We will describe them in terms of regularized step functions~$\Theta_{\delta}(t)$ of width $\delta$, such that in the limit~$\delta\to0$ we recover the discontinuous Heaviside step function. The function~$\Theta_{\delta}(t)$ interpolates between~$0$ and~$1$ for~\mbox{$t\in(-\delta/2,\delta/2)$} and it is constant outside. We choose for the interpolating function 
\begin{equation}
    \Theta_\delta(t) = \frac{1}{2}\left(1+\tanh\left\{\cot\left[\pi\left(\frac{1}{2}-\frac{t}{\delta}\right)\right]\right\}\right), \qquad t\in\left(-\frac{\delta}{2},\frac{\delta}{2}\right),
    \label{eq:RegularizedStep}
\end{equation}
although the results will not qualitatively depend on this particular selection. Then, we can write the smearing functions as
\begin{equation} 
    f(t)^2=\frac{1}{2}\left[\Theta_{\delta}\left(t+T-\frac{\delta}{2}\right)+\Theta_{\delta}\left(-t+T-\frac{\delta}{2}\right)\right].
    \label{eq:SmearingFunction}
\end{equation}
We fix a small step width of $\delta=10^{-4}\sigma$ for all the figures in this chapter. For supports smaller than this width (i.e., $T<\delta$), we readapt the parameter by setting $\delta = T/2$ so that it is still smooth.

\begin{block}[note]
For simplicity, we are choosing to maintain the shape of the test function, considering only the effects of changing its support. In principle its shape may also be relevant to the resulting SLE. However, for sufficiently large supports, the SLEs should be fairly insensitive to the form of the test function, as long as it is reasonably behaved, as is indeed corroborated in \cite{MartinBenito2021}. Furthermore, even when the form of the test function may be relevant, different shapes would simply translate to more or less weight being given to specific time periods when computing the smeared energy density. Therefore, we may understand the physics behind the consequences of different shapes by understanding the physical interpretation of the support first. Besides, one may also argue that more intricate shapes are less natural choices that would require additional motivation.
\end{block}

We saw in~\autoref{chap:ChoiceVacuum} that the freedom in the choice of vacuum is parametrized by $W_{\bfk}(t_0)$ and~$Y_{\bfk}(t_0)$, which define via~\eqref{eq:ParametrizationICphi} and~\eqref{eq:ParametrizationICDotphi} the initial conditions of the selected basis of solutions at time $t_0$. We fix $t_0=0$, the instant at which the Sauter-type electric field reaches its maximum, and consider the smearing functions~\eqref{eq:SmearingFunction}, varying~$T$. In addition, in this section we focus on modes whose wavevectors~$\bfk$ are parallel to the direction of the electric field. Anisotropies will be analysed in detail in~\autoref{sec:Anisotropies}.

\begin{figure}
    \centering
    \begin{subfigure}{0.49\textwidth}
        \centering
        \includegraphics[width=\textwidth]{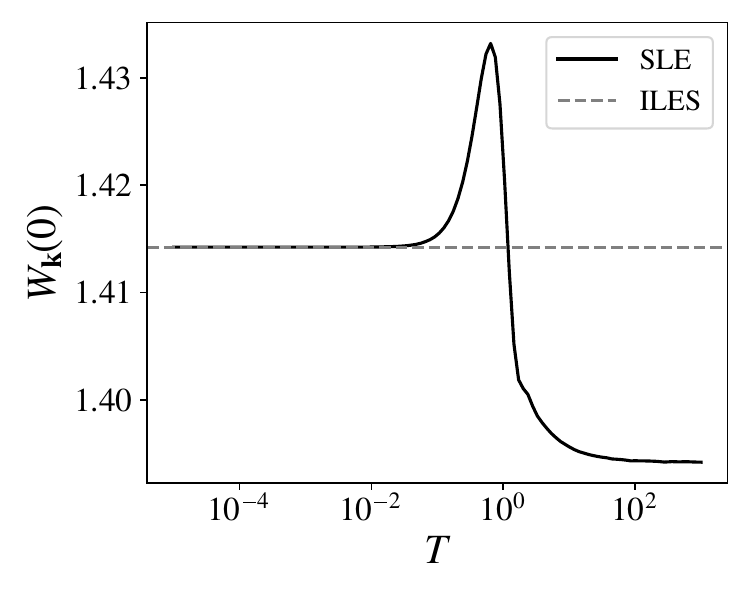}
        \caption{}
        \label{fig:WIR}
    \end{subfigure}
    \begin{subfigure}{0.49\textwidth}
        \centering
        \includegraphics[width=\textwidth]{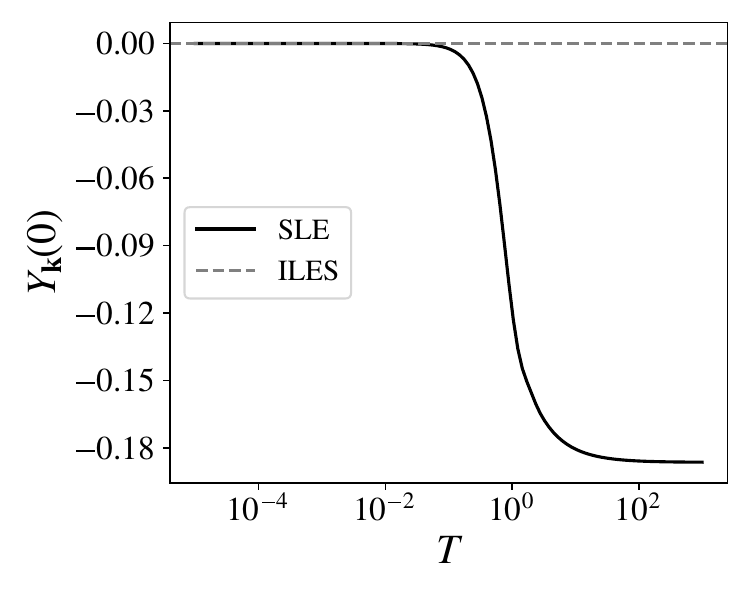}
        \caption{}
        \label{fig:YIR}
    \end{subfigure}
    \caption{Dependence of (a) $W_{\bfk}(t_0=0)$ and (b) $Y_{\bfk}(t_0=0)$ defining the SLEs on the support $[-T,T]$ of the smearing functions \eqref{eq:SmearingFunction}. We show the infrared mode whose wavevector~$\bfk$ is parallel to the electric field and $k=10^{-5}\sigma^{-1}$. 
 We use units $\sigma=1$.}
    \label{fig:WYIR}
\end{figure}

In~\autoref{fig:WYIR} we show~$W_{\bfk}(t_0)$ and~$Y_{\bfk}(t_0)$ for an infrared mode  with~$k=10^{-5}\sigma^{-1}$ as functions of the support of the smearing functions. We identify a transition regime around the time scale $\sigma$, which is the characteristic length of the Sauter-type electric pulse, where the dependence on the support is not monotonic. It separates the behaviours of the SLEs for small and large supports. We have verified that this happens independently of the strength of the electric field.

\begin{itemize}
\item When the support is small (\mbox{$0<T\ll \sigma$}), the SLEs asymptotically approach the values~\eqref{eq:ILES} that characterize the ILES at~$t_0$. The physical justification of this fact resides in the definition of the ILES at $t_0$, which minimizes the instantaneous energy density in~\eqref{eq:EnergyILESWY} obtained identifying the smearing function~$f(t)^2$ with the Dirac delta~\mbox{$\delta(t-t_0)$} in~\eqref{eq:Ek}. We might then say that the ILES at~$t_0$ is the limit for small supports around~$t_0$ of the SLEs. However, note that this limit is singular in the sense that the Dirac delta is not a smooth compact support function, so ILESs are not a particular example of a SLE. These conclusions are also valid for other times different from~$t_0$, as we have verified numerically.
\item We also find an asymptotic constant behaviour for large supports~\mbox{($T \gg \sigma$)}. This is consistent with the fact that the leading contributions to the smeared energy density are for times in the interval $[-\sigma,\sigma]$, and that the electric pulse decreases asymptotically. This limit defines a precise vacuum with a well-defined interpretation: the state which minimizes the energy density when it is smeared over the entire pulse. 
\end{itemize}

For other values of~$k$ we also distinguish analogue behaviours of~$W_{\bfk}(t_0)$ and~$Y_{\bfk}(t_0)$ for small and large supports. However, as we increase~$k$, the dependence on the support decreases. Indeed, the limit~\mbox{$k\rightarrow \infty$} corresponds to local flat spacetime with no electric field, thus all vacua tend to the Minkowski vacuum defined in~\eqref{eq:WYMinkowski}. Nevertheless, how fast or slow we reach the Minkowski vacuum strongly depends on each particular vacuum. We will analyse this in more detail in~\autoref{sec:NumberSLEs} when studying the number of created particles.

Finally, one might wonder why the noticeable dependence of the SLEs on the support of the test function for supports of the order of the characteristic length of the electric pulse seems absent in the case of loop quantum cosmology~\cite{MartinBenito2021}. Indeed, in that work, SLEs are described as independent of the support as long as it is large enough, which agrees with the large support convergent behaviour we observe. Let us then clarify that, although in loop quantum cosmology the equivalent to our potential is different, it also has a characteristic time scale (around the bounce), in which the variations of the potential are most important. This scale plays the same role as our $\sigma$, and there it should be less than a hundredth of a Planck time.\footnote{If we approximate the time-dependent mass in the equation of motion of cosmological perturbations in loop quantum cosmology by a P{\"o}schl-Teller potential, the equivalent to $\sigma$ is easily found as the time after the bounce at which the potential reduces to half its maximum.} Therefore, the considered supports in~\cite{MartinBenito2021} were already quite larger than this scale and the dependence of the SLE on them was minimal, and achieved convergence quickly. In general, the behaviour of SLEs in loop quantum cosmology most likely displays an intermediate regime as we observe in the Schwinger effect, though it corresponds to very small supports around the bounce, which are not physically interesting within the context of cosmology.

\section{Anisotropic power spectrum}\label{sec:Anisotropies}

In this section we consider the extension of the common notion of power spectrum in cosmological scenarios to the Schwinger effect. In addition, we are interested in studying in detail the anisotropies present in this electric background. Motivated by the works in anisotropic cosmologies as Bianchi I \cite{Agullo2022}, we introduce an expansion of the power spectrum in Legendre polynomials and analyse its multipolar contributions.

The Hadamard's elementary function is defined as the vacuum expectation value:
\begin{equation}
    \label{eq:Wightman} G(t,\bfx;t',\bfx')=\langle 0|\{ \hat{\Phi}^{\dagger}(t,\bfx), \hat{\Phi}(t',\bfx') \} |0\rangle= 2\int \frac{\text{d}^3\bfk}{(2\pi)^3} \ e^{-i\bfk\cdot (\bfx  -\bfx')} \Re \left[\phi_{\bfk}(t)\phi^*_{\bfk}(t')\right].
\end{equation} 
In the last equality we used the definition of the quantum field operator~$\hat{\Phi}(t,\bfx)$ in terms of the chosen modes $\phi_{\bfk}(t)$ given by \eqref{eq:FieldOperatorSchwinger}. Note that the Hadamard's elementary function only depends on the position vectors through the difference $\bfx-\bfx'$ because the electric field is spatially homogeneous. Writing the integral in \eqref{eq:Wightman} in spherical coordinates, we can integrate out the azimuthal angle. Indeed, we are assuming that the electric field is applied in the $z$ direction and thus it introduces anisotropy only in the polar angle $\theta$. In addition, taking the limit of coincidence $t\rightarrow t'$ yields 
\begin{equation}
    \lim_{t\rightarrow t'} G(t,\bfx;t',\bfx')=\int \frac{\text{d}k}{k} \int \text{d}(\cos{\theta}) \ e^{-i\bfk\cdot (\bfx  - \bfx')} \mathcal{P}(t,\bfk),
\end{equation}
where we defined the \gls{PowerSpectrum} as
\begin{equation}
    \mathcal{P}(t,\bfk)=\frac{k^3}{2\pi^2}|\phi_{\bfk}(t)|^2.
    \label{eq:PS}
\end{equation}

The power spectrum~\eqref{eq:PS} is determined by the choice of solutions~$\phi_{\bfk}(t)$ used to construct the quantum theory. More specifically, at a given time~$t_0$, it encodes the ambiguities associated with the selection of~$W_{\bfk}(t_0)$ in~\eqref{eq:ParametrizationICphi}, while remaining independent of~$Y_{\bfk}(t_0)$. Thus, the power spectrum at a fixed time does not encode all the information about the quantum vacuum. Furthermore, compared with the spectrum of $W_{\bfk}(t_0)$, the infrared power spectrum blurs the differences between different vacua as a consequence of the factor of $k^3$ in its definition~\eqref{eq:PS}.

\begin{figure}
\centering
    \includegraphics[width=0.7\textwidth]{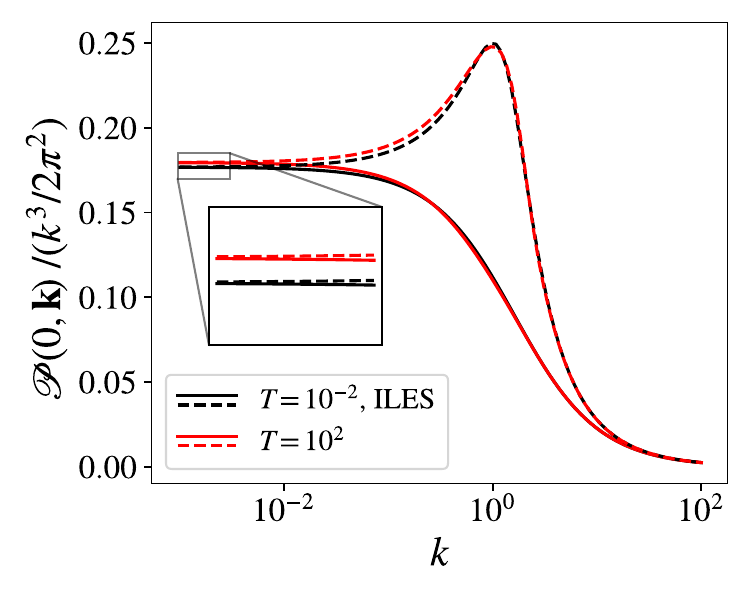}
    \caption{Power spectrum divided by $k^3/(2\pi^2)$ at $t= 0$ for a mode parallel (solid) and antiparallel (dashed) to the electric field and for SLEs with different supports. The power spectra for ILES coincide with those for the SLE of smallest support.}
    \label{fig:PS}    
\end{figure}

We show in~\autoref{fig:PS} the power spectrum~$\mathcal{P}(t_0,\bfk)$ divided by the factor~$k^3/2\pi^2$. This magnitude is computed for SLEs with smearing functions of the type~\eqref{eq:SmearingFunction} of sufficiently small ($T=10^{-2}\sigma$) and sufficiently large ($T=10^2\sigma$) supports.\footnote{These are chosen according to figure~\autoref{fig:WYIR}. This figure refers to a particular infrared mode, but we have verified that the two supports considered here are also sufficiently small and sufficiently large for intermediate and ultraviolet modes as well.} 
\begin{itemize}
    \item We see that all SLEs have the same infrared behaviour except for a constant. This is in agreement with reference~\cite{Banerjee2020}.
    \item In the ultraviolet, all vacua see a vanishing electric field at sufficiently short scales. Accordingly, they all converge to the same Minkowski vacuum at all times.
\end{itemize} 

To investigate the anisotropies we represent modes parallel and antiparallel to the direction of the electric field (i.e., $\theta=0$ and~$\theta=\pi$, respectively). Both the infrared and ultraviolet behaviours are oblivious to the direction of~$\bfk$. This is rooted in the angular dependence of the frequency in~\eqref{eq:Frequency}, \mbox{$\omega_\bfk(t)^2 = k^2 + 2qA(t)\cos{\theta}k + q^2A(t)^2 + m^2$}. Indeed, for 
\begin{equation}
    k \ll \frac{q^2 A(t)^2 + m^2}{2 |q A(t)|} \qquad \text{and for} \qquad k \gg 2 |q A(t)|,
\end{equation}
the angular contribution is negligible. Conversely, this defines an intermediate regime where the dependence on $\theta$ is important. Accordingly, in~\autoref{fig:PS} the difference between parallel and antiparallel modes is significant at these intermediate scales. Note that in this regime the effects of the anisotropy are much more relevant than that of different choices of SLE. Furthermore, the curves for  $\theta = \pi$ are non-monotonic in contrast with those for $\theta = 0$. Indeed, for positive $\cos \theta$, $\omega_\bfk(t_0)^2$ grows monotonously as $k$ increases,  leading to a power spectrum that monotonously decreases. On the other hand, for negative $\cos \theta$, ${\omega_\bfk}(t_0)^2$ presents a minimum at $k=q A(t_0) |\cos \theta|$, which translates into a maximum in the power spectrum around that point (in our case, $k = \sigma^{-1}$). Note that for our computations we have chosen $q$ and $A(t)$ to have same sign. Had we chosen them with opposite signs, the roles of $\theta = 0$ and $\theta = \pi$ would have been interchanged. 

We now expand the power spectrum \eqref{eq:PS} in the Legendre polynomials, $P_{\ell}(\cos{\theta})$, which form an orthonormal basis of square-integrable functions in $[-1,1]$:
\begin{equation}
    \mathcal{P}(t,\bfk)=\sum_{\ell=0}^{\infty} \mathcal{P}_{\ell}(t,k)P_{\ell}(\cos{\theta}),
\end{equation}
where the multipoles are given by
\begin{equation}
    \mathcal{P}_{\ell}(t,k)=\frac{2\ell+1}{2}\int_0^{\pi} \text{d}{(\cos{\theta})} \ \mathcal{P}(t,\bfk)P_{\ell}(\cos{\theta}).
\end{equation} 
Let us consider the multipolar contributions~$\ell\geq 1$ with respect to the isotropic monopole \mbox{$\ell=0$}, i.e., the coefficients
\begin{equation}
    g_{\ell}(t,k)=\frac{\mathcal{P}_{\ell}(t,k)}{\mathcal{P}_0(t,k)}.
\end{equation}

\begin{figure}
    \centering
    \includegraphics[width=0.7\textwidth]{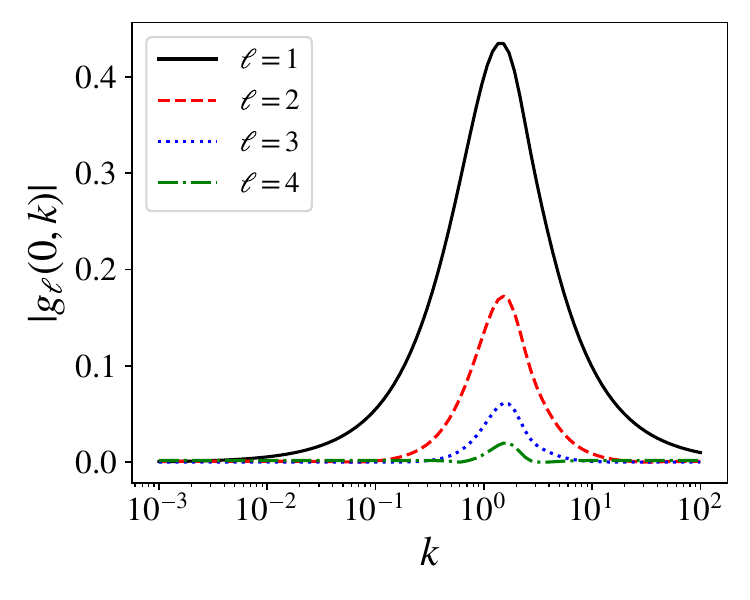}
    \caption{Absolute value of the contributions $g_{\ell}$ of the multipoles $\ell$ with respect to the monopole at $t = 0$ for a SLE with support $T=10^2\sigma$. Note that $g_{\ell}$ are negative for odd values of $\ell$ and positive for even~$\ell$. We use units $\sigma=1$.}
    \label{fig:gl}    
\end{figure}

We show in~\autoref{fig:gl} how the coefficients~$g_l(t,k)$ depend on the wavenumber~$k$ for the SLE with large support $T=10^2\sigma$. We verified that similar behaviours are obtained for smearing functions with different supports. We observe that the maximum contribution of all the multipoles with respect to the monopole happens precisely for the same scale, which is in the aforementioned intermediate regime identified also in~\autoref{fig:PS}. In addition, we confirm that the contribution of multipoles decreases asymptotically in both the infrared and the ultraviolet.

\section{Number of created particles} \label{sec:NumberSLEs}

As mentioned in the previous section, the power spectrum at a fixed time does not fully encapsulate all the information on the vacuum. In cosmology, this is usually the only relevant quantity as it is the only one that can be related with observations of the CMB. However, in general and especially in the context of the Schwinger effect, this can be complemented with the number of created particles in one vacuum with respect to a reference one. 

We take as the reference vacuum the state defined by the `in' solution~$\phi_{\bfk}^{\textrm{in}}(t)$ in~\eqref{eq:InSolution}, which behaves as a positive frequency plane wave in the asymptotic past. Any other choice of a basis of solutions~$\phi_{\bfk}(t)$ leads to a different quantum theory with its own notion of vacuum. The number of excitations per mode in a given vacuum state with respect to the `in' vacuum is determined by~\eqref{eq:NkSchwinger}, and can be written as
\begin{equation}
    \mathcal{N}_\bfk = |\beta^+_\bfk|^2 + |\beta^-_{\bfk}|^2 = 2\left| \phi_{\bfk}(t) \dot{\phi}_{\bfk}^{\textrm{in}}(t) - \phi_{\bfk}^{\textrm{in}}(t) \dot{\phi}_{\bfk}(t)\right|^2 .
\end{equation}
Noticeably, at each time~$t$ this depends on~$\phi_{\bfk}(t)$ as well as its derivative, and therefore encodes information on both~$W_{\bfk}(t)$ and~$Y_{\bfk}(t)$ of the parametrization~\eqref{eq:ParametrizationICphi} and~\eqref{eq:ParametrizationICDotphi}. However, it is still not fully descriptive of the vacuum, as it only depends on a combination of these two functions. As such, it may be used in addition to the power spectrum in order to characterize a given vacuum at a given time.

\begin{figure}
    \centering
    \includegraphics[width=0.7\textwidth]{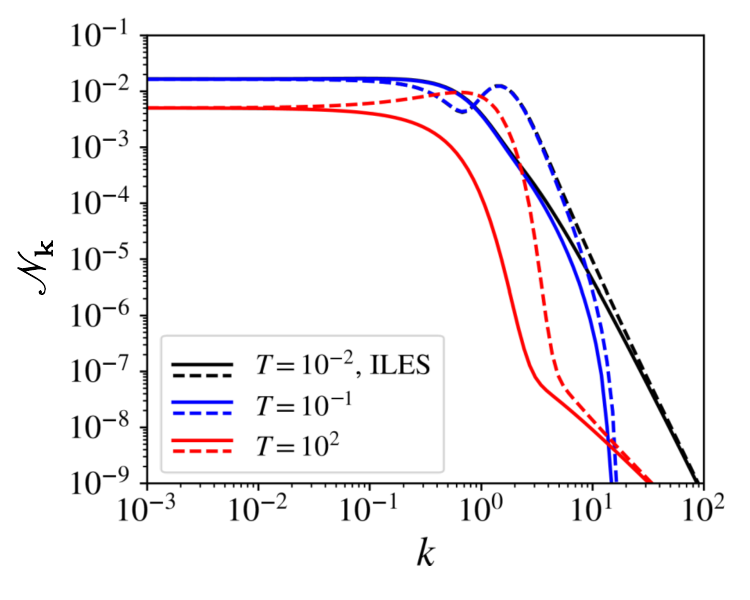}
    \caption{Number of created particles $\mathcal{N}_{\bfk}$ as a function of the module $k$ of the wavevector for SLEs with small and large supports, for modes parallel (solid lines) and antiparallel (dashed lines) to the electric field. For ILES at $t_0$, $\mathcal{N}_{\bfk}$ coincides with that of the SLE with the smallest support considered. We use units $\sigma=1$.}
    \label{fig:Nk}
\end{figure}

\autoref{fig:Nk} shows the behaviour of the number of created particles~$\mathcal{N}_{\bfk}$ for modes parallel and antiparallel to the electric field, as a function of the wavenumber $k$ and for SLEs of sufficiently small ($T = 10^{-2}\sigma$) and sufficiently large ($T = 10^{2}\sigma$) supports around $t_0$. Again, we identify the same infrared behaviour for all vacua, which are distinguished by a constant contribution. In the ultraviolet, however, each vacuum tends to the Minkowski state at a different rate. 

For small supports, the spectral particle number~$\mathcal{N}_{\bfk}$ of the SLE seems to agree with that of the ILES (see \autoref{sec:RoleSmearing}). However, for small enough scales, these states behave differently. To illustrate this separation, we have also represented a SLE with $T=10^{-1}\sigma$, whose~$\mathcal{N}_{\bfk}$ departs from that of the ILES at a lower (numerically achievable)~$k$. This behaviour is compatible with SLEs being of Hadamard type, while the ILES is not. In fact, Hadamard states are infinite-order adiabatic vacua~\cite{Pirk1993,Junker2002}, whose~$\mathcal{N}_{\bfk}$ decays with a power of $k$ proportional to its adiabatic order. Thus, for the ILES the~$\mathcal{N}_{\bfk}$ is not exponentially suppressed, decaying more slowly than SLEs for sufficiently ultraviolet modes, not depicted in~\autoref{fig:Nk}. Along these lines, the~$\mathcal{N}_{\bfk}$ for the SLE with large support $T = 10^2\sigma$ must also decay faster than that for the ILES, for sufficiently ultraviolet modes.

Finally, \autoref{fig:Nk} also shows the intermediate regime where anisotropies are important. As motivated in the previous section, we verify that in the infrared and ultraviolet, the particle number is isotropic. For intermediate scales, modes parallel to the electric field show a monotonic particle number, in contrast to antiparallel modes.

\begin{block}[note]
Solutions to the equation of motion are oscillatory, with increasing frequency after the maximum of the electric pulse, as well as for increasing $k$. Thus, the computation of the SLE becomes computationally demanding for large supports and large~$k$, as it requires the integration of oscillations with very short periods.
\end{block}

\section{Conclusions}
\label{sec:ConclusionsSLEs}

In~\cite{Olbermann2007}, SLEs were introduced in general cosmological spacetimes as the states that minimize the energy density, smeared along the trajectory of an isotropic observer. They were shown to be Hadamard states, and later proven to be good candidates for the vacuum of cosmological perturbations in models with a period of kinetic dominance prior to inflation~\cite{Banerjee2020}. Since then, they have been applied in the context of loop quantum cosmology \cite{MartinBenito2021,MartinBenito2021a}, where it was found that they heavily depend on the choice of smearing function only in regards to whether its support includes or excludes the bounce of loop quantum cosmology. Recently, they have also been applied to fermionic fields in a radiation-dominated CPT-invariant universe \cite{NadalGisbert2023}.

In this chapter, we have extended the construction of SLEs to general spatially homogeneous settings, with the emphasis on the Schwinger effect. To investigate the dependence of these SLEs on the choice of smearing function, we have considered regularized step-like smearing functions with a wide range of supports centred at the maximum of a Sauter-type electric pulse. We discern two asymptotic behaviours of SLEs. In the limit of small supports they behave as ILESs, which instantaneously minimize the energy density (although ILESs are not a particular case of SLEs, just a limiting behaviour). For very large supports the dependence on the support of the smearing function gradually disappears, thus determining in the limit a vacuum which minimizes the smeared energy density over the entire electric pulse. For supports of the order of the characteristic time scale of the electric pulse there is a non-trivial dependence. We have been able to draw parallels with what is observed in \cite{MartinBenito2021}, and conclude that the sizes of the support considered in that work already corresponded to the large support regime, which is why convergence is obtained quickly there and no non-trivial dependence on the smearing function is observed.

We have also calculated the power spectrum in the Schwinger effect, analogously to the usual definition in cosmology. We have shown that all SLEs have the same infrared behaviour except for a constant contribution, in agreement with \cite{Banerjee2020}. In the ultraviolet, all vacua tend to the Minkowski vacuum although at different rates. As the power spectra only depend on the configuration of the state, they all converge for large wavenumbers. However, as the particle number encodes information not only on the configuration of the state but also on its velocity, each vacuum leads to different decay rates when approaching short scales. In particular, we observe that the particle number for all SLEs decays faster than that for the ILES. This might be an indication of SLEs being Hadamard in the Schwinger effect.

Finally, we have analysed the anisotropy of the system. We find that in both the ultraviolet and the infrared regions, the anisotropies do not contribute to either the power spectrum or the number of created particles. An intermediate regime where they are most important has been identified.

\part[Quantum dynamics]{Quantum dynamics}
\label{part:QuantumDynamics}

\plainblankpage

\vspace*{\fill}
\begin{minipage}{0.8\textwidth}

Having established the framework for constructing a quantum theory of a scalar field in the presence of a homogeneous electric field in flat spacetime, we now explore how quantum time evolution can be implemented. Not all quantum theories admit a unitary evolution operator, but those that do define a physically relevant family of quantum vacua that allow for the implementation of unitary dynamics. This is the focus of \autoref{chap:QuantumUnitary}.

In the literature, the time evolution of the number of created particles is often described by an integro-differential equation: the standard quantum Vlasov equation. However, what is rarely emphasized is that using this equation implicitly assumes a specific choice of quantum vacuum with a well-defined physical meaning. In reality, many other physically relevant choices exist, each leading to a different evolution of the particle number. In \autoref{chap:GQVE}, we generalize the standard quantum Vlasov equation to account for ambiguities in the canonical quantization. In addition, we analyze its ultraviolet behavior for the family of quantum vacua allowing for unitary dynamics.

These chapters build upon the study in~\cite{AlvarezGQVE}, while their foundations are primarily based on~\cite{Garay2020} and~\cite{AlvarezUnitary}. The formalism has been significantly reformulated to simplify it and ensure consistency with the rest of the thesis.

\end{minipage}
\vspace*{\fill}
\plainblankpage
\chapter[Quantum unitary dynamics]{Quantum unitary dynamics}
\label{chap:QuantumUnitary}

\chaptermark{Quantum unitary dynamics}

In~\autoref{chap:ChoiceVacuum}, we saw that quantizing a field in flat spacetime in the presence of an electric field leads to a breakdown of time-translational invariance, resulting in an infinite number of possible quantizations compatible with the classical symmetries. To reduce this inherent ambiguity, it is crucial to impose additional physically motivated constraints on the quantization procedure. A well-established approach in the context of fields propagating in homogeneous cosmologies allows for the selection of a unique family of unitarily equivalent Fock representations, thereby ensuring physically equivalent quantizations~\cite{Corichi2006,Cortez2012,Cortez2015,Cortez2020,Cortez2020c}. This approach is based on the unitary implementation of the quantum field dynamics at all times.

If a classical system exhibits symmetry under a given transformation, then in the quantum theory, provided the vacuum remains invariant, it is possible to implement the transformation unitarily. If the symmetry is broken or the vacuum is not invariant, a weaker condition can still be imposed: requiring the transformation to be unitarily implementable~\cite{Cortez2021}. This, in particular, applies to time translations.

Enforcing unitary evolution at all finite times ensures that different quantizations remain physically equivalent throughout the evolution. This is analogous to the situation in quantum mechanics, where the Schrödinger, Heisenberg, and interaction pictures are all related by a unitary transformation, preserving physical observables. In the context of particle production, such as the Schwinger effect, this requirement is crucial: unitary implementation of the dynamics ensures that the number of created particles remains finite at all times.

Unitary dynamics at all finite times is a stronger condition than the usual one found in the literature \cite{Gavrilov1996,Wald1979}, which only requires that the $S$-matrix unitarily connects the asymptotic past and future states, once the external interaction has ceased. However, such approaches fail to eliminate the quantization ambiguities. Moreover, while an S-matrix formalism exists for general backgrounds~\cite{Wald1979}, it does not guarantee the existence of asymptotic free-particle states in non-trivial backgrounds~\cite{Wald1994}. This underscores the need for a more refined approach to quantization in the presence of time-dependent external fields.

Much of the analysis in this chapter originates from~\cite{Garay2020}, but it has been thoroughly reformulated to simplify the formalism and align the notation with the rest of the thesis. Some of these changes were already introduced in~\cite{AlvarezGQVE}. We will focus on the unitary dynamics criterion for a scalar field, as this will provide a solid foundation for understanding the subsequent chapters, which are also developed in the scalar case. A detailed study of the fermionic case can be found in~\cite{AlvarezUnitary}.

In~\autoref{sec:TimeEvolution}, we present the time evolution as a Bogoliubov transformation. In~\autoref{sec:UID}, we study the unitary implementation of the dynamics, which uniquely selects a physically meaningful equivalence class of quantizations. Finally, \autoref{sec:ConclusionsQuantumUnitary} summarises the main results of the chapter.

\section{Time evolution as a Bogoliubov transformation}
\label{sec:TimeEvolution}

In this section, we formalize a concept that frequently appears in the literature on the Schwinger effect: the time-dependent number of particles. Previously, in~\eqref{eq:Num}, we introduced a notion of particle number that compares two different quantizations, where the $\beta$-Bogoliubov coefficient is independent of time. However, this approach does not capture the fact that, in a time-dependent background, the very definition of particles can evolve dynamically.

To account for this, we adopt a time-dependent quantization scheme, defining a quantum theory at each instant~$\tau$ during the evolution. Instead of working with a single, fixed set of annihilation and creation variables~$a_\bfk$ and~$b_\bfk^*$, we introduce a new set~$a_\bfk(\tau)$ and~$b_\bfk^*(\tau)$ for each time~$\tau$. These variables, defined at time~$\tau$, can be compared to those at a reference time~$\tau_0$, denoted~$a_\bfk(\tau_0)$ and~$b_\bfk^*(\tau_0)$. They are related by a Bogoliubov transformation:
\begin{equation}
    \mqty(a_\bfk(\tau)\\b_\bfk^*(\tau))= \mqty(\alpha_\bfk(\tau) & \beta_\bfk^*(\tau) \\ \beta_\bfk(\tau) & \alpha_\bfk^*(\tau)) \mqty(a_\bfk(\tau_0) \\ b_\bfk^*(\tau_0) ).
    \label{eq:BogoliubovVariablestautau0}
\end{equation}
While this transformation formally resembles any other Bogoliubov transformation relating two distinct quantizations, its conceptual significance is different: here, we associate the transformation with a dynamical evolution in time. Varying~$\tau$ allows us to track how the quantization changes as the system evolves. The Bogoliubov transformation now explicitly depends on time, as each instant~$\tau$ defines a new transformation between successive quantizations.

As discussed in~\autoref{chap:ChoiceVacuum}, the choice of annihilation and creation variables at each time~$\tau$ is equivalent to selecting a set of mode functions~\mbox{$\phi^\tau_\bfk = \phi^\tau_\bfk(t)$}, which are solutions to the harmonic oscillator equations~\eqref{eq:Harmonic}. This, in turn, is equivalent to specifying initial conditions at time~$\tau$:
\begin{equation}
    \phi^\tau_\bfk(t=\tau) = \zeta_\bfk(\tau), \qquad \dot{\phi}^\tau_\bfk(t=\tau) = \rho_\bfk(\tau).
    \label{eq:ICtaut}
\end{equation}
The functions~$\zeta_\bfk(\tau)$ and~$\rho_\bfk(\tau)$ act as \gls{InitialConditionDistributors}, determining the values of the modes and their derivatives at each time~$\tau$. These functions fully specify the quantum theory constructed at each~$\tau$, including the associated notions of particles, antiparticles, and quantum vacuum~$|0\rangle^\tau$.

At the reference time~$\tau_0$, we select a particular set of modes~$\phi_\bfk^{\tau_0}$, which determine the corresponding annihilation and creation variables~$a_\bfk(\tau_0)$ and~$b_\bfk^*(\tau_0)$. The ambiguity in choosing these reference modes can often be resolved by imposing physically motivated conditions. For instance, as discussed in~\autoref{sec:ParametrizationVacua}, if the electric field is switched off in the asymptotic past, the system locally recovers Poincar\'{e} symmetry as~$t\rightarrow -\infty$. In this case, a preferred choice is to set~$\tau_0\rightarrow -\infty$ and select the `in' quantum vacuum, where~$\phi_\bfk^{\tau_0}$ behaves as a positive frequency plane wave in the asymptotic past. This selection ensures the preservation of classical Poincar\'{e} symmetry in that regime and uniquely determines~$\phi_{\bfk}^{\tau_0}(t)$ for all $t$.

At a later time~$\tau$, we choose another set of modes~$\phi_\bfk^\tau$, defining new variables~$a_\bfk(\tau)$ and~$b_\bfk^*(\tau)$. The Bogoliubov coefficients~\eqref{eq:BogoliubovVariablestautau0} that relate these two sets at~$\tau_0$ and~$\tau$ are given by~\eqref{eq:BogoliubovCoeffsphis}:
\begin{equation}
    \alpha_\bfk(\tau) = i\left[ \phi_\bfk^{\tau *}(t)\dot{\phi}^{\tau_0}_\bfk(t) - \phi^{\tau_0}_\bfk(t) \dot{\phi}_\bfk^{\tau *}(t) \right], \qquad \beta_\bfk(\tau) = i\left[ \phi^{\tau_0}_\bfk(t) \dot{\phi}_\bfk^{\tau}(t) - \phi_\bfk^{\tau}(t) \dot{\phi}^{\tau_0}_\bfk(t) \right].
\end{equation}
Since these coefficients are independent of the specific time~$t$ at which they are evaluated, we can conveniently compute them at~$t=\tau$. Using the initial conditions~\eqref{eq:ICtaut}, we obtain:
\begin{equation}
    \alpha_\bfk(\tau) = i\left[ \zeta_\bfk^*(\tau) \dot{\phi}_\bfk^{\tau_0}(\tau) - \phi_\bfk^{\tau_0}(\tau)\rho_\bfk^*(\tau) \right], \qquad \beta_\bfk(\tau) = i\left[ \phi_\bfk^{\tau_0}(\tau) \rho_\bfk(\tau) - \zeta_\bfk(\tau) \dot{\phi}_\bfk^{\tau_0}(\tau) \right].
    \label{eq:Bogoliubovtau0tau}
\end{equation}
The \gls{SpectralNumberOfCreatedExcitationsBetweenTau0AndTau} is then defined by
\begin{equation}
    \mathcal{N}_\bfk(\tau) = 2|\beta_\bfk(\tau)|^2.
    \label{eq:Numtau0tau}
\end{equation}
This quantity depends crucially on both the reference vacuum, determined by the functions~$\phi_\bfk^{\tau_0}$, and the choice of initial condition distributors~$\zeta_{\bfk}$ and~$\rho_{\bfk}$.

Once the Bogoliubov coefficients~\eqref{eq:Bogoliubovtau0tau}, which relate the quantizations associated with times~$\tau_0$ and~$\tau$, are defined, we can reinterpret~$\tau$ as an arbitrary time variable. Under this interpretation, these coefficients acquire an explicit dependence on~$\tau$. Consequently, while their derivative with respect to~$t$ remains zero, their derivative with respect to~$\tau$ does not. The same applies to the particle number~\eqref{eq:Numtau0tau}. In~\autoref{chap:GQVE}, we explicitly derive an integro-differential equation governing the evolution of the particle number~$\mathcal{N}_\bfk(\tau)$ as a function of~$\tau$.

\begin{block}[note]
While we have used a dot to denote differentiation with respect to time~$t$, as in~$\dot{\phi}^{\tau_0}_\bfk(t)$ or~$\dot{\omega}_\bfk(t)$, this notation simply indicates differentiation with respect to the argument itself. When these quantities are evaluated at~$t=\tau$, as in $\dot{\phi}_\bfk(\tau)$ or~$\dot{\omega}_\bfk(\tau)$, and~$\tau$ subsequently treated as an independent variable, the dot notation continues to denote differentiation, now with respect to~$\tau$.
\end{block}

We can parametrize the initial condition distributors using the reference parametrization provided in~\eqref{eq:ParametrizationICphi} and~\eqref{eq:ParametrizationICDotphi}. Explicitly:
\begin{equation}
    \zeta_\bfk(\tau) = \frac{1}{\sqrt{2W_\bfk(\tau)}}e^{-i\varphi_\bfk(\tau)}, \qquad
    \rho_\bfk(\tau) = \sqrt{\frac{W_\bfk(\tau)}{2}}\left[ Y_\bfk(\tau) - i \right]e^{-i\varphi_\bfk(\tau)}.
    \label{eq:ParametrizationICDistributors}
\end{equation}
However, it is important to note that~$\zeta_{\bfk}$ is not necessarily a solution to the harmonic oscillator equation~\eqref{eq:Harmonic}. Instead, the actual solutions to the equation of motion are the family of functions~$\phi_\bfk^\tau$, where each~$\phi_\bfk^\tau$ corresponds to a specific time~$\tau$, determined by the initial conditions~\eqref{eq:ICtaut}.

\begin{block}[example]
In the literature of the quantum kinetic approach in the Schwinger effect, which we will study in \autoref{chap:GQVE}, a particular choice for the initial condition distributors is implicitly made~\cite{Kluger1998,Schmidt1998,Aleksandrov2022,Anderson2014,Ruffini2010,Roberts2000,Dunne2009,Dumlu2011,Hebenstreit2009}: at each time~$\tau$, the set of modes~$\phi_\bfk^\tau$ is chosen as the ILES at time~$\tau$~\eqref{eq:ILES}:
\begin{equation}
    \zeta_\bfk(\tau) = \frac{1}{\sqrt{2\omega_\bfk(\tau)}}, \qquad \rho_\bfk(\tau) = -i\sqrt{\frac{\omega_\bfk(\tau)}{2}}.
    \label{eq:ICdistributorsILES}
\end{equation}
Note that the function~$\zeta_\bfk$, when viewed as a function of~$\tau$, does not satisfy the harmonic oscillator equation~\eqref{eq:Harmonic}. This choice leads to the particular expression for the time-dependent number of created particles and antiparticles found, for example, in~\cite{Fedotov2011a}:
\begin{equation} 
    \mathcal{N}^{\text{ILES}}_{\bfk}(\tau)=\omega_{\bfk}(t)|\phi_{\bfk}^{\tau_0}(\tau)|^2+\frac{1}{\omega_{\bfk}(\tau)}|\dot{\phi}_{\bfk}^{\tau_0}(\tau)|^2-1.
\end{equation}
It is essential to understand that this last definition of the number of created excitations is assuming a precise definition of what we call particles and antiparticles throughout the evolution of the system: the notions determined by the ILES at each time of the evolution. Our formalism allows us to write the alternative version of this equation, \eqref{eq:Numtau0tau}, when we select any other initial condition distributors, such as, for example, any adiabatic order initial conditions.
\end{block}

\section{Unitary implementation of the dynamics}
\label{sec:UID}

In this section, we characterize the quantizations that unitarily implement the quantum field dynamics. We review and adapt the results from~\cite{Garay2020} to our present formalism. The problem of unitary implementation has also been extensively studied in cosmological settings~\cite{Corichi2006,Cortez2012,Cortez2015,Cortez2020,Cortez2021,Cortez2020c}. Unlike standard approaches where~$Y_{\bfk}$ is often set to zero, we will also examine the restrictions imposed on this function. This is particularly relevant because the number of excitations explicitly depends on it (see~\eqref{eq:NumWY}).

The time-dependent Bogoliubov transformation~\eqref{eq:BogoliubovVariablestautau0} encodes the evolution of the quantum field. To ensure unitary implementation, we seek a unitary operator~$\hat{B}(\tau)$ acting on the whole Fock space that satisfies:
\begin{equation}
    \begin{pmatrix} \hat{B}(\tau) \hat{a}_\bfk(\tau) \hat{B}(\tau)^{-1} \\ \hat{B}(\tau) \hat{b}^\dagger_\bfk(\tau) \hat{B}(\tau)^{-1} \end{pmatrix}
    =
    \begin{array}{c}
        \begin{pmatrix} \alpha_\bfk(\tau) & \beta_\bfk^*(\tau) \\ \beta_\bfk(\tau) & \alpha_\bfk^*(\tau) \end{pmatrix}
        \begin{pmatrix} \hat{a}_\bfk(\tau_0) \\ \hat{b}_\bfk^\dagger(\tau_0) \end{pmatrix}
    \end{array}.
\end{equation}
However, achieving unitary implementation is non-trivial, and only specific choices of initial condition distributors~$\zeta_{\bfk}$ and~$\rho_{\bfk}$ ensure that the Bogoliubov transformation is unitarily implementable at the quantum level.

As discussed in~\autoref{sec:Bogoliubov}, a necessary and sufficient condition for a Bogoliubov transformation to be unitarily implementable is that the total number of excitations~\eqref{eq:Num} remains finite. This condition is equivalent to requiring that the following integral:
\begin{equation}
    \int \text{d}^3\bfk \ \mathcal{N}_\bfk(\tau) = 2\int_0^{2\pi}\text{d}\varphi \int_0^{\pi} \text{d}\theta  \ \sin{\theta} \int_0^{\infty}\text{d}k \ k^2|\beta_\bfk(\tau)|^2,
    \label{eq:NumUID}
\end{equation}
remains finite at each finite time~$\tau$. Since we are dealing with massive scalar fields, this integral does not suffer from infrared divergences. The integrability of~$|\beta_\bfk(\tau)|^2$ is ensured if and only if, in the ultraviolet limit (\mbox{$k \rightarrow \infty$}), $\beta_\bfk(\tau)$ decays strictly faster than $k^{-3/2}$. For polynomial decays, the asymptotic behaviour must satisfy
\begin{equation} 
    \beta_{\bfk}(\tau)=\order{k^{-\lambda}},\qquad \text{for some} \qquad \lambda>3/2,
    \label{eq:UnitaryCondition}
\end{equation}
at all finite times~$\tau$ and for all directions~$(\theta,\varphi)$.

Due to the anisotropy of the Schwinger effect, the ultraviolet behaviour of~$\beta_{\bfk}(\tau)$ depends on the direction in which we take the large~$k$ limit. From~\eqref{eq:Frequency}, the time derivative of the frequency yields the leading-order contribution:
\begin{itemize}
    \item For generic directions with constant~\mbox{$\theta\neq \pi/2$}: ~\mbox{$\dot{\omega}_{\bfk}(\tau)=\order{k^0}$}.
    \item In the direction orthogonal to the vector potential ($\theta=\pi/2$): \mbox{$\dot{\omega}_{\bfk}(\tau)=\order{k^{-1}}$}.
\end{itemize}

Remember that~$\beta_{\bfk}(\tau)$ depends both on the  particular reference modes~$\phi_{\bfk}^{\tau_0}$ and the initial condition distributors~$\zeta_{\bfk}$ and~$\rho_{\bfk}$. As we said before, in the most realistic case in which the electric field is switched off in the asymptotic past, only the selection of~$\phi_{\bfk}^{\tau_0}$ as the associated with the `in' quantum vacuum preserves Poincar\'{e} symmetry locally in the past. Furthermore, assuming general mild conditions on the time dependence of the frequencies\footnote{In the scalar Schwinger effect, a sufficient condition to satisfy this mild condition is that $\dot{\omega}_{\bfk}(t)/\omega_{\bfk}(t)$ both remains finite and does not change its sign an infinite number of times in each closed interval of time~\cite{Garay2020,AlvarezUnitary}.}, reference~\cite{Garay2020} proves that this `in'  solution behaves in the ultraviolet as
\begin{equation} 
    |\phi_{\bfk}^{\tau_0}(t)|^2=\order{k^{-1}}, \quad \dot{\phi}^{\tau_0}_{\bfk}(t)=i\left[-\omega_{\bfk}(t)+\Lambda_{\bfk}(t)\right]\phi_{\bfk}^{\tau_0}(t),
    \label{eq:zUV}
\end{equation}
where $\Lambda_{\bfk}(t)$ converges to zero at least as fast as $\order{k^{-1}}$ for all directions with $\theta = \text{constant}$.

Once~$\phi_{\bfk}^{\tau_0}$ is fixed, let us characterize the functions~$\zeta_{\bfk}$ and~$\rho_{\bfk}$ which verify the unitary dynamics condition~\eqref{eq:UnitaryCondition}. Using~\eqref{eq:Bogoliubovtau0tau} and~\eqref{eq:zUV}, as well as the parametrization of the initial condition distributors~\eqref{eq:ParametrizationICDistributors}, we can write~$\beta_{\bfk}(\tau)$ as
\begin{equation} 
    \beta_{\bfk}(\tau)=\left\{ \sqrt{\frac{W_{\bfk}(\tau)}{2}}\left[1+iY_{\bfk}(\tau)\right]
    +\frac{1}{\sqrt{2W_{\bfk}(\tau)}}\left[-\omega_{\bfk}(\tau)+\Lambda_{\bfk}(\tau)\right] \right\} e^{-i\varphi_{\bfk}(\tau)}\phi^{\tau_0}_{\bfk}(\tau).
    \label{eq:betaLO}
\end{equation}
We see that both its real and its imaginary parts are $\order{k^{-\lambda}}$ if and only if $W_{\bfk}(\tau)$ and $Y_{\bfk}(\tau)$ behave in the ultraviolet as
\begin{equation} 
    W_{\bfk}(\tau)=\omega_{\bfk}(\tau)\big[1+\order{k^{-\gamma}}\big],   \qquad Y_{\bfk}(\tau)=\order{k^{-\eta}}, \qquad  \gamma,\eta>3/2,
    \label{eq:UID}
\end{equation}
for each finite time $\tau$ and for almost all $\bfk$. These two conditions characterize the choice of $(\zeta_{\bfk}(\tau),\rho_{\bfk}(\tau))$ that allow for a unitary implementation of the dynamics. 

We now analyse whether the examples of quantum vacua introduced in \autoref{chap:ChoiceVacuum} allow for a unitary implementation of the dynamics:
\begin{itemize}
    \item The ILESs, defined by the initial conditions at time~$t_0$ given in~\eqref{eq:ILES}, allow for unitary dynamics, since
\begin{equation}
    W^{\text{ILES}}_{\bfk}(\tau)=\omega_{\bfk}(\tau), \qquad Y^{\text{ILES}}_{\bfk}(\tau)=0.
\end{equation}
    \item All adiabatic vacua of any order allow for a unitary implementation of the dynamics, as they exhibit the ultraviolet behaviour:
\begin{equation}
    W^{(n)}_{\bfk}(\tau)=\omega_{\bfk}(\tau)\left[1+\order{k^{-3}}\right], \qquad Y^{(n)}_{\bfk}(\tau)=\order{k^{-2}}.
\end{equation}
    \item The standard Minkowski plane waves characterized by~\eqref{eq:WYMinkowski} fail to ensure a unitary implementation of the dynamics, since:
\begin{equation}
    W^{\text{M}}_\bfk(\tau) = \sqrt{k^2 + m^2} = \omega_{\bfk}(\tau)\big[1+\order{k^{-1}}\big], \qquad Y^{\text{M}}_\bfk(\tau) = 0.
    \label{eq:WYMinkowskiVacuum}
\end{equation}
However, this Minkowski quantum vacuum is recurrently used in the literature when an electric field is switched on (see, for instance, references~\cite{Kim2011,Huet2014}). While it is true that using Minkowski modes in the Schwinger effect yields finite values of~$\mathcal{N}_{\bfk}(\tau)$ when the electric field is switched on, their total sum diverges~\cite{Ruijsenaars1977a}.
\end{itemize} 

\subsection*{Uniqueness of the quantization}

To what extent do the requirements of symmetry preservation and unitary implementation of time evolution reduce the ambiguity in the selection of the complex structure? In particular, we aim to determine whether quantum representations that admit a unitary implementation of the dynamics are unitarily equivalent.

To this end, let~$(a_\bfk(\tau),b_\bfk(\tau))$ and $(\widetilde{a}_\bfk(\tau),\widetilde{b}_\bfk(\tau))$ denote two sets of time-dependent annihilation and creation operators, each admitting a unitary implementation of the dynamics. These two sets are related by a Bogoliubov transformation of the form: 
\begin{equation}
    \mqty(a_\bfk(\tau)\\b_\bfk^*(\tau))= \mqty(\mathcal{A}_\bfk(\tau) & \mathcal{B}_\bfk^*(\tau) \\ \mathcal{B}_\bfk(\tau) & \mathcal{A}_\bfk^*(\tau)) \mqty(\widetilde{a}_\bfk(\tau) \\ \widetilde{b}_\bfk^*(\tau) ),
\end{equation}
where the time-dependent Bogoliubov coefficients~$\mathcal{B}_\bfk(\tau)$ can be expressed in terms of the initial condition distributors \mbox{$(\zeta_\bfk(\tau),\rho_\bfk(\tau))$} and \mbox{$(\widetilde{\zeta}_\bfk(\tau),\widetilde{\rho}_\bfk(\tau))$} associated with non-tilde and tilde complex structures, respectively:
\begin{equation}
\mathcal{B}_\bfk(\tau) = i\left[ \widetilde{\zeta}_\bfk(\tau) \rho_\bfk(\tau) - \zeta_\bfk(\tau) \widetilde{\rho}_\bfk(\tau) \right].
\end{equation}

By substituting the parametrizations of the initial condition distributors~\eqref{eq:ParametrizationICDistributors} in terms of the real pairs $(W_\bfk(\tau),Y_\bfk(\tau))$ and $(\widetilde{W}_\bfk(\tau),\widetilde{Y}_\bfk(\tau))$, we obtain:
\begin{equation}
|\mathcal{B}_\bfk(\tau)|^2 = \frac{W_\bfk(\tau)}{4\widetilde{W}_\bfk(\tau)}\left[ Y_\bfk(\tau)^2 +1 \right] + \frac{\widetilde{W}_\bfk(\tau)}{4W_\bfk(\tau)}\left[ \widetilde{Y}_\bfk(\tau)^2 +1 \right] - \frac{1}{2}Y_\bfk(\tau) \widetilde{Y}_\bfk(\tau) - \frac{1}{2}.
\label{eq:lambdatau}
\end{equation}
By assumption, both quantizations allow for unitary dynamics. Consequently, the functions \mbox{$(W_\bfk(\tau),Y_\bfk(\tau))$} and \mbox{$(\widetilde{W}_\bfk(\tau),\widetilde{Y}_\bfk(\tau))$} converge in the ultraviolet according to~\eqref{eq:UID}. Substituting this asymptotic behaviour into~\eqref{eq:lambdatau}, we find that for large values of~$k$:
\begin{equation}
    |\mathcal{B}_\bfk(\tau)| = \order{k^{-\lambda}}, \qquad \text{with} \qquad \lambda > 3/2.
\end{equation}
According to~\eqref{eq:UnitaryCondition}, this implies that $|\mathcal{B}_\bfk(\tau)|^2$ is integrable with respect to~$\bfk$ for any~$\tau$, and therefore, both quantizations are unitarily equivalent. 

Consequently, the quantizations that both preserve the symmetries of the classical theory and allow for a unitary implementation of the dynamics form a unique, unitarily equivalent family.

\subsection*{Fermions and cosmological backgrounds}

In~\cite{AlvarezUnitary}, we also study the unitary implementation of the dynamics in the Schwinger effect, but for fermionic fields instead of scalars. Although the Dirac formalism introduces some differences, the underlying principles remain essentially the same. In particular, we derive a characterization of the fermionic quantizations that allow for a unitary implementation of the dynamics, analogous to the scalar case result in~\eqref{eq:UID}. These quantizations also form a unique unitarily equivalent family of vacua. As in the scalar case, the Minkowski quantum vacuum does not belong to this family when an electric field is present.

In isotropic cosmological spacetimes such as FLRW backgrounds, the vacua that allow for a unitary implementation of the dynamics also form a unique unitarily equivalent family~\cite{Corichi2006,Cortez2012,Cortez2015,Cortez2020,Cortez2020c}. However, unlike in the presence of a background electric field, the Minkowski quantum vacuum \textit{does} belong to this family in the cosmological case. This distinction can be justified as follows. As we will discuss in~\autoref{chap:InOut}, once the scalar field is rescaled, the mode equations again reduce to harmonic oscillators with time-dependent frequencies. In cosmological settings, these frequencies typically take the form $\sqrt{k^2 + m(\eta)^2}$, where~$\eta$ denotes conformal time and~$m(\eta)$ is independent of~$\bfk$. This contrasts with the Schwinger case~\eqref{eq:Frequency}, where the frequency includes an anisotropic term $2qA(t)k \cos{\theta}$, introducing a linear $k$-dependence absent in the isotropic case. Consequently, in the FLRW case, the Minkowski quantum vacuum can be expressed in terms of its characteristic time-dependent frequency as:
\begin{equation}
W_\bfk^{\text{M}}(\eta) = \sqrt{k^2 + m(\eta)^2} \left[ 1 + \order{k^{-2}} \right], \qquad Y_\bfk^{\text{M}}(\eta)=0,
\label{eq:WYMinkowskiVacuumCosmology}
\end{equation}
which differs from the ultraviolet behaviour of the Minkowski vacuum in the Schwinger effect, given in~\eqref{eq:WYMinkowskiVacuum}. Still, a similar asymptotic analysis shows that the functions $W_\bfk(\eta)$ and $Y_\bfk(\eta)$ satisfy the ultraviolet conditions in~\eqref{eq:UID} in both scenarios. The faster decay of $W_\bfk^{\text{M}}(\eta)$ in the cosmological case ensures that the Minkowski vacuum allows for a unitary implementation of the dynamics---unlike in the electric field case.

\section{Conclusions}
\label{sec:ConclusionsQuantumUnitary}

In the study of a massive charged scalar field coupled to a spatially homogeneous electric field, we have dealt with the reduction of the ambiguities in the process of canonical quantization. In particular, for those complex structures that preserve the symmetries of the system (the translational invariance due to the homogeneity of the external field and the decoupling between the modes in the equations of motion), we have required that they allow for a unitary implementation of the dynamics.  This requirement serves two main purposes: ensuring the physical equivalence of the quantizations throughout the evolution of the vacuum, and guaranteeing a finite total number of created particles at any finite time. 

The unitary implementation of the quantum dynamics restricts the behaviour of the functions $W_{\bfk}(\tau)$ and $Y_{\bfk}(\tau)$, which have to decay sufficiently fast in the ultraviolet regime. The infinite possibilities for the selection of this function generates a family of unitarily equivalent complex structures characterized by a well-defined total number of created particles (i.e., the sum over all~$\bfk$ contributions) at finite times. Thus, in this chapter we do not propose a unique candidate of this observable, but a selection of unitarily equivalent ones. However, it is crucial to note that each specific selection within this family yields a different total particle number.

\plainblankpage
\chapter[Generalized quantum Vlasov equation]{Generalized quantum Vlasov equation}
\label{chap:GQVE}

\chaptermark{Generalized quantum Vlasov equation}

In the study of classical non-equilibrium physical systems, kinetic theory  has been a very successful tool~\cite{Liboff2003}. In particular, when describing a system composed by identical particles, the starting point in this theory is the Liouville equation for the joint probability distribution of the entire system. If we assume that particles are weakly correlated, we can deduce a closed equation of motion for the probability distribution of each individual particle: the so-called classical Vlasov equation. This equation does not consider collisions between particles. This can be accomplished with a more general but complicated approximation: the Boltzmann kinetic equation. 

A generalization to QFT of the classical Vlasov equation should contemplate particle creation. This is done in the context of the quantum kinetic approach. The widely accepted proposal, based on incorporating a particle creation term, is the so-called \acrfull{QVE}: an integro-differential equation for the time-dependent number of particles and antiparticles~\eqref{eq:Numtau0tau}. In the context of the Schwinger effect, this equation was first presented in \cite{Kluger1998} for scalar charged fields under a spatially homogeneous and time-dependent external electric field. Later, its extension to fermionic quantum fields was proposed in \cite{Schmidt1998}. This equation and its formalism has been used in a wide range of frameworks, including continuum strong quantum chromodynamics \cite{Roberts2000}, electron-positron pair creation in QED (from nuclei phenomena to black hole physics)~\cite{Ruffini2010}, laser technology~\cite{Dunne2009a,Dumlu2011,Hebenstreit2009}, or in cosmology considering a de Sitter spacetime \cite{Anderson2014,Habib2000}. 

By using this QVE, one implicitly adopts a specific prescription for the initial condition distributors---namely, the one defined by the ILES at each moment in the evolution (as illustrated in~\autoref{sec:TimeEvolution}). Moreover, in the literature on the quantum kinetic approach, this QVE is often presented as \textit{the} equation governing the evolution of \textit{the} number of created particles. However, this can be misleading, as we have repeatedly emphasized that ambiguities exist in the canonical quantization. The quantity~$\mathcal{N}_\bfk(\tau)$ is not uniquely defined but depends on the choice of the initial condition distributors.

The primary goal of this chapter is to extend the quantum kinetic approach framework to accommodate arbitrary vacua, thereby deriving a generalized QVE. This generalization will allow us to analyse particle creation beyond the specific case of ILES. Later, we will restrict this generalized QVE by particularizing it to adiabatic vacua~\cite{Habib2000}. For definiteness, we will consider a charged scalar field in the presence of a spatially homogeneous but time-dependent electric field, although extensions to other homogeneous systems, such as quantum matter fields in FLRW spacetimes, follow straightforwardly.

We will then restrict our generalized QVE to the unique family of vacua associated with the quantizations that unitarily implement the dynamics. We will see that there is an interesting connection between the usual QVE and its generalization to modes unitarily implementing the dynamics: under certain conditions, the former is precisely the leading order of the latter in the ultraviolet regime.  This will allow us to propose a more strict criterion for reducing the ambiguity in the quantization based on the ultraviolet behaviour of the generalized QVE. 

This chapter is primarily based on the publication~\cite{AlvarezGQVE}, with several adjustments made to the notation to ensure consistency with the rest of the thesis and to simplify the formalism. Its structure is as follows. In~\autoref{sec:GQVE}, we obtain the generalization to arbitrary quantizations of the QVE. In~\autoref{sec:GQVEUID}, we specialize our generalized QVE to modes satisfying the unitary dynamics criterion. We also propose an additional criterion for reducing the quantization ambiguities. Finally, we summarize the results in~\autoref{sec:ConclusionsGQVE}.

\section{Generalized quantum Vlasov equation}
\label{sec:GQVE}

In the following, we deduce a differential equation for the time-dependent number of created particles for which, unlike~\eqref{eq:Numtau0tau}, there is no need to solve the harmonic oscillator equation with time-dependent frequency first. Of course, this equation, just like~\eqref{eq:Numtau0tau}, will strongly depend on the particular choices of the initial condition distributors~$\zeta_{\bfk}$ and~$\rho_{\bfk}$.  

The evolution of the time-dependent number of excitations is governed by the dynamics of the Bogoliubov coefficients. Therefore, it is useful to derive explicit time evolution equations (with respect to~$\tau$) for both~$\alpha_{\bfk}(\tau)$ and~$\beta_{\bfk}(\tau)$. To obtain these equations, we differentiate~\eqref{eq:Bogoliubovtau0tau} with respect to~$\tau$ and use the harmonic oscillator equation~\eqref{eq:Harmonic}. Finally, by inverting the relations in~\eqref{eq:Bogoliubovtau0tau}, we arrive at
\begin{equation}
    \frac{\text{d}}{\text{d}\tau} \mqty(\alpha_{\bfk}(\tau)\\\beta_{\bfk}(\tau))=
    i\mqty(-s_{\bfk}(\tau)+\frac{\text{d}\varphi_{\bfk}}{\text{d}\tau} &r_{\bfk}(\tau)^*e^{2i\varphi_{\bfk}(\tau)}\\-r_{\bfk}(\tau)e^{-2i\varphi_{\bfk}(\tau)}& s_{\bfk}(\tau)-\frac{\text{d}\varphi_{\bfk}}{\text{d}\tau})\mqty(\alpha_{\bfk}(\tau)\\\beta_{\bfk}(\tau)).
     \label{eq:BogoliubovEvolution}
\end{equation}
Here, the real time-dependent function~$s_{\bfk}(\tau)$ is given by the functions~$W_\bfk(\tau)$ and~$Y_\bfk(\tau)$, which parametrize the initial condition distributors via~\eqref{eq:ParametrizationICDistributors}:
\begin{equation} \label{eq_s}
    s_{\bfk}=\frac{\omega_{\bfk}^2}{2W_{\bfk}}+\frac{1}{2}\left[\frac{\text{d}Y_\bfk}{\text{d}\tau}+W_{\bfk}\left(1+Y_{\bfk}^2\right)\right]+\frac{Y_{\bfk}}{2W_{\bfk}}\frac{\text{d}W}{\text{d}\tau}.
\end{equation}
The time-dependent function $r_{\bfk}$ is determined by its real and imaginary parts, $\mu_{\bfk}$ and $\nu_{\bfk}$, respectively:
\begin{equation} \label{eq_AB}
    \mu_{\bfk}=W_{\bfk}-s_{\bfk}, \qquad
    \nu_{\bfk}=\frac{1}{2W_{\bfk}}\frac{\text{d}W_\bfk}{\text{d}\tau}+W_{\bfk}Y_{\bfk}, \qquad r_{\bfk}=\mu_{\bfk}+i\nu_{\bfk}.
\end{equation}
We have deliberately eliminated the dependence on the phase $\varphi_{\bfk}$ in $s_{\bfk}$ and $r_{\bfk}$, extracting it explicitly in~\eqref{eq:BogoliubovEvolution}. Equations~\eqref{eq:BogoliubovEvolution} match the results found in~\cite{Habib2000}, up to an appropriate change of variables.

Once we have derived the evolution equations, we can generalize the approach presented in~\cite{Kluger1998}. By differentiating~$|\beta_{\bfk}(\tau)|^2$ and applying~\eqref{eq:BogoliubovEvolution}, it follows that
\begin{equation} 
    \frac{\text{d}}{\text{d}\tau} \mathcal{N}_{\bfk}(\tau)=4\Im{e^{-2i\varphi_{\bfk}(\tau)}r_{\bfk}(\tau)\mathcal{M}_{\bfk}(\tau)},
    \label{eq:dotN}
\end{equation}
where we have taken advantage of the real character of $s_{\bfk}$ and introduced the auxiliary function
\begin{equation}
    \mathcal{M}_{\bfk}(\tau)=\alpha_{\bfk}(\tau)\beta_{\bfk}^*(\tau).
\end{equation}
Similarly, an equation for~$\mathcal{M}_{\bfk}(\tau)$ can be obtained in an analogous manner:
\begin{equation} 
    \frac{\text{d}}{\text{d}\tau} \mathcal{M}_{\bfk}(\tau)=ir_{\bfk}^*(\tau)e^{2i\varphi_{\bfk}(\tau)}\left[1+\mathcal{N}_{\bfk}(\tau)\right]+2i\left[-s_{\bfk}(\tau)+\frac{\text{d}\varphi_\bfk}{\text{d}\tau}\right]\mathcal{M}_{\bfk}(\tau),
    \label{eq:dotM}
\end{equation}
which follows from~\eqref{eq:BogoliubovEvolution} and the relation~\eqref{eq:|alpha||beta|} between the Bogoliubov coefficients.

Note that neither equation~\eqref{eq:dotN} nor~\eqref{eq:dotM} depend explicitly on the particular solution~$\phi_{\bfk}^{\tau_0}$ of the harmonic oscillator equation with time-dependent frequency. However, the residual ambiguity in the choice of reference vacuum $|0\rangle^{\tau_0}$ has not disappeared but has been transformed from the freedom in the selection of $\phi_{\bfk}^{\tau_0}$ to the freedom in the initial conditions for $\mathcal{N}_{\bfk}(\tau)$ and $\mathcal{M}_{\bfk}(\tau)$. A natural choice is to set the initial conditions as 
\begin{equation}
    \phi_{\bfk}^{\tau_0}(t=\tau_0)=\zeta_{\bfk}(\tau_0), \qquad \dot{\phi}_{\bfk}^{\tau_0}(t=\tau_0)=\rho_{\bfk}(\tau_0),
\end{equation}
which ensures that the annihilation and creation operators coincide at~$\tau_0$. As a result, the initial production vanishes, i.e., $\beta_{\bfk}(\tau_0)=0$, which implies \mbox{$\mathcal{N}_{\bfk}(\tau_0)=\mathcal{M}_{\bfk}(\tau_0)=0$}. 

To facilitate a direct comparison with results from the quantum kinetic approach~\cite{Kluger1998,Schmidt1998,Fedotov2011a}, it is useful to rewrite equations~\eqref{eq:dotN} and~\eqref{eq:dotM} as a single integro-differential equation for~$\mathcal{N}_{\bfk}(\tau)$, eliminating explicit dependence on the auxiliary function~$\mathcal{M}_{\bfk}(\tau)$. With this objective, we solve~\eqref{eq:dotM} by the method of variation of constants, treating~$\mathcal{N}_{\bfk}(\tau)$ as a fixed function and imposing the initial condition~$\mathcal{M}_{\bfk}(\tau_0)=0$. Then,
\begin{equation}
    \mathcal{M}_{\bfk}(\tau)=e^{2i\varphi_{\bfk}(\tau)}\int^\tau_{\tau_0} \text{d}\tau^\prime \  ir^*_{\bfk}(\tau^\prime)[1+\mathcal{N}_{\bfk}(\tau^\prime)]e^{-i\theta_{\bfk}(\tau,\tau^\prime)},
\end{equation}
where 
\begin{equation} \label{eq_theta}
    \theta_{\bfk}(\tau,\tau^\prime)=2\int_{\tau^\prime}^\tau \text{d}\tau^{\prime\prime} \ s_{\bfk}(\tau^{\prime\prime}).
\end{equation}
Substituting this expression into \eqref{eq:dotN}, we finally arrive at the \gls{GeneralizedQuantumVlasovEquation}, expressed in terms of the real and imaginary parts of $r_{\bfk}=\mu_{\bfk}+i\nu_{\bfk}$:
\begin{align} 
    \frac{\text{d}}{\text{d}\tau} \mathcal{N}_{\bfk}(\tau)=\int^\tau_{\tau_0} \text{d}\tau^\prime \ 4[1+\mathcal{N}_{\bfk}(\tau^\prime)]
    \big\{ \big[&\mu_{\bfk}(\tau)\mu_{\bfk}(\tau^\prime)+\nu_{\bfk}(\tau)\nu_{\bfk}(\tau^\prime)\big]\cos[\theta_{\bfk}(\tau,\tau^\prime)] \nonumber \\ 
    +\big[&\mu_{\bfk}(\tau)\nu_{\bfk}(\tau^\prime)-\nu_{\bfk}(\tau)\mu_{\bfk}(\tau^\prime)\big]\sin[\theta_{\bfk}(\tau,\tau^\prime)]\big\}.
\label{eq:QVE2}
\end{align} 
Note that the time derivative of~$\mathcal{N}_\bfk$ does not depend on the arbitrary phase $\varphi_{\bfk}$, but only on~$W_{\bfk}$ and~$Y_{\bfk}$, as we already deduced in~\autoref{sec:ParametrizationVacua}. This equation is exact and completely general for any given quantization characterized by the initial condition distributors~$\zeta_\bfk$ and~$\rho_\bfk$, determined by~$W_\bfk$ and~$Y_\bfk$ as dictated by~\eqref{eq:ParametrizationICDistributors}.

The equation above reflects the non-local nature and memory effects of pair creation over time: the evolution of~$\mathcal{N}_{\bfk}$ is influenced by its past values through the bosonic enhancement factor~$1+\mathcal{N}_{\bfk}$.\footnote{In fermionic systems, this factor is replaced by the Pauli blocking factor \mbox{$1-\mathcal{N}_{\bfk}$}~\cite{Schmidt1998}.} It is said that the Schwinger effect is non-Markovian. This phenomenon arises due to the coherence between successive particle creation events in the presence of intense external fields. In contrast, when external fields are weak, particle creation events become sufficiently spaced apart in time, making a local approximation of the equation feasible~\cite{Kluger1998,Schmidt1999}. These memory effects in pair creation will play a crucial role when modelling the dynamical collapse of light into a black hole and analysing the energy dissipation via the Schwinger effect, which we will explore in detail in~\autoref{chap:Kugelblitz}.

The integro-differential equation~\eqref{eq:QVE2} may initially appear challenging to solve. However, the canonical approach provides an indirect method to address it. Indeed, the expression \eqref{eq:Numtau0tau} for $\mathcal{N}_{\bfk}$ already constitutes a solution to the equation. The difficulty in solving an integro-differential equation thus translates into calculating a particular solution $\phi_\bfk^{\tau_0}$ of the harmonic oscillator equation with time-dependent frequency \eqref{eq:Harmonic}. As discussed earlier, this can only be done analytically in specific cases, such as when the external field follows a Pöschl-Teller electric pulse (see~\autoref{sec:ParametrizationVacua}).

When we choose $(\zeta_{\bfk}(\tau),\rho_{\bfk}(\tau))$ as the ILES at each time~$\tau$~\eqref{eq:ILES}, the time-dependent functions taking part in the previous equation reduce to
\begin{equation} 
   \mu_{\bfk}^{\text{ILES}}(\tau)=0, \qquad \nu_{\bfk}^{\text{ILES}}(\tau)=\frac{\dot{\omega}_{\bfk}(\tau)}{2\omega_{\bfk}(\tau)}, \qquad \theta_{\bfk}^{\text{ILES}}(\tau,\tau^\prime)=2\int_{\tau^\prime}^\tau \text{d}\tau^{\prime\prime} \ \omega_{\bfk}(\tau^{\prime\prime}),
\label{ABthetaad}
\end{equation}
leading to the standard integro-differential \gls{QuantumVlasovEquation} found in the literature \cite{Kluger1998}:
\begin{equation}
    \frac{\text{d}}{\text{d}\tau} \mathcal{N}_{\bfk}^{\text{ILES}}(\tau)=\frac{\dot{\omega}_{\bfk}(\tau)}{\omega_{\bfk}(\tau)}\int^\tau_{\tau_0} \text{d}\tau^\prime \ \frac{\dot{\omega}_{\bfk}(\tau^\prime)}{\omega_{\bfk}(\tau^\prime)}\left[1+\mathcal{N}_{\bfk}^{\text{ILES}}(\tau^\prime)\right] \cos\left[ 2\int^\tau_{\tau^\prime} \text{d}\tau^{\prime\prime} \ \omega_{\bfk}(\tau^{\prime\prime}) \right].
 \label{eq:QVEILES}
 \end{equation} 
Thus, \eqref{eq:QVE2} is the generalized QVE for an arbitrary choice of functions $(\zeta_{\bfk}(\tau),\rho_{\bfk}(\tau))$. This generalization enables us to express the QVE corresponding to the zeroth-order adiabatic quantum vacuum given by~\eqref{eq:0thWKB}. Indeed, it is straightforward to verify that this choice is characterized by the functions
\begin{equation} 
     {\mu}_{\bfk}^{(0)}(\tau)=\frac{1}{4}\left[ \frac{\ddot{\omega}_{\bfk}(\tau)}{\omega_{\bfk}(\tau)^2}-\frac{3}{2}\frac{\dot{\omega}_{\bfk}(\tau)^2}{\omega_{\bfk}(\tau)^3} \right], \qquad  {\nu}^{(0)}_{\bfk}(\tau)=0, \qquad
    {\theta}_{\bfk}^{(0)}(\tau)=2\int_{\tau^\prime}^{\tau} \text{d}\tau^{\prime\prime} \ W_{\bfk}^{(2)}(\tau^{\prime\prime}).
\label{eq:QVEad0}
\end{equation}
While the only non-vanishing contribution to the standard QVE~\eqref{eq:QVEILES}, $\nu_{\bfk}^{\text{ILES}}(\tau)$, is of first adiabatic order, the generalized QVE characterized by~\eqref{eq:QVEad0} has a nonzero contribution only from~$\mu_{\bfk}^{(0)}(\tau)$, which is of second adiabatic order. This results in $\text{d}\mathcal{N}^{\text{(0)}}_{\bfk}/\text{d}\tau$ being two adiabatic orders higher than $\text{d}\mathcal{N}^{\text{ILES}}_{\bfk}/\text{d}\tau$. Consequently, the generalized QVE for the zeroth-order adiabatic vacuum provides a good balance between accuracy and simplicity when compared to the standard QVE~\eqref{eq:QVEILES}.

Finally, to perform explicit calculations it is more convenient to rewrite the integro-differential equation~\eqref{eq:QVE2}, whose numerical resolution is not generally easy~\cite{Schmidt1999}, as a real linear system of ordinary differential equations.  This was first done in~\cite{Bloch1999} for the standard QVE. To that end, we define two auxiliary time-dependent functions:
\begin{align}
    \mathcal{M}_{1{\bfk}}(\tau)&=\int^\tau_{\tau_0} \text{d}\tau^\prime \ 2[1+\mathcal{N}_{\bfk}(\tau^\prime)]
    \left\{ \mu_{\bfk}(\tau^\prime)\cos[\theta_{\bfk}(\tau,\tau^\prime)]-\nu_{\bfk}(\tau^\prime)\sin[\theta_{\bfk}(\tau,\tau^\prime)] \right\}, \nonumber\\
    \mathcal{M}_{2{\bfk}}(\tau)&=\int^\tau_{\tau_0} \text{d}\tau^\prime \ 2[1+\mathcal{N}_{\bfk}(\tau^\prime)]
    \left\{ \mu_{\bfk}(\tau^\prime)\sin[\theta_{\bfk}(\tau,\tau^\prime)]+\nu_{\bfk}(\tau^\prime)\cos[\theta_{\bfk}(\tau,\tau^\prime)] \right\},
\end{align}
such that
\begin{equation}
    \frac{\text{d}}{\text{d}\tau} \mathcal{N}_{\bfk}(\tau)=2\mu_{\bfk}(\tau) \mathcal{M}_{1{\bfk}}(\tau)-2\nu_{\bfk}(\tau) \mathcal{M}_{2{\bfk}}(\tau).
\end{equation}
Differentiating these auxiliary functions we obtain the linear differential system: 
\begin{equation} \label{eq_difsyst}
    \frac{\text{d}}{\text{d}\tau} \mqty(1+\mathcal{N}_{\bfk}\\\mathcal{M}_{1{\bfk}}\\\mathcal{M}_{2{\bfk}})=2\mqty(0&\mu_{\bfk}&-\nu_{\bfk}\\\mu_{\bfk}&0&s_{\bfk}\\-\nu_{\bfk}&-s_{\bfk}&0) \mqty(1+\mathcal{N}_{\bfk}\\\mathcal{M}_{1\bfk}\\\mathcal{M}_{2\bfk}).
\end{equation}
These real differential equations are also equivalent to the complex differential system composed by \eqref{eq:dotN} and \eqref{eq:dotM}. We have verified that this system of equations is equivalent to the one derived in \cite{Habib2000}, that carries out an analogue analysis focusing on adiabatic modes of arbitrary order.

\section{Generalized QVE and unitary quantum dynamics}
\label{sec:GQVEUID}

Next, we analyse the asymptotic ultraviolet behaviour of the generalized QVE~\eqref{eq:QVE2} for canonical quantizations that unitarily implement the dynamics. This study will provide us with an additional physical criterion, stronger than the unitary implementation of the dynamics, to further constrain the ambiguities inherent in the canonical quantization.

In the ultraviolet regime, the system should asymptotically behave like a free field in flat spacetime, independent of the effects of the curvature or the external fields. This suggests a kind of generic ultraviolet behaviour for the generalized QVE, independent of the specifics of the canonical quantization, at leading order. Such details should certainly play a role in subleading terms. 

Indeed, let us consider a canonical quantization defined by functions $W_{\bfk}$ and $Y_{\bfk}$ that behave in the ultraviolet according to the unitary dynamics requirement~\eqref{eq:UID} but with the stronger condition
\begin{equation} 
    W_{\bfk}(\tau)=\omega_{\bfk}(\tau)\big[1+\order{k^{-\gamma}}\big],   \qquad Y_{\bfk}(\tau)=\order{k^{-\eta}}, \qquad \gamma,\eta>2.
    \label{eq:NewCriterion}
\end{equation}
This faster ultraviolet decay ensures that the leading order of the generalized QVE~\eqref{eq:QVE2} matches that of the standard QVE~\eqref{eq:QVEILES}, as can be verified through direct calculation. Consequently, for general functions~$W_{\bfk}$ and~$Y_{\bfk}$ within this subfamily of quantizations that allow for a unitary implementation of the dynamics, the leading-order ultraviolet behaviour of~$\text{d}\mathcal{N}_{\bfk}/\text{d}\tau$ becomes independent of the specific mode functions chosen for quantization. Instead, it only depends on the properties of the external electric field through~$\omega_\bfk(\tau)$.

On the other hand, when canonical quantizations allow for a unitary implementation of the dynamics but do not satisfy the previous stronger condition \eqref{eq:NewCriterion}, their generalized QVE provides particle creation rates $\text{d}\mathcal{N}_{\bfk}/\text{d}\tau$ whose ultraviolet behaviours at leading order strongly depend on functions $W_{\bfk}$ and $Y_{\bfk}$ themselves. This dependence can lead to slower ultraviolet decay compared to the usual QVE for the ILESs. More precisely, under these conditions, the leading-order terms in the expansions in~$k$ of the functions~\eqref{eq_AB} defining the generalized QVE are
\begin{align}
    \mu_{\bfk}|_{\text{L.O.}}=&2k\left.\left(1-\sqrt{\frac{\omega_{\bfk}}{W_{\bfk}}}\right)\right|_{\text{L.O.}}=\order{k^{1-\gamma}}, \nonumber\\
    \nu_{\bfk}|_{\text{L.O.}}=&\frac{1}{2}k^{-1}q\dot{A}\cos{\theta}+kY_{\bfk}|_{\text{L.O.}}=\order{k^{-1}}+\order{k^{1-\eta}}.
\end{align}
If either~$\gamma<2$ or~$\eta<2$, one of these terms decays more slowly than $\nu_{\bfk}^{\text{ILES}}=\order{k^{-1}}$ in the case of the usual QVE for the ILES (see~\eqref{ABthetaad}). In the limiting cases where either~$\gamma=2$ and~$\eta\geq 2$ or vice versa, the ultraviolet decay rate matches the general case (\mbox{$\gamma,\eta >2$}), but in a state-dependent manner.

Note that this analysis remains valid as long as the leading order of the generalized QVE is of the same adiabatic order as the standard QVE. However, certain exceptions arise, such as canonical quantizations based on higher-order adiabatic approximations, where the generalized QVE is of higher order. In such cases, the~$n$th-adiabatic approximation systematically cancels lower-order contributions, including those from the usual QVE. Consequently, the leading order follows that of the~$n$th-adiabatic approximation, which exhibits a faster decay in the ultraviolet, significantly suppressing particle production. More explicitly, the leading orders of~$\mu^{(n)}_{\bfk}$ and~$\nu^{(n)}_{\bfk}$ for~$n$th-order adiabatic modes (with~$n\geq 2$) are:
\begin{align}
    \mu^{(n)}_{\bfk}|_{\text{L.O.}}&=W_{\bfk}^{(n)}-W_{\bfk}^{(n+2)}=\mathcal O (k^{-(n+2)}), \nonumber\\
    \nu^{(n)}_{\bfk}|_{\text{L.O.}}&=k\big(Y_{\bfk}^{(n)}-Y_{\bfk}^{(n+2)}\big)=\mathcal O (k^{-(n+3)}).
\end{align}

For all these reasons, we consider that the physically reasonable choices for general $W_{\bfk}$ and $Y_{\bfk}$ should satisfy~\eqref{eq:NewCriterion}. This ensures not only a unitary implementation of the dynamics but also guarantees that the particle creation rate is independent of the details of the quantization at leading order in the ultraviolet, decaying at least as fast as for the ILES. Other selections not satisfying this criterion but such that they cancel the contribution for the usual QVE (e.g., higher order-adiabatic approximations, which have $Y^{(n)}_{\bfk}=\order{k^{-2}}$), lead to particle creation rates which converge even faster to zero than all the others, and are therefore good candidates as well.

Furthermore, one could impose even more restrictive criteria to further reduce the ambiguity in quantization, based on the generalized QVE for higher adiabatic orders. A motivation for these criteria may come from the fact that, in cosmological settings and within the strict family of adiabatic vacua, it is necessary to consider higher adiabatic orders to obtain a well-defined renormalized stress-energy tensor~\cite{Parker1974}. 

\section{Conclusions}
\label{sec:ConclusionsGQVE}

In this chapter we have written a generalized version of the usual quantum Vlasov equation~\cite{Kluger1998}, which is an integro-differential equation for the number of created particles throughout time for the Schwinger effect, extending it to arbitrary canonical quantizations. Moreover, we have specialized it for arbitrary $n$th-order adiabatic modes, calculating its leading order in an adiabatic expansion.

Focusing on the quantizations that allow for a unitary implementation of the dynamics, we have proved that there is a wide family of them whose generalized QVE behaves, at leading order in the ultraviolet asymptotic expansion, exactly as the standard QVE for the ILES. Namely, the time dependence of such leading order is only due to the characteristics of the external agent (electric field) responsible for the creation of particles, and not to the specific modes used to quantize our field.

On the other hand, we have also proved that there is another family of quantizations that, while also allowing for a unitary implementation of the dynamics, yields a generalized QVE whose leading order in the ultraviolet limit depends explicitly on the quantization (via a time dependent term that is not simply determined by the time dependence of the external agent). In view of this last result we have proposed a new criterion which, together with the unitary implementation of the dynamics, restricts even more the quantizations that we consider acceptable: those for which the leading order of the generalized QVE is just that of the ILES (except when this leading order vanishes, e.g. for the higher order adiabatic vacua). This criterion guarantees that the particle creation rate is independent of the details of the quantization at leading order in the ultraviolet, and which decays at least as fast as for the ILES.

\part[Connection with the experiment]{Connection with the experiment}
\label{part:ConnectionExperiment}

\plainblankpage

\vspace*{\fill}
\begin{minipage}{0.8\textwidth}

This part of the thesis aims to bridge our theoretical results with potential experiments. For instance, we have extensively discussed the quantum ambiguities that arise in the choice of quantum vacuum. But is this discussion purely theoretical, or can these ambiguities manifest in a tangible Schwinger effect experiment? In \autoref{chap:OperationalRealization}, based on~\cite{AlvarezOperational}, we demonstrate that quantum ambiguities are not mere theoretical artifacts but have an intrinsic physical nature, providing an operational realization of them: each way of measuring the number of created particles selects a particular quantum vacuum. This point of view gives a clear and physical meaning to the time evolution of the number of particles produced as the counts in a specific detector and, at the same time, relates commonly used quantization prescriptions to particular measurement setups.

\autoref{chap:InOut}, based on~\cite{AlvarezInOut}, moves beyond the Schwinger effect to explore homogeneous cosmologies. Analogue gravity experiments, such as those realized in Bose-Einstein condensates, often aim at simulating cosmological pair production within a specific time window due to the dynamical expansion of the universe. However, these experiments have a start and an end, which introduces unavoidable transitions out of and into static regimes that alter the intended expansion profile. We show that the resulting particle spectra can be overwhelmingly dominated by these transition periods. Consequently, it becomes impossible to faithfully isolate the effects of the background dynamics during the targeted time window alone---without the transitions---, and one is forced to carefully interpret experimental outcomes. We also study the importance of these transition regimes in prospective Schwinger effect experiments. In contrast to the cosmological case, although the electric field must also be switched on and off, transition effects do not dominate particle production, and such a reinterpretation may not be required.

\end{minipage}
\vspace*{\fill}
\plainblankpage
\chapter[Operational realization of quantum vacuum ambiguities]{Operational realization of \\ quantum vacuum ambiguities}
\label{chap:OperationalRealization}

\chaptermark{Operational realization of quantum vacuum ambiguities}

In~\autoref{chap:QuantumUnitary}, we introduced the concept of the time-dependent spectral number of created pairs, $\mathcal{N}_{\bfk}(\tau)$. This quantity reflects the fact that the notions of particles and antiparticles evolve over time, from the initial time to the time~$\tau$. Crucially, this evolution depends not only on the physical characteristics of the system but also on the choices made during quantization. Specifically, defining the time evolution of~$\mathcal{N}_\bfk(\tau)$ requires selecting a (global) notion of vacuum at each time~$\tau$. This clearly poses questions about the physical interpretation of $\mathcal{N}_\bfk(\tau)$, and the discussion in the literature is still open~\cite{Ilderton2022,Dabrowski2014,Dabrowski2016a,Yamada2021,Domcke2022,Diez2023}.

Recent works have experimentally implemented gravitational particle production in black hole \cite{Steinhauer2016,MunozdeNova2019} and cosmological \cite{Eckel2018, Wittemer2019, Steinhauer2022, Viermann2022} analogue systems, where by means of two-point correlation functions of the density contrast, the number of produced particles after the expansion was measured. Motivated by the experimental accessibility of this quantity, we provide a way of understanding the physical meaning of the possible definitions of $\mathcal{N}_\bfk(\tau)$ in terms of the number of particles measured well after the time $\tau$ at which the interaction between the external agent and the detector has been switched off.

This chapter follows the findings of~\cite{AlvarezOperational}. In~\autoref{sec:SetupOperational}, we describe the theoretical experiment setup, where we vary the way in which we switch off the interaction to measure the number of created particles and antiparticles. Since the electric field vanishes asymptotically in our setting, we analyse in~\autoref{sec:MeasuredParticleNumber} the particle number that would be detected in the experiment by comparing the `in' and `out' quantum vacua. Then, in~\autoref{sec:RelationMeasuredTheoretical}, we establish the connection between this measured particle number and the theoretical particle number---subject to the quantum ambiguities extensively discussed throughout this thesis---when the electric field is still on. Finally, we summarize the conclusions of the chapter in~\autoref{sec:ConclusionsOperational}.

\section{Setup}
\label{sec:SetupOperational}

We consider a simple setup in which an electric field is switched on smoothly from zero, so that there is no ambiguity in the choice of initial vacuum: the `in' quantum vacuum, presented in~\autoref{chap:ChoiceVacuum}, is preferred. In order to measure the actual number of particles at a certain time~$\tau_1$, one would need to instantaneously disconnect the interaction between the detector and the external agent, here the background electric field, and measure afterward. However, instantaneous processes are unfeasible, and thus we cannot have experimental access to that magnitude. Instead, one possibility is to start switching the interaction off smoothly at that time, wait some time until the interaction is completely switched off, and finally measure. We denote this outcome as~$N^{\text{exp},\tau_1}_\bfk$. In order to measure the number of particles at a later time~$\tau_2$, we would need to repeat the experiment switching the interaction off at that new instant. In this way, we would obtain a set of   measurement  results $N^{\text{exp},\tau_1}_\bfk$, $N^{\text{exp},\tau_2}_\bfk$, ..., which tells us what is the number of particles measured in our experiment if we start to switch off the interaction at $\tau_1$, $\tau_2$, ... Nevertheless, this procedure and the results of the measurement will depend on how we switch the interaction off, which might be conditioned by the particular characteristics of the detector that we are using.

Here, we propose to relate the different ways in which we can switch the interaction off and measure the number of particles on the one hand, with the theoretical ambiguities in the choice of the quantum vacuum on the other. For each measurement setup, leading to a family of results~$\{N^{\text{exp},\tau_i}_\bfk\}$, we can find among all possible quantizations at each $\tau_i$ a notion of vacuum such that 
\begin{equation}
    \mathcal{N}_\bfk(\tau_i)=N^{\text{exp},\tau_i}_\bfk.
\end{equation}
The meaning of $\mathcal{N}_\bfk(\tau)$ becomes clear in this case: it is the resulting number of particles that would be measured, following our particular measurement process, if we switched off our experiment at time~$\tau$. Thus, canonical quantum ambiguities are inherently physical in the sense that they are intimately related to the infinitely many different ways of measuring.

As a working case, we focus on the scalar Schwinger effect in ($1+1$) dimensions. However, our analysis can be extrapolated to higher dimensions and to similar particle creation scenarios due to an external time-dependent agent or to other matter fields (e.g., Dirac fields). 

\begin{figure}
    \centering
    \includegraphics[width=0.7\textwidth]{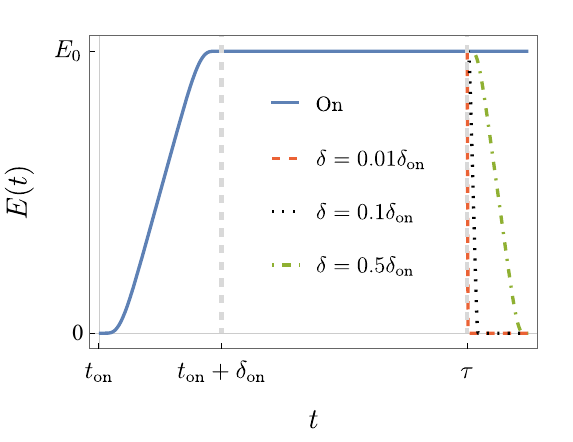}
    \caption{Time evolution of the electric field (solid line) with different switch off profiles (dashed/dotted lines) corresponding to different values of $\delta$, starting at $\tau$.}
    \label{fig:E(t)}
\end{figure}

We want to account for realistic (non-instantaneous) switch-ons and -offs of the electric field (see~\autoref{fig:E(t)}) to mimic the smooth interaction between the background field and the detector. We start the experiment switching the electric field on at a time $t_{\text{on}}$, and after a time $\delta_{\text{on}}$ the electric field smoothly reaches the constant value $E_0$. Then, in order to measure the number of created particles at a given time $\tau$, we start switching it off at that time---effectively disconnecting the detector from the field---, and after a lapse~$\delta$, the electric field vanishes. This entire switching process is characterized by the properties of the experimental setup. 

The finite duration of the switch-on and switch-off phases is modelled using regularized step functions~$\Theta_{\delta_{\text{on}}}(t)$ and~$\Theta_{\delta}(t)$, with respective widths~$\delta_{\text{on}}$ and~$\delta$. These~$C^{\infty}$ functions, previously introduced in~\eqref{eq:RegularizedStep}, were used in~\autoref{chap:SLEs} to model the smearing functions that define the SLEs. The specific form of these functions does not qualitatively affect our results.

In the following, times are given in units of the switch on duration~$\delta_{\text{on}}$, while we parametrize different switch offs by varying~$\delta$. In all figures we fix~$m=\delta_{\text{on}}^{-1}$ and \mbox{$qE_0=\delta_{\text{on}}^{-2}$}, so that the electric field reaches the critical Schwinger limit $m^2/q$~\cite{Schwinger1951}. For lower field strengths, the probability of pair production becomes negligible. In addition, we set $t_{\text{on}}=0$, and $A(t_{\text{on}})=0$. The numerical calculations were performed using \textsc{Mathematica} by solving the initial value problem corresponding to the harmonic oscillator equations with time-dependent frequency~\eqref{eq:Harmonic}, for each value of~$k$, with the appropriate initial conditions in each case (`in', `out', or zeroth-order adiabatic).

\section{Measured particle number}
\label{sec:MeasuredParticleNumber}

Given a particular experimental setting, we can compute the asymptotic number of created particles~$N^{\text{exp},\tau}_\bfk$ that would be measured by our detector when we start the switch off at time~$\tau$. As discussed in~\autoref{chap:ChoiceVacuum}, we are here in a situation where the `in'  and `out'  quantum vacua are preferred. Indeed, for~$t < t_{\text{on}}$, before the electric field is switched on, the system is in the `in' region; and from~$t_{\text{off}}=\tau+\delta$, when the electric field is switched off, the system enters the `out'  region. In both regions, Poincar\'{e} symmetry is locally restored, and the `in' and `out' quantum vacua preserve this symmetry in the asymptotic past and future into the quantum theory, respectively. 

The quantity~$N^{\text{exp},\tau}_\bfk$ measures how excited is the `in' vacuum with respect to the `out'  vacuum. To compute this, we evaluate the $\beta$-Bogoliubov coefficient~\eqref{eq:BogoliubovCoeffsphis} from the transformation relating these vacua. Since the $\beta$-coefficient is time-independent, we can compute it at any time. For example, we can evolve the `in' solution from the initial time and compare it with the `out'  solution at $t_{\text{off}}$, yielding:
\begin{equation}
    N^{\text{exp},\tau}_\bfk= 2 \left| \phi^{\text{in}}_\bfk(t_{\text{off}})\dot{\phi}^{\text{out}}_\bfk(t_{\text{off}})-\phi^{\text{out}}_\bfk(t_{\text{off}}) \dot{\phi}^{\text{in}}_\bfk(t_{\text{off}}) \right|^2.
    \label{eq:numas}
\end{equation}

\begin{figure}
    \centering
    \includegraphics[width=0.9\textwidth]{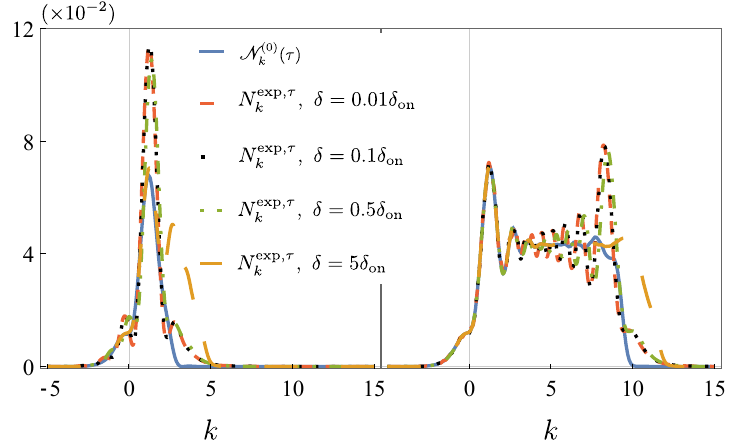}
    \caption{Spectra of the asymptotic number of created particles~$N^{\text{exp},\tau}_k$ for different switch offs, starting at times $\tau=3\delta_{\text{on}}$ (left) and $\tau=10\delta_{\text{on}}$ (right), for different switch off durations $\delta$ (dashed lines). We also represent the computed number of created particles $\mathcal{N}^{(0)}_k(\tau)$ in the zeroth-order adiabatic vacuum at those times (solid line). We use units~$\delta_{\text{on}}=1$.}
    \label{fig:numk}
\end{figure} 

It is important to remark that for a different measurement process starting at the same~$\tau$, here characterized by another duration of the switch-off $\delta$, the outcome $N^{\text{exp},\tau}_\bfk$ will change. In \autoref{fig:numk}, we show the spectra of asymptotically produced particles $N^{\text{exp},\tau}_\bfk$ for $\tau=3\delta_{\text{on}}$ and $\tau=10\delta_{\text{on}}$ for different switch-off durations $\delta$. Since we are working in the ($1+1$)-dimensional case, the wavevector reduces to $\bfk = k \ \textbf{e}_3$, with $k$ taking any real value. Observe that the slower the switch-offs, the longer the electric field can accelerate particles, and thus modes with larger $k$ become excited. This behaviour is in agreement with that of~\cite{Adorno2018}, where they thoroughly analyse the role played by~$\delta$ and~$\tau$ in particle production for a similar profile of the electric field. Note that the oscillations in the spectral distribution have already been observed in analogue experiments by means of two-point correlation functions~\cite{Steinhauer2022,Viermann2022}. For convenience, we also show in~\autoref{fig:numk} the theoretical number of created particles $\mathcal{N}^{(0)}_\bfk(\tau)$ for the zeroth-order adiabatic quantum vacuum at time~$\tau$, determined by the initial conditions~\eqref{eq:0thAdiabaticVacuum}.

\section{Relation between measured and theoretical particle numbers}
\label{sec:RelationMeasuredTheoretical}

In \autoref{chap:QuantumUnitary} we formalized the notion of the time-dependent particle number~$\mathcal{N}_\bfk(\tau)$. For reference, we rewrite this magnitude here, using the~$\beta$-Bogoliubov coefficient~\eqref{eq:Bogoliubovtau0tau} and following~\eqref{eq:Numtau0tau}:
\begin{equation}
    \mathcal{N}_\bfk(\tau) = 2 \left| \phi^{\text{in}}_\bfk(\tau)\rho_\bfk(\tau)-\zeta_\bfk(\tau) \dot{\phi}^{\text{in}}_\bfk(\tau) \right|^2.
    \label{eq:numt}
\end{equation}
We emphasize that, once a preferred `in' quantum vacuum has been chosen, the ambiguity in defining this time-dependent particle number arises from the choice of initial condition distributors~$\zeta_\bfk$ and~$\rho_\bfk$. Our objective now is to establish a connection between this theoretically defined but ambiguous quantity and the measured particle number~\eqref{eq:numas}.

Each measurement procedure selects a particular vacuum for which the theoretical number of particles $\mathcal{N}_\bfk(\tau)$ has a well-defined physical meaning. Indeed, among all possibilities for choosing the initial condition distributors~$\zeta_\bfk$ and~$\rho_\bfk$, there is one for which the number of predicted particles at time~$\tau$ coincides with the outcome~$N^{\text{exp},\tau}_\bfk$ that a particular measurement device would yield. More explicitly, for each time~$\tau$, we choose
\begin{equation}
    \zeta_\bfk(\tau) = \phi_\bfk^{\text{out}}(t = \tau), \qquad \rho_\bfk(\tau) = \dot{\phi}_\bfk^{\text{out}}(t = \tau),
\end{equation}
where $\phi^{\text{out}}_\bfk$ is the `out'  vacuum associated with the switch-off starting at time~$\tau$. In fact, replacing~$t_{\text{off}}$ by~$\tau$ in~\eqref{eq:numas} does not change the resulting $N^{\text{exp},\tau}_\bfk$, since this magnitude does not depend on the instant at which it is evaluated. Consequently, this choice for the initial condition distributors makes~\eqref{eq:numt} equal to~\eqref{eq:numas}; i.e., \mbox{$\mathcal{N}_\bfk(\tau) = N_\bfk^{\text{exp},\tau}$}.

The set of corresponding vacua $\{\ket{0}^{\text{exp},\tau}\}$ defined in this way, one for each $\tau$, allows one to construct $N^{\text{exp}}_\bfk(\tau)$, viewed as function of the time at which we start switching the interaction off. At each time $\tau$, the `in' vacuum is an excited state with respect to the vacuum $\ket{0}^{\text{exp},\tau}$: its excitations correspond precisely to the quanta that would be measured by our detector. Note that $\tau$ denotes the time at which we want to calculate the particles produced by the electric field and not the starting point of a programmed switch off.

This prescription defines a family of physical vacua: those for which there exists a switch-off giving the same particle number as the one predicted by the vacuum. Furthermore, all these vacua unitarily implement the dynamics, as they are associated with a finite number of particles by construction.

\begin{figure}
    \centering
    \includegraphics[width=0.9\textwidth]{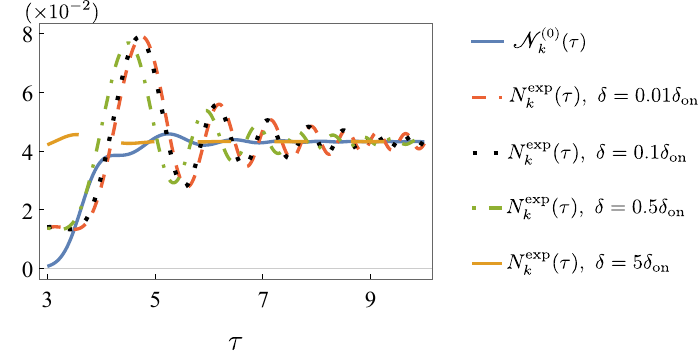}
    \caption{Evolution of the number of created particles $N_k^{\text{exp}}(\tau)$ with $k=3$ for different switch off durations~$\delta$. The solid line corresponds to the zeroth-order adiabatic prescription. We use units~$\delta_{\text{on}}=1$.}
    \label{fig:numt}
\end{figure}

In our simple setup, the measurement device is characterized by $\delta$, and for each of its values we have different notions of~$N_\bfk^{\text{exp}}(\tau)$. This is illustrated in \autoref{fig:numt}, where one can see the time evolution of $N_\bfk^{\text{exp}}(\tau)$ for $k=3$, for different durations of the switch off~$\delta$. For each time~$\tau$, we compute the asymptotic number of particles $N_{\bfk}^{\text{exp},\tau}$ when we start switching the interaction off at $\tau$. The observed oscillations in~$\tau$ were already present in~\cite{Kluger1998,Schmidt1998,Dabrowski2014,Dabrowski2016a}, but now we can provide them with a full physical meaning, as they follow from a measurement-based notion of particle. Moreover, recent works try to implement experimental setups that make use of this behaviour to enhance particle production (see the recent study~\cite{Aleksandrov2025} or other references on the dynamically assisted Schwinger effect~\cite{Schutzhold2008,Bulanov2010}). In \autoref{fig:numt} we also show the time evolution of the theoretical particle number when we choose zeroth-order adiabatic vacua at each time~$\tau$, $\mathcal{N}_\bfk^{(0)}(\tau)$.

The amplitudes of the fluctuations are smaller as we increase the value of~$\delta$. Note that as $\tau$ increases, these amplitudes decrease and the number of particles become more independent of~$\delta$. This result is compatible with~\cite{Adorno2018}, where it was proved that for a sufficiently large time~$\tau$ (larger than the ones considered here) the switch-on and -off effects only affect as next-to-leading corrections to the contribution to the constant part of the electric field.

As an aside, for each measurement process we could have reassigned the asymptotic outcome $N_{\bfk}^{\text{exp},\tau}$ to any other time different from the time at which the switch off starts (e.g., results in~\cite{Ilderton2022} are obtained by taking $\tau+\delta/2$ as the reference time instead). However, note that this would lead to a simple relabelling of the $\tau$~axis in \autoref{fig:numt}, shifting each curve proportionally to its value of~$\delta$.

One may wonder whether it is possible to find, given a choice of vacuum, a particular switch-off starting at~$\tau$ such that the measurement of the associated asymptotic number of particles~$N_{\bfk}^{\text{exp},\tau}$ coincides with the value of~$\mathcal{N}_\bfk(\tau)$, computed using said vacuum prescription. This requirement does not unequivocally determine the time evolution of the `in' mode~$\phi_\bfk^{\text{in}}$ after time $\tau$. Therefore, each function~$\phi_\bfk^{\text{in}}$ compatible with the previous condition would lead to a different mode equation. However, it is definitely non-trivial that one can find a time-dependent frequency~$\omega_\bfk(t)$ of the form of~\eqref{eq:Frequency} fulfilling this requirement for all values of~$\bfk$.

Finally, we illustrate the application of our operational notion of particles by interpreting two usual notions of vacuum in terms of measurements:
\begin{itemize}
    \item First, we consider the ILES at time~$\tau$, defined by the initial conditions~\eqref{eq:ILES}. In~\cite{Ilderton2022}, it was proved that the theoretical particle number calculated using initial condition distributors based on this vacuum (see~\eqref{eq:ICdistributorsILES}) coincides with the asymptotic particle number measured in the unfeasible situation in which the interaction between the electric field and detector is instantaneously switched off at~$\tau$, i.e., $\delta=0$. Indeed, this setting can be easily implemented in the electric potential with a continuous but non-differentiable step function at~$\tau$. The most regular solution to the harmonic oscillator equation~\eqref{eq:Harmonic} has a continuous but non-differentiable second derivative and corresponds precisely to the ILES~\eqref{eq:ILES}. In addition, in agreement with~\cite{Ilderton2022}, the particle number in the instantaneous case~$\delta=0$ coincides with the limit~$\delta \to 0$.
    \item Another vacuum prescription that is commonly used is precisely choosing initial condition distributors based on the zeroth-order adiabatic quantum vacuum at each time~$\tau$ (see~\eqref{eq:0thAdiabaticVacuum}). We infer from \autoref{fig:numk} and \autoref{fig:numt} that the particle number spectrum of the adiabatic vacuum deviates from that of an arbitrarily fast switch-off. Indeed, in the latter, there appear fluctuations with larger amplitudes both in the spectrum and in the time evolution.
\end{itemize}

\section{Conclusions}
\label{sec:ConclusionsOperational}

Quantum vacuum ambiguities are inherent to QFT in the presence of an external, time-dependent agent. In this chapter, we show that knowing the particularities of how we measure the particle number allows us to identify a particular quantum vacuum with clear physical meaning: its associated notion of particle is that which would be measured by our detector in a potential experiment.

This operational procedure can be used to interpret the notion of particle associated with usual vacuum prescriptions. This is the case, for example, of the ILES at a certain time, which provides the same particle number as an instantaneous switch-off at that time or of the zeroth-order adiabatic vacuum, which departs from this behaviour.

In conclusion, we select a family of vacua---those related to a realistic switch off of the interaction between the detector and the external agent---that are physical in the sense that they accommodate information about real outcomes. These vacua are intrinsically well-behaved as they allow for a unitary implementation of the dynamics.
\chapter[The relevance of `on' and `off' transitions in quantum pair production experiments]{The relevance of `on' and `off' transitions in quantum pair production experiments}
\label{chap:InOut}

Particle creation phenomena are typically difficult to access experimentally. Instead, motivated by the original idea by Unruh~\cite{Unruh1981}, analogue gravity experiments \cite{Barcelo2011,Jacquet2020,Almeida2023} have been used as a tool to explore the dynamics of quantum fields in non-trivial backgrounds or effective curved spacetimes. In recent years, numerous experiments have been carried out in hydrodynamical, condensed matter, optical systems, and others \cite{Philbin2008,Weinfurtner2011,Euve2016,MunozdeNova2019,Drori2019,Shi2023,Torres2017,Eckel2018,Wittemer2019,Banik2022,Giacomelli2021,Braidotti2022,Steinhauer2022,Jacquet2022,Viermann2022}, demonstrating the potential of such platforms for the study of quantum fields.

Cosmological analogue experiments often aim to measure the analogue of particle production caused by the expansion of the universe over a specific cosmological time interval~\cite{Eckel2018,Wittemer2019,Banik2022,Steinhauer2022,Viermann2022,Sparn2024}. Similarly, in Schwinger effect experiments, one might wonder what is the production of particles due to an electric field that is switched on during a certain period of time \cite{Schuetzhold2009,Bulanov2010,Aleksandrov2022,Aleksandrov2025,Ilderton2022}. However, in all these situations, one cannot avoid the existence of transitions from and to static regimes in which the cosmological expansion ceases or the electric field vanishes. Analogue experiments have a beginning and an end, and electric fields must be switched on and off to implement specific pair production processes in the laboratory. Therefore, a very natural question arises: How do these `on' and `off' transitions impact the results of the experiments? Are they negligible, or do they affect the particle production process? If the latter is true, one has to be careful when interpreting the results of such experiments, as the particle production occurring during the particular time window that one is trying to simulate could be overshadowed by the production taking place during the transitions.

To address these questions, we first derive fundamental insights from the more general case of cosmological pair production in homogeneous and isotropic cosmologies. Indeed, these transition regimes are also present in early universe scenarios, where the computation of cosmologically produced particles typically relies on the fact that the universe's expansion becomes sufficiently slow at very early and late times---such as at the onset of inflation and well into the reheating epoch, respectively. Our analysis is therefore also interesting in actual cosmological scenarios, and characterizes which regions of spacetime are more relevant regarding cosmological pair production.

Specifically, we study the impact of `on' and `off' transitions on particle production in $D$-dimensional FLRW expanding universes. We will consider a massive spectator scalar field, allowing for a non-minimal coupling to the geometry. Our results demonstrate that the coupling parameter between the field and the geometry has a strong influence on the resulting pair production spectrum. In particular, we find that particle production during the targeted time window is inevitably affected by `on' and `off' transition periods. Furthermore, in the case of a non-conformal coupling, production during abrupt transitions dominates pair creation, significantly overshadowing the contributions from the intermediate region. However, when the coupling is conformal, the `on' and `off' transitions do not substantially enhance pair creation, resulting in a much lower overall density of produced particles.

We will then apply these fundamental results to two experimental setups, highlighting the need for extreme caution when interpreting the physical results of these experiments. On the one hand, we will discuss analogue pair production in (1+2)-dimensional \acrfull{BECs}, which simulates the problem of a non-conformally coupled field in an FLRW universe. On the other hand, we will discuss the Schwinger effect due to a switchable electric field in $(1+3)$ dimensions. Its behaviour regarding `on' and `off' transitions is for the most part equivalent to that of a conformally coupled field in an FLRW universe. However, unlike the cosmological case, the anisotropic nature of the electric field causes the contribution from the intermediate regime to increasingly dominate over that of the abrupt transitions when the field remains switched on for a sufficiently long duration, leading to enhanced particle production as the field duration grows.

This chapter is based on~\cite{AlvarezInOut} and is structured as follows. In \autoref{sec:Cosmo}, we review the framework for describing cosmological pair production for a real scalar field in FLRW in $D$ spatial dimensions. In \autoref{sec:OnOffTransitions}, we analyse the impact of the transitions between static and dynamic regimes of the scale factor on particle production. Then, in \autoref{sec:BEC}, we apply these ideas to the case of analogue pair production in BECs. We discuss these matters in the context of the Schwinger effect in \autoref{sec:Schwinger}. Finally, we elaborate our conclusions in \autoref{sec:ConclusionsInOut}. 

\section{Cosmological pair production}
\label{sec:Cosmo}

Let us consider a $(1+D)$-dimensional FLRW expanding universe with vanishing spatial curvature~\cite{Friedman1922, Friedman1924, Lemaitre1931, Robertson1935, Robertson1936a, Robertson1936b, Walker1937}, 
\begin{equation}
\dd s^2 = a^2(\eta)\left(-\text{d} \eta^2+\text{d} \bfx^2 \right),
 \label{eq:GeneralLineElement}
\end{equation}
where~$\eta$ is the conformal time and $a(\eta)$ the scale factor. The dynamics of a real, non-minimally coupled to gravity scalar field~$\Phi(\eta,\bfx)$ with mass~$m$ is described by the KG equation
\begin{equation}
    \Phi^{\prime\prime} + (D-1)\mathcal{H}\Phi^\prime - a^2(\Delta + m^2 + \xi R) \Phi = 0,
\end{equation}
where ${}^\prime = \partial/\partial\eta$, $\Delta$ is the Laplace operator and $\mathcal{H}=a^\prime / a$ the conformal Hubble parameter. The field is coupled via the parameter~$\xi$ to the Ricci curvature scalar~$R$.

We follow a procedure similar to the one in \autoref{chap:ChoiceVacuum}, developed for the Schwinger effect. In this case, however, we propose solutions to the Klein-Gordon equation of the form
\begin{equation}
    \Phi_\bfk(\eta,\bfx) = (2\pi)^{-\frac{D}{2}} a(\eta)^{\frac{1-D}{2}} \phi_\bfk(\eta) e^{i\bfk\cdot \bfx},
    \label{eq:AnsatzPhiCosmology}
\end{equation}
were we need to introduce the factor~$a(\eta)^{\frac{1-D}{2}}$ to ensure that the modes~$\phi_\bfk(\eta)$ satisfy decoupled harmonic oscillator equations of the form 
\begin{equation}
    \phi^{\prime\prime}_{\bfk}(\eta) + \omega_{k}(\eta)^2 \phi_{\bfk}(\eta) = 0,
    \label{eq:ModeEquation}
\end{equation}
with the time-dependent frequency given by
\begin{equation}
    \omega_{k}(\eta)^2 = k^2 + m^2a(\eta)^2 +\frac{1+(4\xi-1)D}{4} \left\{2 \frac{a^{\prime\prime}(\eta)}{a(\eta)}+(D-3) \left[ \frac{a^\prime(\eta)}{a(\eta)} \right]^2 \right\}.
    \label{eq:MasterFrequency}
\end{equation}
This frequency encodes all the relevant information about the gravitational background through the evolution of the scale factor. Because the setting is isotropic, $\omega_k(\eta)$ depends only on the magnitude of the wavevector, $k = |\bfk|$. To reflect this isotropy at the quantum level---unlike in the Schwinger case, which is anisotropic---we choose solutions such that~\mbox{$\phi_\bfk(\eta) = \phi_k(\eta)$}, i.e., solutions that only depend on the modulus of~$\bfk$, not its direction.

The quantum field operator is then defined as
\begin{equation}
\hat{\Phi}(\eta,\bfx) = a(\eta)^{\frac{1-D}{2}} \int \frac{\text{d}^D\bfk}{\left(2\pi\right)^{\frac{D}{2}}} \ \left[\hat{b}_{\bfk} \phi_\bfk(\eta)e^{i\bfk\cdot\bfx}
+ \hat{b}_{\bfk}^\dagger \phi_\bfk^*(\eta)e^{-i\bfk\cdot\bfx}\right].
\label{eq:FieldExpansion}
\end{equation}
The operators satisfy the standard commutation relations \mbox{$[\hat{b}_{\bfk},\hat{b}_{\bfk'}^{\dagger}]=\delta(\bfk-\bfk')$}, whereas all the other commutators vanish. 

In the following, we consider an initially static universe that begins expanding at some finite time~$\eta_{\text{on}}$ until it halts expansion and returns to a static state from a later time~$\eta_{\text{off}}$. This scenario is sensible in the context of the inflationary universe, at the beginning of which the geometry expands slowly. After inflation, the universe thermalizes, and the expansion becomes again very adiabatic. This defines `in' and `out' regions in which the universe is static, and the frequency \eqref{eq:MasterFrequency} becomes constant. Then, we adopt again an `in-out' formalism, where the `in' solutions are defined as
\begin{equation}
    \phi^{\text{in}}_k(\eta)=\frac{1}{\sqrt{2\omega_k^{\text{in}}}} e^{-i\omega_k^{\text{in}}\eta}, \quad \eta\leq \eta_{\text{on}},
\label{eq:vin}
\end{equation}
with $\omega_k^{\text{in}}=\sqrt{k^2+ m^2a^2(\eta_{\text{on}})}$, setting the `in' quantum vacuum. Similarly, the `out' solutions, 
\begin{equation}
    \phi^{\text{out}}_k(\eta)= \frac{1}{\sqrt{2\omega_k^{\text{out}}}} e^{-i\omega_k^{\text{out}}\eta}, \quad \eta\geq \eta_{\text{off}},
\label{eq:vout}
\end{equation}
with frequencies \mbox{$\omega_k^{\text{out}}=\sqrt{k^2+m^2a^2(\eta_{\text{off}})}$}, define the `out' vacuum. The $\beta$-coefficient of the Bogoliubov transformation relating both quantizations can be computed via
\begin{equation}
    \beta_k = i\left\{ \phi^{\text{in}}_k(\eta) \left[\phi^{\text{out}}_k(\eta)\right]^\prime - \phi^{\text{out}}_k(\eta) \left[\phi^{\text{in}}_k(\eta)\right]^\prime \right\}.
    \label{eq:betakvinvout}
\end{equation}
The total number density of produced particles and antiparticles is then obtained by summing over all modes:
\begin{equation}
    \mathcal{N} = 2\int \text{d}^D\bfk \ |\beta_k|^2.
    \label{eq:n}
\end{equation}

\section{`On' and `off' transitions}
\label{sec:OnOffTransitions}

We aim to model the expansion of the universe occurring between two times, $\eta_{\text{on}}$ and~$\eta_{\text{off}}$. However, the way in which we model the transition between the `in' regime (\mbox{$\eta \leq \eta_{\text{on}}$}) and the intermediate region (\mbox{$\eta_{\text{on}} \leq \eta \leq \eta_{\text{off}}$}), as well as the transition between this intermediate region and the `out' regime (\mbox{$\eta \geq \eta_{\text{off}}$}), inevitably influences particle production. Whether these transitions are abrupt or adiabatic can greatly affect production. This raises a critical question: To what extent can the impact of these transitions be considered negligible? Are there cases where particle production during these phases becomes so pronounced that it masks the effects of the expansion we intend to simulate? Our analysis shows that transition effects are inevitable in all cases and have a significant impact on pair creation.

\begin{figure}
    \centering
    \includegraphics[width=0.6\textwidth]{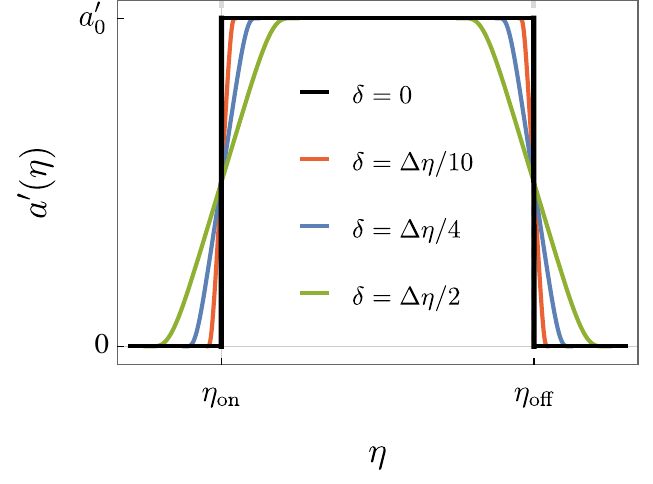}
    \caption{Expansion rate in~\eqref{eq:DerivativeScale} as function of conformal time, for different transition durations $\delta$, with $\Delta\eta = \eta_{\text{off}} - \eta_{\text{on}}$.}
    \label{fig:SwitchOnAndOff}
\end{figure}

Our goal is to identify whether particle production during the transitions is more or less significant than the production during the intermediate expansion---the regime in which we are actually interested in the case of analogue experiments, for example. To illustrate these ideas, we consider a universe undergoing a constant expansion rate in terms of conformal time; i.e., \mbox{$a^\prime(\eta)=a_0^\prime$}, for~\mbox{$\eta_{\text{on}} \leq \eta \leq \eta_{\text{off}}$}. We model the transitions between this intermediate region and the static regimes by the regularized interpolation function $\Theta_{\delta}(\eta)$ already introduced in~\eqref{eq:RegularizedStep}, and used in \autoref{chap:SLEs} and \autoref{chap:OperationalRealization}. We can then write the expansion rate during the entire expansion, including the transitions, as
\begin{equation} 
\label{eq:DerivativeScale}
    a^{\prime}(\eta)=a_0^\prime\left[\Theta_{\delta}(\eta-\eta_{\text{on}}-\delta/2)-\Theta_{\delta}(\eta-\eta_{\text{off}}+\delta/2)\right].
\end{equation}
In \autoref{fig:SwitchOnAndOff}, we represent this expansion rate for various transition durations~$\delta$.\footnote{Note that, in contrast to the convention used in \autoref{chap:OperationalRealization}, here we denote by~$\delta$ the duration of both the `on' and `off' transitions, which we take to be equal.} In the following, we will say that the scale factor undergoes an `abrupt transition' when it is continuous but not differentiable. During these abrupt transitions, the expansion rate~$a^\prime(\eta)$ involves discontinuous (but finite) step functions, as depicted in \autoref{fig:SwitchOnAndOff} for~$\delta=0$.

As in the previous chapter, we perform the numerical calculations using \textsc{Mathematica}, solving the initial value problem for each~$k$ associated with the harmonic oscillator equations~\eqref{eq:ModeEquation}. In this case, however, we consider the homogeneous time-dependent frequency~\eqref{eq:MasterFrequency} and restrict to the `in-out' formalism.

\begin{figure}
    \centering
    \includegraphics[width=0.75\textwidth]{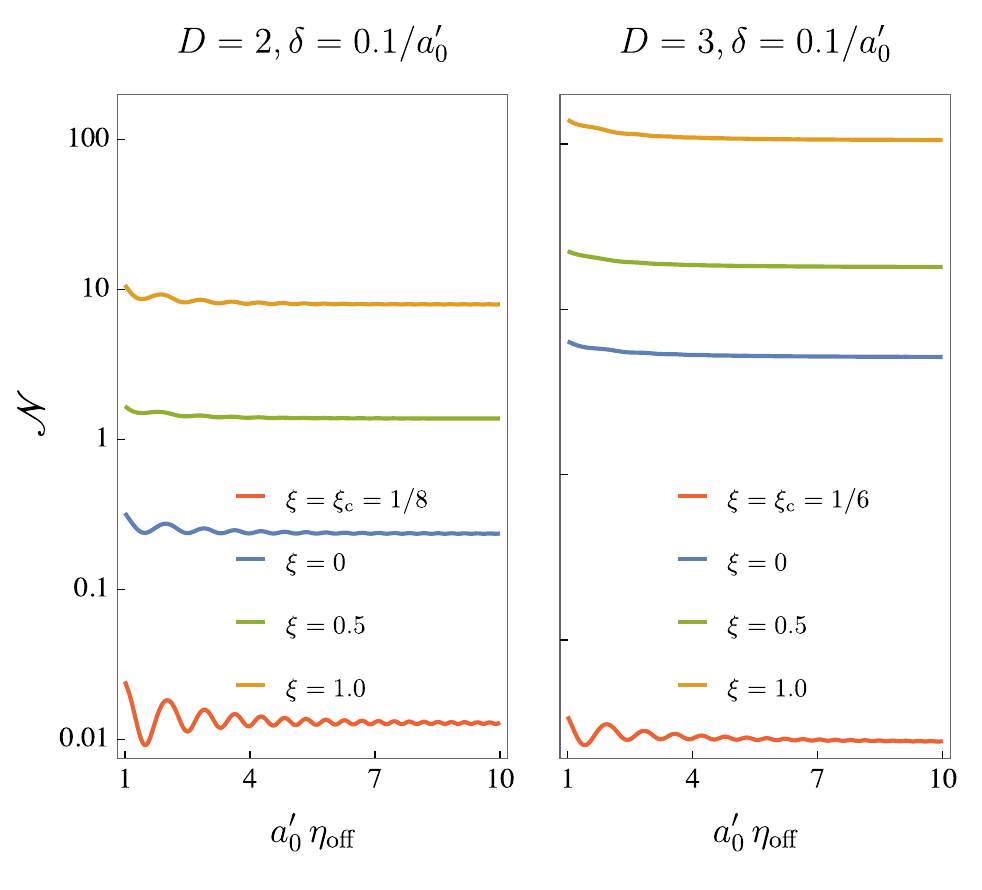}
    \caption{Total number density of produced particles $\mathcal{N}$ as a function of $\eta_{\text{off}}$ ($\eta_{\text{on}}=0$) for fast transitions ($\delta=0.1/a_0^\prime$), in the case of two (left) and three (right) spatial dimensions. Results are shown for different values of the coupling~$\xi$, where the red lines correspond to the conformal coupling case.}
    \label{fig:TotalDensity}
\end{figure}

In \autoref{fig:TotalDensity}, we calculate the total number density of produced pairs, as defined in~\eqref{eq:n}, for fast transitions ($\delta=0.1/a_0^\prime$) and various durations of the intermediate expansion, $\eta_{\text{on}} \leq \eta \leq \eta_{\text{off}}$. Specifically, we fix $\eta_{\text{on}} = 0$ and numerically compute the $\beta$-Bogoliubov coefficient using~\eqref{eq:betakvinvout} for each process, with its duration parametrized by a particular value of~$\eta_{\text{off}}$. We then integrate over all modes~$k$ to obtain the total density. We ensured that the numerical upper limit cutoff was sufficiently large to guarantee convergence of the integration. This procedure is repeated for each value of~$\eta_{\text{off}}$. We examine how these particle densities depend on the coupling parameter $\xi$ and present results for $D=2$ and $D=3$ spatial dimensions. Our analysis reveals that \mbox{\emph{a)} the} total number density of produced pairs in the conformal coupling case, where \mbox{$\xi=(D-1)/(4D)$}, is significantly lower---by several orders of magnitude---than in the non-conformal case; and that \mbox{\emph{b)} particle} production stabilizes for sufficiently prolonged expansions. 

\begin{figure*}
    \centering
    \includegraphics[width=0.9\textwidth]{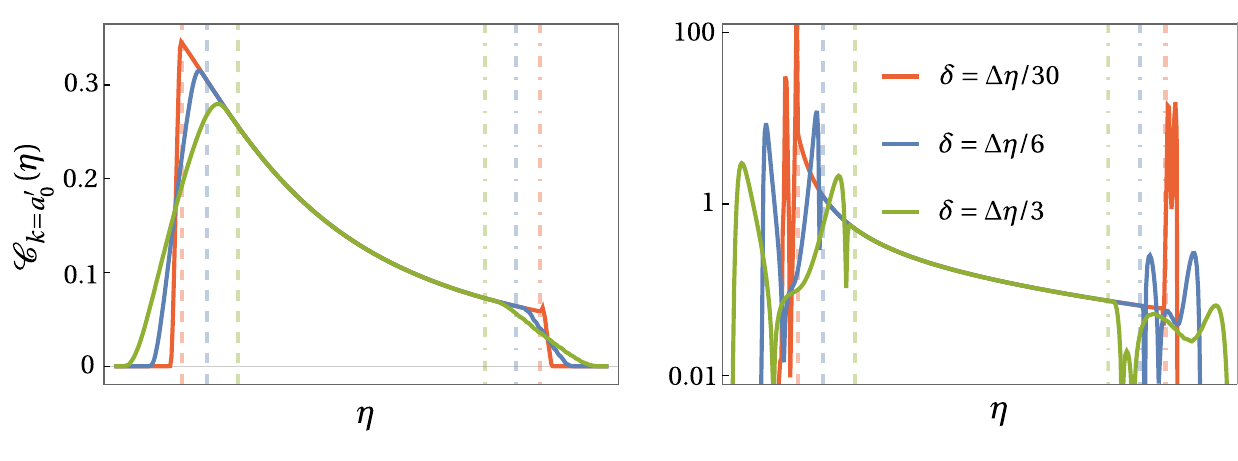}
    \caption{Function $\mathcal{C}_k$ for $D=3$ dimensional universe expansion with \mbox{$\Delta \eta=3/a_0$}, for different transition durations~$\delta$, fixed \mbox{$k= a_0$}. The left panel illustrates the conformal coupling case ($\xi=1/6$) while the right panel corresponds to a non-conformal coupling case ($\xi= 1$).}
    \label{fig:Ck}
\end{figure*}

\begin{itemize}
    \item \textit{Production in the conformal case is suppressed with respect to the non-conformal case.} The difference between these two cases can be understood by examining the behaviour of the frequency \eqref{eq:MasterFrequency}, since pair production is dictated by the dynamics of the field modes. It is when the frequency rapidly varies over time, that particle production is enhanced. During the intermediate expansion phase ($\eta_{\text{on}} \leq \eta \leq \eta_{\text{off}}$), the frequency remains bounded in both cases, provided the expansion is sufficiently smooth. However, a crucial distinction arises during the abrupt transitions: while~$a^\prime$ remains bounded, $a^{\prime\prime}$ approaches a Dirac delta. In the conformal coupling case, the mode frequency simplifies to~\mbox{$\omega_k^2 = k^2 + m^2a^2$}, meaning that both $\omega_k$ and its time-derivative remain bounded even during abrupt transitions. Conversely, for non-conformal coupling, the frequency explicitly depends on the first and second derivatives of the scale factor. As a result, the frequency and its time-derivatives sharply diverge at the transition points, leading to significant enhancement of particle production. The crucial point is that this enhancement arises not from the intermediate expansion, but rather from the `on' and `off' transitions. This is particularly evident in our example, as in our computations we employ a constant expansion rate during the intermediate regime, ensuring that the derivatives of the scale factor entering the frequency~\eqref{eq:MasterFrequency} vanish except at the transitions.
    \item \textit{Production stabilizes for sufficiently prolonged intermediate expansions.} This behaviour reinforces the conclusion that the dominant contribution to particle creation arises from the abrupt `on' and `off' transitions. Even when the intermediate expansion lasts significantly longer, the particle number density remains effectively constant, indicating that the intermediate phase contributes minimally to the overall production. The resulting horizontal asymptote can thus be interpreted as capturing the average effect of the abrupt transitions. We will see that, in the case of the Schwinger effect, anisotropies lead to a drastically different asymptotic behaviour.
\end{itemize}

In order to characterize the rate of change of the mode frequency (for a general expansion rate), we define the dimensionless function
\begin{equation} 
    \mathcal{C}_k(\eta)=\left|\frac{\omega_k^{\prime}(\eta)}{\omega_k^2(\eta)}\right|.
    \label{eq:adCk}
\end{equation}
This provides a straightforward-to-compute magnitude, as numerical methods are not required---unlike in the evaluation of the number density of produced particles. From the expression of $\omega_k$ in~\eqref{eq:MasterFrequency}, it follows that $\mathcal{C}_k$ is a strictly decreasing function of $k$, reflecting that the time variation of the frequency always decreases as one considers larger wavenumbers. This behaviour results in a suppression of particle creation in the ultraviolet. Regarding its dependence on the conformal time~$\eta$, we saw in \autoref{chap:GQVE} that the function~$\mathcal{C}_k(\eta)$ is directly linked to the fluctuations of the particle density per mode $k$, as captured by the quantum Vlasov equation~\cite{Kluger1998,Schmidt1998,AlvarezGQVE}. When $\mathcal{C}_k(\eta)$ becomes large during the transition phases, the particle number density experiences rapid oscillations, leading to enhanced particle production compared to the intermediate expansion period. On the other hand, if $\mathcal{C}_k(\eta)$ remains small during the transitions, the time variation of particle production is comparatively less oscillating, and the overall production rate is significantly lower. 

The frequency~\eqref{eq:MasterFrequency} depends in general on the derivatives of the scale factor, and 
\begin{equation}
\mathcal{C}_k =\omega_k^{-3} \left|m^2a a^{\prime} + \frac{1+(4\xi-1)D}{4} \left[ \frac{a^{\prime\prime\prime}}{a} + (D-4)\frac{a^{\prime}a^{\prime\prime}}{a^2} - (D-3)\left( \frac{a^{\prime}}{a} \right)^3\right]\right|.
\label{eq:Cknonconformal}
\end{equation}
For abrupt transitions, $a^{\prime\prime\prime}$ approaches the derivative of a Dirac delta, which strongly dominates over the terms proportional to $a'$ and $a^{\prime\prime}$. In the frequency~\eqref{eq:MasterFrequency}, the term with~$a^{\prime\prime}$ dominates. From~\eqref{eq:Cknonconformal}, this leads to $\mathcal{C}_k$ exploding during abrupt transitions, allowing production for a broad range of modes. Therefore, abrupt transitions in an expanding universe dramatically impact the spectra of produced particles, masking the contributions from the actual expansion process itself without such transitions.

In the particular case of conformal coupling, \eqref{eq:adCk} simplifies to
\begin{equation}
    \mathcal{C}_k=\left|\frac{m^2a a^{\prime}}{\left( k^2+m^2a^2 \right)^{3/2}}\right|,
    \label{eq:adCk16}
\end{equation}
which is bounded from above by $\mathcal{C}_{k=0}=\mathcal{H}/m$. Even in cases where the scale factor undergoes abrupt transitions, the function $\mathcal{C}_k$ remains finite, as it depends only on the first derivative of the scale factor and not on higher orders. Nevertheless, higher-order time derivatives of the frequency involve higher derivatives of the scale factor, which, in the limit of abrupt transitions, tend to Dirac delta distributions and their derivatives. Consequently, although still relevant, the impact of transitions in pair production is smaller in this case than in the non-conformal scenario.

For the particular shape given in~\eqref{eq:DerivativeScale}, \autoref{fig:Ck} shows the function~$\mathcal{C}_k$ for different values of~$\delta$ in $D=3$ dimensions. In the conformal coupling case $\xi = 1/6$, $\mathcal{C}_k$ remains bounded throughout the entire expansion, even during abrupt transitions. However, in the non-conformal coupling case (for instance, \mbox{$\xi = 1$}), $\mathcal{C}_k$ exhibits sharp oscillations during the `on' and `off' transitions. The amplitude of these oscillations increases by several orders of magnitude as the transitions become shorter, highlighting the sensitivity of the system to more rapid transitions. It is clear in this case that the primary contribution to the particle excitation number arises predominantly from the transitions, overshadowing the effects of the linear expansion in the intermediate region.

Note that in scenarios where the scale factor varies rapidly---particularly involving abrupt decelerations---such variations act as effective `on' and `off' transitions. Oscillatory or cyclic cosmologies, with alternating phases of expansion and contraction as discussed in~\cite{Schmidt2024,Sparn2024,Agullo2024}, exemplify this behaviour. In these cases, particle production is significant throughout the entire evolution.

In \autoref{sec:BEC} and \autoref{sec:Schwinger} we present two illustrative examples from laboratory settings where we apply the results just developed. The first involves gravitational analogue experiments with BECs that mimic the dynamics of a non-conformally coupled scalar field in an FLRW expanding universe. The second focuses on the Schwinger effect, whose anisotropic nature introduces important nuances to our analysis that we discuss below.

\section{BEC analogue experiment}
\label{sec:BEC}

We focus on the problem of analogue particle production in a quasi two-dimensional, \mbox{spin-0} BEC~\cite{TolosaSimeon2022,Viermann2022,Sparn2024,Schmidt2024}. Low-energy excitations on top of the condensate's ground state behave as a massless scalar field propagating in a curved spacetime defined by the so-called acoustic metric. This acoustic metric, determined by the properties of the condensate, can be experimentally controlled to emulate a two-dimensional FLRW metric. As a result, the system provides an analogue for cosmological particle production in a $(1+2)$-dimensional spacetime.

The role of the scale factor in this analogue setup is played by the scattering length, whose time dependence can be precisely controlled using Feshbach resonances~\cite{Stwalley1976,Cornish2000,Chin2010}. The expansion process is implemented across various stages: an initial `in' region where the scattering length remains constant, followed by an `on' transition into an intermediate region designed to mimic the desired cosmological scenario, and finally an `off' transition leading to a final `out' region where the scattering length returns to a constant value.

In analogue experiments, the `on' and `off' transitions are unavoidable, and typically modelled as instantaneous. This approach was adopted in, e.g., references~\cite{Viermann2022,Sparn2024}, where the abrupt transition model was shown to align well with experimental data. However, understanding the impact of these transitions on particle production is crucial. While the primary focus of such experiments lies in the intermediate region, where the desired scale factor behaviour is replicated, it is essential to analyse how these transitions influence the dynamics to properly isolate and interpret the physical effects of interest.

In this analogue platform, we do not have the freedom to select the value of $\xi$, which is zero in this case. For flat spatial sections, the mode equation corresponds to taking $D=2$ and \mbox{$m=0$} in the time-dependent frequency~\eqref{eq:MasterFrequency}, yielding~\cite{TolosaSimeon2022,Viermann2022,Schmidt2024,Sparn2024}
\begin{equation}
    \omega_k^2 = k^2 - \frac{a''}{2a} + \left( \frac{a'}{2a} \right)^2 .
    \label{eq:BECFrequency}
\end{equation}
In this situation, the coupling is minimal and therefore non-conformal, and second derivatives of the scale factor appear in the frequency. Regarding the density of produced particles, this BEC experiment corresponds to the scenario described by the non-conformal coupling curve $\xi=0$ in the left panel of \autoref{fig:TotalDensity}.
\begin{figure}
    \centering
    \includegraphics[width=0.55\textwidth]{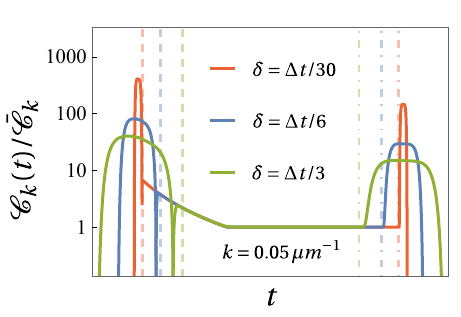}
    \caption{Function $\mathcal{C}_k$ for a typical analogue gravity experiment in quasi two-dimensional BECs for $k=0.05 \, \mu\text{m}^{-1}$ and transitions of different abruptness characterized by $\delta$. The values are normalized with respect to~$\bar{\mathcal{C}}_k \simeq 5 \times 10^{-8}$, assuming an intermediate region of duration~$\Delta t = 3 \, \text{ms}$. The BEC parameters correspond to the expansion linear in $t$ presented in~\cite{Viermann2022}. 
    }
    \label{fig:CkBEC}
\end{figure}

We computed the function~$\mathcal{C}_k$ corresponding to the frequency~\eqref{eq:BECFrequency}, using the same functional form of the scale factor as in~\eqref{eq:DerivativeScale}, but replacing conformal time~$\eta$ with laboratory time $t$. In \autoref{fig:CkBEC}, we replicate the laboratory conditions reported in~\cite{Viermann2022} for a scale factor linear in $t$ and observe that~$\mathcal{C}_k$ increases by several orders of magnitude during the `on' and `off' transitions compared to its lowest value during the expansion, $\bar{\mathcal{C}}_k$. This reinforces our earlier conclusion: The effects of abrupt transitions overshadow the contributions from the intermediate expansion process, effectively masking the dynamics we aim to analyse. One must, therefore, be aware of the role of transitions concerning particle production when performing such experiments, as the outcome stems from the `on' and `off' transitions rather than from the background time-dependence in the intermediate region.

Under laboratory conditions, it is more realistic to assume a nonzero initial occupation number~$n_k^0$, such as that of a thermal state, which results in stimulated particle production from the beginning. In this scenario, the expression for the expected particle number density is modified to~\mbox{$n_k = n_k^{0} + |\beta_k|^2 (1+2n_k^{0}$)}. This adjustment merely introduces an affine transformation. Therefore, the stimulated production of particles and antiparticles remains primarily dictated by the `on' and `off' transitions.

\section{Schwinger effect}
\label{sec:Schwinger}

\begin{figure}
    \centering
    \includegraphics[width=0.55\textwidth]{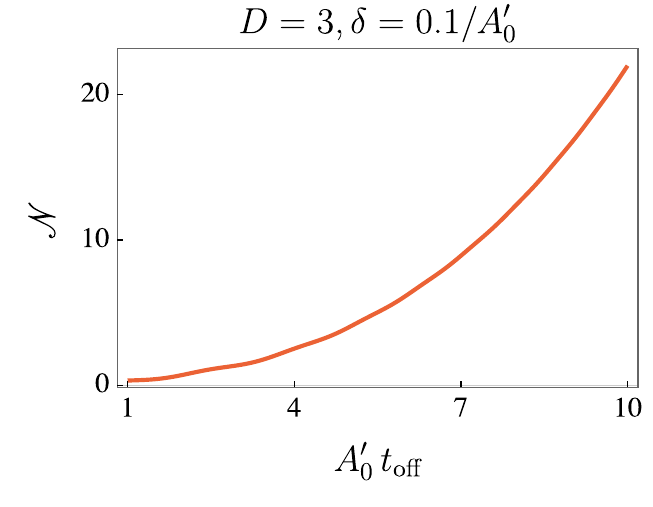}
    \caption{Total number density of produced particles $n$ as function of $t_{\text{off}}$ ($t_{\text{on}}=0$) in the Schwinger effect for electric potentials of the form~\eqref{eq:DerivativeScale}, with an intermediate electric strength of~$-A_0^\prime$ and fast switch-on and switch-off transitions ($\delta=0.1/A_0^\prime$).}
    \label{fig:TotalDensitySchwinger}
\end{figure}

In the Schwinger effect, the role of the scale factor in the cosmological case is replaced by the electromagnetic potential~$A_\mu$. Since the frequency~$\omega_{\bfk}(t)$ in~\eqref{eq:Frequency} is independent of any time derivatives of the potential, the analysis regarding its time variation yields conclusions similar to those in the case of a cosmologically conformally coupled scalar field. However, the intermediate regime has a more significant impact on the particle spectrum in the Schwinger effect than in the cosmological case. As shown in \autoref{fig:TotalDensitySchwinger}, the total number of produced particles in the Schwinger effect continues to increase for large values of~$t_{\text{off}}$, in contrast to the cosmological case (\autoref{fig:TotalDensity}), where particle production eventually converges. 

This is a consequence of the anisotropic nature of the Schwinger effect, which causes the intermediate regime to become increasingly dominant over abrupt transitions as the electric field remains switched on for a sufficiently long duration. Indeed, the linear dependence on~$k$ in the frequency~\eqref{eq:Frequency} causes higher~$k$ modes to become excited as the duration of the electric field increases, particularly for modes aligned with the electric field direction. This is consistent with reference~\cite{Adorno2018}, where it is demonstrated that, when the electric field remains on for sufficiently long durations, the dominant contribution to particle production comes from the intermediate regime rather than the switch-on and switch-off transitions. Nonetheless, even in this case---and even more so when the electric field is switched on for shorter durations---the effects of `on' and `off' transitions remain unavoidable.

\section{Conclusions}
\label{sec:ConclusionsInOut}

We highlight the importance of carefully accounting for `on' and `off' transitions when interpreting quantum pair production due to the expansion of the universe or a strong electromagnetic field. In experiments designed for simulating production within some time window, such regimes are inevitable---experiments have a beginning and an end---, and always influence particle production. Here, we have distinguished when these effects simply influence particle production without dominating it, and when they overwhelmingly dictate the outcome, necessitating a fundamental reinterpretation of the resulting spectra.

This issue is particularly critical in analog gravity experiments that simulate a non-conformally coupled field to an expanding FLRW universe, such as~\cite{Hung2013,Steinhauer2022,Viermann2022,Sparn2024}. We showed that transitions dominate particle production, effectively overshadowing the contributions from the intermediate dynamics. Therefore, one has to be careful when interpreting the outcomes of such experiments, as the main contribution to pair production does not come from the specific expansion during the intermediate phase that the setup is intended to simulate. In scenarios involving alternating periods of expansion and contraction~\cite{Schmidt2024,Sparn2024,Agullo2024}, rapid changes in the scale factor significantly affect the particle spectrum, effectively acting as `on' and `off' transitions within the intermediate regime.

The Schwinger effect presents a notably different scenario. While the `on' and `off' transitions still influence the outcome, in a similar way as in the conformally coupled cosmological case, their relative impact diminishes as the electric field remains switched on for longer times. In this regime, the intermediate period becomes increasingly dominant in determining the particle spectrum. This behaviour stems from the intrinsic anisotropy of the electromagnetic background and stands in sharp contrast to the isotropic cosmological case, where extending the duration of the intermediate expansion has little effect on the spectrum, which remains dominated by the abrupt transitions. 

Studying how `on' and `off' transitions affect production is also of interest in the context of the early universe, where particle production is typically computed from the onset of inflation until the expansion of spacetime slows down significantly, well into the reheating epoch. These periods behave as approximate `in' and `out' regions, where the mode frequency evolves very slowly, and between which `on' and `off' transitions occur.

Analogue experiments inherently incorporate such transitions, and, if appropriately tuned, they could even simulate the \textit{actual} cosmological scenario---including the transition from inflation to reheating, where most particles are known to be produced~\cite{Ema2016,Markkanen2017,Ema2019,Chung2019,BasteroGil2019,Yu2023,Cembranos2023}. However, it is crucial to abandon the idea of isolating the contribution of a specific intermediate region to pair production, as `on' and `off' transitions in such experiments remain unavoidable. As such, the measured particle spectra must be appropriately interpreted.

Finally, one might theoretically consider computing particle creation in a fully dynamical setting without relying on asymptotic `in' and `out' regions. Such an approach could, in principle, isolate the intermediate region of interest. However, as shown repeatedly throughout this thesis, the absence of asymptotic static regimes introduces ambiguities in the definition of the quantum vacuum. In the end, as discussed in \autoref{chap:OperationalRealization}, the effects of `in' and `out’ transitions are not only unavoidable but are intrinsic to particle creation phenomena, just as quantum vacuum ambiguities are an inherent feature of quantum field theory in curved spacetimes~\cite{AlvarezOperational}.

\plainblankpage
\part[Charged fields and black holes]{Charged fields and black holes}
\label{part:BlackHoles}

\plainblankpage

\vspace*{\fill}
\begin{minipage}{0.8\textwidth}

In this part, we explore the profound implications of quantum effects in nature. In line with the central theme of this thesis, we examine how pair creation in strong electromagnetic fields can lead to intriguing phenomena related to black holes.

In \autoref{chap:Kugelblitz}, we show that it is not possible to concentrate enough light to precipitate the formation of an event horizon. We argue that the dissipative quantum effects coming from the Schwinger effect are enough to prevent any meaningful buildup of energy that could create a black hole in any realistic scenario. The content of this chapter is mostly drawn from~\cite{AlvarezKugelblitz,AlvarezKugelblitzComment}.

In \autoref{chap:Superradiance}, we cover another particle creation phenomenon occuring in charged black holes due again to strong electric fields: charge superradiance. Unlike a classical charged bosonic field, a classical charged fermion field on a static charged black hole does not exhibit superradiant scattering. Based on~\cite{AlvarezSuperradiance}, we demonstrate that the quantum version of this classical process is however present for fermions. We construct a vacuum state for the fermion field which has no incoming particles from past null infinity, but which contains, at future null infinity, a nonthermal flux of particles. This state describes both the discharge and energy loss of the black hole, and we analyze how the interpretation of this phenomenon depends on the choice of quantum vacuum. 

\end{minipage}
\vspace*{\fill}
\plainblankpage
\chapter[No black holes from light]{No black holes from light}
\label{chap:Kugelblitz}

\chaptermark{No black holes from light}

One of the consequences of the fact that energy---and not mass---is the one responsible for the curvature of spacetime is the a priori possibility of having massless fields being held together by gravity. These exotic structures (known as \gls{geons}) were first considered by Wheeler~\cite{Wheeler1955,Misner1957,Wheeler1981} for electromagnetic fields. The cases of the (almost massless) neutrinos~\cite{Brill1957} and the gravitational field itself~\cite{Brill1964,Anderson1997} were subsequently studied. These objects are found to be unstable under perturbations~\cite{Perry1999}, leading to either a \textit{leakage} of the massless field~\cite{Wheeler1955} or its collapse into a black hole~\cite{Gundlach2003}. In this context, the term \gls{kugelblitz} (German for \textit{ball lightning}) has become popular as a way to refer to any hypothetical black hole formed by the gravitational collapse of electromagnetic radiation.\footnote{The term \textit{kugelblitz} was initially used by Wheeler in an unpublished reference preceding~\cite{Wheeler1955}, to be later substituted by the term \textit{geon}. While the electromagnetic and gravitational geons considered by Wheeler and his collaborators did not have singularities nor horizons, ``kugelblitz'' has spread in popular science to refer to black holes formed by the gravitational collapse of electromagnetic radiation. Examples of this use include \href{https://www.youtube.com/watch?v=gNL1RN4eRR8}{The Kugelblitz: A Black Hole Made From Light}, and \href{https://www.youtube.com/watch?v=EzZGPCyrpSU}{5 REAL Possibilities for Interstellar Travel}, by the YouTube channels SciShow Space and PBS Space Time, respectively, as well as the Wikipedia entry for \href{https://en.wikipedia.org/wiki/Kugelblitz_(astrophysics)}{Kugelblitz (astrophysics)}.}

Kugelblitze are allowed by general relativity: there are exact solutions to Einstein-Maxwell equations describing black holes generated by the collapse of electromagnetic energy~\cite{Robinson1962,Senovilla2014}. Kugelblitze have been studied in the context of the cosmic censorship hypothesis~\cite{Lemos1992,Lemos1999,Senovilla2014}, the evaporation of white holes~\cite{Senovilla2014}, dark matter~\cite{Guiot2020}, and have even been proposed as the engine of a really speculative option for interstellar travel~\cite{Crane2009,Crane2011,Lee2015}. However, none of these works take into account quantum effects, which should play an important role in determining whether a kugelblitz can form or not. This is especially so if we are interested in black holes of small sizes such as the artificial ones required in~\cite{Crane2009,Crane2011,Lee2015}.

The hypothetical formation of a kugelblitz---even one with as little energy as to be just a few orders of magnitude above the Planck length---would involve electromagnetic field strengths larger than the threshold above which the Schwinger effect stops being exponentially suppressed~\cite{Sauter1931,Heisenberg1936}. This phenomenon hinders the formation of the kugelblitz, since the created particles can scatter out of the region where the radiation is collapsing, carrying their energy with them.

In this chapter, we show that the dissipation of energy via Schwinger effect alone is enough to prevent the formation of kugelblitze with radii ranging from $10^{-29}$ to $10^8$~m. Specifically, we consider the scenario where an external flux of electromagnetic radiation is being focused on a spherical region until there is enough energy to form a Schwarzschild black hole. However, our analysis takes into account that a significant fraction of the energy leaks out of the region of formation due to the Schwinger effect: electron-positron pairs are created inside the region, accelerated by the existing electric field, and subsequently expelled with ultrarelativistic velocities. Our analysis strongly suggests that the formation of black holes solely from electromagnetic radiation is highly implausible under the conditions of the present-day universe, either by concentrating light in a hypothetical laboratory setting or in naturally occurring astrophysical phenomena.

In~\autoref{sec:Setup}, we describe the setup of our model, where we attempt to concentrate a sufficiently strong electric field within a spherical region to create a kugelblitz. To make the scenario more realistic, we consider the creation of electron-positron pairs. While heavier charged particle-antiparticle pairs could also be produced, their production is exponentially suppressed with increasing particle mass, as indicated by~\eqref{eq:ExponentialSuppression}. Before calculating the dissipation rate due to the Schwinger effect, we first extend in~\autoref{sec:HomogeneousElectricFlatFermions} our previous analysis of charged scalar fields in flat spacetime with a homogeneous electric background (\autoref{sec:HomogeneousElectricFlat}) to Dirac fields. Then, in~\autoref{sec:DissipatedEnergy}, we compute the energy dissipated by the Schwinger effect through adiabatic regularization of the time-time component of the Dirac field’s stress-energy tensor. In \autoref{sec:BackOfTheEnvelope}, we present a simplified back-of-the-envelope calculation. While it does not introduce new results beyond the technical calculations of the previous two sections, it serves as an intuitive approach to grasp the core of the phenomenon without requiring a deep dive into the detailed calculations. In \autoref{sec:NoBlackHoles}, we revisit our model, where we demonstrate that Schwinger dissipation is so overwhelming that it prevents black hole formation. In \autoref{sec:ValidityResults}, we examine the validity of our approximations, ensuring that our results hold within the regime of radii studied. Finally, we summarize the main conclusions of this chapter in~\autoref{sec:ConclusionsKugelblitz}.

This chapter builds on the work presented in~\cite{AlvarezKugelblitz} and~\cite{AlvarezKugelblitzComment}. However, \autoref{sec:HomogeneousElectricFlatFermions} and \autoref{sec:DissipatedEnergy} provide a more detailed explanation of the QFT calculations to ensure a consistent comparison with previous results for scalar fields.

In contrast to the rest of this thesis, in this chapter we will explicitly display the dependence on the constants~$\hbar$, $c$, and~$G$. However, in~\autoref{sec:HomogeneousElectricFlatFermions} and~\autoref{sec:DissipatedEnergy}, we will revert to natural units for convenience.

\section{Setup}
\label{sec:Setup}

\begin{figure}
    \centering
    \includegraphics[width=0.65\textwidth]{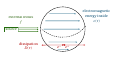}
    \caption{Schematic representation of the kugelblitz formation setup, illustrating the physical interpretation of the terms appearing in~\eqref{Eq: ODE for the change of energy}.}
    \label{fig:Kugelblitz}
\end{figure}

In order to generate a spherically symmetric kugelblitz, it would be necessary to concentrate a critical amount of energy
\begin{equation}
    \epsilon_{\textsc{bh}} = \frac{Rc^4}{2G}
\end{equation}
in a sphere of radius~$R$. To achieve this with focused radiation, we assume a constant influx~$f$ of electromagnetic energy into the sphere. However, the rate at which the energy focused inside the sphere grows is limited by the dissipation due to the scattering of the radiation. The change of energy~$\epsilon(\tau)$ inside the sphere is then governed by
\begin{equation}\label{Eq: ODE for the change of energy}
\frac{\text{d}}{\text{d}\tau} \epsilon(\tau) = 4 \pi R^2 f - D(\tau),
\end{equation}
where $D(\tau)$ is the dissipation rate. As a lower bound on the scattered energy, we only consider the Schwinger effect, since, as we will see, to reach energies close to $\epsilon_\textsc{bh}$ solely with electromagnetic energy, we will need electric field strengths above the Schwinger limit \cite{Sauter1931,Heisenberg1936}, \mbox{$1.3 \cdot 10^{18} \text{ V/m}$}. A schematic illustration of the setup is shown in \autoref{fig:Kugelblitz}.

Estimating~$D(\tau)$ is challenging because, in principle, the Schwinger effect and the kugelblitz formation model we are proposing have fundamentally different natures:
\begin{itemize}
    \item As discussed in \autoref{chap:GQVE}, the generalized QVE reveals that the Schwinger effect exhibits memory effects, making it inherently non-Markovian. Specifically, the generalized QVE is an integro-differential equation for the particle number per mode that depends on the number of particles produced in the past. This arises due to the backreaction of the created pairs, which influences the ability of the background electric field to generate more pairs.
    \item Conversely, the differential equation~\eqref{Eq: ODE for the change of energy}, which describes the potential formation of a kugelblitz, assumes Markovianity, as it is purely local in nature and does not account for past particle production events.
\end{itemize}
To ensure the validity of our analysis, we must find a way to reconcile these two fundamentally different aspects.

The production of particles is restricted to the sphere of radius~$R$ where the electromagnetic energy is confined. These particles are scattered in all directions and eventually leave the sphere in some average \textit{exiting} time~$\sigma_\text{x}$, after which they stop influencing the pair creation. Hence, the fermion production in the sphere after that time can be considered to be \textit{reset}. Another way of understanding this approximation is to consider that the correlations between pairs of fermions produced in the sphere at different instants $\tau_1$ and $\tau_2$ are negligible whenever $|\tau_1-\tau_2| \gg \sigma_\text{x}$. Thus, as long as $\sigma_\text{x}$ is much smaller than the timescale $T$ of the formation of the kugelblitz, it is safe to describe the process in a coarse-grained way that relegates the memory effects to the timescales below $\sigma_\text{x}$. The continuous process is hence discretized into a sequence of non-Markovian processes of typical duration~$\sigma_\text{x}$, and because $\sigma_\text{x} \ll T$, the discrete evolution can be approximated by a continuous one. We will see that this is indeed a good approximation, as we can estimate $\sigma_\text{x}$ to be half the light-crossing time of the sphere, $R/c$, and the timescales predicted by.~\eqref{Eq: ODE for the change of energy} for the formation of a kugelblitz are consistent with $R \ll cT$. 

\begin{figure}
    \centering
    \includegraphics[width=0.9\textwidth]{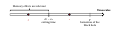}
    \caption{Schematic illustration of the relevant timescales in the setup.}
    \label{fig:Timescales}
\end{figure}

At every instant $\tau$, we model this process by considering an electric field pulse of maximum strength $E(\tau)$ that is switched on and off adiabatically, is homogeneous in the sphere, and stays on for a characteristic time~$\sigma_{\text{x}}$. Adiabaticity ensures that the tails of the electric field profile contribute negligibly to the particle production~\cite{Adorno2018,AlvarezOperational,Ilderton2022,Diez2023}. To make the calculation concrete, let us consider an adiabatic P{\"o}schl-Teller pulse~\eqref{eq:PoschlTeller} (for each~$\tau$): 
\begin{equation} \label{eq:EPT(t)}
    E_\tau(t)=\frac{E(\tau)}{\cosh^2\left(t/\sigma_\text{x}\right)}.
\end{equation}
$E_{\tau}(t)$ reaches its maximum amplitude~$E(\tau)$ at~$t=0$, vanishes asymptotically for $t\rightarrow \pm \infty$, and has a characteristic duration $\sigma_\text{x}$. Here, for a fixed instant~$\tau$, $t$ denotes a time variable at scales comparable to the exiting time~$\sigma_\text{x}$, where memory effects become significantly relevant (see \autoref{fig:Timescales}). 

To compute the energy density carried by the particle-antiparticle pairs, we employ the standard adiabatic regularization and renormalization of the stress-energy tensor. However, before proceeding with this step, we must extend the framework developed in \autoref{sec:HomogeneousElectricFlat}---formulated for charged scalar fields in flat spacetime under a homogeneous electric background---to the case of charged fermions. Our approach will primarily follow the procedure outlined in~\cite{BeltranPalau2020}, which examines Dirac fields in (1+3)-dimensions, while adapting the formalism and notation to align with our conventions. In the following two sections, we omit the~$\tau$ label for simplicity in the notation.

\section{Charged fermions in flat spacetime with homogeneous electric background}  
\label{sec:HomogeneousElectricFlatFermions}

Let us consider a Dirac field~$\Psi(t,\bfx)$ propagating in Minkowski spacetime coupled to an external electromagnetic potential~$A_{\mu}(t,\bfx)$. The dynamics of the fermionic field is governed by the Dirac equation~\eqref{eq:Dirac}, which in flat spacetime takes the form\footnote{In this thesis, we have chosen a different representation for the flat-space Dirac matrices~\eqref{eq:FlatGammaMatrices} than that used in~\cite{AlvarezKugelblitz,AlvarezKugelblitzComment}. As justified in~\autoref{chap:QuantizationScalars}, this results in a different explicit expression for the Dirac equation compared to the one presented in~\cite{AlvarezKugelblitz,AlvarezKugelblitzComment}.}:
\begin{equation} 
    \left[\widetilde{\gamma}^{\mu}\left(\partial_{\mu}+iqA_{\mu}\right)-m\right]\Psi=0.
    \label{eq:DiracMinkowski}
\end{equation}
Similarly to the scalar case discussed in \autoref{chap:ChoiceVacuum}, the temporal gauge $A_{\mu}(t,\bfx)=(0,\textbf{A}(t))$ is preferred, as it explicitly preserves homogeneity in the equations of motion. We fix this gauge, as well as the direction of the electric field so that \mbox{$\textbf{E}(t)=-\dot{A}(t) \ \textbf{e}_3$}. 

In order to proceed with the adiabatic regularization, it will be useful to work with the unitarily transformed field
\begin{equation}
    \Psi^{\prime}=U\Psi, \qquad \text{with} \qquad U=\frac{1}{\sqrt{2}}\widetilde{\gamma}^0(i-\widetilde{\gamma}^3),
    \label{eq:UnitaryTransformationPsi}
\end{equation}
leading to the following reformulation of the Dirac equation~\eqref{eq:DiracMinkowski}:
\begin{equation} 
    \left[ \widetilde{\gamma}^0\partial_0-\widetilde{\gamma}^1\partial_1-\widetilde{\gamma}^2\partial_2-i\partial_3+qA(t)-im\widetilde{\gamma}^3 \right]\Psi^{\prime}(t,\bfx)=0.
    \label{eq:NewDirac}
\end{equation}
Exploiting the homogeneity of the equation of motion, we look for solutions of the form:
\begin{equation}
    \Psi^{\prime}_{\bfk L}(t,\bfx) = (2\pi)^{-\frac{3}{2}} u_{\bfk L}(t)e^{i\bfk\cdot\bfx}.
\end{equation}
This choice parallels the form of the solutions used for the scalar case in~\eqref{eq:AnsatzPhi}, where the plane-wave factor~$e^{i\bfk\cdot \bfx}$ ensures that different wavenumbers~$\bfk$ remain dynamically decoupled. However, in the fermionic case, the time dependence is carried by a basis of orthonormal spinors~$u_{\bfk L}(t)$. On the other hand, $L$ denotes the quirality of the fermion: $L=1$ for positive quirality and $L=-1$ for negative, such that
\begin{equation}
\left( 1 - L \gamma ^{5} \right) \Psi  = 0,
\label{eq:lefthanded}
\end{equation}
where~$\gamma^5$ is the chirality matrix
\begin{equation}
\gamma ^{5} =
i { \widetilde {\gamma }}^{0} { \widetilde {\gamma }}^{1} { \widetilde {\gamma }}^{2}
{ \widetilde {\gamma }}^{3} =
\left(
\begin{array}{cc}
0 & I_{2}  \\
I_{2} & 0
\end{array}
\right).
\label{eq:gamma5}
\end{equation}

Substituting this ansatz into the Dirac equation~\eqref{eq:NewDirac}, we derive a differential equation for the spinors:
\begin{equation}
    \dot{u}_{\bfk L}(t) + \left\{ i k_1 \widetilde{\gamma}^0\widetilde{\gamma}^1 + i k_2 \widetilde{\gamma}^0\widetilde{\gamma}^2 - \left[ k_3 + qA(t) \right] \widetilde{\gamma}^0 + i m\widetilde{\gamma}^0 \widetilde{\gamma}^3 \right\} u_{\bfk L}(t) = 0.
    \label{eq:ukL}
\end{equation}
From this equation, we extract the time-dependence of the spinor into two functions~$h_\bfk^I(t)$ and~$h_\bfk^{II}(t)$ such that
\begin{equation}
    u_{\bfk L}(t) = \begin{pmatrix} h_\bfk^I(t) \eta_{\bfk L} \\ h_\bfk^{II}(t) L\eta_{\bfk L}\end{pmatrix},
\end{equation}
where~$\eta_{\bfk L}$ is a two-spinor.

The solutions~$\Psi_{\bfk L}^\prime$ are orthonormal with respect to the Dirac product~\eqref{eq:DiracProduct}---that is, \mbox{$(\Psi_{\bfk L}^\prime, \Psi_{\bfk^\prime L^\prime}^\prime) = 0$}---, if and only if
\begin{equation}
    u_{\bfk L}^\dagger(t) u_{\bfk L^\prime} (t) = \delta_{L,L^\prime}.
\end{equation}
This condition is equivalent to the normalization constraints:
\begin{equation}
    |h_{\bfk}^{I}(t)|^2+|h_{\bfk}^{II}(t)|^2=1 \qquad \text{and} \qquad \eta_{\bfk L}^\dagger \eta_{\bfk L^\prime} = \delta_{L,L^\prime}.
    \label{eq:Normalizationhk}
\end{equation}

Substituting this ansatz into the equation of motion~\eqref{eq:ukL} for the spinor~$u_{\bfk L}(t)$, we find that it is particularly useful to choose the two-spinor basis such that:
\begin{equation}
    \left( k_1\sigma_1 + k_2 \sigma_2 + m\sigma_3 \right) \eta_{\bfk L} = L \kappa \eta_{\bfk L},
\end{equation}
where~$\kappa = \sqrt{k_1^2 +k_2^2 + m^2}$. This condition is satisfied by the following choice of spinors:
\begin{equation}
    \eta_{\bfk,L=+1}=\frac{1}{\sqrt{2\kappa (\kappa+m)}} \mqty(\kappa+m \\k_1+ik_2), \qquad \eta_{\bfk, L=-1}=\frac{1}{\sqrt{2\kappa (\kappa+m)}} \mqty(-k_1+ik_2 \\ \kappa+m),
    \label{eq:2DSpinors}
\end{equation}
The prefactor ensures normalization according to~\eqref{eq:Normalizationhk}.

The time-dependent functions~$h_{\bfk}^{I}(t)$ and~$h_{\bfk}^{II}(t)$ satisfy the coupled differential equations:
\begin{align}
    \dot{h}_{\bfk}^{I}(t)&-i[k_3+qA(t)]h_{\bfk}^{I}(t)-i\kappa h_{\bfk}^{II}(t)=0, \nonumber \\
    \dot{h}_{\bfk}^{II}(t)&+i[k_3+qA(t)]h_{\bfk}^{II}(t)-i\kappa h_{\bfk}^{I}(t)=0.
    \label{eq:hk}
\end{align}

To ensure that the chosen complex structure preserves the symmetries of the equations of motion, we consider two key aspects:
\begin{itemize}
    \item Spatial homogeneity is already explicitly incorporated through the plane-wave factor~$e^{i\bfk\cdot\bfx}$.
    \item If the pair~$(h_\bfk^I,h_\bfk^{II})$ is solution to the system of equations~\eqref{eq:hk}, then~$(-(h_\bfk^{II})^*,(h_\bfk^I)^*)$ is also a solution. 
\end{itemize}
To reflect these symmetries in the complex structure---and consequently in the quantum theory---we propose:
\begin{equation}
    \Psi^{\prime +}_{\bfk L}(t,\bfx) = (2\pi)^{-\frac{3}{2}} u_{\bfk L}^+(t)e^{i\bfk\cdot\bfx}, \qquad \Psi^{\prime -}_{\bfk L}(t,\bfx) = (2\pi)^{-\frac{3}{2}} u^-_{\bfk L}(t)e^{i\bfk\cdot\bfx},
    \label{eq:ComplexStructureSchwingerDirac}
\end{equation}
where the spinors are of the form:
\begin{equation}
    u^+_{\bfk L}(t) = \begin{pmatrix} h_\bfk^I(t) \eta_{\bfk L} \\ h_\bfk^{II}(t) L\eta_{\bfk L}\end{pmatrix}, \qquad u^-_{\bfk L}(t) = \begin{pmatrix} -[h_\bfk^{II}(t)]^* \eta_{\bfk L} \\ [h_\bfk^I(t)]^* L\eta_{\bfk L}\end{pmatrix}.
    \label{eq:u+u-}
\end{equation}
The time-dependent functions $h_{\bfk}^{I}$ and $h_{\bfk}^{II}$ parametrize the ambiguities in the choice of the complex structure.

We now proceed with the canonical quantization procedure described in~\autoref{sec:Fermions}. The specific form of the quantum Dirac field operator~\eqref{eq:DiracOperator}, corresponding to the complex structure~\eqref{eq:ComplexStructureSchwingerDirac}, is given by:
\begin{equation}
\label{eq:DiracOperatorSchwinger}
    \hat{\Psi}^{\prime}(t,\bfx) = \sum_{L=\pm 1} \int \frac{\text{d}^3\bfk}{(2\pi)^{\frac{3}{2}}} \ \left[ \hat{c}_{\bfk L} u_{\bfk L}^+ (t)
    + \hat{d}_{\bfk L}^{\dagger}u_{\bfk L}^- (t) \right] e^{i\bfk\cdot\bfx}
\end{equation}
with the annihilation and creation operators~$\hat{c}_{\bfk L}$ and~$\hat{d}_{\bfk L}^{\dagger}$ satisfying the canonical anticommutation relations 
\begin{equation} 
    \{\hat{c}_{\bfk L},\hat{c}_{\bfk^\prime L^\prime}^{\dagger}\}=\{\hat{d}_{\bfk L},\hat{d}_{\bfk^\prime L^\prime}^{\dagger}\}=\delta(\bfk-\bfk')\delta_{L,L^\prime},
    \label{eq:anticommutatorsvariables}
\end{equation}
and all other anticommutators vanish.

Regarding Bogoliubov transformations, the explicit spatial homogeneity ensures that the Bogoliubov coefficients are diagonal, analogous to the scalar case (see \eqref{eq:BogoliubovDiagonal}):
\begin{equation}
    \alpha^\pm_{\bfk\bfk^\prime} = \alpha^\pm_\bfk \delta(\bfk-\bfk^\prime), \qquad \beta^\pm_{\bfk\bfk^\prime} = \beta^\pm_\bfk \delta(\bfk-\bfk^\prime).
\end{equation}
The relations between the coefficients given in~\eqref{eq:BogoliubovConstraintsFermions} allow us to write two of them in terms of the others: \mbox{$\alpha_\bfk^- = (\alpha_\bfk^+)^* \equiv \alpha_\bfk^*$} and \mbox{$\beta_\bfk^- = -(\beta_\bfk^+)^* \equiv -\beta_\bfk^*$}. These coefficients can be computed in terms of the time-dependent functions~$h_\bfk^I$ and~$h_\bfk^{II}$:
\begin{equation}
    \alpha_\bfk = (h_\bfk^I)^*\widetilde{h}_\bfk^I + (h_\bfk^{II})^* \widetilde{h}_\bfk^{II}, \qquad \beta_\bfk = h_\bfk^I \widetilde{h}_\bfk^{II} - \widetilde{h}_\bfk^I h_\bfk^{II},
    \label{eq:betahk}
\end{equation}
and they satisfy:
\begin{equation}
    |\alpha_\bfk|^2 + |\beta_\bfk|^2 = 1.
\end{equation}
This last relation is structurally similar to the constraint in the scalar case~\eqref{eq:|alpha||beta|}, but with a plus sign instead of a minus sign, reflecting the underlying difference in bosonic and fermionic statistics, i.e., the anticommutators instead of commutators.

\subsection*{P{\"o}schl-Teller pulse}

From now on, we consider the analytically solvable Sauter electric potential~\eqref{eq:Sauter}\footnote{This potential differs from the one used in the original reference~\cite{AlvarezKugelblitz}, where it was defined with an opposite sign and a constant shift: \mbox{$A(t) = -E_0\sigma \tanh(t/\sigma)$}, to maintain consistency with the convention adopted in previous sections.}
\begin{equation}
    A(t)=E_0\sigma\left[\tanh \left(\frac{t}{\sigma}\right) + 1\right],
\end{equation}
which corresponds to the P{\"o}schl-Teller electric pulse in~\eqref{eq:EPT(t)}, after omitting the~$\tau$ label and replacing~$E(\tau)$ with~$E_0$, and identifying~$\sigma$ with the exiting time~$\sigma_\text{x}$. Our goal is to reformulate the `in' and `out' formalism developed in~\autoref{chap:ChoiceVacuum} for charged scalars, but now applied to charged fermions.

In the asymptotic past there is still no electric field on, thus the `in' solutions $h_{\bfk}^{I}(\tau)$, $h_{\bfk}^{II}(\tau)$ to Eq.~\eqref{eq:hk} are the particular solutions that asymptotically behave as  Minkowski positive frequency plane waves, i.e., 
\begin{equation}
    h_{\bfk}^{\text{in},I/II}(t) \sim \pm\sqrt{\frac{\omega_\bfk^{\text{in}} \mp k_3
    }{2\omega_\bfk^{\text{in}}}}e^{-i\omega_\bfk^{\text{in}} t}, 
    \qquad \text{when}\quad t\rightarrow -\infty.
\end{equation}
The evolution of these in-solutions can be written in terms of hypergeometric functions \cite{BeltranPalau2019}:
\begin{equation}
    h_\bfk^{\text{in},I/II}(t) = \pm \sqrt{\frac{\omega_\bfk^{\text{in}}\mp k_3}{2\omega_\bfk^{\text{in}}}} e^{-i\omega_\bfk^{\text{in}}t} \left( 1+e^{\frac{2t}{\sigma}} \right)^{-i\sigma\omega^-_\bfk} {}_2F_1\left( \lambda_\bfk^\pm, 1+\lambda_\bfk^{\mp}, 1-i\sigma\omega_\bfk^{\text{in}}; \frac{1}{2}\left[1+\tanh\left(\frac{t}{\sigma}\right)\right] \right),
\end{equation}
where
\begin{equation}
    \omega_\bfk^-=\frac{1}{2}\left( \omega_\bfk^{\text{out}}-\omega_\bfk^{\text{in}} \right), \qquad \lambda_\bfk^\pm = i\sigma\left( \omega_\bfk^- \pm qE_0\sigma \right),
\end{equation}
and the `in' and `out' frequencies are given in~\eqref{eq:InFrequency} and~\eqref{eq:OutFrequency}.

The `out' solutions are defined by their asymptotic behaviour as positive frequency plane waves, when the electric field is asymptotically switched off:
\begin{equation}
    h_{\bfk}^{\text{out},I/II}(t) \sim \pm \sqrt{\frac{\omega_\bfk^{\text{out}} \mp [k_3+2qE_0\sigma]}{2\omega_\bfk^{\text{out}}}}e^{-i\omega_\bfk^{\text{out}} t}, \qquad \text{when}\quad t\rightarrow +\infty.
\end{equation}

In the `out' region the `in' solutions become a linear combination of positive and negative frequency `out' plane waves:
\begin{align}
\label{eq:hkSauter}
 h_{\bfk}^{\text{in},I/II}(t) \sim  &\pm \alpha_{\bfk}\sqrt{\frac{\omega_\bfk^{\text{out}} \mp [k_3+2qE_0\sigma]}{2\omega_\bfk^{\text{out}}}}e^{-i\omega_\bfk^{\text{out}} t} \nonumber \\ & 
 +\beta_{\bfk}\sqrt{\frac{\omega_\bfk^{\text{out}} \pm [k_3+2qE_0\sigma]}{2\omega_\bfk^{\text{out}}}}e^{i\omega_\bfk^{\text{out}} t},
 \qquad \text{when}\quad t\rightarrow +\infty.
\end{align}

To compute the $\beta$-Bogoliubov coefficient, we evaluate~\eqref{eq:betahk} in the asymptotic future:
\begin{equation}
    \beta_\bfk = \lim_{t\rightarrow +\infty} \left[h_\bfk^{\text{out},I}(t) h_\bfk^{\text{in},II}(t) - h_\bfk^{\text{in},I}(t) h_\bfk^{\text{out},II}(t)\right].
\end{equation}
Taking its squared modulus, we obtain the fermionic counterpart of the scalar result~\eqref{eq:nkSauterScalar} for the number of created particles and antiparticles:
\begin{equation} \label{eq:nkSauter}
    \mathcal{N}_\bfk = 2|\beta_\bfk|^2=\frac{\cosh{\left[2\pi qE_0\sigma^2\right]}-\cosh{\left[\pi\left(\omega_\bfk^{\text{out}}-\omega_\bfk^{\text{in}}\right)\sigma\right]}}{\sinh{\left[\pi\omega_\bfk^{\text{in}}\sigma\right]}\sinh{\left[\pi\omega_\bfk^{\text{out}}\sigma\right]}}.
\end{equation}

\section{Dissipated energy}
\label{sec:DissipatedEnergy}

Here, we calculate the energy density of particles created via Schwinger effect that are scattered out of the sphere where a potential kugelblitz is forming. To achieve this, we compute the energy density of fermions generated by a homogeneous electric field pulse defined in the entire space, which matches the energy density within the region of formation.
As the vacuum expectation value of the energy-momentum tensor involves ultraviolet divergences, we will resort to its adiabatic regularization and renormalization. The adiabatic regularization used in gravitational scenarios in~\cite{Suen1987,Suen1987a} can be readapted in the presence of a homogeneous time-dependent electric field, both in Minkowski~\cite{Cooper1989,Kluger1991,Kluger1992,Barbero2018} and FLRW spacetimes~\cite{Barbero2018,Ferreiro2018,Ferreiro2018a,Ferreiro2019,BeltranPalau2020}. We mainly follow again the reference~\cite{BeltranPalau2020}, where an analogue computation is done for the adiabatic regularization of the charge current.

From the action~\eqref{eq:ActionFermion}, the energy-momentum tensor of a Dirac field $\Psi(t,\bfx)$ in a flat and electromagnetic background is given by
\begin{equation}
T_{\mu \nu }= \frac {i}{2} \left( {\overline {\Psi }} \widetilde{\gamma} _{(\mu }
\partial _{\nu )} \Psi - \left( \partial _{( \mu }{\overline {\Psi }}\right)
\widetilde{\gamma} _{\nu )} \Psi + 2iqA_{(\mu} {\overline {\Psi }} \widetilde{\gamma}_{\nu) } \Psi \right),
\label{eq:Tmunuflat}
\end{equation}
where parentheses are used to denote symmetrization of indices. In terms of the unitarily transformed field~$\Psi^\prime = U\Psi$ given in~\eqref{eq:UnitaryTransformationPsi}, the time-time component of this tensor can be written as
\begin{equation}
T_{00 }= -i \left( \Psi ^{\prime \dagger}
\partial_t \Psi^\prime - \left( \partial_t {\Psi^{\prime \dagger}}\right)
\Psi^\prime \right).
\end{equation}

To define a quantum operator that takes into account both contributions coming from the flux of particles and antiparticles, we need to introduce commutators.\footnote{In the original reference~\cite{AlvarezKugelblitz}, these commutators were not introduced, resulting in only the contribution to the dissipated energy from antiparticles being considered. This omission leads to a discrepancy by a factor of~$2$ in the estimated dissipation rate via the Schwinger effect. However, as we will see, due to the robustness of our results, this discrepancy does not affect our outcomes in any way.} However, in order to preserve the spinorial structure, these commutators can only act on the annihilation and creation operators and not on the Dirac spinors. This quantum operator is then defined as
\begin{equation}
\hat{T}_{00 }= -\frac{i}{2} \left\{ \left[\hat{\Psi} ^{\prime \dagger},
\partial_t \hat{\Psi}^\prime \right] - \left[\left( \partial_t {\hat{\Psi}^{\prime \dagger}}\right),
\hat{\Psi}^\prime \right] \right\}.
\end{equation}
Its vacuum expectation value in the quantum vacuum~$|0\rangle$, determined by the chosen quantization scheme, can be written using~\eqref{eq:u+u-}--\eqref{eq:anticommutatorsvariables}) as
\begin{equation} \label{eq:ExpT}
    \langle 0 |\hat{T}_{00}(t,\bfx)|0 \rangle= \int \text{d}^3 \bfk \ \rho_\bfk(t),
\end{equation}
where the contribution from each mode~$\bfk$ is
\begin{equation}
\rho_\bfk(t)=\frac{4}{(2\pi)^3} \Im{h_{\bfk}^{I}(t)\dot{h}_{\bfk}^{I*}(t)+h_{\bfk}^{II}(t)\dot{h}_{\bfk}^{II*}(t)}.
\label{eq:rhok}
\end{equation}

This result requires renormalization. To proceed with its adiabatic regularization \cite{Barbero2018,Ferreiro2019,BeltranPalau2020}, we compute the energy density $\rho_{\bfk}^{\text{ad}}(\tau)$ for the zeroth-order adiabatic quantum vacuum\footnote{Here, we assign a zero adiabatic order to the electric potential~$A(t)$. It is noteworthy that when the background includes both gravitational and electromagnetic contributions, references~\cite{Ferreiro2018,Ferreiro2018a} state that the adiabatic regularization scheme must be performed treating $A(t)$ as a variable of first adiabatic order. However, in a pure electromagnetic background in Minkowski spacetime, the adiabatic order of the electromagnetic potential is irrelevant for the computation of physical observables.}, determined by the choice of functions:
\begin{equation}
    h_{\bfk}^{\text{ad},I/II}(t)=\pm\sqrt{\frac{\omega_{\bfk}(t)\mp [k_3+qA(t)]}{2\omega_{\bfk}(t)}}e^{-i\int^t \text{d}t^\prime \ \omega_{\bfk}(t^\prime) }.
\end{equation}
This expression for the zeroth-order adiabatic modes is analogous to the expression given in~\autoref{sec:ParametrizationVacua} for scalar fields. Up to zeroth-order in the adiabatic expansion, this results in an energy density per mode
\begin{equation}
    \rho_{\bfk}^{\text{ad}}(t)=\frac{4}{(2\pi)^3} \omega_{\bfk}(t).
    \label{eq:rhoadiabatic}
\end{equation}

We will be interested in evaluating the renormalized expectation value~$\langle \hat{T}_{00} \rangle_{\text{ren}}$ when the electric field is asymptotically switched off. We need to evaluate this quantity for the `in' quantum vacuum in the asymptotic future. Substituting the asymptotic behaviour~\eqref{eq:hkSauter} of the $h_\bfk^{\text{in},I/II}$ functions in the energy density per mode~\eqref{eq:rhok} we obtain:
\begin{equation} \label{eq:rhofuture}
    \lim_{t\rightarrow +\infty} \rho_\bfk^{\text{in}}(t) = \frac{4}{(2\pi)^3}  \left(1-\mathcal{N}_\bfk\right)\omega_\bfk^{\text{out}},
\end{equation}
where $1-\mathcal{N}_{\bfk}$ is the usual Pauli-blocking factor determined by~\eqref{eq:nkSauter}. Subtracting the contribution at~$t\rightarrow +\infty$ from the zeroth-order adiabatic vacuum~\eqref{eq:rhoadiabatic}:
\begin{equation}
\label{eq:T00ren}
    \lim_{t\rightarrow +\infty} \langle \hat{T}_{00} \rangle_{\text{ren}} =\lim_{t\rightarrow +\infty}\left[\langle \text{in}|\hat{T}_{00} |\text{in}\rangle - \langle\text{ad}|\hat{T}_{00}|\text{ad}\rangle \right]
    =-\int \frac{\text{d}^3\bfk}{(2\pi)^3} \ 4 \mathcal{N}_{\bfk}\omega_\bfk^{\text{out}}.
\end{equation}
The factor of $4$ comes from the contribution of each function $h_{\bfk}^{\text{in},I}(t)$ and $h_{\bfk}^{\text{in},II}(t)$, being each one the same for the two different spins.

In the regime where~$qE_0\sigma^2\gg 1$, the particle number~\eqref{eq:nkSauter} for the Sauter-type electric potential yields
\begin{equation}
\mathcal{N}_{\bfk} \approx
\begin{cases}
     \ 2\exp\left( -\pi\frac{k_{\perp}^2+m^2}{qE_0} \frac{1}{1-[k_3/(qE_0\sigma)+1]^2} \right) & \text{for} -2qE_0\sigma \leq k_3 \leq 0, \\
     \ 0 & \text{otherwise},
\end{cases}
\end{equation}
where~$k_\perp = \sqrt{k_1^2 + k_2^2}$ is the transverse momentum. The negative exponent suppresses the contribution of modes with large~$k_\perp^2+m^2$ (with respect to $qE_0$) in the integral~\eqref{eq:T00ren}. Thus, when~$qE_0\sigma^2\gg 1$, we can approximate
\begin{equation}
\omega_\bfk^{\text{out}}=\sqrt{(k_3+2qE_0\sigma)^2+k_\perp^2 + m^2}\approx k_3 + 2qE_0\sigma
\end{equation}
in the domain of the integral. As a result, we obtain
\begin{equation}
    \lim_{t\rightarrow +\infty} \langle\hat{T}^{00}\rangle_{\textrm{ren}} \approx -\frac{2}{\pi^2}\int_{-2qE_0\sigma}^{0} \text{d}k_3  \int_{0}^{+\infty}\text{d}k_{\perp} k_{\perp} \ (k_3 + 2qE_0\sigma) \exp\left[ -\frac{\pi\frac{k_{\perp}^2+m^2}{qE_0}}{1-\left(\frac{k_3}{qE_0\sigma}+1\right)^2} \right].
\end{equation}
The integration in the transverse momentum $k_{\perp}$ is straightforward. After simple parity arguments and the change of variables $u=k_3/(qE_0\sigma)+1$, it leads to
\begin{equation}\label{Eq: final integral}
    \lim_{t\rightarrow +\infty} \langle\hat{T}^{00}\rangle_{\textrm{ren}} \approx -\frac{2}{\pi^3}(qE_0)^3\sigma^2 \int^1_0 \text{d}u \ (1-u^2)\exp\left( -\pi\frac{m^2}{qE_0} \frac{1}{1-u^2} \right).
\end{equation}
The values of~$u$ around~$1$ give negligible contributions to the integral. For other values of~$u$, assuming the strong field regime~$qE_0\gg m^2$, we can approximate the exponential term as one, resulting in
\begin{equation}
    \lim_{t\rightarrow +\infty} \langle\hat{T}^{00}\rangle_{\text{ren}}\approx -\frac{4}{3\pi^3}(qE_0)^3\sigma^2.
\end{equation}

Finally, notice that in our setup the particle production does not happen in free space, but inside a bounded region of diameter $2R$. However, since the calculation was performed for a homogeneous scenario, the number of produced particles is also homogeneous. Therefore, the only effect that the bounded region has is the removal of the contribution from the modes with $k_3 \lesssim \pi/R$. After the change of variables preceding~\eqref{Eq: final integral}, this corresponds to $u \lesssim \pi/[e E(t) \tau_\text{x} R] \approx \pi/[e E(t) \tau_\text{x}^2] \ll 1$, where we recovered the notation from~\autoref{sec:Setup}. Hence, the contribution from this range of frequencies is negligible in comparison with the result of the integral in $u$ of~\eqref{Eq: final integral}, which is $\mathcal{O}(1)$.

\section{No black holes from light}
\label{sec:NoBlackHoles}

To summarize the key results from the previous two sections, we focus on the regime
\begin{equation}\label{Eq: approx}
    qE(\tau)\hbar \gg m^2 c^3 \quad \text{and} \quad qE(\tau)\sigma_\text{x}^2 c \gg \hbar.
\end{equation}
The first inequality simply requires that the electric field strength is much larger than the Schwinger limit, \mbox{$1.3 \cdot 10^{18} \text{ V/m}$}, above which pair production effectively takes place. As for the second inequality, we will verify its consistency with our results later.

In this regime, we have shown that the energy dissipated via the Schwinger effect at time~$\tau$ is given by
\begin{equation}\label{Eq: approx energy density particle production}
    \lim_{t\rightarrow \infty} \langle\hat{T}^{00}(t,\bfx)\rangle_{\text{ren}}\approx -\frac{4 q^3}{3\pi^3\hbar^2}\sigma_\text{x}^2 E(\tau)^3.
\end{equation}

Dividing this result by the duration~$\sigma_\text{x}$ and multiplying by the volume of the sphere yields the energy dissipation rate
\begin{equation}\label{Eq: estimated Schwinger dissipation}
D(\tau) \approx  \frac{16 q^3}{9\pi^2\hbar^2} \sigma_\text{x} [R E(\tau)]^3 . 
\end{equation}
Since the electromagnetic energy in the sphere is 
\begin{equation}
    \epsilon(\tau)=\left[\frac{4}{3}\pi R^3\right] \left[\frac{1}{2}\varepsilon_0E(\tau)^2\right],
    \label{eq:EMEnergy}
\end{equation}
we can rewrite~\eqref{Eq: ODE for the change of energy} as a first order differential equation for the electric field~$E(\tau)$:
\begin{equation}\label{Eq: model}
\varepsilon_0 E(\tau) \frac{\text{d}}{\text{d}\tau} E(\tau) = \frac{3}{R}f - \frac{4q^3\sigma_\text{x}}{3\pi^3\hbar^2} E(\tau)^3.
\end{equation}
This equation has a fixed point
\begin{equation}\label{Eq: E infinity}
    E_\infty=\bigg( \frac{9 \pi^3 \hbar^2 f}{4 q^3 R \sigma_\text{x}} \bigg)^{1/3},
\end{equation}
and all its solutions are monotonic and convergent to this fixed point as $\tau\rightarrow +\infty$. Because the electric field needs to build up for the kugelblitz to form dynamically, the monotonicity of the solutions implies that, for the kugelblitz to be viable, the fixed point $E_\infty$ must be above the electric field $E_\textsc{bh}$ required to form an electromagnetic black hole. Setting~$q=e$, the elementary charge of the electron:
\begin{align}\label{Eq: Electric field BH}
E_\infty > E_\textsc{bh} & = \sqrt{\frac{3}{4\pi R^3} \frac{2\epsilon_\textsc{bh}}{ \varepsilon_0 }} = \sqrt{3} \frac{c^4 \sqrt{\alpha}}{e G} \frac{\ell_\textsc{p}}{R} \sim \frac{10^{27} \; \text{V}}{R},
\end{align}
where
\begin{equation}
    \alpha = \frac{e^2}{4\pi \epsilon_0 \hbar c} \sim \frac{1}{137} \qquad \text{and} \qquad \ell_\textsc{p} = \sqrt{\frac{\hbar G}{c^3}} \sim 10^{-35} \; \text{m}
\end{equation}
are the fine structure constant and the Planck length, respectively. Otherwise the electric field in the sphere would stabilize before reaching the critical value~$E_\textsc{bh}$, and the black hole would never form. Notice for reference that the strongest electromagnetic fields in nature are found in magnetars~\cite{Mereghetti2015,Turolla2015,Kaspi2017}, a kind of neutron star. Magnetars display magnetic strengths of $10^{11} \text{ T}$, corresponding to electric strengths of $10^{19} \text{ V/m}$. Meanwhile, the strongest electric field achieved so far in the laboratory is of the order of $10^{15} \text{ V/m}$~\cite{Yoon2021}.

Since $\sigma_\text{x}$ is bounded from below by $R/c$, it can be checked that electric field strengths close to $E_\textsc{bh}$ for 
\begin{equation}
    10^{-29} \text{ m} \lesssim R \lesssim 10^{8} \text{ m}
    \label{eq:Rvalid}
\end{equation}
fall well within the regime of approximation given by~\eqref{Eq: approx}. Indeed, the higher bound on the radii considered is imposed by the first approximation, which assures that we are in the regime where Schwinger particle production happens. The lower bound assures the fulfilment of the second inequality, which can be rewritten as
\begin{equation}
E(\tau) \sigma_\text{x}^2 c^2 \approx E(\tau) R^2 \gg \frac{\hbar c}{e} \sim 10^{-7} \; \textrm{V m}.
\end{equation}
From~\eqref{Eq: Electric field BH}, \mbox{$E_\textsc{bh} R \sim 10^{27} \text{ V}$}. Thus, both inequalities are satisfied for the interval of radii given in~\eqref{eq:Rvalid}, and therefore need to be satisfied by $E(\tau)$ from some instant forward, since it needs to approach $E_\textsc{bh}$ for the kugelblitz to form. 

In this regime, the scattered particles are ultrarelativistic  and we can estimate the exiting time by half the light-crossing time of the sphere, \mbox{$\sigma_\text{x} \approx R/c$}. Then, $E_\infty > E_\textsc{bh}$ implies that 
\begin{equation}\label{Eq: power needed}
f R > \frac{4}{\sqrt{3}\pi^3} \frac{c^5 \alpha^{\frac{3}{2}}}{G} \frac{1}{\ell_\textsc{p}} \sim 10^{83} \; \textrm{W/m}.
\end{equation}
The intensity required to form a laboratory-scale kugelblitz ($R \lesssim 1 \text{ m}$) would be of approximately $10^{83} \text{ W/m}^2$, more than 50 orders of magnitude above state-of-the-art laser pulse intensities, which reach \mbox{$10^{27} \text{ W/m}^2$}~\cite{Yoon2021}. For astrophysical sources, the intensity required is still many orders of magnitude above the highest-intensity sources in the universe, including quasars~\cite{Hopkins2007,Sulentic2014,Wolf2024} and supernovae~\cite{Nicholl2020}. Moreover, from~\eqref{Eq: power needed}, the total power input must satisfy  
\begin{equation}\label{Eq: total power needed}
4 \pi R^2 f > \frac{16}{\sqrt{3} \pi^2} \frac{c^5 \alpha^{\frac{3}{2}}}{G} \frac{R}{\ell_\textsc{p}} \sim R \cdot (10^{84} \; \textrm{W/m}),
\end{equation}
which is far from the bolometric luminosity of the brightest quasars, $10^{41}$ W~\cite{Hopkins2007,Sulentic2014,Wolf2024}, for any kugelblitz radius above the Planck length. This shows that the formation of a kugelblitz requires energy levels that are not achievable either naturally or artificially. 

\section{A back-of-the-envelope calculation}
\label{sec:BackOfTheEnvelope}

To give some additional intuition on the competition between the attractive effect of gravity and the dissipation via Schwinger effect that prevents the formation of kugelblitze, we offer the following back-of-the-envelope estimation that illustrates how the effects scale with the size of the (potential) black hole. This is not intended to replace the previous rigorous analysis, but rather to provide an intuitive understanding of why the Schwinger effect dominates over gravitational effects for the aforementioned window of black hole sizes, thereby preventing the formation of kugelblitze. This simplified approach can help grasp the basic core of the phenomenon without requiring a deep dive into the detailed calculations.

The energy density of a pair produced by Schwinger effect in a strong homogeneous electric field is given by~\eqref{Eq: approx energy density particle production}, with the energy density being proportional to the third power of the field strength~$E$. Seeing that this is the correct scaling is easy: the standard calculation for the particle density creation rate via Schwinger effect above the Schwinger limit~\cite{Schwinger1951} gives $\text{d}n/\text{d}t\propto E^2$. On the other hand, the energy of each particle in the strong field limit behaves as $\omega\sim \sigma_\text{x}E$, since $\sigma_\text{x}$ represents the effective time each particle is subjected to the electric field. Therefore, the power is proportional to~$\sigma_\text{x}E^3$, and the energy is proportional to $\sigma_\text{x}^2E^3$, which is the scaling that the energy density dissipated in~\eqref{Eq: approx energy density particle production} shows.

Now, we can rewrite this scaling in terms of the radius $R$ and the total electromagnetic energy \mbox{$\epsilon=(4\pi R^3/3) \varepsilon_0 E^2/2$}. Since the time that it takes for the (ultrarelativistic)  particles to leave the region is \mbox{$\sigma_\text{x}\approx R/c$}, the energy density dissipated scales as
\begin{equation}\label{Eq: 2}
    \lim_{t\rightarrow \infty} \langle\hat{T}^{00}(t,\bfx)\rangle_{\text{ren}} \sim \sigma_\text{x}^2E^3 \sim \epsilon^{\frac32}R^{-\frac{5}{2}}.
\end{equation}
This provides intuition as to why in our calculations the energy density lost via Schwinger effect scales with a negative power of the radius. Now notice that in this estimation there is no a priori relationship between the total energy $\epsilon$ and the radius, as there would be in a black hole, for which $R=2GM/c^2$. If we want to compare with the mass scaling of the effective energy density of a black hole, we could look at a situation where we have electromagnetic radiation at a point just about enough to produce a black hole of mass~$M$ and just naively take $\epsilon\sim Mc^2$, $R\sim 2G M/c^2$, and then one sees from~\eqref{Eq: 2} that the Schwinger effect decays slower ($M^{-1}$) than the effective energy density for a black hole ($M^{-2}$) as the mass increases.  This shows that as long as one is past the Schwinger limit there are regimes (precisely the aforementioned range between $10^{-29}$ and $10^8$ m) where the energy density dissipated by the Schwinger effect scales favourably as compared to the energy density necessary to form a black hole, and can therefore overpower gravitational confinement.

\section{Validity of the results} 
\label{sec:ValidityResults}

In order to reach our conclusions, we used an admittedly simple description of the formation of a kugelblitz. Here, we justify that we should not expect significant deviations between the extreme orders of magnitude predicted in our simple setup and those of a more sophisticated one. 

First, let us examine in greater detail the broad range of radii~\eqref{eq:Rvalid} for which our results apply. On one hand, $10^{-29} \text{ m}$ is more than ten orders of magnitude below the smallest focus spot size achieved for a laser~\cite{Self1983,Siegman1986}, and it is close enough to the Planck length to consider it outside of any naturally occurring phenomena. On the other hand, the amount of energy necessary to form a black hole with $R \gtrsim 10^9 \text{ m}$ is approximately $10^{53} \text{ J}$, which is the energy output of a bright quasar for over $10^4 \text{ years}$~\cite{Hopkins2007,Sulentic2014,Wolf2024}. Nevertheless, black holes of these sizes and larger do exist. These are the so-called supermassive black holes (with masses $M \gtrsim 10^{6} M_{\odot}$)~\cite{Kormendy1995,Ferrarese2005,Volonteri2010,Kormendy2013}. However, the proposed mechanisms for their formation involve the merger of smaller black holes and/or the evolution from an intermediate-mass black hole through the accretion of matter~\cite{Loeb1994,Ebisuzaki2001,Bromm2003,Volonteri2003,Gurkan2004,Volonteri2007,Tanaka2009,Kulier2015,Latif2016,Pacucci2020}, rather than its direct collapse~\cite{Begelman2006}. Thus, the formation of a kugelblitz of these characteristics seems extremely implausible, except for maybe the exceptionally conditions of the very early universe.

Let us analyse the main approximations made throughout our argument, namely: \mbox{a) estimating} the dissipation using the specific electric pulse~\eqref{eq:EPT(t)}, \mbox{b) considering} a uniform electric field, \mbox{c) assuming a Minkowski background,} \mbox{d) estimating} the exiting time $\sigma_\text{x}$ by half the light-crossing time of the region of formation, \mbox{e) modelling} dissipation as a Markovian sequence of short non-Markovian processes, and \mbox{f) not} considering the possibility of confining the scattered fermions.

\subsection*{a) Electric field pulse profile}

As we discussed above, the adiabaticity of the process implies that the tails of the pulse have a negligible impact on the particle production. Any smooth peaked pulse would therefore yield the same order of magnitude of particle production. If the adiabatic approximation is not fulfilled then the dissipation effects are even stronger due to extra particle production~\cite{Adorno2018,AlvarezOperational,Ilderton2022,Diez2023}, further hindering the formation of a kugelblitz.

\subsection*{b) Uniform electric field approximation} 

To evaluate the suitability of this approximation, we need to understand how fermion pair production is affected by the time and space dependence of the electric field, as well as the related presence of a magnetic field. Although the general case is not analytically tractable, many authors have studied how the spatio-temporal dependence of the electromagnetic pulse affects particle production~\cite{Hebenstreit2010,Kohlfurst2022,Gies2005,Dunne2005,Dunne2006,Schutzhold2008,Amat2022,Nikishov1970,Ritus1985,Ruf2009,Bulanov2010b,Jiang2014,Karabali2019}.

First, there is evidence that the time dependence of the electric field enhances pair production~\cite{Dunne2005,Dunne2006,Schutzhold2008}, while the space dependence suppresses it~\cite{Gies2005,Dunne2005,Dunne2006,Amat2022}. The latter, however, is only significant at scales below the pair formation length~\cite{Ritus1985,Nikishov1970,Dunne2005a,Dunne2005,Dunne2006,DiPiazza2012,Gies2016}, 
\begin{equation}
    \ell = \frac{m c^2}{e E}.
\end{equation}
In the hypothetical formation of a kugelblitz of radius $R$, the electric field would have to get increasingly closer to $E_\textsc{bh}$, for which the associated pair formation length would be
\begin{equation}
\frac{\ell_\textsc{bh}}{R} = \frac{m c^2}{e E_\textsc{bh} R} = \frac{m G}{\sqrt{3 \alpha} c^2 \ell_\textsc{p}} \sim 10^{-22}. 
\end{equation}
Even if we managed to devise a setup for which the Schwinger effect was suppressed for weaker electric fields, suppressing it until the formation of the black hole would require inhomogeneities with typical length scales of the order of $\ell_\textsc{bh}$ or below. In laboratory setups, where $R \lesssim 1 \text{ m}$, reaching this regime would require radiation with wavelengths of the order of $10^{-22} \text{ m}$ or below, which is more than ten orders of magnitude below the current shortest laser wavelength~\cite{Yoneda2015}---in fact, $10^{-21} \text{ m}$ is the order of magnitude of the shortest wavelength ever measured for radiation coming from an astrophysical source~\cite{Amenomori2019,Cao2021,Kar2022}. For $\ell_\textsc{bh}$ to approach, for instance, the wavelengths of $\gamma$-ray bursts~\cite{Klebesadel1973,Kouveliotou1993,Bonnell1996,Gendre2013} ($10^{-12}$ m or below~\cite{Ohmori2019}), we would require \mbox{$R \gtrsim 10^9 \,\textrm{m}$}. Outside this regime, the suppression of the Schwinger effect due to spatial inhomogeneities of the electric field is negligible. 

Regarding the magnetic field that would unavoidably exist in a dynamical scenario, its presence increases pair production~\cite{Nikishov1970,Dunne2005a,Harko2006,Karabali2019}, and this effect is increased by curvature and strong gravitational fields~\cite{Calucci1999,DiPiazza2002,DiPiazza2006,Karabali2019,Siahaan2023}. While it is true that a single plane electromagnetic wave cannot lead to pair production~\cite{Dunne2005a}, the formation of a kugelblitz would require focusing radiation, and possibly multiple sources. This makes things even worse for the formation of a kugelblitz: it has been shown that both focused radiation~\cite{Narozhny2004,Bulanov2005,Narozhny2000} and light from multiple sources~\cite{Brown1964,Brezin1970,Fried2001,Dunne2009,Bulanov2010b,Bulanov2010,Mocken2010} are more efficient at pair production than a constant electric field.

Overall, the approximation by a uniform electric field should lead to an underestimation of the dissipation via Schwinger effect, since the neglected effects either enhance pair production, or are irrelevant in the regimes where a kugelblitz could form~\cite{Ritus1985,Dittrich2000,DiPiazza2012}.

\subsection*{c) Assumption of a Minkowski background} 

One could argue that to analyse the formation of kugelblitze one should work with QFT in a dynamically curved background. However, we prove here that the energy densities that one can realistically reach before and after Schwinger dissipation dominates are not only not forming an event horizon, but are also well within the weak (gravitational) field approximation compatible with a flat spacetime (except for the very late stages, which are unreachable anyway). To support this claim, we analyse the escape velocity of the created fermions.

The setup under analysis falls within the regime of approximation~\eqref{Eq: approx}. In this regime, the energy of fermions produced via Schwinger effect (in the centre-of-mass reference frame of the collapsing light) can be approximated by
\begin{equation}\label{Eq: energy of a fermion}
\gamma m c^2 \approx e E \sigma_\text{x} c,
\end{equation}
where $\gamma=1/\sqrt{1-\beta^2}$ is the Lorentz factor (with $\bm\beta=\bm v/c$, and $\beta = |\bm\beta|$), and we have omitted the dependence of $E$ on $\tau$ for the sake of a lighter notation. From this, the velocity of the fermions can be approximated by
\begin{equation}\label{Eq: estimated velocity}
v \approx c \,\sqrt{1 - \bigg( \frac{m c}{e E \sigma_\text{x}} \bigg)^{\!\!2}}\approx c,
\end{equation}
since in this regime of approximation $e E \gg m^2 c^3/\hbar$ and $eE \sigma_\text{x}^2 \gg \hbar/c$, and multiplying both inequalities yields $(e E \sigma_\text{x})^2 \gg m^2 c^2$. This means that fermions produced via Schwinger effect are ultrarelativistic. 

We check now that the estimated fermion velocity $v$ from~\eqref{Eq: estimated velocity} is much larger than  the necessary velocity $v_\text{esc}$ to escape the region of formation of the kugelblitz. This escape velocity can be easily estimated as
\begin{equation}
v_\text{esc} \approx \sqrt{\frac{2G\epsilon}{R c^2}} = c \,\sqrt{\frac{\epsilon}{\epsilon_\textsc{bh}}} = \frac{c E}{E_\textsc{bh}},  
\end{equation}
where recall that $\epsilon$ is the electromagnetic energy in the region of formation, $\epsilon_\textsc{bh}$ is the threshold electromagnetic energy needed to form the black hole, and $E_\textsc{bh}$ is its associated electric field strength. Then, from~\eqref{Eq: estimated velocity}, 
\begin{equation}
\frac{v_\text{esc}}{v} \approx \frac{E}{E_\textsc{bh}},
\end{equation}
which means that the escape velocity is only comparable to the velocity of the scattered ultrarelativistic particles when the electric field strength $E$ is comparable with $E_\textsc{bh}$. This implies that the gravitational influence of the radiation on the exiting fermions is negligible except in the very final stages where the formation would be imminently taking place. However, the calculations made in the main text show that it is not realistically possible to reach neither~$E_\textsc{bh}$ nor any significant fraction of it. Thus, to arrive at our conclusions it is never necessary to work outside of the regime where we can safely assume that the velocity estimated in~\eqref{Eq: estimated velocity} is always significantly bigger than the escape velocity of the region where the kugelblitz is forming.

\subsection*{d) Estimation of the exiting time} 

To say that $\sigma_\text{x} \approx R/c$, we neglected the gravitational attraction that the confined radiation exerts on the scattered fermions. However, as we proved above, in the regime of approximation where the electric field is strong enough to produce particle-antiparticle pairs, the fermions produced by the field quickly become ultrarelativistic, reaching significantly larger velocities than those required to escape the collapsing region. Hence, for the ensemble of particles produced at random positions and with random directions of movement within the sphere of radius $R$, spherical symmetry implies that the average exiting time is
\begin{equation} \label{Eq: estimation taue}
\sigma_\text{x} \approx \frac{R}{v} \approx \frac{R}{c}.
\end{equation}

\subsection*{e) Markovian approximation}

We described the dissipation via Schwinger effect as a sequence of independent non-Markovian dissipation processes of negligibly small duration as compared with the time $T$ of formation of the kugelblitz, i.e., $\sigma_\text{x} \ll T$. Hence, it is necessary to examine whether this is actually the case. We can bound $T$ from below by the time it would take to form the kugelblitz in the absence of dissipation:
\begin{equation}
T \gtrsim \frac{\epsilon_\textsc{bh}}{4 \pi R^2 f} = \frac{c^4}{8 \pi G R f}.
\end{equation}
On the other hand, we have seen that $\sigma_\text{x} \approx R/c$, and therefore
\begin{equation}
\frac{\sigma_\text{x}}{T} \lesssim 4 \pi R^2 f \frac{2 G}{c^5}\sim  \frac{4\pi R^2 f}{10^{52}\text{ W} } .
\end{equation}
Hence, for $\sigma_\text{x}$ to be non-negligible in comparison with~$T$, the total power input needs to be \mbox{$4\pi R^2 f \sim   10^{51} \text{ W}$}. This lower bound is already ten orders of magnitude above the power output of quasars~\cite{Hopkins2007,Sulentic2014,Wolf2024}. 

\subsection*{f) Confinement of fermions}

We could even consider the contrived scenario where one is able to use some external mechanism to confine the scattered fermions. However, this does not eliminate the dissipation, it just changes the form in which it happens. This is so because the deceleration required to stop the particles leads to bremsstrahlung that quickly scatters off the region where the kugelblitz is forming. This radiation, as we will see, carries a significant enough fraction of the energy that the fermions initially had, hence yielding a \textit{corrected} estimated dissipation that is still large enough for our conclusions to hold.

To see that this is the case, first recall that the power radiated by an accelerating charge is given by~\cite{Jackson1999}
\begin{equation}\label{Eq: radiated power}
\dot{\epsilon}_\text{rad} = \frac{2}{3} r_e m c \gamma_t^6 \big[\dot{\bm\beta}_t^2 - (\bm\beta_t\times\dot{\bm\beta}_t)^2\big],
\end{equation}
where $\bm\beta_t = \bm v(t)/c$ is the velocity of the particle at time $t$ (normalized by $c$), $\gamma_t = 1 / \sqrt{1-\beta_t^2}$ is its Lorentz factor, and 
\begin{equation}
    r_e = \frac{e^2}{4 \pi \varepsilon_0 m c^2}
\end{equation}
is the classical electron radius. We will consider two cases: one where the charge is \textit{completely} stopped before reaching some radius $R' \geq R$, and another where the confinement is achieved by forcing the charge to orbit inside the region of radius~$R'$. 

\subsubsection*{Confinement by slowing down particles}

We can decompose the acceleration as $\bm{\dot{\beta}}_t = \bm{\dot{\beta}}_t^{\parallel} + \bm{\dot{\beta}}_t^{\perp}$, where $\bm{\dot{\beta}}_t^{\parallel}$ and $\bm{\dot{\beta}}_t^{\perp}$ are, respectively, the tangent and normal components. Then, the power radiated can be bounded from below neglecting the contribution that comes from the normal component:
\begin{equation}\label{Eq: first bound}
\dot{\epsilon}_{\text{rad}} = \frac{2}{3} r_e m c \gamma_t^6 \big[(\dot{\bm\beta}_t^{\parallel})^2 + (\dot{\bm\beta}_t^{\perp})^2(1-\bm\beta_t^2) \big] \geq \frac{2}{3} r_e m c \gamma_t^6 (\dot{\bm\beta}_t^{\parallel})^2=\frac{2}{3} r_e m c \frac{\gamma_t^2 \dot\gamma_t^2}{\gamma_t^2 - 1}.
\end{equation}

We can estimate the energy radiated by the charge after it has been slowed down to a certain $\beta_\text{b} < \beta$, where $\beta$ denotes the velocity that the charge initially had when it left the region of formation of the kugelblitz. Taking that initial time to be $t=0$, and denoting with $t_\text{b}$ the time that it takes to decelerate the charge, we can use the Cauchy-Schwarz inequality to obtain a lower bound for the total radiated energy: 
\begin{equation}
    \epsilon_\text{rad} \geq \frac{2r_emc}{3t_\text{b}} \bigg( \int_0^{t_\text{b}}\text{d}t\, \frac{\gamma_t\dot\gamma_t}{\sqrt{\gamma_t^2-1}} \bigg)^{\!\!2} = \frac{2r_e mc}{3t_\text{b}} \bigg(\sqrt{\gamma_\text{b}^2-1}-\sqrt{\gamma^2-1}\bigg)^{\!\!2},
\end{equation}
where $\gamma_\text{b} = 1 / \sqrt{1-\beta_\text{b}^2}$ is the Lorentz factor at $t = t_\text{b}$. Since we are assuming that the charge is confined in some region of radius $R'\geq R$, $t_\text{b}$ must be bounded from above by the time it would take for the particle to leave the region of radius $R'$ with a velocity greater or equal than~$\beta_\text{b}$, which at the same time can be upper-bounded by $R'/(c\beta_\text{b})$, yielding
\begin{equation}
    \epsilon_\text{rad} \geq \frac{2r_e mc^2}{3R'} \beta_\text{b} \bigg(\sqrt{\gamma_\text{b}^2-1}-\sqrt{\gamma^2-1}\bigg)^{\!\!2} \approx \frac{2r_e mc^2}{3R'} \beta_\text{b} \bigg(\frac{\beta_\text{b}}{\sqrt{1-\beta_\text{b}^2}}-\gamma\bigg)^{\!\!2} ,
\end{equation}
where we used that the charge is initially ultrarelativistic,~$\gamma \gg 1$. 

Now, to estimate the fraction~$\chi$ of the initial energy of the charge that leaves the region of formation of the kugelblitz, we can bound it from below by only taking into account the energy radiated until~$t=t_\text{b}$, and assuming that the remaining energy stays with the charge, which itself remains inside the region of radius $R'$. Notice that the charge can stay \textit{anywhere} inside the region of radius $R'$, of which the region of formation of the kugelblitz is only a subregion of radius $R$. This means that the energy of the fermions that have been stopped is \textit{diluted} by a factor that accounts for their different volumes, namely $(R/R')^3$. Therefore, the energy dissipated from the region of formation of the kugelblitz is the initial energy carried by the charge, $\gamma m c^2$, minus the energy that remains, $(\gamma m c^2 - \epsilon_\text{rad})$, multiplied by the \textit{dilution factor} $(R/R')^3$. From this, the fraction of dissipated energy is
\begin{equation}\label{Eq: fraction that stays}
    \chi = 1 - \frac{R^3}{R'^3}\bigg( 1 - \frac{\epsilon_\text{rad}}{\gamma m c^2} \bigg) \geq 1 - \frac{R^3}{R'^3} \bigg[ 1 - \frac{2r_e}{3\gamma R'}\beta_\text{b}\bigg( \frac{\beta_\text{b}}{\sqrt{1-\beta_\text{b}^2}} - \gamma \bigg)^{\!\!2}  \bigg].
\end{equation}
Notice that to derive this inequality we only required $\beta_\text{b} < \beta$, and therefore~\eqref{Eq: fraction that stays} actually represents an infinite set of inequalities. Although we could optimize over $\beta_\text{b}$ to obtain the stricter lower bound of $\chi$, for the sake of simplicity we will just use $\beta_\text{b}=1/2$. In this case,
\begin{equation}\label{Eq: fraction particular case}
    \chi \geq 1 - \frac{R^3}{R'^3}\bigg[ 1 - \frac{r_e}{3\gamma R'} \bigg(\frac{1}{\sqrt{3}}-\gamma\bigg)^{\!\!2} \bigg] \approx 1 - \bigg( \frac{R}{R'} \bigg)^{\!\!3} \bigg[  1 - \frac{r_e \gamma}{3 R} \bigg(\frac{R}{R'}\bigg) \bigg].
\end{equation}

For $R/R'>3R/(r_e \gamma)$, equation~\eqref{Eq: fraction particular case} implies that $\chi > 1$, which is impossible. What actually happens in this case is that $R'$ is \textit{too small}, and the exiting charges cannot be stopped before they leave the region of radius $R'$ (since to do that they would need to dissipate more energy than the charged particle has). Therefore, the confinement of the electron-positron pairs can only be attempted for $R'$ such that $R/R'\leq 3R/(r_e \gamma)$, and only then the bound given by~\eqref{Eq: fraction particular case} applies. Moreover, since $\gamma \approx e E \sigma_\text{x}/ (m c) \geq e E R / (m c^2)$,
\begin{equation}
    \frac{r_e \gamma}{3 R} \gtrsim \frac{r_e e}{3 m c^2} E \sim \frac{E}{10^{21} \text{ V/m}}.
\end{equation}
We can therefore distinguish two regimes: 
\begin{itemize}
    \item $E \gg 10^{21} \text{ V/m}$. In this case, we have $r_e \gamma / (3 R) \gg 1$, which can lead to two different cases. First, if $R/R'>3R/(r_e\gamma)$, the charges cannot be stopped before leaving $R'$. Second, if \mbox{$R/R' \leq 3R/(r_e\gamma) \ll 1$}, and \mbox{$\chi \geq 1 - (R/R')^3 \sim 1$}, most of the energy leaves the region of formation of the kugelblitz.
    \item $E \lesssim 10^{21} \text{ V/m}$, which leads to $r_e \gamma / (3 R) \lesssim 1$. In order to form a kugelblitz in the range of radii under analysis, the electric field has to be bigger than the Schwinger limit, \mbox{$E \gtrsim 10^{18} \text{ V/m}$}, implying that $r_e\gamma / (3R) \gtrsim 10^{-3}$. In this case,
\begin{equation}
    \chi \geq  1 - \bigg( \frac{R}{R'} \bigg)^{\!\!3} \bigg[  1 - \frac{r_e \gamma}{3 R} \bigg(\frac{R}{R'}\bigg) \bigg] \gtrsim \frac{r_e \gamma}{3 R} \gtrsim 10^{-3}, 
\end{equation}
where in the second inequality we have used that, as a function of $R/R'$, the right hand side of~\eqref{Eq: fraction particular case} attains its minimum at $(R/R')^* = \min\{1,9R/(4r_e \gamma)\} \approx 1$. This estimation would mean that with this setup the dissipation is, in the best case, three orders of magnitude below the one used originally, given by~\eqref{Eq: estimated Schwinger dissipation}. However, from~\eqref{Eq: E infinity} we see that a correction factor of $10^{-3}$ for the dissipation term $D$ would only increase an order of magnitude the value of $E_\infty$. Even in this extremely optimistic scenario, the resulting reduction is far from being enough to close the huge gap between the estimations of the required conditions to form a kugelblitz and what seems realistically achievable. 
\end{itemize}

\subsubsection*{Confinement by making particles orbit}

Finally, let us analyse the case where the charges are confined in the region of radius $R' \geq R$ not by slowing them down before they leave, but by making them orbit inside the region instead. In that scenario, the main contribution to the radiated energy comes from the normal component of the acceleration. To give a lower bound on the energy radiated by the confined charges, we look at the contribution of the normal component in~\eqref{Eq: radiated power}, which yields
\begin{equation}\label{Eq: power radiated synchrotron}
    \dot{\epsilon}_\text{rad} \geq \frac{2}{3} r_e m c \gamma_t^6 (\dot{\bm\beta}_t^\perp)^2 (1 - \bm\beta_t^2) \geq \frac{2 r_e m c^3}{3 R'^2} \gamma_t^6 \beta_t^4 (1-\beta_t^2) = \frac{2 r_e m c^3}{3 R'^2} (\gamma_t^2-1)^2, 
\end{equation}
where in the last inequality we have used that the radius of the orbit must be less or equal than~$R'$, and therefore we have that $|\dot{\bm\beta_t^\perp}| \geq \bm\beta_t^2 c / R'$. Now, because the energy radiated is energy lost by the particle, we have that $\dot{\epsilon}_\text{rad} = - \dot\gamma_t m c^2$, and thus
\begin{equation}\label{Eq: dot gamma}
\dot{\gamma}_t \leq - \frac{2 r_e c}{3 R'^2} (\gamma_t^2-1)^2.
\end{equation}

In order to compare the power dissipated in this setup with the estimated dissipation~$D$ given in~\eqref{Eq: estimated Schwinger dissipation}, we can compute the fraction~$\chi$ of the initial energy of the charge that still dissipates in this setup over a period of time equivalent to the timescale $\sigma_\text{x}$ of the dissipation in the original setup, 
\begin{equation}\label{Eq: fraction synchrotron}
    \chi = 1 - \bigg( \frac{R}{R'} \bigg)^{\!\!3} \frac{\gamma_\text{x} m c^2}{\gamma m c^2},
\end{equation}
where $\gamma_\text{x}$ is the final Lorentz factor. To bound $\gamma_\text{x}$ from above, we can integrate~\eqref{Eq: dot gamma} from the time $t=0$, when the charge left the region of formation of the kugelblitz, to $t=\sigma_\text{x}$, obtaining
\begin{equation}
   \int_{0}^{\sigma_\text{x}} \text{d}t \  \frac{\dot\gamma_t}{(\gamma_t^2-1)^2} = \left[ -\frac{\gamma_t}{2(\gamma_t^2-1)} + \frac{1}{4}\log\frac{\gamma_t+1}{\gamma_t-1} \right]_{\gamma_t=\gamma}^{\gamma_t=\gamma_\text{x}} \leq  -\frac{2 r_e c}{3 R'^2}\sigma_\text{x}.
\end{equation}
Since $\sigma_\text{x} \geq R/c$, and $\gamma_\text{x} < \gamma$, the previous inequality can be rewritten as
\begin{equation}
    \frac{2 r_e R}{3 R'^2} \leq \left[ -\frac{\gamma_t}{2(\gamma_t^2-1)} + \frac{1}{4}\log\frac{\gamma_t+1}{\gamma_t-1} \right]_{\gamma_\tau=\gamma_\text{x}}^{\gamma_t=\gamma} \leq \frac{1}{2}\bigg( \frac{\gamma_\text{x}}{\gamma_\text{x}^2-1} - \frac{\gamma}{\gamma^2-1}  \bigg), 
\end{equation}
where in the last step we used that $\log[(\gamma_t+1)/(\gamma_t-1)]$ is a decreasing function of $\gamma_t$. From here, and since, as discussed, $\gamma\gg 1$, we get
\begin{equation}
   \frac{\gamma_\text{x}}{\gamma_\text{x}^2-1} - \frac{1}{\gamma} \geq \frac{4 r_e R}{3 R'^2}.
\end{equation}
Solving the corresponding quadratic inequality, we arrive at
\begin{equation}\label{Eq: inequality for gamma x}
    \gamma_\text{x} \lesssim \frac{1}{2}\bigg( \frac{1}{\gamma} + \frac{4 r_e R}{3 R'^2}   \bigg)^{\!\!-1} +\frac{1}{2}\sqrt{1 + \bigg( \frac{1}{\gamma} + \frac{4 r_e R}{3 R'^2}   \bigg)^{\!\!-2}}. 
\end{equation}
If $4r_e R / (3 R'^2) \gtrsim 1$, then~\eqref{Eq: inequality for gamma x} implies
\begin{equation}
    \gamma_\text{x} \lesssim \frac{1}{2}\bigg(\frac{3 R'^2}{4 r_e R}   \bigg) +\frac{1}{2}\sqrt{1 + \bigg(\frac{3 R'^2}{4 r_e R}   \bigg)^{\!\!2}} \lesssim \frac{1+\sqrt{2}}{2},
\end{equation}
where we used again that $\gamma \gg 1$. In this case, $\gamma_\text{x}/\gamma \ll 1$, and from~\eqref{Eq: fraction synchrotron} we conclude that \mbox{$\chi\sim 1$}, i.e., most of the energy leaves the region of formation of the kugelblitz. If, on the contrary, $4r_e R / (3 R'^2) \ll 1$, then~\eqref{Eq: inequality for gamma x} reduces to
\begin{equation}
    \gamma_\text{x} \lesssim \bigg( \frac{1}{\gamma} + \frac{4 r_e R}{3 R'^2}   \bigg)^{\!\!-1},
\end{equation}
yielding
\begin{equation}\label{Eq: fraction synchrotron case 2}
    \chi = 1 - \bigg( \frac{R}{R'} \bigg)^{\!\!3} \frac{\gamma_\text{x} m c^2}{\gamma m c^2} \gtrsim 1 - \bigg(\frac{R}{R'}\bigg)^{\!\!3} \bigg[1 + \frac{4 r_e e}{3 m c^2} E \bigg( \frac{R}{R'} \bigg)^{\!\!2}\bigg]^{\!-1},
\end{equation}
where as before we used that $\gamma \approx e E \sigma_\text{x} / (m c) \geq e E R / (m c^2)$. Since, as a function of $R/R'$, the right hand side of~\eqref{Eq: fraction synchrotron case 2} is minimized at $(R/R')^* = 1$, we can bound $\chi$ from below as
\begin{equation}\label{Eq: fraction bound}
    \chi \geq 1 - \bigg[ 1 + \frac{4 r_e e}{3 m c^2} E \bigg]^{\!-1} \sim 1 - \bigg( 1 + \frac{E}{10^{20} \text{ V/m}} \bigg)^{\!\!-1}.
\end{equation}
Then, as in our first analysis, we can distinguish two regimes: 
\begin{itemize}
    \item $E \gg 10^{20} \text{ V/m}$. The second term of~\eqref{Eq: fraction bound} becomes negligible, and, again, $\chi \gtrsim 1$, i.e., the dissipation is essentially the same as in the original setup.
    \item $E \lesssim 10^{20} \text{ V/m}$. We can use once more that $E \gtrsim 10^{18} \text{ V/m}$, in which case $\chi \gtrsim 10^{-2}$. However, as we argued before, even this best-case-scenario correction factor of $10^{-2}$ for the dissipated power $D$ does not modify our conclusions, and would not allow any realistic scenario to satisfy the necessary conditions for the formation of a kugelblitz. 
\end{itemize}

\section{Conclusions}
\label{sec:ConclusionsKugelblitz}

We showed that it is not possible to generate a black hole out of the gravitational collapse of electromagnetic radiation in the range of length scales comprised between $10^{-29} \text{ m}$ and $10^8 \text{ m}$. 

To reach this conclusion, we studied the rate at which electromagnetic energy can be focused on a spherical region of a certain radius when a constant inward flux is applied, while part of it is leaked away by the particle-antiparticle pairs created in the process via Schwinger effect. 

Our analysis indicates that the power needed to form a kugelblitz is tens of orders of magnitude above what can be achieved in any realistic scenario, both in the laboratory and in astrophysical setups. Moreover, we showed that the  approximations incurred in this analysis do not affect the regimes where our conclusions apply. Furthermore, even if one only trusts the estimations of the model to some extent, the predicted orders of magnitude are so vastly unrealistic as to make this study a very compelling argument against the viability of kugelblitze, both artificially or as a naturally occurring phenomenon.
\chapter[Quantum fermion superradiance on charged black holes]{Quantum fermion superradiance \\ on charged black holes}
\label{chap:Superradiance}

\chaptermark{Quantum fermion superradiance on charged black holes}

Classical \gls{Superradiance} is a phenomenon whereby a field wave is amplified during a scattering process.
In black hole physics, superradiance arises when low-frequency bosonic field waves are scattered on a rotating black hole \cite{Misner1972,Press1972,Teukolsky1974,Chandrasekhar1985}. 
For bosonic fields, superradiance is a consequence of the area theorem and first law of black hole mechanics \cite{Bekenstein1973}, the former holding for matter fields satisfying the weak energy condition. 
Superradiance does not occur for classical fermion fields on rotating black hole backgrounds \cite{Unruh1973,Chandrasekhar1976,Iyer1978} because they do not satisfy the weak energy condition, and so the area law no longer holds \cite{Unruh1973}.

Both bosonic and fermionic fields do however exhibit the quantum analogue of classical superradiance \cite{Starobinsky1973,Unruh1974,Dai2023,Dai2023a}.
Particles are spontaneously emitted in low-frequency field modes, in precisely those frequencies which correspond to classically superradiant modes for bosonic fields. 
The radiation is nonthermal in nature and is in addition to the usual Hawking radiation \cite{Hawking1975} emitted by the black hole. 

Classical superradiance also occurs on static, nonrotating black holes when both the black hole and the scattered field have a nonzero charge (`charge superradiance')~\cite{Bekenstein1973,Nakamura1976,DiMenza2015,Benone2016,DiMenza2020}.
As with the classical superradiance of neutral fields on rotating black holes, charge superradiance only exists for bosonic and not fermionic fields~\cite{Maeda1976}.
A natural question is whether there is a quantum analogue of this classical charge superradiance. 
For a  massless charged scalar field, this process was studied many years ago  \cite{Gibbons1975} and revisited more recently~\cite{Balakumar2020}. 
In~\cite{Balakumar2020}, `in' and `out' vacuum states are constructed for the charged quantum scalar field on a \acrfull{RN} black hole. 
The `in' state is devoid of particles at past null infinity, but contains an outgoing flux of particles at future null infinity. 
This flux is present only in those modes which exhibit classical charge superradiance. 

Our purpose in this chapter is to investigate whether the quantum analogue of charge superradiance also occurs for a massless charged fermion field. We construct analogues of the `in' and `out' states defined for a charged scalar field in~\cite{Balakumar2020}. These quantum vacua describe the discharge and energy loss of the charged black hole, leading to a dissipative phenomenon that is significantly more intense than in the scalar case.

However, we show that quantum superradiance is not exhibited by all fermionic quantum states that can be defined in RN black holes. In particular, we construct a candidate `Boulware' state---originally introduced for scalar fields on Schwarzschild black holes~\cite{Boulware1975}---which exhibits no particle flux at either past or future null infinity. 
This marks a significant distinction from scalar fields on RN black holes, where, due to additional restrictions in the canonical quantization process, a direct analogue of the `Boulware' state does not exist~\cite{Balakumar2022}.

This chapter is based on~\cite{AlvarezSuperradiance}. In~\autoref{sec:ClassicalFermionsRN}, we particularize the Dirac formalism detailed in~\autoref{sec:Fermions} to the case of massless fermions on a RN black hole. In \autoref{sec:InOutQuantum} we discuss the ambiguities in the definition of quantum vacuum, and we construct the `in' and `out' states that describe the phenomenon of quantum superradiance. 
In \autoref{sec:QuantumSuperradiance} we examine this effect, calculating the number density of created particles as well as the black hole discharge and energy loss. \ref{sec:conc} contains further discussion and our conclusions.

\section{Massless charged fermions on a charged black hole}
\label{sec:ClassicalFermionsRN}

We consider a massless charged Dirac field propagating on a \gls{ReissnerNordstromBlackHole}.
The spacetime is described by the line element
\begin{equation}
\text{d}s^{2} = - f(r) \, \text{d}t^{2} + \left[ f(r) \right] ^{-1} \text{d}r^{2}+ r^{2} \text{d}\theta ^{2} + r^{2}\sin ^{2} \theta \, \text{d}\varphi ^{2} ,
\label{eq:RNmetric}
\end{equation}
where the metric function $f(r)$ is given by 
\begin{equation}
f(r) = 1 - \frac{2M}{r} + \frac{Q^{2}}{r^{2}} ,
\label{eq:fr}
\end{equation}
with $M$ being the mass and $Q$ the electric charge of the black hole.
We restrict attention to the situation where $M > |Q| > 0 $,
in which case
the metric function $f(r)$ has two zeros at $r=r_{\pm }$, where
\begin{equation}
r_{\pm } = M \pm {\sqrt {M^{2}-Q^{2}}}.
\label{eq:rpm}
\end{equation}
The larger root~$r_{+}$ is the location of the black hole event horizon 
and $r_{-}$ is the location of the inner horizon.
We will be interested only in the region of spacetime exterior to the event horizon, $r>r_+$. 

A massless fermionic field~$\Psi$ with charge $q$ propagating on the RN black hole satisfies the Dirac equation~\eqref{eq:Dirac}. To specialize this equation, we must take the following key points into account:
\begin{itemize}
    \item To reflect the spherical symmetry of the system in the equations of motion, we fix the gauge
\begin{equation}
A_\mu = \left(-\frac{Q}{r}, 0, 0, 0\right).
\label{eq:gaugepot}
\end{equation}
    \item A suitable basis of $\gamma ^{\mu }$ matrices for the RN metric (\ref{eq:RNmetric}) is given by~\cite{Unruh1974}
\begin{equation}
\gamma ^{t}   = 
\frac {1}{{{\sqrt {f(r)}}}} { \widetilde {\gamma }}^{0},
\qquad 
\gamma ^{r}  = 
{\sqrt {f(r)}} { \widetilde {\gamma }}^{3}, \qquad
\gamma ^{\theta } = 
\frac {1}{r} { \widetilde {\gamma }^{1}},
\qquad
\gamma ^{\varphi }  = 
\frac {1}{r\sin \theta } { \widetilde {\gamma }}^{2},
\label{eq:gamma}
\end{equation}
where our choice for the representation for the flat-space Dirac matrices~$\widetilde{\gamma}^a$ is provided in~\eqref{eq:flatspacegamma}.
    \item The spinor connection matrices, $\Gamma_{\mu}$, which appear in the definition~\eqref{eq:SpinorCovariantDerivative} of the spinor covariant derivatives~$\nabla_\mu$ in the Dirac equation, can be computed using~\eqref{eq:SpinorConnectionMatrices}. In the RN background, they are given by:
\begin{eqnarray}
\Gamma _{t} & = &
\frac {1}{4} \frac{\text{d} f}{\text{d} r}
{ \widetilde {\gamma }}^{0} { \widetilde {\gamma }}^{3},
\nonumber \\
\Gamma _{r}& =  & 0,
\nonumber \\
\Gamma _{\theta} & = &
-\frac{1}{2} \sqrt{f(r)} { \widetilde {\gamma }}^{1}
{ \widetilde {\gamma }}^{3}
,
\nonumber \\
\Gamma _{\varphi}& = &
-\frac{1}{2} \left[ \sqrt{f(r)} \sin\theta \, { \widetilde {\gamma }}^{2} { \widetilde {\gamma }}^{3} + \cos\theta \,
{ \widetilde {\gamma }}^{2} { \widetilde {\gamma }}^{1} \right].
 \label{eq:Gamma}
\end{eqnarray}
\end{itemize}
The resulting form for the Dirac equation \eqref{eq:Dirac} is then given by 
\begin{equation}
\left\{  \gamma^t\left(\partial_t-i\frac{qQ}{r}\right)
    +  \gamma^r\left[\partial_r+\frac{1}{4f(r)}\frac{\text{d}f}{\text{d}r} +\frac{1}{r}\right] 
    +  \gamma^\theta\left[ \partial_\theta + \frac{\cot\theta}{2}\right]
    +  \gamma^\varphi\partial_\varphi \right\} \Psi = 0.
    \label{eq:DiracCoord}
\end{equation}

We search for a separable orthonormal basis of solutions~$\{\Psi_\Lambda\}$ of the Dirac equation~\eqref{eq:DiracCoord} with respect to the Dirac product~\eqref{eq:DiracProduct}. For a given chirality $L=\pm 1$ and a set of quantum indices \mbox{$\Lambda=\{m,l,\omega\}$}, we look for solutions of the form~\cite{Unruh1973,Unruh1974,Vilenkin1978}
\begin{equation}
    \Psi_\Lambda (t,r,\theta,\varphi) = \frac{1}{\sqrt{8\pi^2}\mathcal{F}(r,\theta)}e^{-i\omega t}e^{im\varphi} 
    \left(
\begin{array}{c}
\eta_\Lambda(r,\theta) \\
L\eta_\Lambda(r,\theta)
\end{array}
\right),
\label{eq:ansatz}
\end{equation}
where 
\begin{equation}
\mathcal{F}(r,\theta)=r[f(r)\sin^2\theta]^{1/4}.
\end{equation}
Each component of the two-spinors~$\eta_\Lambda(r,\theta)$ is separable into radial and angular functions:
\begin{equation}
    \eta_\Lambda(r,\theta) = 
    \left(
\begin{array}{c}
R_{1,\Lambda}(r) S_{1,\Lambda}(\theta) \\
R_{2,\Lambda}(r) S_{2,\Lambda}(\theta)
\end{array}
\right).
\end{equation}

Introducing this ansatz into the Dirac equation~\eqref{eq:DiracCoord}, we find two linearly independent equations for the angular functions: 
\begin{align}
\label{eq:S12}
    \left[ \frac{\text{d}}{\text{d}\theta} - \frac{m}{\sin\theta} \right] S_{1,\Lambda}(\theta) & = \left(l+\frac{1}{2}\right) S_{2,\Lambda}(\theta), \nonumber \\
    \left[ \frac{\text{d}}{\text{d}\theta} + \frac{m}{\sin\theta} \right] S_{2,\Lambda}(\theta) & = -\left(l+\frac{1}{2}\right) S_{1,\Lambda}(\theta),
\end{align}
as well as for the radial functions:
\begin{align}
\label{eq:R12}
    r\sqrt{f(r)}\left[ \frac{\text{d}}{\text{d}r} -\frac{iL}{f(r)}\left( \omega+\frac{qQ}{r} \right)\right] R_{1,\Lambda}(r) & = \left(l+\frac{1}{2}\right) R_{2,\Lambda}(r), \nonumber \\
    r\sqrt{f(r)}\left[ \frac{\text{d}}{\text{d}r} + \frac{iL}{f(r)}\left( \omega+\frac{qQ}{r} \right)\right] R_{2,\Lambda}(r) & = \left(l+\frac{1}{2}\right) R_{1,\Lambda}(r).
\end{align}
These functions have a discrete spectrum, with \mbox{$l=\frac{1}{2}, \frac{3}{2}, ...$}; and \mbox{$m= -l, -l+1, ..., l-1, l$}. 

\subsection*{Angular functions}

We outline some key properties of the angular functions that will be crucial for proving essential results in the remainder of this chapter. The angular functions~$S_{1,\Lambda}(\theta)$ and~$S_{2,\Lambda}(\theta)$, which we take to be real, are closely related to the well-known spin-weighted spherical harmonics~$_sY_l^m(\theta,\varphi)$~\cite{Newman1966,Goldberg1967}:
\begin{equation}
    S_{1,\Lambda}(\theta) = \sqrt{\sin\theta} _{\frac{1}{2}}Y_l^{-m}(\theta,\varphi) e^{im\varphi}, \qquad S_{2,\Lambda}(\theta) = \sqrt{\sin\theta} _{-\frac{1}{2}}Y_l^{-m}(\theta,\varphi) e^{im\varphi}.
\end{equation}
They  are normalized according to
\begin{equation}
    \int_0^\pi \text{d}\theta \ S_{1,\Lambda}(\theta) = \int_0^\pi \text{d}\theta \ S_{2,\Lambda}(\theta) = 1
\end{equation}
and satisfy the following addition relations, which can easily be deduced from those for the spin-weighted spherical harmonics~\cite{Monteverdi2024}:
\begin{align}
    \sum_{m=-l}^l S_{1,\Lambda}(\theta)^2 =& \sum_{m=-l}^l S_{2,\Lambda}(\theta)^2  = \frac{2l+1}{4\pi} \sin\theta, \nonumber \\
    \sum_{m=-l}^l m S_{1,\Lambda}(\theta)^2 =& -\sum_{m=-l}^l m S_{2,\Lambda}(\theta)^2  = \frac{2l+1}{8\pi} \sin\theta \cos\theta, \nonumber \\
    \sum_{m=-l}^l S_{1,\Lambda}(\theta) S_{2,\Lambda}(\theta) =& \sum_{m=-l}^l m S_{1,\Lambda}(\theta) S_{2,\Lambda}(\theta)  = 0, \nonumber \\
    \sum_{m=-l}^l S_{j,\Lambda}(\theta)\frac{\text{d}}{\text{d}\theta} S_{k,\Lambda}(\theta) =& ~ 0,
    \label{eq:angularproperties}
\end{align}
for $j,k=1,2$. Another useful property concerns the symmetries of the angular functions under the transformation~\mbox{$m \rightarrow -m$}. 
From their governing equations~\eqref{eq:S12} we have
\begin{equation}
    S_{1,(-m,l,\omega)}=\pm S_{2,(m,l,\omega)}, \qquad S_{2,(-m,l,\omega)}=\mp S_{1,(m,l,\omega)}.
    \label{eq:symmetriesm}
\end{equation}

\subsection*{Radial functions}

Writing the radial equations~\eqref{eq:R12} in terms of the tortoise coordinate~$r_*$, defined by
\begin{equation}
    \frac{\text{d}r_*}{\text{d}r} = \frac{1}{f(r)},
\end{equation}
one can verify that the radial functions behave as plane waves asymptotically far from the black hole ($r_*\rightarrow +\infty$):
\begin{equation}
    R_{1,\Lambda}(r_*) \propto e^{iL\omega r_*}, \qquad R_{2,\Lambda}(r_*) \propto e^{-iL\omega r_*}.
    \label{eq:R-infty}
\end{equation}
This is as expected, since the RN spacetime is asymptotically flat. At the event horizon~$r_+$ ($r_* \rightarrow -\infty$), these functions also behave as plane waves,
\begin{equation}
    R_{1,\Lambda}(r_*) \propto e^{iL\widetilde{\omega} r_*}, \qquad R_{2,\Lambda}(r_*) \propto e^{-iL\widetilde{\omega} r_*},
    \label{eq:R+infty}
\end{equation}
but now with a shifted wave number
\begin{equation}
    \widetilde{\omega}=\omega+\frac{qQ}{r_+}.
    \label{eq:omegatilde}
\end{equation}
For positive chirality $L=1$, the plane waves of $R_{1,\Lambda}(r_*)$ are outgoing at both future null infinity~$\mathscr{I}^+$ and past the event horizon~$\mathscr{H}^-$, while those of $R_{2,\Lambda}(r_*)$ are ingoing at both past null infinity~$\mathscr{I}^-$ and the future event horizon~$\mathscr{H}^+$. 
For negative chirality $L=-1$, the roles of $R_{1,\Lambda}(r_*)$ and $R_{2,\Lambda}(r_*)$ are reversed. 
Therefore, in what follows, we will restrict our attention to the case of positive chirality, $L=1$.

\begin{block}[note]
We could have considered massive fermions. However, when the mass is nonzero, the two chiralities become classically coupled, making the formalism slightly more complex (see, for instance, the discussion in~\cite{McKellar1993} for rotating black holes). For simplicity, we restrict our analysis to massless fermions.

Nonetheless, it is important to recognize the limitations of this assumption. While positive and negative chiralities remain decoupled at the classical level for massless fermions, quantum effect---particularly the axial anomaly---are expected to introduce couplings between them. In this work, we focus on particle creation processes that preserve chirality as a first approximation, but it would be worthwhile in future studies to investigate whether quantum mixing between chiralities indeed arises.

Moreover, in the case of massive fermions, we anticipate quantum superradiance to be exponentially suppressed, based on our analysis of pair creation in the Schwinger effect. This suppression would be analogous to that observed in the Schwinger effect, where the excitation probability is proportional to~$e^{-\pi m^2 / (qE)}$~\cite{Schwinger1951}.
\end{block}

\section{`In-up' and `out-down' bases and classical superradiance}

We now define two well-known orthonormal bases of solutions to the Dirac equation. 
Elements of the bases will be of the form given by~\eqref{eq:ansatz}, with certain choices of the radial functions. 

\begin{figure}
    \centering
    \includegraphics[width=\textwidth]{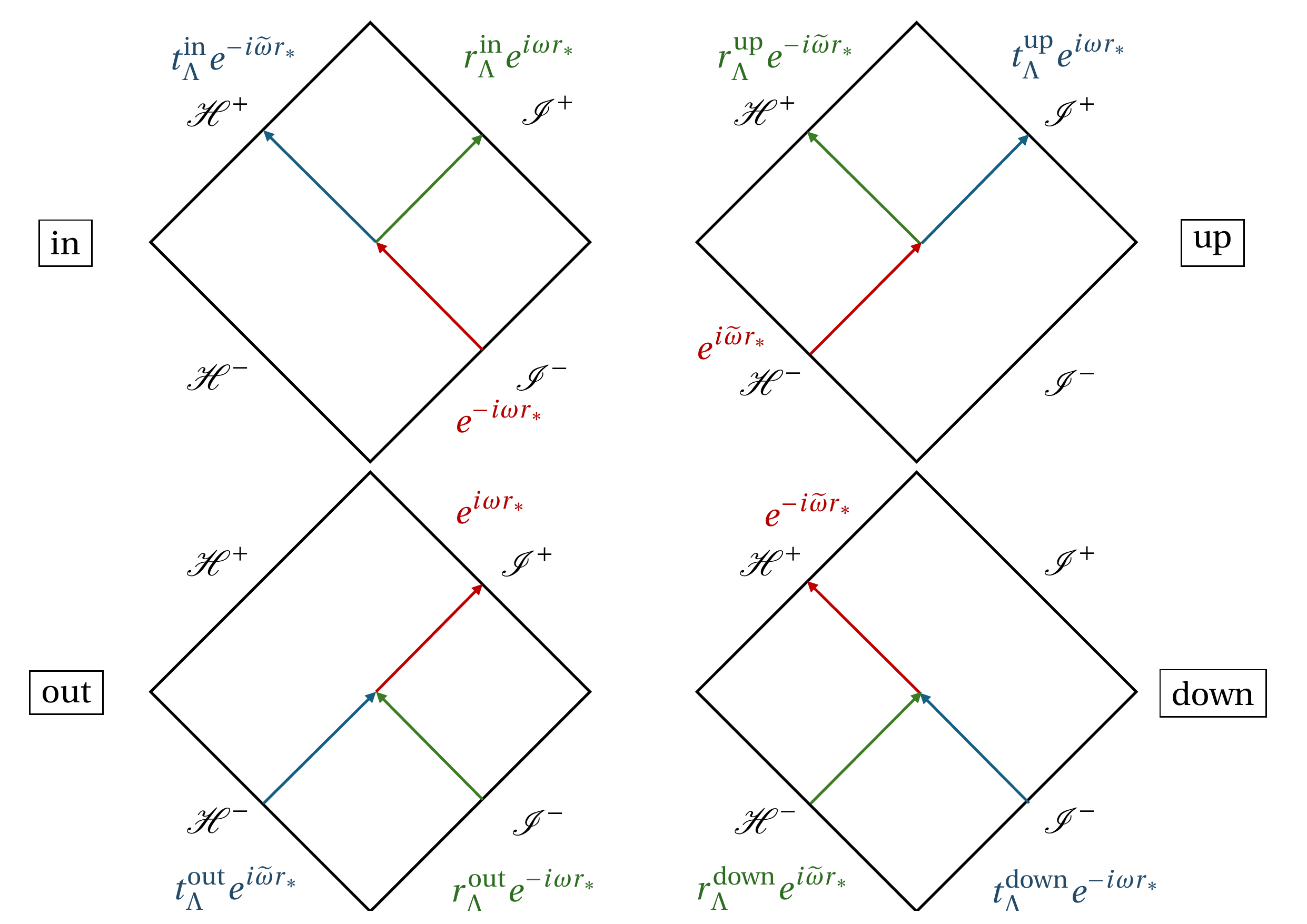}
    \caption{Asymptotic behaviour of the `in', `up', `out' and `down' modes in the Penrose diagram corresponding to the exterior region of the RN event horizon.}
    \label{fig:RN}
\end{figure}

The so-called \gls{InUpBasis} is determined by imposing initial conditions for the radial functions on the past hypersurface~$\mathscr{H}^- \cup \mathscr{I}^-$~\footnote{Strictly speaking, we are choosing a Cauchy surface close to~$\mathscr{H}^- \cup \mathscr{I}^-$.}. The `in' modes represent unit flux of incoming plane waves from~$\mathscr{I}^-$, with no contribution coming from~$\mathscr{H}^-$:
\begin{equation}
    R_{2,\Lambda}^{\text{in}}(r_*) \sim e^{-i\omega r_*}, \ \ \ r_*\to +\infty; \qquad R_{1,\Lambda}^{\text{in}}(r_*) \sim 0, \ \ \ r_*\to -\infty.
    \label{eq:RinA}
\end{equation}
These ingoing plane waves are partly transmitted to the future horizon~$\mathscr{H}^+$ and, since this is a scattering problem, partly reflected back to~$\mathscr{I}^+$. According to~(\ref{eq:R-infty}, \ref{eq:R+infty}), this translates into the asymptotic behaviour of the radial functions given by
\begin{equation}
    R_{2,\Lambda}^{\text{in}}(r_*) \sim t_{\Lambda}^{\text{in}} e^{-i\widetilde{\omega} r_*}, \ \ \ r_*\to -\infty; \qquad R_{1,\Lambda}^{\text{in}}(r_*) \sim r_{\Lambda}^{\text{in}}e^{i\omega r_*}, \ \ \ r_*\to +\infty. 
    \label{eq:RinB}
\end{equation}
The factors~$t_\Lambda^{\text{in}}$ and~$r_{\Lambda}^{\text{in}}$ are called the \gls{TransmissionAndReflectionCoefficients}, respectively. In \autoref{fig:RN}, we present a schematic illustration of the behaviour of these modes, along with all the modes that will be introduced in the following.

On the other hand, the `up' modes correspond to unit flux of outgoing plane waves at~$\mathscr{H}^-$, with no contribution from~$\mathscr{I}^-$:
\begin{equation}
    R_{1,\Lambda}^{\text{up}}(r_*) \sim e^{i\widetilde{\omega} r_*}, \ \ \ r_*\to -\infty; \qquad R_{2,\Lambda}^{\text{up}}(r_*) \sim 0, \ \ \ r_*\to +\infty.
    \label{eq:RupA}
\end{equation}
Part of this outgoing flux is transmitted to~$\mathscr{I}^+$ while the other part is reflected back down $\mathscr{H}^+$, that is,
\begin{equation}
    R_{1,\Lambda}^{\text{up}}(r_*) \sim t_\Lambda^{\text{up}} e^{i\omega r_*}, \ \ \ r_*\to +\infty; \qquad R_{2,\Lambda}^{\text{up}}(r_*) \sim r_\Lambda^{\text{up}} e^{-i\widetilde{\omega}r_*}, \ \ \ r_*\to -\infty.
    \label{eq:RupB}
\end{equation}

The second basis is the so-called \gls{OutDownBasis}. 
In particular, the `out' and `down' solutions are the time reversals of the `in' and `up' modes, respectively, so that
\begin{equation}
    R_{1,\Lambda}^{\text{out/down}}=\left( R_{2,\Lambda}^{\text{in/up}} \right)^*, \qquad R_{2,\Lambda}^{\text{out/down}}= \left( R_{1,\Lambda}^{\text{in/up}} \right)^*.
    \label{eq:timereverse}
\end{equation}
In this case, the chosen hypersurface on which we impose the initial behaviour is formed by future null infinity and the future event horizon, $\mathscr{H}^+\cup\mathscr{I}^+$. 
The `out' solutions are outgoing plane waves at~$\mathscr{I}^+$, vanishing at~$\mathscr{H}^+$, such that when they are evolved to the past, part of the wave emanates from~$\mathscr{H}^-$ and part is incoming from~$\mathscr{I}^-$:
\begin{align}
    R_{1,\Lambda}^{\text{out}}(r_*) \sim e^{i\omega r_*}, \ \ \ \ \ \ \ \ \ \ r_*\to +\infty; & \qquad R_{2,\Lambda}^{\text{out}}(r_*) \sim 0, \ \ \ \ \ \ \ \ \ \ \ \ \ \ \ \ \ \ \ \ \ r_*\to -\infty; \nonumber \\
    R_{1,\Lambda}^{\text{out}}(r_*) \sim t_{\Lambda}^{\text{out}} e^{i\widetilde{\omega} r_*}, \ \ \ r_*\to -\infty; & \qquad R_{2,\Lambda}^{\text{out}}(r_*) \sim r_{\Lambda}^{\text{out}}e^{-i\omega r_*}, \ \ \ r_*\to +\infty.
    \label{eq:Rout}
\end{align}

Analogously, the `down' modes are ingoing solutions near~$\mathscr{H}^+$, vanishing at~$\mathscr{I}^+$, such that in the past, part of the flux is incoming from~$\mathscr{I}^-$ and the other part is outgoing from~$\mathscr{H}^-$, that is,
\begin{align}
    R_{2,\Lambda}^{\text{down}}(r_*) \sim e^{-i\widetilde{\omega} r_*}, \ \ \ \ \ \ \ \ \ \ r_*\to -\infty; & \qquad R_{1,\Lambda}^{\text{down}}(r_*) \sim 0, \ \ \ \ \ \ \ \ \ \ \ \ \ \ \ \ \ \ \ \ \ \ r_*\to +\infty; \nonumber \\
    R_{2,\Lambda}^{\text{down}}(r_*) \sim t_\Lambda^{\text{down}} e^{-i\omega r_*}, \ \ \ r_*\to +\infty; & \qquad R_{1,\Lambda}^{\text{down}}(r_*) \sim r_\Lambda^{\text{down}} e^{i\widetilde{\omega}r_*}, \ \ \ r_*\to -\infty.
    \label{eq:Rdown}
\end{align}
The constants $t_\Lambda^{\text{out/down}}$ and $r_\Lambda^{\text{out/down}}$ are, respectively, the transmission and reflection coefficients for the `out' and `down' modes.

To determine whether classical superradiance exists for charged fermions on RN, it is useful to find relations between the transmission and reflection coefficients of the different modes. We define
\begin{equation}
    W_{1,\Lambda}  =  \widetilde{R}_{1,\Lambda}R_{2,\Lambda} -  \widetilde{R}_{2,\Lambda}R_{1,\Lambda}, \qquad
    W_{2,\Lambda}  =  \widetilde{R}_{1,\Lambda}^*R_{1,\Lambda} -  \widetilde{R}_{2,\Lambda}^*R_{2,\Lambda},
    \label{eq:W1W2}
\end{equation}
for any two pairs of solutions $(R_{1,\Lambda},R_{2,\Lambda})$ and $( \widetilde{R}_{1,\Lambda}, \widetilde{R}_{2,\Lambda})$. It is straightforward to verify that these quantities do not depend on $r$. As a consequence: \mbox{$|r_{\Lambda}^{\text{in}}|=|r_{\Lambda}^{\text{up}}|=|r_{\Lambda}^{\text{out}}|= |r_{\Lambda}^{\text{down}}|$} and \mbox{$t_{\Lambda}^{\text{in}}=t_{\Lambda}^{\text{up}}=( t_{\Lambda}^{\text{out}} )^* = ( t_{\Lambda}^{\text{down}} )^*$}, with
\begin{equation}
    |r_{\Lambda}^{\text{in}}|^2+|t_{\Lambda}^{\text{in}}|^2=1.
    \label{eq:rtrelations}
\end{equation}
From~\eqref{eq:rtrelations}, we deduce that all reflection coefficients satisfy the condition~$|r_{\Lambda}|\leq 1$, which confirms the absence of classical superradiance for Dirac fields~\cite{Maeda1976}, similarly to the case of fermions on rotating black holes~\cite{Chandrasekhar1985}. 
Nonetheless, as we will demonstrate below, due to the wave number shift~\eqref{eq:omegatilde} experienced by an observer near the event horizon, quantum superradiance will occur for fermions in RN backgrounds.

\begin{block}[note]
Classical superradiance does occur in charged black hole for massless charged scalar fields. Indeed, the relations~\eqref{eq:rtrelations} between the reflection and transmission coefficients for a fermion field are different in the scalar case. Specifically, one of the key relations takes the form
\begin{equation}
    \omega \left[ 1-|r_\Lambda^{\text{in}}|^2 \right] = \tilde{\omega} |t_\Lambda^{\text{in}}|^2.
\end{equation}
From this, we deduce that scalar modes satisfying~$\omega\tilde{\omega} <0$ lead to~$|r_\Lambda^{\text{in}}| > 1$, indicating superradiant amplification. This corresponds to low frequencies within the interval \mbox{$0<\omega<qQ/r_+$} when the field and the black hole share the same charge sign ($qQ>0$), and within \mbox{$-|qQ|/r_+<\omega<0$} when they have opposite charges ($qQ<0$). Surprisingly, this will turn out to be exactly the same frequency interval in which quantum superradiance for fermions occurs.
\end{block}

\section{Choice of quantum vacuum}
\label{sec:InOutQuantum}

We now proceed with the canonical quantization of a fermionic field~$\Psi$ on a classical RN background. 
First, we need to choose an orthonormal basis of solutions to the Dirac equation~\eqref{eq:Dirac} with respect to the inner product~\eqref{eq:DiracProduct}. 
In the last section, we reviewed two well-known possibilities: the `in-up' and the `out-down' bases. 
Second, we need to choose our complex structure, splitting our chosen basis into two subsets, $\{\Psi_\Lambda^+\}$ and $\{\Psi_\Lambda^-\}$. 
As we have discussed in depth in previous chapters, this procedure is full of ambiguities.

Let us start with the `in-up' basis. Since RN admits a globally timelike Killing vector~$\partial_t$, it provides a natural way to do the splitting with respect to the proper time~$t$ of a static observer asymptotically far from the black hole (at~$\mathscr{I}^-$), whose frequency is given by
\begin{equation}
    i\partial_t \Psi_\Lambda^{\text{in}} = \omega \Psi_\Lambda^{\text{in}}.
\end{equation}
We then define~$\Psi_\Lambda^{+\text{in}}$ as the modes with positive frequency, $\omega > 0$, and~$\Psi_\Lambda^{-\text{in}}$ as those with negative frequency, $\omega < 0$. 

While we could proceed with the splitting of the `up' modes analogously by splitting them into positive and negative values of~$\omega$---and we will return to this choice later---there is no fundamental obstruction to adopting a different splitting. In fact, as shown in~\eqref{eq:R+infty}, the relevant wave number for a static (and hence accelerated) observer at~$\mathcal{H}^-$ is not~$\omega$ but the shifted wave number~$ \widetilde{\omega}$~\eqref{eq:omegatilde}. Accordingly, we can define~$\Psi_\Lambda^{+\text{up}}$ as the modes with~$ \widetilde{\omega} > 0$, and~$\Psi_\Lambda^{-\text{up}}$ as those with~$ \widetilde{\omega} < 0$. 
Finally, following the canonical quantization procedure described above, the quantum field operator~\eqref{eq:DiracOperator} translates in this case into
    \begin{multline}
        \hat{\Psi}_{|\text{in}\rangle}=\sum_{l=\frac{1}{2}}^{\infty} \sum_{m=-l}^l \left[ \int_0^{\infty} \text{d}\omega \ \hat{c}^{\text{in}}_\Lambda \Psi_\Lambda^{+\text{in}} + \int_{-\infty}^{0} \text{d}\omega \ \hat{d}^{\text{in}\dagger}_\Lambda \Psi_\Lambda^{-\text{in}}  
        \right. \\   \left. 
        +\int_0^{\infty} \text{d} \widetilde{\omega} \ \hat{c}^{\text{up}}_\Lambda \Psi_\Lambda^{+\text{up}} + \int_{-\infty}^{0} \text{d} \widetilde{\omega} \ \hat{d}^{\text{up}\dagger}_\Lambda \Psi_\Lambda^{-\text{up}} \right].
        \label{eq:expansionPsiinup}
    \end{multline}
These annihilation and creation coefficients satisfy the anticommutation relations~\eqref{eq:DiracOperator}, and define the `in' quantum vacuum, denoted here as~$|\text{in}\rangle$.
This state, by construction, has no flux of particles coming from the past null infinity~$\mathscr{I}^-$. 

Following similar criteria, we can construct a quantization scheme for the `out-down' basis. 
We choose modes~$\Psi_\Lambda^{+\text{out}}$ with positive frequency with respect to a static observer at  future null infinity~$\mathscr{I}^+$ (so that $\omega > 0$), and~$\Psi_\Lambda^{+\text{down}}$ to have $ \widetilde{\omega} > 0$. Modes~$\Psi_\Lambda^{-\text{out}}$ and~$\Psi_\Lambda^{-\text{down}}$ are defined analogously. 
These choices define the `out' and `down' annihilation and creation operators via the field expansion
    \begin{multline}
        \hat{\Psi}_{|\text{out}\rangle}=\sum_{l=\frac{1}{2}}^{\infty} \sum_{m=-l}^l  \left[ \int_0^{\infty} \text{d}\omega \ \hat{c}^{\text{out}}_\Lambda \Psi_\Lambda^{+\text{out}} + \int_{-\infty}^{0} \text{d}\omega \ \hat{d}^{\text{out}\dagger}_\Lambda \Psi_\Lambda^{-\text{out}} \right. \\ 
        +  \left. \int_0^{\infty} \text{d} \widetilde{\omega} \ \hat{c}^{\text{down}}_\Lambda \Psi_\Lambda^{+\text{down}} + \int_{-\infty}^{0} \text{d} \widetilde{\omega} \ \hat{d}^{\text{down}\dagger}_\Lambda \Psi_\Lambda^{-\text{down}} \right],
        \label{eq:expansionPsioutdown}
    \end{multline}
which in turn determine the `out' quantum vacuum~$|\text{out}\rangle$. 
In this vacuum there are no particles outgoing to the future null infinity~$\mathscr{I}^+$.

\begin{block}[note]
Although the `in-up' and `out-down' bases are explicitly invariant under time translations due to the factor~$e^{-i\omega t}$ in~\eqref{eq:ansatz}, the complex structure adopted here is not. This is because~$\widetilde{\omega}$ does not represent a true frequency of the system, unlike~$\omega$, which does. As a result, the `in' and `out' quantum states constructed here are not invariant under the classical symmetries of the background, as they are non-static despite the classical background being static.

Nonetheless, such an approach can be advantageous, as seen in other contexts. For instance, in the case of the Unruh vacuum for a static Schwarzschild black hole, the Unruh state is itself non-static. Yet, it effectively captures features of gravitational collapse and leads to the prediction of Hawking radiation. In a similar spirit, the `in' and `out' states defined in this setup may encode dynamical physics, thereby providing a static framework to study quantum superradiance.
\end{block}

\begin{block}[note]
In the case of a charged scalar field~$\Phi$, the splitting criterion employed here for both the `in-up' and `out-down' bases is not optional but mandatory~\cite{Balakumar2022}. 
This requirement arises from the necessity for modes~$\Phi_\Lambda^{+\text{in/up/out/down}}$ to have positive KG norm (and~$\Phi_\Lambda^{-\text{in/up/out/down}}$ to have negative KG norm) in order to ensure the standard bosonic commutation relations for the creation and annihilation operators presented in~\autoref{chap:QuantizationScalars}. Indeed, the KG norms of the `in' and `out' modes~$\Phi_\Lambda^{\text{in/out}}$ have the same sign as~$\omega$, whereas the KG norms of the `up' and `down' modes~$\Phi_\Lambda^{\text{up/down}}$ have the same sign as~$\widetilde{\omega}$. In contrast, for fermionic fields, a natural notion of inner product~\eqref{eq:DiracProduct} exists without requiring any additional constraints and all modes have positive norm. This provides greater flexibility in the choice of mode splitting.
\end{block}

Alternatively, we can define the splitting of the `up' modes according to a static observer at~$\mathscr{I}^+$ (instead of with respect to a static observer close to~$\mathscr{H}^-$). 
In this case, the modes~$\Psi_\Lambda^{+\text{up}}$ would be those with~$\omega >0$ (instead of $ \widetilde{\omega} >0$), while~$\Psi_\Lambda^{-\text{up}}$ would have~$\omega <0$ (and not $ \widetilde{\omega} <0$). By keeping the splitting of the `in' modes with respect to the static observer  at~$\mathscr{I}^-$, this new definition of positive frequency leads to a different quantum vacuum, a `Boulware' state~$|B\rangle$, which has no particles at either~$\mathscr{I}^-$ or~$\mathscr{I}^+$. 
Similarly, we can apply this approach to the `out-down' basis, such that modes~$\Psi_\Lambda^{+\text{out/down}}$ are those with~$\omega > 0$, and modes~$\Psi_\Lambda^{\text{out/down}}$ are those with~$\omega < 0$. 
This results in another, possibly distinct, `Boulware'-like state~$|B'\rangle$. Note that, in the scalar case, due to the constraints on the quantization imposed by the sign of the KG norm, a `Boulware'-like state---characterized by the absence of particle flux at both $\mathscr{I}^-$ and $\mathscr{I}^+$---cannot be defined~\cite{Balakumar2022}.

\section{Quantum superradiance} 
\label{sec:QuantumSuperradiance}

In this section, we will show that while the `out' state $|\text{out}\rangle$ is empty at future null infinity~$\mathscr{I}^+$, the `in' state $|\text{in}\rangle$ contains an outgoing flux of particles at~$\mathscr{I}^+$. This particle production phenomenon is known as quantum superradiance. We will quantify the number of particles per unit time created during this process. 

The Bogoliubov transformation relating the `in' and `out' quantum theories can be written, following the structure of~\eqref{eq:BogoliubovPsis}, as:
\begin{equation}
    \begin{pmatrix} 
    \Psi^{+\text{out}}_\Lambda \\ 
    \Psi^{+\text{down}}_\Lambda \\
    \Psi^{-\text{out}}_\Lambda \\
    \Psi^{-\text{down}}_\Lambda 
    \end{pmatrix}
    =
    \sum_{\Lambda^\prime} 
    \begin{array}{c}
        \begin{pmatrix} \alpha_{\Lambda\Lambda^\prime}^{+\text{in/out}} & \alpha_{\Lambda\Lambda^\prime}^{+\text{up/out}} & \beta_{\Lambda\Lambda^\prime}^{+\text{in/out}} &
        \beta_{\Lambda\Lambda^\prime}^{+\text{up/out}} \\
        \alpha_{\Lambda\Lambda^\prime}^{+\text{in/down}} & \alpha_{\Lambda\Lambda^\prime}^{+\text{up/down}} & \beta_{\Lambda\Lambda^\prime}^{+\text{in/down}} &
        \beta_{\Lambda\Lambda^\prime}^{+\text{up/down}} \\
        \beta_{\Lambda\Lambda^\prime}^{-\text{in/out}} &
        \beta_{\Lambda\Lambda^\prime}^{-\text{up/out}} &
        \alpha_{\Lambda\Lambda^\prime}^{-\text{in/out}} & \alpha_{\Lambda\Lambda^\prime}^{-\text{up/out}} \\
        \beta_{\Lambda\Lambda^\prime}^{-\text{in/down}} &
        \beta_{\Lambda\Lambda^\prime}^{-\text{up/down}} &
        \alpha_{\Lambda\Lambda^\prime}^{-\text{in/down}} & \alpha_{\Lambda\Lambda^\prime}^{-\text{up/down}} \\
        \end{pmatrix}
        \begin{pmatrix} 
        \Psi^{+\text{in}}_{\Lambda^\prime} \\ 
        \Psi^{+\text{up}}_{\Lambda^\prime} \\
        \Psi^{-\text{in}}_{\Lambda^\prime} \\
        \Psi^{-\text{up}}_{\Lambda^\prime}
    \end{pmatrix}
    \end{array}.
\end{equation}
Following~\eqref{eq:Num}, the total number of created particles and antiparticles in the `out' state with respect to the `in' state (and vice versa) can be calculated from the $\beta$-Bogoliubov coefficients:
\begin{align}
    \mathcal{N}_{|\text{in}\rangle} 
    = \sum_{\Lambda,\Lambda^\prime} &\big( |\beta_{\Lambda\Lambda^\prime}^{+\text{in/out}}|^2 + |\beta_{\Lambda\Lambda^\prime}^{+\text{up/out}}|^2 + |\beta_{\Lambda\Lambda^\prime}^{+\text{in/down}}|^2 +
    |\beta_{\Lambda\Lambda^\prime}^{+\text{up/down}}|^2 \nonumber \\
     +& |\beta_{\Lambda\Lambda^\prime}^{-\text{in/out}}|^2 + |\beta_{\Lambda\Lambda^\prime}^{-\text{up/out}}|^2 + |\beta_{\Lambda\Lambda^\prime}^{-\text{in/down}}|^2 +
    |\beta_{\Lambda\Lambda^\prime}^{-\text{up/down}}|^2 \big)
    \label{eq:NSuperradiance}
\end{align}

The next step is to determine the values of the $\beta$-coefficients. From the asymptotic behaviours of the radial functions given in~(\ref{eq:RinA}--\ref{eq:Rdown}), and from the relations between the transmission and reflection coefficients in~\eqref{eq:rtrelations}, we obtain 
\begin{equation}
    R_{j,\Lambda}^{\text{out}} = r_\Lambda^{\text{out}} R_{j,\Lambda}^{\text{in}} + t_\Lambda^{\text{out}} R_{j,\Lambda}^{\text{up}}, \qquad
    R_{j,\Lambda}^{\text{down}} = t_\Lambda^{\text{down}} R_{j,\Lambda}^{\text{in}} + r_\Lambda^{\text{down}} R_{j,\Lambda}^{\text{up}},
\end{equation}
for $j=1,2$. Then, the `in-up' basis is related to the `out-down' basis by the linear combinations
\begin{equation}
    \Psi_{\Lambda}^{\text{out}} = r_\Lambda^{\text{out}} \Psi_{\Lambda}^{\text{in}} + t_\Lambda^{\text{out}} \Psi_{\Lambda}^{\text{up}}, \qquad
    \Psi_{\Lambda}^{\text{down}} = t_\Lambda^{\text{down}} \Psi_{\Lambda}^{\text{in}} + r_\Lambda^{\text{down}} \Psi_{\Lambda}^{\text{up}}.
    \label{eq:relbasisBogoliubov}
\end{equation}
According to \eqref{eq:omegatilde}, when~$qQ>0$, modes with~$\omega>0$ also have ${\widetilde {\omega}}>0$. Therefore, for positive frequencies~$\omega$, the relations~\eqref{eq:relbasisBogoliubov} connect only solutions within the one-particle sector of the Hilbert space:
\begin{equation}
    \Psi_{\Lambda}^{+\text{out}} = r_\Lambda^{\text{out}} \Psi_{\Lambda}^{+\text{in}} + t_\Lambda^{\text{out}} \Psi_{\Lambda}^{+\text{up}}, \qquad
    \Psi_{\Lambda}^{+\text{down}} = t_\Lambda^{\text{down}} \Psi_{\Lambda}^{+\text{in}} + r_\Lambda^{\text{down}} \Psi_{\Lambda}^{+\text{up}}.
\end{equation}
Since there is no mixing between particle and antiparticle states, $\beta_{\Lambda\Lambda^\prime}^{+\text{in/out}}$ and $\beta_{\Lambda\Lambda^\prime}^{+\text{up/out}}$ vanish for~$\omega>0$. For modes with~${\widetilde {\omega}}<0$, which also satisfy~\mbox{$\omega<-qQ/r_+$}, we reach similar conclusions, where now~\eqref{eq:relbasisBogoliubov} only relates antiparticle states. However, this does not hold for modes in the range~\mbox{$-qQ/r_+<\omega<0$}, for which ${\widetilde {\omega }}>0$ and we have
\begin{equation}
    \Psi_{\Lambda}^{-\text{out}} = r_\Lambda^{\text{out}} \Psi_{\Lambda}^{-\text{in}} + t_\Lambda^{\text{out}} \Psi_{\Lambda}^{+\text{up}}, \qquad
    \Psi_{\Lambda}^{+\text{down}} = t_\Lambda^{\text{down}} \Psi_{\Lambda}^{-\text{in}} + r_\Lambda^{\text{down}} \Psi_{\Lambda}^{+\text{up}}.
    \label{eq:relbasisBogoliubov+-}
\end{equation}
For this specific frequency interval, there is a mixing of particle and antiparticle states. 
Therefore, for $qQ>0$, the only non-vanishing $\beta$-coefficients are
\begin{equation}
    \beta_{\Lambda\Lambda^\prime}^{+\text{in/down}} =
    \beta_{\Lambda\Lambda^\prime}^{-\text{up/out}} = t_\Lambda^{\text{out}} \delta(\omega-\omega^\prime)\delta_{l,l^\prime} \delta_{m,m^\prime} \qquad \text{for} \ \ -qQ/r_+<\omega<0.
    \label{eq:BetasqQpos}
\end{equation}
Furthermore, since $t_\Lambda^{\text{out}}=t_\Lambda^{\text{down}}$, particles and antiparticles are created in equal proportions. 
This is consistent with the fact that particle production is fundamentally a pair creation process, ensuring that the total charge of the produced fermions remains neutral. 

Similarly, for the case where~$qQ<0$, analogous conclusions are drawn, with the only non-vanishing contributions now coming from modes with frequencies within the interval~\mbox{$0< \omega < -qQ/r_+$}:
\begin{equation}
    \beta_{\Lambda\Lambda^\prime}^{+\text{up/out}} =
    \beta_{\Lambda\Lambda^\prime}^{-\text{in/down}} = t_\Lambda^{\text{out}} \delta(\omega-\omega^\prime)\delta_{l,l^\prime} \delta_{m,m^\prime}, \qquad \text{for} \ \ 0< \omega < -qQ/r_+.
    \label{eq:BetasqQneg}
\end{equation}

Substituting~\eqref{eq:BetasqQpos} and~\eqref{eq:BetasqQneg} into~\eqref{eq:NSuperradiance}, we obtain the total number of excitations in terms of the transmission coefficients, accounting for both signs of the product~$qQ$:
\begin{equation}
    \mathcal{N}_{|\text{in}\rangle} = \delta(0) \frac{1}{16\pi^3} \sum_{l=\frac{1}{2}}^{+\infty} (2l+1) \int_{\min\{-\frac{qQ}{r_+},0\}}^{\max\{-\frac{qQ}{r_+},0\}} \text{d}\omega \ |t_\Lambda^{\text{out}}|^2.
    \label{eq:numberparticles}
\end{equation}
The factor~$\delta(0) = (2\pi)^{-1} \int \text{d}t$ arises from squaring the Dirac delta function~$\delta(\omega - \omega^\prime)$ that appears in~\eqref{eq:BetasqQpos}, reflecting the fact that we are computing the number of particles created over an infinite time interval in a static spacetime. To handle this, we redefine~$\mathcal{N}_{|\text{in}\rangle}$ as the total number of excitations per unit time, identifying it with the term accompanying~$\delta(0)$ in~\eqref{eq:numberparticles}. On the other hand, the factor of~$2l+1$ arises from the summation over~$m$, as the transmission coefficients are independent of this quantum number, given that the radial functions are also independent of~$m$~\eqref{eq:R12}. 
Additionally, we have included a prefactor of $1/16\pi^3$, which arises from the normalization of the modes. 

Although we previously saw that classical superradiance does not occur for fermions on a RN background, this result shows that quantum superradiance is indeed present.
From henceforth, we will call modes with $\omega {\widetilde {\omega }}<0$ `superradiant' modes, since these modes give rise to quantum superradiance.

To numerically compute the transmission coefficients~$t_\Lambda^{\text{out}}$, we solved the system of radial equations~\eqref{eq:R12} with the asymptotic boundary conditions
\begin{equation}
R_{1,\Lambda}^{\text{out}}(r_*) \overset{r_*\to +\infty}{\sim} e^{i\omega r_*}, 
\qquad 
R_{2,\Lambda}^{\text{out}}(r_*) \overset{r_*\to -\infty}{\sim} 0.
\end{equation}
By then calculating the constant~$W_{1,\Lambda}$ in~\eqref{eq:W1W2} for the `out' and `down' radial functions, it follows that 
\begin{equation}
    t_\Lambda^{\text{out}} = \lim_{r_*\rightarrow -\infty} R_{1,\Lambda}^{\text{out}}(r_*) e^{-i \widetilde{\omega}r_*}.
    \label{eq:toutnumerics}
\end{equation}
To numerically solve this boundary value problem for each pair \mbox{$(l,\omega)$}, we used the function \texttt{scipy.integrate.solve\_bvp} in \textsc{Python}, which implements a collocation method based on polynomial interpolation. The tolerance level, the number of nodes in the discretization of $r$, and the numerical cutoffs for \mbox{$r_* \rightarrow \pm\infty$} were chosen carefully to ensure convergence of the solution. Due to the computational complexity and the large volume of data involved, we made use of the High Performance Computing cluster resources provided by the Universidad Complutense de Madrid.

\begin{figure}
    \centering
    \includegraphics[width=0.8\textwidth]{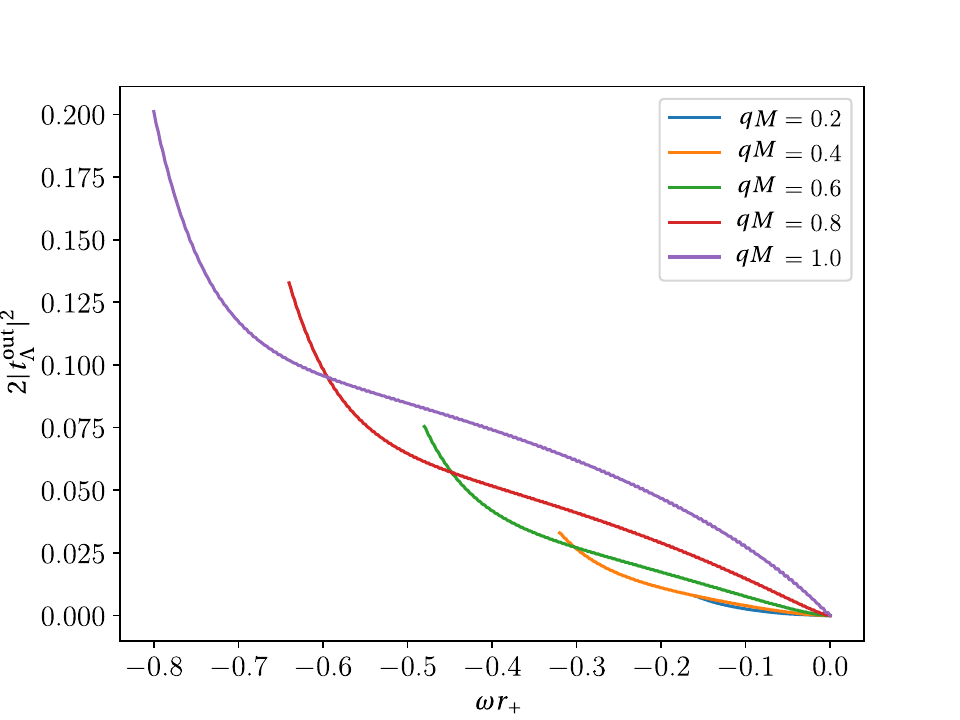}
    \caption{Number of created particles per mode, \mbox{$(2l+1)|t_{\Lambda}^{\text{out}}|^2$}, as a function of the frequency~$\omega$, for modes with $l=1/2$ and various positive fermion charges~$q$. The black hole charge is fixed at $Q=0.8M$. Outside the depicted interval~$[-qQ/r_+,0]$, no particle production occurs, and the particle number drops to zero.}
    \label{fig:densityomega}
\end{figure}

In figure~\ref{fig:densityomega}, we illustrate the dependence of the particle number density on the frequency~$\omega$ for fixed~$l=1/2$, when $qQ>0$. 
Modes with~$\omega = 0$ lack the energy required to cross the transmission barrier, resulting in the reflection of all fermions. 
Consequently, no particle production occurs at vanishing frequencies. 
As the frequency increases in absolute value, the modes gain enough energy to be partially transmitted down the event horizon. 
The particle production peaks at the threshold frequency~$\omega = -qQ/r_+$. 
Beyond this point, the quantum superradiance effect ceases, as described by~\eqref{eq:numberparticles}. 
Our analysis shows that, across all values of~$q$ and~$Q$ studied, the particle density contribution decreases by several orders of magnitude with each increasing value of~$l$. 
For instance, at the threshold~\mbox{$\omega = -qQ/r_+$}, where particle production is most significant,  the contribution of modes with~\mbox{$l>7/2$} is more than ten orders of magnitude smaller than that of the dominant~$l=1/2$ mode. 
Finally, increasing the fermion charge~$q$ broadens the spectrum, leading to an overall enhancement in particle production.
This will be discussed further in~\autoref{sec:EnergyLoss}.

In contrast, consider the `Boulware'-type states~$|B\rangle$ and~$|B'\rangle$, defined in \autoref{sec:InOutQuantum} according to the criterion that $\Psi_\Lambda^{+\text{in/up/out/down}}$ have~\mbox{$\omega > 0$} and $\Psi_\Lambda^{-\text{in/up/out/down}}$ have \mbox{$\omega < 0$}. 
From~\eqref{eq:relbasisBogoliubov}, it is clear that positive and negative frequencies are not mixed. 
This results in vanishing $\beta$-coefficients across the entire spectrum, meaning that these two states, initially defined using two different bases of field modes, represent the same exact quantization: one that is empty at both~$\mathscr{I}^-$ and~$\mathscr{I}^+$. Therefore, no quantum superradiance occurs with respect to these vacua. 
In the remainder of this chapter, we therefore focus on the study of the `in' and `out' quantum vacua.

\section{Black hole discharge}
\label{sec:Discharge}

For a Dirac spinor~$\Psi$, we define the classical charge current as
\begin{equation}
J^{\mu} = -q {\overline {\Psi }} \gamma ^{\mu } \Psi.
\label{eq:current}
\end{equation}
This quantity is conserved, $\nabla_\mu J^\mu = 0$. The \gls{QuantumChargeCurrentOperator} is defined as
\begin{equation}
    \hat{J}^\mu = -\frac{q}{2} \left[ \hat{{\overline {\Psi }}}, \gamma ^{\mu } \hat{\Psi} \right],
\end{equation}
where the commutator acts only on the annihilation and creation operators in order to preserve the spinorial structure of products of the form~$-q{\overline {\Psi }} \gamma ^{\mu } \Psi$, as justified in \autoref{sec:DissipatedEnergy}. Given a quantization scheme with a particular choice of complex structure defined by the modes~$\Psi_\Lambda^+$ and~$\Psi_\Lambda^-$, the expectation value of~$\hat{J}^\mu$ is given by
\begin{equation}
    \langle \hat{J}^\mu \rangle = \frac{1}{2}\sum_\Lambda \left( j_\Lambda^{-\mu} - j_\Lambda^{+\mu} \right), \qquad j_\Lambda^{\pm\mu} = -q \overline{\Psi}_\Lambda^\pm\gamma^\mu\Psi_\Lambda^\pm.
    \label{eq:expcurrent}
\end{equation}
The expressions for the components of the current~\mbox{$j_\Lambda^\mu = -q\overline{\Psi}_\Lambda\gamma^\mu\Psi_\Lambda$} in terms of the functions appearing in the mode ansatz~\eqref{eq:ansatz} are
\begin{align}
    j_\Lambda^t & = \frac{q}{4\pi^2r^2f\sin\theta} \left( |R_{1,\Lambda}|^2S_{1,\Lambda}^2 + |R_{2,\Lambda}|^2S_{2,\Lambda}^2 \right), \nonumber \\
    j_\Lambda^r & = \frac{qL}{4\pi^2r^2\sin\theta} \left( |R_{1,\Lambda}|^2S_{1,\Lambda}^2 - |R_{2,\Lambda}|^2S_{2,\Lambda}^2 \right), \nonumber \\
    j_\Lambda^\theta & = \frac{qL}{2\pi^2r^3\sqrt{f}\sin\theta} \Re \left( R_{1,\Lambda}^* R_{2,\Lambda} \right) S_{1,\Lambda}S_{2,\Lambda}, \nonumber \\
    j_\Lambda^\varphi & = \frac{qL}{2\pi^2r^3\sqrt{f}\sin^2\theta} \Im \left( R_{1,\Lambda}^* R_{2,\Lambda} \right) S_{1,\Lambda}S_{2,\Lambda},
    \label{eq:currentcomp}
\end{align}
where $\Re $ denotes the real part and $\Im $ the imaginary part.

From the properties of the angular functions in \autoref{sec:ClassicalFermionsRN}, when performing the finite sum over~$m$ to compute the expectation values~\eqref{eq:expcurrent} we obtain a vanishing contribution from the angular components, so that
\begin{equation}
    \langle \hat{J}^\theta \rangle = \langle \hat{J}^\varphi \rangle = 0,
    \label{eq:angularcurrent}
\end{equation}
independently of the quantum state under consideration. 
This is expected from the spherical symmetry of the configuration. 

Since we want to quantify quantum superradiance, we are interested in the components of the charge current leading to different expectation values for `in' and `out' states. 
Using that the `out-down' basis is the time reverse of the `in-up' basis (see~\eqref{eq:timereverse}) and the symmetries of the angular functions under the transformation~$m\rightarrow -m$ given in \autoref{sec:ClassicalFermionsRN} (the radial functions are independent of~$m$), we arrive at the results
\begin{equation}
    \langle \text{in} | \hat{J}^t | \text{in} \rangle = \langle \text{out} | \hat{J}^t | \text{out} \rangle, \qquad
    \langle \text{in} | \hat{J}^r | \text{in} \rangle = - \langle \text{out} | \hat{J}^r | \text{out} \rangle.
    \label{eq:currentB-B+}
\end{equation}
Thus, we will now focus on the computation of the expectation value of the radial component of the charge current. 

From the semiclassical Maxwell equation \mbox{$\nabla_\mu F^{\mu\nu} = \langle \hat{J}^\nu \rangle$}, where \mbox{$F^{\mu\nu}=\partial^\mu A^\nu - \partial^\nu A^\mu$} is the antisymmetric electromagnetic tensor, we deduce that the expectation value of the current density operator is conserved for all quantum vacua: $\nabla_\mu \langle \hat{J}^\mu \rangle = 0$. Taking into account the vanishing angular components~\eqref{eq:angularcurrent} and the fact that the expectation values~$\langle \hat{J}^t \rangle$ in the quantum states considered here do not depend on time, integrating this conservation equation leads to
\begin{equation}
    \langle \hat{J}^r \rangle = -\frac{\mathcal{K}}{r^2},
\end{equation}
where~$\mathcal{K}$ is an integration constant independent of $r$. 
From~\eqref{eq:expcurrent}, the sign of~$\mathcal{K}$ matches that of the contribution to the charge current from the particle states, $j_\Lambda^{+\mu}$. Consequently, $\mathcal{K}$ represents the net charge flux emitted by the black hole, defined as the charge flux of particles minus that of antiparticles. 

To compute~$\mathcal{K}$ for the `in' vacuum we only need to evaluate~$\langle \hat{J}^r \rangle$ at~$r_*\rightarrow +\infty$. 
Using the asymptotics of the radial functions~(\ref{eq:RinA}--\ref{eq:RupB}), this results in
\begin{equation}
    \mathcal{K}_{|\text{in}\rangle} = \frac{q}{16\pi^3} \sum_{l=\frac{1}{2}}^{+\infty} (2l+1) \int_{-\frac{qQ}{r_+}}^0 \text{d}\omega \ |t_\Lambda^{\text{out}}|^2.
    \label{eq:K}
\end{equation}
Note that when $qQ<0$, the lower limit of the integral is larger than the upper limit, introducing a negative sign when the order is reversed. 
Only the superradiant modes contribute to the charge flux, with the absolute value of $\mathcal{K}_{|\text{in}\rangle}$ given by the total particle number per unit time in~\eqref{eq:numberparticles} multiplied by the charge of the fermionic field~$q$. 
This result is independent of the chirality~$L$ due to the invariance of $\sum_\Lambda j_\Lambda^r$~\eqref{eq:currentcomp} under the transformation \mbox{$R_{1,\Lambda} \leftrightarrow R_{2,\Lambda}$}.

The contribution of particles to the total charge current is equal in magnitude but opposite in sign to that of antiparticles;  in other words \mbox{$\sum_\Lambda j_\Lambda^{+r} = - \sum j_\Lambda^{-r}$}. 
In addition, from~\eqref{eq:K}, we observe that the sign of the charge flux~$\mathcal{K}_{|\text{in}\rangle}$ matches the sign of the black hole charge~$Q$. 
This implies that when the black hole is positively (negatively) charged, positive (negative) charges are emitted outwards while an equal number of negative (positive) charges are absorbed, resulting in the discharge of the black hole due to quantum superradiance.

Finally, the expression for~$\mathcal{K}$ in~\eqref{eq:K} is only valid for the `in' state (for `out' state, according to~\eqref{eq:currentB-B+}, we have \mbox{$\mathcal{K}_{|\text{out}\rangle}=-\mathcal{K}_{|\text{in}\rangle}$}), and each quantum vacuum has its own value of~$\mathcal{K}$. 
For the `Boulware'-type quantum state~$|B\rangle$, we find that $\mathcal{K}_{|B\rangle}=0$ and there is no charge current in the radial direction.

\section{Black hole energy loss}
\label{sec:EnergyLoss}

The classical stress-energy momentum tensor for a Dirac field~$\Psi$ in background gravitational and electromagnetic fields is
\begin{equation}
T_{\mu \nu }= \frac {i}{2} \left( {\overline {\Psi }} \gamma _{(\mu }
\nabla _{\nu )} \Psi - \left( \nabla _{( \mu }{\overline {\Psi }}\right)
\gamma _{\nu )} \Psi + 2iqA_{(\mu} {\overline {\Psi }} \gamma_{\nu) } \Psi \right).
\label{eq:Tmunuclassical}
\end{equation}
With the same caveat as for the current, namely that commutators act only on the annihilation and creation operators and not the Dirac spinors, the associated \gls{QuantumStressEnergyMomentumTensorOperator} is
\begin{equation}
    \hat{T}_{\mu\nu} = \frac{i}{4}  \left( \left[ \hat{\overline{\Psi}}, \gamma_{(\mu} \nabla_{\nu)} \hat{\Psi} \right] - \left[ \nabla_{(\mu} \hat{\overline{\Psi}}, \gamma_{\nu)} \hat{\Psi} \right]  + 2iqA_{(\mu} \left[ \hat{{\overline {\Psi }}}, \gamma_{\nu)} \hat{\Psi} \right] \right).
\end{equation}
Fixing a particular quantization scheme defined by modes~$\Psi_\Lambda^+$ and~$\Psi_\Lambda^-$, we find the expectation value of $\hat{T}_{\mu\nu}$ to be
\begin{equation}
    \langle \hat{T}_{\mu\nu} \rangle = \frac{1}{2} \sum_\Lambda \left( t_{\mu\nu,\Lambda}^- - t_{\mu\nu,\Lambda}^+ \right),
\end{equation}
where $t_{\mu\nu,\Lambda}^\pm$ are the classical stress-energy momentum tensor components~\eqref{eq:Tmunuclassical} for the modes~$\Psi_\Lambda^\pm$. 
Their expressions (omitting~$\Lambda$ and the~$\pm$ signs for simplicity in the notation) in terms of the radial and angular functions  and mode contributions to the charge current~\eqref{eq:currentcomp} are
 \begin{align}
    t_{tt} =& ~ -\left( \omega+\frac{qQ}{r} \right) \frac{j_t}{q}, \nonumber \\
    t_{tr} =& ~ -\frac{1}{2} \left( \omega+\frac{qQ}{r} \right)\frac{j_r}{q} - \frac{1}{8\pi^2r^2\sin\theta}\left[ \Im\left( R_1^*R_1^\prime \right) S_1^2 + \Im\left( R_2^*R_2^\prime \right) S_2^2 \right], \nonumber \\
    t_{t\theta} =& ~ -\frac{1}{2} \left( \omega+\frac{qQ}{r} \right)\frac{j_\theta}{q} - \frac{L}{4r\sin\theta} \left( \frac{r}{2}f^\prime - f \right) \frac{j_\varphi}{q}, \nonumber \\
    t_{t\varphi} =& ~ \frac{m}{2}\frac{j_t}{q} + \frac{L\cos\theta f}{4} \frac{j_r}{q} + \frac{L\sin\theta}{4r} \left( \frac{r}{2}f^\prime - f \right) \frac{j_\theta}{q} - \frac{1}{2}\left( \omega+\frac{qQ}{r} \right)\frac{j_\varphi}{q}, \nonumber \\
    t_{rr} =& ~ \frac{L}{4\pi^2r^2f\sin\theta} \left[ \Im\left( R_1^* R_1^\prime \right) S_1^2 - \Im\left( R_2^* R_2^\prime \right) S_2^2 \right], \nonumber \\
    t_{r\theta} =& ~ \frac{L}{8\pi^2r\sqrt{f}\sin\theta} \left[ \Im\left( R_1^* R_2^\prime \right) + \Im\left( R_2^* R_1^\prime \right)\right] S_1 S_2, \nonumber \\
    t_{r\varphi} =& ~ -\frac{L\cos\theta}{4 f} \frac{j_t}{q} + \frac{m}{2} \frac{j_r}{q} - \frac{L}{8\pi^2r\sqrt{f}} \left[ \Re\left( R_1^* R_2^\prime \right) - \Re\left( R_2^* R_1^\prime \right) \right] S_1 S_2, \nonumber \\
    t_{\theta\theta} =& ~ -\frac{L}{4\pi^2 r\sqrt{f}\sin\theta} \Im\left( R_1^*R_2 \right) \left( S_1^\prime S_2 - S_1 S_2^\prime \right), \nonumber \\
    t_{\theta\varphi} =& ~ \frac{m}{2} \frac{j_\theta}{q} + \frac{L}{8\pi^2r\sqrt{f}} \Re\left( R_1^* R_2 \right) \left( S_1^\prime S_2 - S_1 S_2^\prime \right), \nonumber \\
    t_{\varphi\varphi} =& ~ m \frac{j_\varphi}{q}.
    \label{eq:tmunu}
\end{align}

Taking into account the properties of the angular functions in \ref{sec:ClassicalFermionsRN}, substituting the modes \eqref{eq:ansatz} in the stress-energy momentum tensor components in~\eqref{eq:tmunu}, and performing the finite sum over $m$, all components of the stress-energy momentum tensor expectation value vanish except for~$\langle \hat{T}_{tt} \rangle$, $\langle \hat{T}_{rr} \rangle$ and~$\langle \hat{T}_{tr} \rangle$. 
However, while the first two coincide for the `in' and `out' states, this is not the case for the $tr$-component, which satisfies
\begin{equation}
    \langle \text{in} | \hat{T}_{tr} | \text{in} \rangle = - \langle \text{out} | \hat{T}_{tr} | \text{out} \rangle.
\end{equation}
In order to quantify the quantum superradiance phenomenon we focus now on calculating this radial energy flux expectation value.

Due to the electromagnetic background, the expectation value of the stress-energy momentum tensor is not conserved: $\nabla^\mu \langle \hat{T}_{\mu\nu} \rangle = 4\pi F_{\mu\nu} \langle \hat{J}^{\mu} \rangle$. Taking into account the fact that all the expectation values are time-independent, we integrate the equation for~$\nu=t$, resulting in 
\begin{equation}
    \langle \hat{T}_t^r \rangle = -\frac{\mathcal{L}}{r^2} + \frac{4\pi \mathcal{K}Q}{r^3},
    \label{eq:TtrL}
\end{equation}
where~$\mathcal{L}$ does not depend on $r$, but does depend on the particular quantum state considered. 
Physically ${\mathcal{L}}$ is the flux of energy from the black hole. 
To evaluate~$\langle \hat{T}_t^r \rangle$ for the `in' vacuum at~$r_* \rightarrow +\infty$, we use the asymptotic behaviour of the radial functions~(\ref{eq:RinA}--\ref{eq:RupB}), as well as the properties of the angular functions given in \autoref{sec:ClassicalFermionsRN}. 
Identifying this result with~\eqref{eq:TtrL}, we obtain 
\begin{equation}
    \mathcal{L}_{|\text{in}\rangle} = -\frac{1}{16\pi^3} \sum_{l=\frac{1}{2}}^{+\infty} (2l+1) \int_{-\frac{qQ}{r_+}}^0 \text{d}\omega \ \omega |t_\Lambda^{\text{out}}|^2.
\end{equation}
When $qQ<0$, we again need to reverse the order of the integral limits and introduce a negative sign. Each superradiant mode contributes to the energy flux in the radial direction with an energy proportional to its particle number, $(2l+1)|t_\Lambda^{\text{out}}|^2$, and its frequency~$\omega$. 
Due to the invariance of $\sum_\Lambda t_{tr,\Lambda}$ under the exchange of $R_{1,\Lambda}$ and $R_{2,\Lambda}$~\eqref{eq:tmunu}, the energy flux~$\mathcal{L}_{|\text{in}\rangle}$ is the same for positive and negative chiralities. 

\begin{figure}
    \centering
    \includegraphics[width=0.8\textwidth]{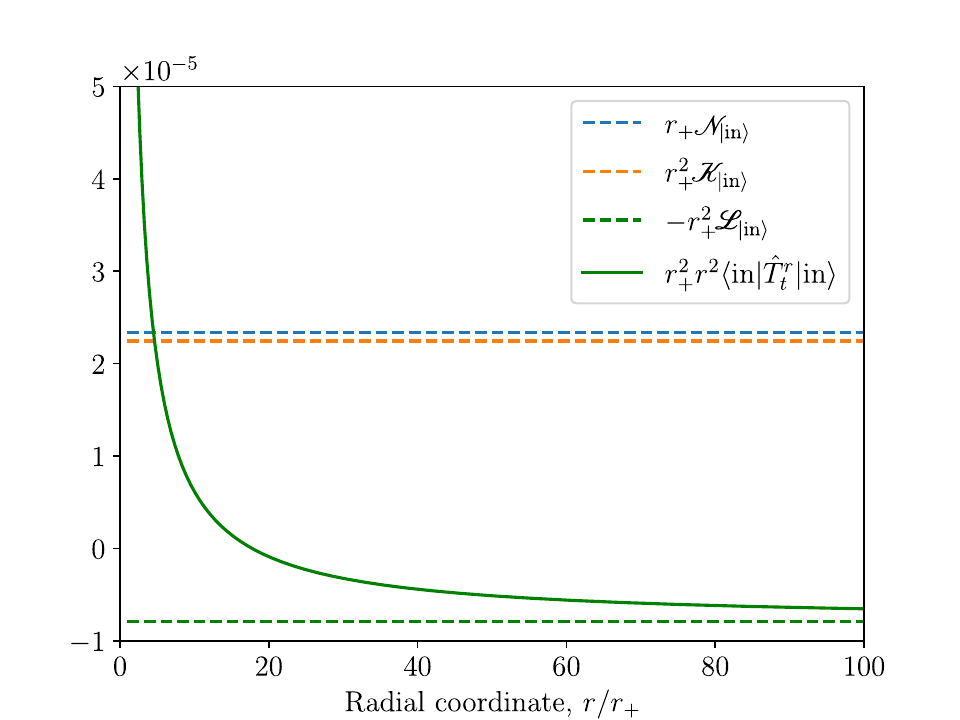}
    \caption{Expected energy density flux dissipated by the black hole, $r^2 \langle \text{in} | \hat{T}_t^r | \text{in} \rangle$, as a function of the radial coordinate, for $q M = 0.6$ and $Q = 0.8 M$. 
    Far from the black hole, it approaches a horizontal asymptote at~$-\mathcal{L}_{|\text{in}\rangle}$. 
    The total number of particles created per time, $\mathcal{N}_{|\text{in}\rangle}$, and the charge flux constant, $\mathcal{K}_{|\text{in}\rangle}$, are also shown.}
    \label{fig:NKLr}
\end{figure}

In particular, $\mathcal{L}_{|\text{in}\rangle}$ is always positive, and due to the black hole discharge studied above, we have $\mathcal{K}_{|\text{in}\rangle}Q > 0$. 
As a result, there is a spherical surface with radius
\begin{equation}
    r_0=\frac{4\pi \mathcal{K}_{|\text{in}\rangle} Q}{\mathcal{L}_{|\text{in}\rangle}},
    \label{eq:r0}
\end{equation}
where the expectation value in~\eqref{eq:TtrL} vanishes. Inside this sphere, there is an ingoing flux of energy into the black hole, while outside, there is a net energy loss. This can be seen in figure~\ref{fig:NKLr}, which shows how the energy flux~$r^2\langle \text{in} | \hat{T}_t^r | \text{in} \rangle$ decreases as one moves away from the black hole, and asymptotically approaches the constant $-\mathcal{L}_{|\text{in}\rangle}$. We also show the total particle number per unit time $\mathcal{N}_{|\text{in}\rangle}$~\eqref{eq:numberparticles}, and the charge flux constant $\mathcal{K}_{|\text{in}\rangle}$~\eqref{eq:K}. The behaviour of~$r^2\langle \text{in} | \hat{T}_t^r | \text{in} \rangle$ resembles that observed in the case of a charged scalar field on RN~\cite{Balakumar2020}, where an \gls{EffectiveErgosphere} indicates a sign change in this component of the stress-energy momentum tensor outside the event horizon~\cite{DiMenza2015,Denardo1973,Denardo1974}. 

\begin{figure}
    \centering
    \includegraphics[width=0.8\textwidth]{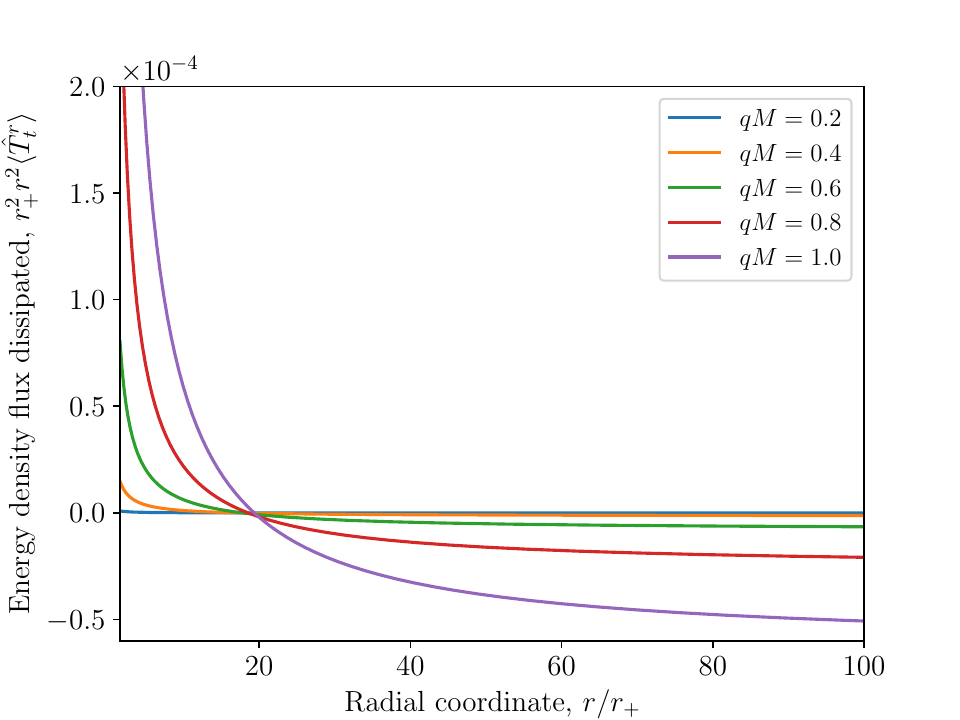}
    \caption{Expected energy density flux dissipated by the black hole, $r^2 \langle \text{in} | \hat{T}_t^r | \text{in} \rangle$, as a function of the radial coordinate, for $Q = 0.8 M$ and various fermion charges $q$.}
    \label{fig:Ttr}
\end{figure}

\begin{figure}
    \centering
    \includegraphics[width=0.8\textwidth]{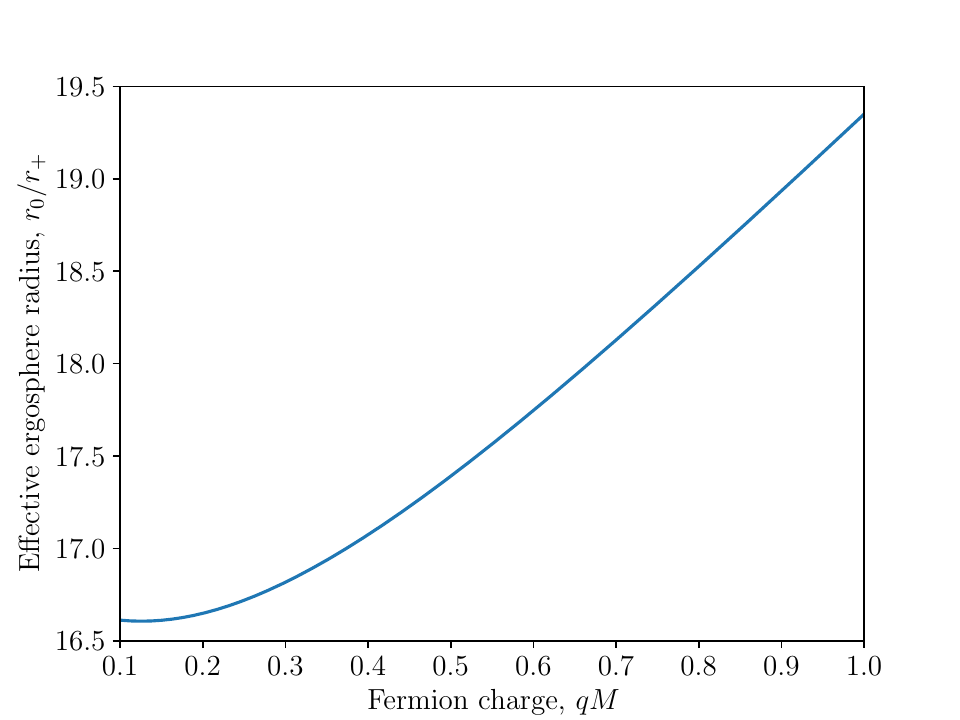}
    \caption{Effective ergosphere radius $r_{0}$, where the expectation value $r^2 \langle \text{in} | \hat{T}_t^r | \text{in} \rangle$ vanishes, as a function of the fermion field charge $q$, with $Q = 0.8M$.}
    \label{fig:ergo}
\end{figure}

\begin{figure}
    \centering
    \includegraphics[width=0.8\textwidth]{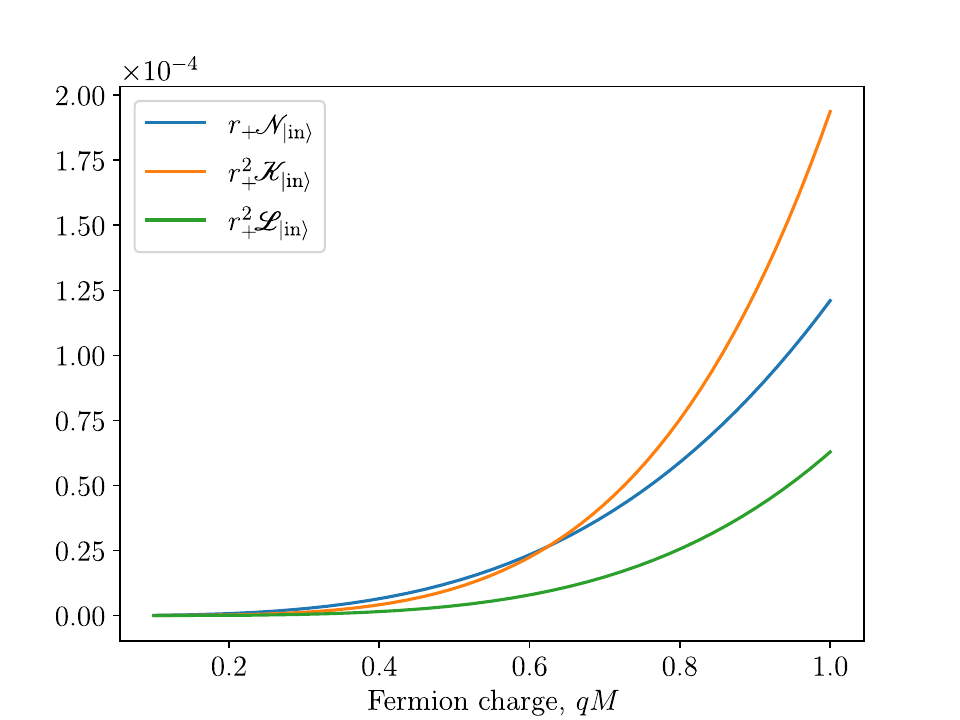}
    \caption{Enhancement of quantum superradiance when increasing the fermion charge~$q$, for a black hole charge \mbox{$Q = 0.8 M$}. We show the quantities $\mathcal{N}_{|\text{in}\rangle}$, $\mathcal{K}_{|\text{in}\rangle}$ and~$\mathcal{L}_{|\text{in}\rangle}$ as functions of $q$.}
    \label{fig:NKLq}
\end{figure}

In figure~\ref{fig:Ttr} we explore how~$r^2\langle \text{in} | \hat{T}_t^r | \text{in} \rangle$ changes as we vary the fermion charge~$q$. As the charge~$q$ increases, the ingoing energy flux inside the effective ergosphere grows, enabling a greater extraction of energy from the black hole, which is expelled outside the ergosphere. This results in a net energy gain at the expense of drawing energy from the black hole. Figure~\ref{fig:ergo} shows how the boundary $r_0$ \eqref{eq:r0} of the effective ergosphere shifts with $q$, revealing a slight expansion of the ergosphere as $q$ increases. This expansion enhances quantum charge superradiance. Indeed, in figure ~\ref{fig:NKLq} we observe that particle creation~$\mathcal{N}_{|\text{in}\rangle}$, charge flux~$\mathcal{K}_{|\text{in}\rangle}$ and energy flux~$\mathcal{L}_{|\text{in}\rangle}$ all increase with larger~$q$, as the electromagnetic interaction between the RN black hole and the charged field intensifies.

Although these results are similar to the scalar case, charge quantum superradiance is considerably more intense for fermions. Notably, the effective ergosphere is one order of magnitude larger for fermions than for charged scalars, for which $r_{0}\sim 2r_{+}$~\cite{Balakumar2020}. This leads to charge and energy fluxes that are two orders of magnitude higher than in the scalar case.

We close our analysis of quantum superradiance for charged fermions by considering how the quantities $\mathcal{N}_{|\text{in}\rangle}$, $\mathcal{K}_{|\text{in}\rangle}$ and~$\mathcal{L}_{|\text{in}\rangle}$
depend on the signs of the fermion charge $q$ and the black hole charge $Q$. In figure~\ref{fig:NKLq} we consider $q>0$ and $Q>0$.
Under the transformation~\mbox{$qQ \rightarrow -qQ$}, both the total number of particles (per unit time)~$\mathcal{N}_{|\text{in}\rangle}$ and the energy flux~$\mathcal{L}_{|\text{in}\rangle}$ remain invariant. In contrast, the charge flux~$\mathcal{K}_{|\text{in}\rangle}$ changes sign when~\mbox{$Q\rightarrow -Q$} but remains invariant when~\mbox{$q\rightarrow -q$}. 
This behaviour follows from the transformation of the radial differential equation~\eqref{eq:R12}  under the mapping~\mbox{$qQ \rightarrow -qQ$}, leading to the transformation~\mbox{$R_{j,(\omega,l,m)}^{\text{out}} \rightarrow (R_{j,(-\omega,l,m)}^{\text{out}})^*$}. 
Consequently, as implied by~\eqref{eq:toutnumerics}, the transmission coefficient transforms as~\mbox{$t_{(\omega,l,m)}^{\text{out}} \rightarrow (t_{(-\omega,l,m)}^{\text{out}})^*$}.
Therefore the behaviour for different signs of $q$ and $Q$ can be deduced from that depicted in figure~\ref{fig:NKLq} by making an appropriate transformation.

\section{Conclusions}
\label{sec:conc}

In this chapter we have studied a massless, charged fermion field propagating on a static, charged RN black hole background.
Classically, the charged fermion field does not exhibit superradiance \cite{Maeda1976}, but we have shown that the quantum analogue of charge superradiance can occur, depending on the quantum state of the field.  
We define an `in' vacuum state which is empty at the past null infinity. 
In this state, quantum superradiance is present: charged fermions are spontaneously emitted into those field modes whose frequency lies in the range for which a bosonic field would exhibit classical superradiance.
As a result, the black hole discharges and also loses energy. 

However, as we know, there is an inherent ambiguity in how the vacuum state is defined; the `in' vacuum is not the only possibility.
For example, we can construct the time-reverse of the `in' state, namely the `out' vacuum,  which is as empty as possible at the future null infinity.
Both these states can be defined analogously for a quantum charged scalar field on RN \cite{Balakumar2020}.
For a quantum charged fermion as considered here, there is a third possibility. 
We can define a state which is as empty as possible at {\em {both}} past and future null infinity.
Such as state can only be defined for fermions; there is no analogue for a charged scalar field \cite{Balakumar2022}.
In this `Boulware'-like state, there is no spontaneous emission of charged fermions, and accordingly this state is that which most closely resembles the `Boulware' state  \cite{Boulware1975} for a neutral scalar or fermion field on a static Schwarzschild or RN black hole. 

The situation for charge superradiance on charged black holes, as studied in this chapter, is somewhat analogous to that for rotational superradiance on rotating Kerr black holes.
In both scenarios, classical superradiance is present for scalars but not for fermions; however both scalars and fermions can exhibit quantum superradiance.
Furthermore, for scalar fields in both setups, it is not possible to define a `Boulware'-like state which is as empty as possible at both past and future null infinity \cite{Ottewill2000,Balakumar2022}.
Considering neutral fermions on a rotating Kerr black hole, as is the case here for charged fermions on a charged black hole, a `Boulware'-like state can be defined \cite{Casals2013}. 
However, while this state on Kerr is a vacuum state asymptotically far from the black hole, it diverges on the stationary limit surface (the boundary of the ergosphere) \cite{Casals2013}. 
It would be interesting to investigate whether the `Boulware'-like state we have defined here in this chapter is regular everywhere outside the event horizon. 
This would require a study of all the components of the charge current and stress-energy momentum tensor, which is beyond the scope of our present study.

\begin{comment}
Finally, it is worth addressing a subtle point: we are discussing particle creation in a static spacetime, whereas such phenomena are typically associated with scenarios where time-translational invariance is broken. At first glance, this may seem contradictory. The key observation is that while the `in-up' and `out-down' bases do exhibit this invariance via a factor~$e^{-i\omega t}$, the choice of complex structure is not invariant under time translations. This is because the quantity~$\widetilde{\omega}$, which guides the splitting of `up' and `down' modes near the horizon, is not a genuine frequency of the system---unlike~$\omega$, which is.

My preliminary interpretation is that the `in' and `out' quantum states introduced here may be effectively mimicking a dynamical process that has yet to be fully identified. In this sense, these quantum states could play a role analogous to that of the Unruh vacuum in a static Schwarzschild black hole: the Unruh state encapsulates features of gravitational collapse and leads to the Hawking radiation effect. Similarly, the `in' and `out' states constructed in this setup may encode the physics of a dynamical setting, offering a static realization of quantum superradiance.
\end{comment}

\plainblankpage
\part[Conclusions]{Conclusions}
\label{part:Conclusions}

\plainblankpage
\chapter[Conclusions]{Conclusions}
\label{chap:Conclusions}

Throughout this thesis, we have explored quantum phenomena arising in a variety of non-trivial backgrounds, with a particular focus on settings involving strong electromagnetic fields and charged fields. Using the Schwinger effect as a central example---where a quantum field in flat spacetime interacts with a spatially homogeneous time-dependent electric field---we identified common features underlying particle creation processes. We extended our analysis to a range of distinct scenarios, including dynamically evolving spacetimes such as cosmologically expanding universes, Bose-Einstein condensate analogue gravity experiments simulating such expansions, and static charged black holes, with spatially inhomogeneous configurations.

One of the central features we have emphasized throughout this thesis is that constructing a quantum theory in non-trivial backgrounds inherently involves certain freedom in the choice of the quantization scheme. These choices are far from trivial: each one leads to distinct notions of particles and antiparticles, a different quantum vacuum, and different values for expectation values of observables. Not all choices are physically meaningful, so identifying those that are well-motivated requires a clear understanding of which physical criteria are reasonable to impose on the quantum theory.

In this context, we aimed to recover the intuitive notion of the vacuum as a state of minimum energy---a notion that holds in flat spacetime in the absence of external fields. However, when the background is dynamical, as is the case whenever an external electric field is present, energy is not conserved, and this idea must be generalized. Previous studies in FLRW cosmologies~\cite{Fewster2000,Olbermann2007} proposed the family of states of low energy, defined as those that minimize the energy density smeared over a compact time window. These states are particularly valuable because they satisfy the Hadamard condition in these cosmologies~\cite{Olbermann2007}. In \autoref{chap:SLEs}, we extended their definition to general, time-dependent homogeneous backgrounds, not necessarily isotropic, particularly in the context of the Schwinger effect. We expect the Hadamard condition to still be satisfied, and we have justified that our construction is, at the very least, consistent with it. It is important to emphasize why we speak of a \textit{family} of quantum vacua: the construction still contains ambiguities, particularly in the choice of smearing function, and each smearing function determines a distinct vacuum state. Notably, we were able to recover several standard vacua in the literature as special cases. This includes, for example, the instantaneous lowest-energy state, which, while not strictly part of the SLE family, is recovered in the limit where the smearing becomes localized at a single instant of time---an approach widely used but where the quantization is not adapted to the dynamics of the full evolution.

Looking ahead, an exciting direction for future work would be to extend the definition of states of low energy to more general curved spacetimes settings such as dynamically collapsing black holes. Such an extension could provide a robust and systematic method for defining Hadamard states in those contexts.

In the literature on the Schwinger effect---particularly within the quantum kinetic approach---it is common to encounter an integro-differential equation describing the time evolution of the number of created pairs: the quantum Vlasov equation~\cite{Kluger1998}. However, this equation is often presented as providing a unique description of particle production, seemingly independent of any ambiguities inherent to the canonical quantization procedure. This apparent uniqueness contrasts with the evidence of quantization ambiguities we have emphasized throughout this work. In \autoref{chap:GQVE}, we demonstrated that the widely used form of the quantum Vlasov equation implicitly assumes a specific choice of quantum vacuum: the instantaneous lowest-energy state. We generalized this equation to incorporate arbitrary choices of quantum vacuum and show that, for a particular family of states, the particle creation rate becomes independent of the quantization details at leading order in the ultraviolet. In this regime, the generalized quantum Vlasov equation naturally reduces to its standard form. This result provides a new, more restrictive criterion for selecting the quantum vacuum, stronger than the requirement that the quantum theory admits a unitary implementation of time evolution. 

The implicit assumption of particular choices of vacuum also happens in other frameworks, such as the Wigner formalism to describe spatially inhomogeneous settings in the Schwinger effect (see, e.g., \cite{Hebenstreit2010,Sheng2019,Kohlfurst2022,Olugh2019}). These preliminary results are based on work that will be presented in an upcoming publication, and that leads to the generalization of this formalism to account for other quantum vacua, and thus, to other characteristics that one wants to imprint on the quantum theory in inhomogeneous configurations.

The physical interpretation of the widely used time-dependent particle number remained an open question in the literature. What does it truly mean to say that a specific number of particles and antiparticles have been created at a particular finite time? Moreover, if this number changes depending on how we quantize, what should we expect to measure in an actual experiment? In \autoref{chap:OperationalRealization}, we provided an operational reinterpretation of this quantity: measuring the number of excitations at a given time would require switching off the interaction between the detector and the background field at that instant. Crucially, how this switch-off is implemented---instantaneously or gradually---determines the resulting outcome. For instance, a sudden switch-off naturally selects the instantaneous lowest-energy state as the quantum vacuum, whereas adiabatic vacua correspond to smoother transitions. This perspective links the ambiguities inherent in canonical quantization with the experimental procedure itself, establishing a bridge between the abstract mathematical framework and physical observables. Our results showed that quantum ambiguities are not merely theoretical artifacts but have operational meaning: they reflect the different possible ways one can interact with and probe the system.

In realistic Schwinger effect experiments, the electric field must be switched on and off at finite times, leading to `on' and `off' transitions. Similarly, gravitational analogue experiments---such as those based on quasi two-dimensional Bose-Einstein condensates simulating FLRW cosmologies~\cite{TolosaSimeon2022,Viermann2022,Sparn2024,Schmidt2024}---exhibit these transitions due to their finite duration. In \autoref{chap:InOut}, we analysed how these inevitable transitions affect particle production. Considering a scalar field non-minimally coupled to the geometry in a homogeneous and isotropic universe, we found that particle spectra are often dominated by these transitions, especially in the non-conformal coupling case---of which BEC analogues are a prime example. For the Schwinger effect, the situation is subtler: while transitions still contribute, the anisotropic structure of the theory leads to a suppression of transition effects when the electric field is kept on for a sufficiently long duration, allowing the intermediate regime to dominate. 

In summary, careful attention must be paid to `on' and `off' transitions when modelling or interpreting quantum pair production experiments. One cannot assume that the observed particle production arises solely from the intended intermediate regime; in many cases, the way the system is switched on and off plays a decisive role in shaping the outcome. Connecting with the discussion in \autoref{chap:OperationalRealization}, this result is nothing but a manifestation of the inherent ambiguities in the canonical quantization. Indeed, one might consider the possibility of bypassing the `in' and `out' static regimes altogether, focusing exclusively on the intermediate region of interest in the model. However, such an approach would eliminate the existence of well-defined `in' and `out' quantum vacuum states, leading to ambiguities in the choice of the quantum vacuum. Ultimately, the effects of the `in' and `out’ transitions are not only unavoidable but are intrinsic to particle creation phenomena, just as quantum vacuum ambiguities are an inherent aspect of QFTCS.

Finally, in the last part of this thesis, we applied the theoretical framework developed throughout the earlier chapters to physically motivated scenarios involving black holes. In \autoref{chap:Kugelblitz}, we investigated the concept of kugelblitze---black holes formed from the gravitational collapse of pure electromagnetic radiation. While these are legitimate classical solutions of general relativity~\cite{Robinson1962,Senovilla2014}, we asked a natural question in light of our understanding of the Schwinger effect: can such objects realistically form in our present-day universe? Our conclusion is negative. The Schwinger effect effectively prevents the formation of an event horizon when attempting to generate a black hole by concentrating intense electromagnetic fields. This holds for all realistic scenarios involving current or conceivable electromagnetic sources---whether natural or artificial---over an enormous range of scales, from $10^{-29}$ to $10^8$~m. Within this range, pair production due to the Schwinger effect dominates well before gravitational collapse can occur. Of course, this does not rule out the formation of kugelblitze in the early universe, where extreme conditions and different dynamics prevail. A separate analysis, incorporating early-universe physics, would be required to address that possibility. Nevertheless, our findings highlight the powerful role that semiclassical phenomena play in high-energy environments, prohibiting the formation of event horizons that would be classically permitted under general relativity. 

Motivated by this research, we are investigating whether a similar situation arises where gravitational waves, rather than electromagnetic radiation, are involved. The formation of black holes and singularities due to the collapse of gravitational waves has been studied as classical solutions in general relativity~\cite{Brill1964,Pretorius2018}, but quantum effects have largely been overlooked. We aim to investigate whether quantum effects play a crucial role in black hole formation during gravitational wave collisions, potentially revealing fundamental discrepancies between general relativity and quantum predictions regarding gravitational horizons and singularities.

Lastly, in \autoref{chap:Superradiance} we studied quantum charge superradiance for fermionic fields propagating on a charged black hole background. Unlike scalar fields, which exhibit classical charge superradiance~\cite{Starobinsky1973,Unruh1974,Dai2023,Dai2023a}, fermions do not show this effect classically. However, we constructed a quantum state that does exhibit quantum superradiance: the black hole discharges and loses energy through the creation of particle-antiparticle pairs in a region near (but outside) the event horizon---an effective ergosphere. Interestingly, the quantization ambiguities allow for the construction of alternative quantum vacua, such as a `Boulware' vacuum that respects the staticity of the spacetime and shows no superradiant behaviour. This reflects the richness of semiclassical theory, where the choice of quantum vacuum leads to physically distinct but consistent predictions.
\plainblankpage 

%%% Bibliography %%%
\newrefcontext[]
\renewcommand*{\bibname}{References} 
\renewcommand*{\refname}{References}

\blankpage\printbibliography[title={References},heading=bibintoc,notkeyword={Alvarez},resetnumbers=true]

\end{refsection}

\begin{comment}
%%% Appendices: Work that *YOU* Developed %%%
\appendix
\input{Matter/06-Appendices}
\input{Chapters/Appendices/00-AppendixA}
\input{Chapters/Appendices/01-AppendixB}

%%% Annexes: Work that *YOU DID NOT* Develop %%%
\input{Matter/07-Annexes}
\input{Chapters/Annexes/00-AnnexA}

%%% Back Page %%%
\input{Matter/08-BPage}
\end{comment}

\end{document}